\documentclass[12pt]{article}
\usepackage{graphicx}
\usepackage{amsmath}

\oddsidemargin  - 5 mm
\evensidemargin - 5 mm
\newtheorem{theorem}{Theorem}

\newtheorem{proposition}[theorem]{Proposition}

\input{tcilatex}

\begin{document}

\date{}
\title{Elliptic Curves, Algebraic Geometry Approach in Gravity Theory and Some
Applications in Theories with Extra Dimensions I.}
\author{Bogdan G. Dimitrov \thanks{%
Electronic mail: bogdan@thsun1.jinr.ru} \\
Bogoliubov Laboratory for Theoretical Physics\\
Joint Institute for Nuclear Research \\
6 Juliot Curie str. \\
Dubna 141980, Russia}
\maketitle

\centerline{\textit{Dedicated to the memory  of  Prof. Sawa S. Manoff /1943 -
27.05. 2005/}}

\vskip .5cm

\begin{abstract}
\ \ Motivated by the necessity to find exact solutions with the elliptic
Weierstrass function of the Einstein's equations (see gr-qc/0105022),the
present paper develops further the proposed approach in hep-th/0107231
,concerning the s.c. cubic algebraic equation for effective parametrization.
Obtaining an ''embedded'' sequence of cubic equations, it is shown that it
is possible to parametrize also a multi-variable cubic curve, which is not
the standardly known case from algebraic geometry. Algebraic solutions for
the contravariant metric tensor components are derived and the
parametrization is extended in respect to the covariant components as well.

\ \ It has been speculated that corrections to the extradimensional volume
in theories with extra dimensions should be taken into account, due to the
non-euclidean nature of the Lobachevsky space.It was shown that the
mechanism of exponential "damping" of the physical mass in the
higher-dimensional brane theory may be more complicated due to the variety
of contravariant metric components for a spacetime with a given constant
curvature. The invariance of the low-energy type I string theory effective
action is considered in respect not only to the known procedure of
compactification to a four-dimensional spacetime, but also in respect to
rescaling the contravariant metric components. As a result, instead of the
simple algebraic relations between the parameters in the string action,
quasilinear differential equations in partial derivatives are obtained,
which have been solved for the most simple case.

\ \ In the Appendix, a new block structure method is presented for solving
the well known system of operator equations in gravity theory in the
N-dimensional case.
\end{abstract}

\tableofcontents

\section{\protect\bigskip INTRODUCTION}

Inhomogeneous cosmological models have been intensively studied in the past
in reference to colliding gravitational \bigskip waves [1] or singularity
structure and generalizations of the Bondi - Tolman and Eardley-Liang-Sachs
metrics [2, 3]. In these models the inhomogeneous metric is assumed to be of
the form [2]
\begin{equation}
ds^{2}=dt^{2}-e^{2\alpha (t,r,y,z)}dr^{2}-e^{2\beta (t,r,y,z)}(dy^{2}+dz^{2})
\tag{1.1}
\end{equation}
(or with $r\rightarrow z$ and $z\rightarrow x$) with an energy-momentum
tensor $G_{\mu \nu }=k\rho u_{\mu }u_{\nu }$ for the irrotational dust and
functions $\alpha (t,r,y,z)$ and $\beta (t,r,y,z)$, determined by the
Einstein's equations. The particular form of the metric and therefore of the
functions $\alpha $ and $\beta $ may restrict seriously the spacetime
symmetries, due to which in some cases, for example when there is a two -
parameter group of motions (i. e. $\beta =\beta (t,y)$) with commuting
Killing vectors, then ''there are no solutions to these equations'' [2].
This represents in fact the serious motivation why the functions $\alpha $
and $\beta $ are chosen in a special form [4]
\begin{equation}
\alpha \equiv log\left[ h(z,\xi ,\overline{\xi })\Phi ^{^{\prime }}+h(z,\xi ,%
\overline{\xi })\Phi \nu ^{^{\prime }}\right] \text{ \ \ ,}  \tag{1.2}
\end{equation}
\begin{equation}
\beta \equiv log\Phi (t,z)+\nu (z,\xi ,\overline{\xi })\text{ \ ,}  \tag{1.3}
\end{equation}
so that the integrations of (some) of the components of the Einstein's
equations is ensured. In (1.2) and (1.3) $^{\prime }\equiv \frac{\partial }{%
\partial z}$, $.\equiv \frac{\partial }{\partial t}$ and $\xi ,\overline{\xi
}$ are a pair of complex (conjugated) variables
\begin{equation}
\xi \equiv x+iy\text{ \ \ \ \ ; \ \ \ }\overline{\xi }\equiv x-iy\text{ \ \
.\ }  \tag{1.4}
\end{equation}
A\ nice feature of the approach is that in the limit $t\rightarrow \infty $
[5] and under a special choice of the pressure as a definite function of
time the metric approaches an isotropic form [4]. Other papers, also
following the approach of Szafron-Szekerez are [6,7]. In [7], after an
integration of one of the components - $G_{1}^{0}$ of the Einstein's
equations, a solution in terms of an elliptic function is obtained.

In different notations, but again in the framework of the Szafron-Szekerez
approach the same integrated in [7] \textbf{nonlinear differential equatio}n
\
\begin{equation}
\left( \frac{\partial \Phi }{\partial t}\right) ^{2}=-K(z)+2M(z)\Phi ^{-1}+%
\frac{1}{3}\Lambda \Phi ^{2}  \tag{1.5}
\end{equation}
was obtained in the paper [8] of Kraniotis and Whitehouse. They make the
useful observation that (1.5) in fact defines a (cubic) algebraic equation
for an elliptic curve, which according to the standard algebraic geometry
prescribtions (see [9] for an elementary, but comprehensive and contemporary
introduction) can be parametrized with the \textbf{elliptic Weierstrass
function}
\begin{equation}
\rho (z)=\frac{1}{z^{2}}+\sum\limits_{\omega }\left[ \frac{1}{(z-\omega )^{2}%
}-\frac{1}{\omega ^{2}}\right]  \tag{1.6}
\end{equation}
and the summation is over the poles in the complex plane. More explicitely
the simple idea about the parametrization procedure shall be explained in
Sect. II of the present paper. \ \ However, two important problems
immediately arise, which so far have remained without an answer\textbf{: }

1. The parametrization procedure with the elliptic Weierstrass function in
algebraic geometry is adjusted for cubic algebraic equations with number
coefficients! Unfortunately, equation (1.5) is not of this type, since it
has coefficient functions in front of the variable $\Phi $, which depend on
the complex variable $z$. In view of this, it makes no sense to define
''Weierstrass invariants'' as
\begin{equation}
g_{2}=\frac{K^{2}(z)}{12}\text{ \ \ ; \ \ \ }g_{3}=\frac{1}{216}K^{3}(z)-%
\frac{1}{12}\Lambda M^{2}(z)\text{ \ \ ,}  \tag{1.7}
\end{equation}
since the above \textbf{functions} have to be set up equal to the complex
numbers $g_{2}$ and $g_{3}$ (the s. c. Eisenstein series)
\begin{equation}
g_{2}=60\sum\limits_{\omega \subset \Gamma }\frac{1}{\omega ^{4}}%
=\sum\limits_{n,m}\frac{1}{(n+m\tau )^{4}}\text{ \ \ \ ,}  \tag{1.8}
\end{equation}
\begin{equation}
g_{3}=140\sum\limits_{\omega \subset \Gamma }\frac{1}{\omega ^{6}}%
=\sum\limits_{n,m}\frac{1}{(n+m\tau )^{6}}\text{ \ \ \ \ }  \tag{1.9}
\end{equation}
and therefore \textbf{additional equations }have to be satisfied in order to
ensure the parametrization with the Weierstrass function.

2. Is the Szekerez - Szafron metric the only case, when the parametrization
with the Weierstrass function is possible? Closely related to this problem
is the following one - is only one of the components of the Einstein's
equation parametrizable with \ $\rho (z)$ and its derivative?

An attempt to give an answer to the first problem was made in the preceeding
paper [10], where it has been proved that the Weierstrass function and its
derivative satisfy a ''parametrizable'' form of a cubic equation of a more
general type
\begin{equation}
\left[ \rho ^{^{\prime }}(z)\right] ^{2}=4\rho ^{3}(z)-g_{2}(z)\rho
(z)-g_{3}(z)\text{ \ \ \ \ \ }  \tag{1.10}
\end{equation}
with convergent coefficients (at $n\rightarrow \infty $) in the Laurent
expansion of the functions $g_{2}(z)$ and $g_{3}(z)$. Note also that still
the more general problem about the validity of the equation

\begin{equation}
(\frac{d\rho }{dz})^{2}=M(z)\rho ^{3}+N(z)\rho ^{2}+P(z)\rho +E(z)
\tag{1.11}
\end{equation}
remains unsolved. In future applications, we shall follow the standard
approach in parametrizing a cubic algebraic equation with number
coefficients - the s.c.''arithmetic theory''.

Concerning the second problem, which is of primary importance in view of
possible applications in multidimensional gravity theories, M-theory and
supergravity theories, it will be treated in the second part of this paper.
In it the developed previously in [10] parametrization procedure for the s.
c. cubic algebraic equation of reparametrization invariance of the
gravitational Lagrangian and as well the proposed in this (first) part
method for parametrization of embedded sequence of cubic algebraic equations
will be applied in respect to the Einstein's \ system of equations in their
general form (i.e. not dependent on any concrete metric and also with an
arbitrary energy - momentum tensor). The treatment will be performed within
the assumption that the components of the metric tensor can be represented
as a factorized product of two vector fields (but not necessarily lying in
the tangent bundle of the given manifold), i. e.
\begin{equation}
g^{\alpha \beta }(\overrightarrow{x})=k^{\alpha }(\overrightarrow{x}%
)k^{\beta }(\overrightarrow{x})\text{ \ \ \ \ ; \ \ \ }\overrightarrow{x}%
=(x^{1},x^{2},.....,x^{n})\text{ \ \ \ \ }\alpha ,\beta =1,2,...n\text{ \ \
\ .}  \tag{1.12}
\end{equation}
Then the Einstein's equations, if written in respect to the $n-$th component
$k^{n}$ of \ the vector field \ $\overrightarrow{k}%
=(k^{1},k^{2},.....,k^{n}) $, can be represented as a system of $n$
(multicomponent) \textbf{cubic} algebraic equations and additionally $\left(
\begin{array}{c}
n \\
2
\end{array}
\right) =\frac{n(n-1)}{2}$ \textbf{quartic} algebraic equations.
Consequently, it is not surprising at all that parametrization with the
Weierstrass function and its derivative can be performed in respect to some
of the components in the Einstein's system of equations not only in the
Szekeres - Szafron case, but also in the general case as well. Note also
that the parametrization with the Weierstrass function can be applied also
in respect to quartic algebraic equations [11, 12].

The first part of the present paper continues and develops further the
approach from the previous paper [10], where a definite choice of the
contravariant metric tensor was made in the form of the factorized product $%
\widetilde{g}^{ij}=dX^{i}dX^{j}$ and $%
X^{i}=X^{i}(x_{1},x_{2},x_{3},.....,x_{n})$ are generalized coordinates.
They can be regarded as an $n-$ dimensional hypersurface, defining a
transition from an \textbf{initially defined} set of coordinates $%
x_{1},x_{2},x_{3},.....,x_{n}$ on a chosen manifold to another set of the
generalized coordinates $X^{1},X^{2},.....,X^{n}$. In section 1I of the
present paper it will be reminded briefly how the cubic algebraic equation
in respect to the differentials $dX^{i}$ was derived in [10], but in fact
the aim will be to show that depending on the choice of variables in the
gravitational lagrangian or in the Einstein's equations, a wide variety of
algebraic equations (of third, fourth, fifth, ninth and tenth degree) in
gravity theory may be treated, if a distinction between the covariant metric
tensor components and the contravariant ones is made. This idea, originally
set up by Schouten and Schmutzer, was further developed in the papers of
S.Manoff, mainly with the purpose of classification of such more general
theories of gravity with two different metrics and affine connections
(covariant and contravariant affine connections), or also theories,
admitting torsion and shear (for a review of the general approach, one may
consult the review article [13]). Further, possible observational
consequences of propagation of signals in spaces with affine connections and
metrics were explored in [14].

The present paper will also be related with the idea about the distinction
between the covariant and contravariant metric components should be made,
although no shear or torsion will be assumed to exist. The physical idea,
which will be exploited will be: can such a gravitational theory with a more
general contravariant tensor be equivalent to the usual and known to us
theory with a contravariant metric tensor, which is at the same time the
inverse one of the covariant one? By ''equivalence''\ it is meant that the
gravitational lagrangian in both approaches should be equal, on the base of
which the s.c. cubic algebraic equation for reparametrization invariance (of
the gravitational Lagrangian) was obtained in [10]. The derivation was based
also on the construction of another connection $\widetilde{\Gamma }%
_{kl}^{s}\equiv \frac{1}{2}dX^{i}dX^{s}(g_{ik,l}+g_{il,k}-g_{kl,i})$, which
\textbf{is the same as the known Christoffell connection }$\Gamma _{kl}^{s}$%
\textbf{, but with the contravariant tensor component replaced with the
factorized product }$dX^{i}dX^{s}$. The connection $\widetilde{\Gamma }%
_{kl}^{s}$ has two very useful properties: 1. It has an affine
transformation law, i.e. it transforms after change of coordinates as the
usual Christoffell connection. For completeness the proof is given in
Appendix A of the present paper, it was not given in the previous paper
[10]. 2. $\widetilde{\Gamma }_{kl}^{s}$ is an \textbf{equiaffine connection }%
(see \ also Appendix A for the elementary proof), which is a typical notion,
introduced in classical \textbf{affine geometry} [15, 16] and meaning that
there exists a volume element, which is preserved under a parallel
displacement of a basic $n-$dimensional vector $e\equiv
e_{i_{1}i_{2}....i_{n}}$. However, this notion turns out to be very
convenient and important, since for such types of connections we can use the
known formulae for the Ricci tensor, but with the connection $\widetilde{%
\Gamma }_{kl}^{s}$ instead of the usual Christoffell one $\Gamma _{kl}^{s}$
and moreover, the Ricci tensor $\widetilde{R}_{ij}$ will again be a
symmetric one, i. e. $\widetilde{R}_{ij}=\widetilde{R}_{ji}=\partial _{k}%
\widetilde{\Gamma }_{ij}^{k}-\partial _{i}\widetilde{\Gamma }_{kj}^{k}+%
\widetilde{\Gamma }_{kl}^{k}\widetilde{\Gamma }_{ij}^{l}-\widetilde{\Gamma }%
_{ki}^{m}\widetilde{\Gamma }_{jm}^{k}$. It should be emphasized that the
derivation of the cubic equation and of all the other algebraic equations in
principle does not depend on the properties of the affine connection. Also,
the theoretical reasoning for the distinction between covariant and
contravariant components comes again from affine geometry, where the
assumption about the existence of inverse contravariant metric components in
fact ''sets up'' the correspondence between ''covectors'' (in our
terminology - these are the ''vectors'') and the ''vectors'' (i.e. the
contravariant vectors''), on which the covariant and the contravariant
tensors are being built.. However, \textbf{such a correspondence is not
necessarily to be established} (see again [15, 16]) and both kinds of
tensors have to be treated as different mathematical objects, defined on one
and the same manifold. For a contemporary introduction in affine geometry,
one may consult also the monographs [17, 18].

Let us point out some important consequences from the distinction between
covariant and contravariant metric components: 1. One can perform a
classification of all the affine connections, resulting in the same
gravitational Lagrangian. In a sense, this is a problem, which has been
solved already in the 60-ties, but here the proposal is to perform this
classification on the base of solving algebraic equations instead of
nonlinear differential equations in partial derivatives. 2. One can find all
the contravariant metric tensor components, satisfying the same
gravitational Lagrangian. The last fact brings some new and important
consequences for gravitational physics in the sense that if one considers
the Einstein's equations, one can speak about \textbf{separate classes of
solutions for the covariant metric tensor components and for the
contravariant ones.} These (i.e. both) components constitute the algebraic
variety, satisfying the Einstein's equations, if \textbf{they are considered
as a set of intersecting multivariable cubic and quartic algebraic surfaces (%
}further instead of cubic surfaces we shall continue to use the terminology
''cubic curves''). But in principle, the notion of ''intersecting algebraic
curves and surfaces'' in algebraic geometry is very complicated and
unfortunately, even though some approaches may be found in older monographs
(see for example vol. II of [19]), they are purely mathematical ones and not
at all applicable to really given algebraic curves. The second volume of the
known monograph [19] of Hodge and Pedoe contains some material about
intersection of quadratic manifolds - these are simply quadratic algebraic
equations, which for our concrete case are for example the equations from
the quadratic system $g_{ij}\widetilde{g}^{jk}=g_{ij}dX^{j}dX^{k}=\delta
_{i}^{k}$, determining the differentials $dX^{i}$ from the defining equality
$g_{ij}\widetilde{g}^{jk}=\delta _{i}^{k}$ for the existence of the inverse
contravariant metric tensor components $\widetilde{g}^{jk}$. By itself, the
knowledge about this system of equations is useful since it helps to make
the clear distinction between a \textbf{linear algebraic system of equations}
(where the unknown variables are in a vector-column) and an \textbf{%
operational system of equations (}see the monograph of Gantmacher [20] for
treatment of such equations), in which the unknown variables constitute a
matrix. The above mentioned system is definitely of the second type and not
of the first type. This is an important point, since it may be thought that
if the determinant $det\mid dX^{i}dX^{j}\mid =0$ (which really happens for
any dimensions), then the system does not have a solution. It has been shown
in Appendix $C$, and in part also in section 1I, on the base of an
originally created approach with the help of the s.c. ''block structure
matrices'' and for the general $N-$dimensional case, that yet the system is
correctly determined in respect to $g_{ij}$.

But now, it is another point \ if we would \ like to treat the same system
as a system of \textbf{quadratic\ algebraic equations}, intersecting the set
of \textbf{cubic and quartic algebraic Einstein's equations}. Therefore,
extending the theory by admitting more general contravariant metric tensor
components and finding the intersection variety (i.e. the algebraic variety
of the variables $dX^{i}$, satisfying both systems of equations), the
\textbf{standardly known solutions of the Einstein's equations should be
recovered}. In principle, the s.c. ''intersection theory'' in algebraic
geometry is very complicated, and it is not very clear how a concrete
complicated system of intersecting algebraic equations can be analyzed. In
this paper we shall investigate \textbf{only} the cubic algebraic equation
for reparametrization invariance of the gravitational Lagrangian. It may be
noted that general theorems for \ intersection of \ algebraic curves of
different (arbitrary)\ degrees are given in [21, 22], but they refer to the
following cases: a) plane curves, b) a system of one - parametric curves of
degree $n$, having a known quantity of common points with another curve of
degree $p$. c) two intersecting curves of different degrees and a \textbf{%
third one, }passing through a \textbf{known number of the intersection points%
} of the first two curves and some other similar cases. As it may be
guessed, not a single of any these (and other)\ cases is adapted for the
treatment of the intersecting systems of cubic, quartic and quadratic
algebraic equations in gravity theory.

Yet one of the serious shortcomings of the algebraic geometry approach is
related to the fact that the standardly known procedure for parametrization
of a cubic algebraic equation with the Weierstrass function is adapted to 1.
plane cubic algebraic curves of two variables. 2. algebraic curves with
constant coefficients. We already commented at the beginning (and also in
[10]) on the second point, so in this paper all efforts will be concentrated
on the problem \textbf{how to parametrize a given non - plane,
multicomponent cubic algebraic curve with the Weierstrass function and its
derivative - in the present case, this is of course the equation of
reparametrization invariance. }Previously \ [10], it has been shown how a
selected variable $dX^{i}$ of the cubic algebraic equation \ and the ratio $%
\frac{a}{c}$ of \ two coefficient functions in the performed linear -
fractional transformation can be parametrized with the Weierstrass function
and its derivative. \textbf{One of the purposes of the present paper will be
to demonstrate how this parametrization can be extended to all the variables
}$dX^{i}$\textbf{\ in the cubic algebraic equation. }It will be demonstrated
in section 3 how an embedded system of cubic algebraic equations can be
obtained, and each equation from this sequence will be for the algebraic
subvariety of the solutions - i.e. the first equation will be for the
algebraic variety of $n$ variables, the second one - for $(n-1)$
variables,...., the last one will be only for the $dX^{1}$ variable\textbf{.
Thus a qualitatively new moment appears in the investigation of elliptic
curves and cubic algebraic equations - it turns out that it is possible to
parametrize not only a plane cubic algebraic curve }$%
y^{2}=4x^{3}-g_{2}x-g_{3}$, \textbf{but also a multi - variable cubic
algebraic equation.} In section 4 it has been proved that yet the situation
is different from the standard case - the obtained ''uniformization''
functions for the differentials $dX^{1}$, $dX^{2}$, $dX^{3}$ are complicated
irrational functions of the Weierstrass function $\rho (z)$, its derivative $%
\rho ^{^{\prime }}(z)$, the (covariant)\ metric tensor and affine connection
components, which depend on the global coordinates $X^{1}$, $X^{2}$, $X^{3}$%
. Therefore, it might seem that due to the dependence on the global
coordinates one cannot assert that the functions $dX^{1}$, $dX^{2}$, $dX^{3}$
are ''uniformization functions'', since they should depend \textbf{only} on
the complex coordinate $z$. But in section 5 it has been proved that the
obtained expressions can be transformed to a system of first - order
nonlinear differential equations in respect to $X^{1}$, $X^{2}$, $X^{3}$ and
since a solution of this system always exists, one may write down $%
X^{1}=X^{1}(z)$, $X^{2}=X^{2}(z)$, $X^{3}=X^{3}(z)$. This solves completely
the problem about the existence of ''uniformization functions''. On the
other side, the generalized coordinates depend also on the s. c. ''initially
given'' coordinates $\mathbf{x}=(x_{1}$, $x_{2}$, $x_{3}$), so the complex
function dependence is more complicated, i. e. $X^{1}=X^{1}(\mathbf{x(}z))$,
$X^{2}=X^{2}(\mathbf{x(}z))$, $X^{3}=X^{3}(\mathbf{x(}z))$. In section 6 it
was shown, introducing special notations and the s.c. ''one - dimensional
Poisson bracket'', that additionally a system of first - order nonlinear
differential equations may be written also in respect to the initial
coordinates $x_{1}$, $x_{2}$, $x_{3}$. Sections 7, 8 and 9 treat an
important issue, which is purely a mathematical one, but can have important
physical applications - how the complex coordinate dependence of \ the
initial and general coordinates can be extended from one to two complex
coordinates $z$, $v$. In section 7 it was shown that this cannot be done by
assuming that $X^{i}=X^{i}(x(z),z,v)$, since then the obtained system of
equations will be conradictable. Intuitively, this conclusion can be
justified because it is unnatural for the complex coordinate $z$ to be in a
''privilleged'' position because of the dependence of the initial
coordinates only on $z$.Therefore, it is much more natural if the complex
coordinate dependence is realized in the following way: $%
X^{i}=X^{i}(x(z,v),z)$. By performing a full analysis of the system of
equations, in Sections 8 and 9 it was shown that this time no contradiction
and incompatibility in the system of equations appear.

Throughout all the sections from 3 to 9, the approach was applied by
assuming that the second differentials of the generalized coordinates are
zero $d^{2}X^{i}=0$. This may seem to be a serious restriction, but \textbf{%
it was necessary to be imposed in order to be able to construct the
algorithm for finding the solutions of the cubic equation}. So it is not
realistic to hope that just at once such an algorithm may be constructed,
without making any assumptions whatsoever. The important point is another -
since the main tool of the proposed approach is the linear - fractional
transformation, will this approach be applicable also for the more
complicated cases of 1. again the cubic algebraic equation of
reparametrization invariance, but with $d^{2}X^{i}\neq 0$ and 2. the system
of \textbf{cubic and quartic Einstein's algebraic equations}. In the
subsequent second and third parts of this paper, it will be shown that the
approach continues to work, although with some modifications and
unfortunately, with considerable (but not unsurmountable) technical
difficulties. In these two cases, the application of this approach will
require: 1. \textbf{enlarging the algebraic variety} of solutions with the
addition of the second differentials (first case) or the first and the
second derivatives (the second case) 2. Applying a \textbf{multi - component
linear - fractional transformation}, which accounts for all the variables in
the extended algebraic variety. This will result in several coefficient
functions in front of the highest degrees, all of which have to be set up to
zero. In other words, if in the presently investigated case we will have
\textbf{only one }cubic equation with an algebraic variety of $\ n$, $(n-1)$%
, $....,2$, $1$ dimensions after \textbf{each} application of the linear -
fractional transformation, then after each case one would have \textbf{at
least} two algebraic equations, one of which may be of third, and the other
- of second order. 3. Since the originally given multi - variable algebraic
equation is transformed by means of a linear - fractional transformation in
respect to all the variables in the algebraic variety, which includes also
first derivatives (first case) and first and second derivatives (second
case), \textbf{additionally, }if one sets up $d^{2}X^{i}=Z^{i}$ and $%
dX^{i}=Y^{i}$, one has to take into account also that $dY^{i}=Z^{i}$, which
further will be called the ''compatibility system of equations''. This would
result in a system of differential equations in respect to the coefficient
functions in the multi - component linear - fractional transformation. Even
for the simpler (first)\ case with $d^{2}X^{i}\neq 0$, the solution of this
system will be connected with considerable difficulties. However, yet it
will turn out for the first case that such a (unique!) solution can found,
and it helps to express \textbf{uniquely} the coefficient functions in the
linear - fractional transformation for $d^{2}X^{i}$ through the ones in the
transformation for $dX^{i}$.

There are a number of physical applications of the algebraic geometry
formalism, concerning \textbf{gravitational theories with extra dimensions},
which will be considered in the last section 10 of this paper. These
applications and most of the corresponding equations are not directly
related to parametrization with the Weierstrass function, but the approach
in principle follows the same main idea of introducing a more general
contravariant tensor. In our opinion, one of the most important applications
is related to the so called ''exponential damping''\ of the mass $m_{0}$ of
the visible three - brane in a fundamental higher - dimensional brane
theory, which has been considered in the pioneering paper of L. Randall and
R. Sundrum and which results \ in an ''exponentially damped''\ physical mass
$m=m_{0}e^{-kr_{-}\pi }$. If one assumes $e^{kr_{-}\pi }\simeq 10^{15}$,
then this in fact turns out to be the mechanism for producing physicall mass
scales not far from the Planck scale $\sim 10^{19}$ GeV. The problem however
is that the contravariant metric components play the role of a ''coupling''\
parameter in the effective action of the visible brane. So it has been shown
in this paper, that for the same scalar (constant and negative)
gravitational curvature of a five - dimensional space with an embedded flat $%
4D$ Minkowski space, other contravariant components $\widetilde{g}^{ij}$ may
be found, which are proportional to the components $g^{ij}$, i.e. $%
\widetilde{g}^{ij}=l(\mathbf{x})\widetilde{g}^{ij}$ ($\mathbf{x\equiv }%
\overrightarrow{\mathbf{x}}=(x_{1}$, $x_{2}$, $x_{3}$, $x_{4}$)) and the
function of proportionality $l(\mathbf{x})$ (it will be called a \textbf{%
length function}) may be found as an \textbf{(exact analytical) solution of
a quasilinear differential equation in partial derivatives.} Thus, it may be
supposed that the physical Higgs mechanism for dynamical mass generation may
turn out to be more complicated, and so in principle, \textbf{there might be
a variety of generated physical masses.} At least, this should strictly
follow from the whole mathematical construction and mostly from the fact
that solution of a differential equation of partial derivatives may be
\textbf{any function, }depending on the first integrals of the
characteristic system of equations. It is perhaps from a physical point of
view interesting to note that in one of found solutions for $l$ (form.
(10.197))
\begin{equation}
l^{2}=\frac{1}{1-const.\text{ }e^{24\text{ }k\text{ }\varepsilon \text{ }y}}%
\text{ \ \ \ \ \ \ }\varepsilon =\pm 1\text{ \ \ ,}  \tag{1.13}
\end{equation}
the ''scale function''\ will indeed be equal to one (i.e. we have the usual
gravitational theory with $\widetilde{g}^{ij}=g^{ij}$ and $l=1$) for $%
\varepsilon =-1$ and $y\rightarrow \infty $ (the s.c. infinite extra -
dimensions), but for $\varepsilon =+1$ there will be even a decrease of the
''length function''\ due to the exponenrtial factor in the denominator. Yet,
the derivation of a justifiable physical result from a solution of the
characteristic equation poses one more interesting problem, this time
\textbf{in reference to the invariance of the low - energy type I string
theory effective action under both compactification from a ten -dimensional
spacetime to five - dimensional one and also (again) rescaling of the
contravariant metric components,} which has been investigated also in
section 10. The peculiar moment here is that together with the
compactification (which is the standard known approach so far), here the
rescaling is also taken into consideration, which again leads to the
construction of \ \textbf{quasilinear differential equations in partial
derivatives in respect to the function }$l(x)$. Here the important fact is
that these equations are non - contradictable in the limit $l=1$, which
means that the usual and known relations between the parameters in the low -
energy type I string action in this limit are obtained. However, from the
solutions of the characteristic equations the limit $l=1$ can be set up and
can give some new relations between the parameters. Thus, if any information
about the string scale $m_{s}$, the string coupling constant $\lambda $ and
the electromagnetic coupling constant $g_{5}$ is available, then in
principle the obtained solutions should result also in the unit lenght scale
$l=1$, if the known theory of gravity is correct. If this does not happen,
then under certain energies, string scales, electromagnetic and string
couplings there might be deviations from the case $l=1$. It turns out, the
standard gravitational theory may be tested, since it is related to the
above mentioned parameters.

\textbf{\ }

\section{\protect\bigskip BASIC \ KNOWLEDGE \ ABOUT \ ALGEBRAIC \ EQUATIONS
\ IN \ GRAVITY \ THEORY}

\subsection*{\protect\bigskip 2.1. THE \ INTRODUCED \ CONNECTION \ $%
\widetilde{\Gamma }_{kl}^{s}\equiv \frac{1}{2}%
dX^{i}dX^{s}(g_{ik,l}+g_{il,k}-g_{kl,i})$}

This section has the purpose to review the possible application of algebraic
geometry and theory of algebraic equations in gravity theory. A part of the
exposition will be based on the paper [10] (see also [23]), but some new
generalizations will be presented.

It is known in gravity theory that the knowledge of the \textbf{metric
tensor }$g_{ij}$ determines the space - time geometry, which means that the
\textbf{Christoffell connection}
\begin{equation}
\Gamma _{ik}^{l}\equiv \frac{1}{2}g^{ls}(g_{ks,i}+g_{is,k}-g_{ik,s})
\tag{2.1}
\end{equation}
and the \textbf{Ricci tensor}
\begin{equation}
R_{ik}=\frac{\partial \Gamma _{ik}^{l}}{\partial x^{l}}-\frac{\partial
\Gamma _{il}^{l}}{\partial x^{k}}+\Gamma _{ik}^{l}\Gamma _{lm}^{m}-\Gamma
_{il}^{m}\Gamma _{km}^{l}\text{ \ \ \ }  \tag{2.2}
\end{equation}
can be calculated. \textbf{Now suppose that one knows the the Ricci tensor }$%
R_{ij}=R_{ij}(x)$\textbf{\ as a function of the space - time coordinates,
but one would like to find the Christoffell connection }$\Gamma _{ij}^{k}$.
Then one can consider the defining equation (2.2)\ in two different ways

1. As a\textbf{\ first - order system of nonlinear differential equations in
partial derivatives}
\begin{equation}
\frac{\partial \Gamma _{ij}^{k}}{\partial x^{l}}=F(R_{ik},\Gamma
_{il}^{m}\Gamma _{km}^{l})  \tag{2.3}
\end{equation}
with a quadratic nonlinearity in the components of the Christoffell
connection.The number of equations is equal to the number of components of
the symmetric tensor $R_{ik}$, which for an $n$-dimensional space - time is $%
\left(
\begin{array}{c}
n \\
2
\end{array}
\right) +n$ and the number of the unknown variables is equal to the number
of the Christoffell connection components, which is $\left[ \left(
\begin{array}{c}
n \\
2
\end{array}
\right) +n\right] n.$ Note that even if $R_{ik}$ have been calculated from
the initially given connection $\Gamma _{ij}^{k}$, the solution of the
system of differential equations (2.3) will not give only the initial
connection, but also annother connections. The reason is that the
corresponding boundary conditions for the system (2.3) have not been fixed,
thus there is no theorem for uniqueness of the solution.

2. As a system of $\left(
\begin{array}{c}
n \\
2
\end{array}
\right) +n$ algebraic equations for an algebraic variety (i.e. the variety
of the unknown variables in the algebraic equation) of $\left[ \left(
\begin{array}{c}
n \\
2
\end{array}
\right) +n\right] n$ variables (the components of the Christoffell
connection) plus $\left[ \left(
\begin{array}{c}
n \\
2
\end{array}
\right) +n\right] n^{2}$ derivatives $\Gamma _{ij}^{k},_{l}$. Thus the total
number of the elements of the algebraic variety is $\frac{n^{2}(n+1)^{2}}{2}$%
. Note that the derivatives in an algebraic equation can also enter as
elements of the algebraic variety.

One might therefore expect that there will be a whole variety of
connections, which will give the same Ricci tensor.

It is useful to remember also from standard textbooks on gravity theory [24]
that the s. c. \textbf{Christoffel connection of the first kind:}
\begin{equation}
\Gamma _{i;kl}\equiv g_{im}\Gamma _{kl}^{m}=\frac{1}{2}%
(g_{ik,l}+g_{il,k}-g_{kl,i})  \tag{2.4}
\end{equation}
is obtained from the expression for the zero covariant derivative $0=\nabla
_{l}g_{ik}=g_{ik,l}-g_{m(i}\Gamma _{k)l}^{m}$ \textbf{without assuming} that
$g_{ij}g^{jk}=\delta _{i}^{k}$ . \textbf{After the derivation} of this
formulae it becomes clear that the consistency of the definition \ (2.1) of
the Christoffell connection (derivable by contracting (2.4) with the
contravariant metric tensor $g^{is}$)\ would require the existence of an
inverse contravariant metric tensor. But then by contraction of (2.4) with
\textbf{another } conravariant tensor field $\widetilde{g}^{is}$ one might
as well define \textbf{another connection:}
\begin{equation}
\widetilde{\Gamma }_{kl}^{s}\equiv \widetilde{g}^{is}\Gamma _{i;kl}=%
\widetilde{g}^{is}g_{im}\Gamma _{kl}^{m}=\frac{1}{2}\widetilde{g}%
^{is}(g_{ik,l}+g_{il,k}-g_{kl,i})\text{ \ \ \ ,}  \tag{2.5}
\end{equation}
not consistent with the initial metric $g_{ij}$. Clearly the connection
(2.5) is defined under the assumption that the metric tensor $g_{ij}$ does
not have on inverse one and therefore $\widetilde{g}^{is}g_{im}\equiv
l_{m}^{s}$. The use of such a convention may seem strange, but it was shown
above that it is fully consistent with the defining equations and as will be
demonstrated further, this more general definition will allow some important
generalization and possibilities for obtaining more general solutions even
of the Einstein's equations.

In fact, the definition $\widetilde{g}^{is}g_{im}\equiv l_{m}^{s}$ turns out
to be inherent to gravitational physics. For example, in the \textbf{%
projective formalism} one decomposes the standardly defined metric tensor
(with $g_{ij}g^{jk}=\delta _{i}^{k}$) as
\begin{equation}
g_{ij}=p_{ij}+h_{ij}\text{ \ \ ,}  \tag{2.6}
\end{equation}
together with the additional assumption that the two subspaces, on which the
\textbf{projective tensor }$p_{ij}$ and the tensor $h_{ij}$ are defined, are
orthogonal. This means that
\begin{equation}
p_{ij}h^{jk}=0\text{ \ \ \ .}  \tag{2.7}
\end{equation}
As a consequence
\begin{equation}
p_{ij}p^{jk}=\delta _{i}^{k}-\frac{1}{e}h_{ij}h^{jk}\neq \delta _{i}^{k}
\tag{2.8 }
\end{equation}
and \textbf{under this condition} the projective connection $\overline{%
\overline{\Gamma }}_{ik}^{l}$ is defined in a similar way
\begin{equation}
\overline{\overline{\Gamma }}_{ik}^{l}\equiv \frac{1}{2}%
p^{ls}(p_{ks,i}+p_{is,k}-p_{ik,s})\text{ \ \ \ \ .}  \tag{2.9}
\end{equation}

Another example of gravitational theories with more than one connection are
the so called \textbf{theories with affine connections and metrics} [13], in
which there is one connection $\Gamma _{\alpha \beta }^{\gamma }$ for the
case of a parallel transport of covariant basic vectors $\nabla _{e_{\beta
}}e_{\alpha }=\Gamma _{\alpha \beta }^{\gamma }$ $e_{\gamma }$ and a \textbf{%
separate connection} $P_{\alpha \beta }^{\gamma }$ for the contravariant
basic vector $e^{\gamma }$, the defining equation for which is $\nabla
_{e_{\beta }}e^{\alpha }=P_{\gamma \beta }^{\alpha }$ $e^{\gamma }$. Since $%
e_{\alpha }e^{\beta }\equiv f_{\alpha }^{\beta }(x)$ for such theories, one
can obtain after covariant differentiation that the two connections are
related in the following way [13]
\begin{equation}
f_{j,k}^{i}=\Gamma _{jk}^{l}\text{ }f_{l}^{i}+P_{lk}^{i}f_{jk}^{l}\text{ \ \
\ \ \ ;\ \ \ \ \ \ (}f_{j,k}^{i}=\partial _{k}f_{j}^{i}\text{) \ \ \ \ . }
\tag{2.10}
\end{equation}
However, the function $f_{j}^{i}(x)$ cannot be determined from any physical
considerations, but similarly to the proposed here approach one can write
down the gravitational Lagrangian in terms of the two connections and make
it equal to the gravitational Lagrangian with only one connection. Then the
defining relation can again be expected to be an algebraic equation (at
least quadratic) in respect to the two connections $\Gamma _{\alpha \beta
}^{\gamma }$ and $P_{\alpha \beta }^{\gamma }$.

In the present case the introduced connection (2.5) should not be identified
with the connection $P_{\alpha \beta }^{\gamma }$, since the connection $%
\widetilde{\Gamma }_{kl}^{s}\equiv \widetilde{g}^{is}\Gamma _{i;kl}$ \ is \
introduced by means of modifying the contravariant tensor and not on the
base of any separate parallel transport. Moreover, the freedom in
determination of $\widetilde{\Gamma }_{kl}^{s}$ \ would contradict the
relation between the two connections (2.10) in case if such an
identification is made.

In [10] a definite choice of the contravariant tensor $\widetilde{g}^{ij}$
was made in the form of the factorized product $\widetilde{g}^{ij}\equiv
dX^{i}dX^{j}$, where $X^{i}=X^{i}(x_{1},x_{2},......,x_{n})$ are some
generalized coordinates, which can be regarded as an $n-$ dimensional
hypersurface, defining a transition from an \textbf{initially} defined set
of coordinates $x_{1},x_{2},......,x_{n}$ on the chosen manifold to \textbf{%
another set of the (generalized) coordinates} $X^{i}$ ($i=1,2.....n$ ; $n$ -
the dimension of spacetime)$.$ If $dX^{i}$ are assumed to be vectors, lying
in the tangent space to the hypersurface, the defined earlier connection
(2.5) assumes the form
\begin{equation}
\widetilde{\Gamma }_{kl}^{s}\equiv dX^{i}dX^{s}\Gamma _{i;kl}=\frac{1}{2}%
dX^{i}dX^{s}(g_{ik,l}+g_{il,k}-g_{kl,i})\text{ \ \ \ \ . }  \tag{2.11}
\end{equation}
The connection \ $\widetilde{\Gamma }_{kl}^{s}$ has two important
properties, already mentioned in the Introduction and proved in Appendix A:
it has an affine transformation law under change of variables and moreoever,
it is an equiaffine connection [15, 16], which means that it can be
represented as a gradient of a scalar quantity, i.e. $\widetilde{\Gamma }%
_{ks}^{s}=\partial _{k}lge$, where $e$ is some scalar quantity. Since by
definition a connection is an equiaffine if and only if the corresponding
Ricci tensor $\widetilde{R}_{ij}$ is a symmetric one in respect to the
indices $i$ and $j$, the usual formulae for the Ricci tensor can be used,
but with the newly defined connection $\widetilde{\Gamma }_{kl}^{s}$ instead
of the usual Christoffell connection $\Gamma _{kl}^{s}$.\ \

In the formal mathematical sense, (2.11) can be treated as a system of
quadratic algebraic equations in respect to the differentials $dX^{i}$ and
under known connection components $\widetilde{\Gamma }_{kl}^{s}$. Such an
interpretation would be incorrect because for fixed indices $k$ and $l$ and
a varying indice $s$ (2.11) represents also a system of \ linear algebraic
equations in respect to the variables $G_{i}\equiv
g_{ik,l}+g_{il,k}-g_{kl,i} $ ($k,l$ - fixed). \textbf{Since the determinant }%
$det\parallel \widetilde{g}^{ij}\parallel =det\parallel
dX^{i}dX^{j}\parallel \equiv 0$\textbf{, for non - zero }$\widetilde{\Gamma }%
_{kl}^{s}$\textbf{\ there will be no solutions of this system of equations
in respect to }$G_{i}$\textbf{. This means that the components of the
connection } $\widetilde{\Gamma }_{kl}^{s}$\textbf{\ cannot be fixed
arbitrarily,} but instead one can choose the differential functions $dX^{i}$
and the derivatives of the metric tensor $g_{ij,k}$. There is one exception
- for certain $k$ and $l$ one can choose all the connection components $%
\widetilde{\Gamma }_{kl}^{s}$ $\ $\ for all $s=1,2......n$ to be zero. Then
the system of equations (2.11) for $\widetilde{\Gamma }_{kl}^{s}=0$ will be
satisfied for \textbf{arbitrary} $g_{ij,k}$ with the given $k$ and $l$.

\subsection*{\protect\bigskip 2.2. BASIC \ ALGEBRAIC \ EQUATIONS \ IN \
GRAVITY \ THEORY}

Now if one applies again the new definition $\widetilde{g}^{ij}\equiv
dX^{i}dX^{j}$ of the contravariant tensor in respect to the Ricci tensor,
then the following \textbf{fourth - degree algebraic equation} can be
obtained
\begin{equation*}
R_{ik}=dX^{l}\left[ g_{is,l}\frac{\partial (dX^{s})}{\partial x^{k}}-\frac{1%
}{2}pg_{ik,l}+\frac{1}{2}g_{il,s}\frac{\partial (dX^{s})}{\partial x^{k}}%
\right] +
\end{equation*}
\begin{equation}
+\frac{1}{2}dX^{l}dX^{m}dX^{r}dX^{s}\left[
g_{m[k,t}g_{l]r,i}+g_{i[l,t}g_{mr,k]}+2g_{t[k,i}g_{mr,l]}\right] \text{ \ \
\ \ \ ,}  \tag{2.12}
\end{equation}
where $p$ is the scalar quantity

\begin{equation}
p\equiv div(dX)\equiv \frac{\partial (dX^{l})}{\partial x^{l}}\text{,}
\tag{2.13}
\end{equation}
which ''measures'' the \textbf{divergency }of the vector field $dX$. The
algebraic variety of the equation consists of the differentials $dX^{i\text{
}}$ and their derivatives $\frac{\partial (dX^{s})}{\partial x^{k}}$.

In the same spirit one can investigate the problem whether the gravitational
Lagrangian in terms of the new contravariant tensor can be equal to the
standard representation of the gravitational Lagrangian. The condition for
the equivalence of the two representations gives a \textbf{cubic algebraic
equation} in respect to the algebraic variety of the first differential $%
dX^{i}$ and the second one $d^{2}X^{i}$ [10]
\begin{equation}
dX^{i}dX^{l}\left( p\Gamma _{il}^{r}g_{kr}dX^{k}-\Gamma
_{ik}^{r}g_{lr}d^{2}X^{k}-\Gamma _{l(i}^{r}g_{k)r}d^{2}X^{k}\right)
-dX^{i}dX^{l}R_{il}=0\text{ \ \ \ \ .}  \tag{2.14}
\end{equation}

From this equation the following important generalization can be made: Let
us assume that the components $\Gamma _{ik}^{l}$ of the standard
Christoffell connection in the above equation are not given, but instead
only the metric tensor $g_{ij}$, its partial derivatives and the scalar
curvature $R$ are given. Then the above written equation assumes the form of
a \textbf{fifth - degree algebraic equation }
\begin{equation}
dX^{i}dX^{l}dX^{r}dX^{s}\left( p\Gamma _{s;il}g_{kr}dX^{k}-\Gamma
_{s;ik}g_{lr}d^{2}X^{k}-\Gamma _{s;l(i}g_{k)r}d^{2}X^{k}\right) -R=0\text{ \
\ \ \ .}  \tag{2.15}
\end{equation}
\textbf{\ }The mathematical treatment of such equations is known since the
time of Felix Klein's famous monograph [25], published in 1884. A way for
resolution of such equations on the base of earlier developed approaches in
by means of reducing the fifth - degree equations to the so called \textbf{%
modular equation} has been presented in the more recent \ monograph of \
Prasolov and Solov'yev [9]. It should be stressed that the presented in [10]
(standardly known) method for parametrization with the Weierstrass function
of a cubic algebraic equation should be considered as \textbf{just one
possibility} for solution of third - order algebraic equations. Some other
methods for solution of third-, fifth- and higher- order \ algebraic
equations, based on resolvents with a group of linear substitutions [26, 27]
date from the end of the previous century (see particularly [28] and also
[29] for a complete and detailed description of the properties of \textbf{%
elliptic, theta and modular functions} from the viewpoint of the earliest
developments more that 100 years ago). It is known as well that solutions of
$n-$ th degree algebraic equations can be presented also in \textbf{theta -
constants} [30] and also in \textbf{special functions} [31]. The first
possibilitty, related also to modular equations, is interesting in view of
the not yet proven hypothesis in the paper by Kraniotis and Whitehouse [8]
that \textbf{''all nonlinear solutions of general relativity are expresed in
terms of theta - functions, associated with Riemann - surfaces''}.

In [10] the Einstein's equations in vacuum \textbf{for the general case}
were presented under the assumption about the newly defined contravariant
metric tensor
\begin{equation*}
0=\widetilde{R}_{ij}-\frac{1}{2}g_{ij}\widetilde{R}=\widetilde{R}_{ij}-\frac{%
1}{2}g_{ij}dX^{m}dX^{n}\widetilde{R}_{mn}=
\end{equation*}
\begin{equation*}
=-\frac{1}{2}pg_{ij}\Gamma _{mn}^{r}g_{kr}dX^{k}dX^{m}dX^{n}+\frac{1}{2}%
g_{ij}(\Gamma _{km}^{r}g_{nr}+\Gamma
_{n(m}^{r}g_{k)r})d^{2}X^{k}dX^{m}dX^{n}+
\end{equation*}
\begin{equation}
+p\Gamma _{ij}^{r}g_{kr}dX^{k}-(\Gamma _{ik}^{r}g_{jr}+\Gamma
_{j(i}^{r}g_{k)r})d^{2}X^{k}\text{ \ \ \ .}  \tag{2.16 }
\end{equation}

\bigskip This equation is again a system of \textbf{cubic equations} or it
can analogously to (2.15) be presented as a system of \textbf{fifth - degree
}algebraic equations. \textbf{In addition, if the differentials }$dX^{i}$%
\textbf{\ and }$d^{2}X^{i}$\textbf{\ are known, but not the covariant tensor
}$g_{ij}$, \textbf{the same equation can be considered also as a \ cubic
algebraic equation in respect to the algebraic variety of the metric tensor
components }$g_{ij}$\textbf{\ and their first derivatives }$g_{ij,k}$\textbf{%
. }

\subsection*{\protect\bigskip 2.3. SOME \ GENERAL \ ANALYSIS \ OF \ THE \ N
- DIMENSIONAL \ SYSTEM \ OF \ EQUATIONS \ $g_{ij}\widetilde{g}^{jk}=\protect%
\delta _{i}^{k}$}

Of course, a more general theory with the definition of the contravariant
tensor as $\widetilde{g}^{ij}\equiv dX^{i}dX^{j}$ should \ contain in itself
the standard gravitational theory with $g_{ij}g^{jk}=\delta _{i}^{k}$. From
a mathematical point of view, this should be performed by \textbf{%
considering the intersection of the cubic algebraic equation (2.16) (or of
its fifth - degree \ counterpart)\ \ with the system of \ }$n^{2}$ \textbf{%
quadratic algebraic equations for the algebraic variety of the }$n$
variables
\begin{equation}
g_{ij}dx^{j}dx^{k}=\delta _{i}^{k}\text{ \ \ .}  \tag{2.17}
\end{equation}
It might seem that the system of equations (2.17) does not have solutions in
respect to $g_{ij\text{ }}$(and thus no solutions of the Einstein's
equations can be found for the standard case), since the determinant $%
det\parallel dx^{i}dx^{j}\parallel _{i,j=1,..n}=0$ equals to zero! Such a
statement would be however \textbf{wrong}, since (2.17) should not be
considered as a system of linear algebraic equations (for which the unknown
variables are contained in a \textbf{vector - column and therefore all the
theorems of linear algebra can be applied), }but rather than that as an
\textbf{operator equation of the kind }$Y_{ij}g^{jk}=\delta _{i}^{k}$
\textbf{with unknown variables }$Y_{ij}\equiv g_{ij}$, \textbf{which
constitute a matrix and not a vector - column! Therefore the linear algebra
theorems cannot be applied. }A general theory of operator equations of the
type $AX=XB$ ($X$- the unknown matrix) is developed in the known monograph
of Gantmaher [20]. In Appendix C it will be shown how the \textbf{operator
system of equations} $Y_{ij}g^{jk}=\delta _{i}^{k}$ can be transformed to a
system $\widetilde{A}_{ij}\widetilde{Y}$ $^{j}=\delta _{i}$ ($\delta _{i}$
is a \textbf{\ vector - column}, consisting only of $1$ and $0;$ $i,j=1,2...,%
\frac{n(n+1)}{2}$). The method\textbf{\ } at first will be demonsrated for
the simple case when $i,j=1,2,3$; subsequently the method will be
generalized to arbitrary $n-$dimensions, i.e. it will be explicitely shown
how the matrix \ $\widetilde{A}_{ij}$ can be constructed \ for the general $%
n-$dimensional case. The advantage of the proposed approach is that it will
allow explicitely to answer the question: will the operator system of
equations $Y_{ij}g^{jk}=\delta _{i}^{k}$ have a solution in the case, when
the contravariant metric tensor $\widetilde{g}^{jk}$ is chosen in the form
of a factorized product $\widetilde{g}^{jk}\equiv dX^{j}dX^{k}$? It will
become evident that this choice, inspite of the zero determinant $\parallel
dx^{i}dx^{j}\parallel _{i,j=1,..n}$, \textbf{does not} put a restriction on
the solvability of the considered \textbf{predetermined} system \ of $\
n^{2} $ \ equations (for different values of the indices $i$ and $k$ ) for
the $\left(
\begin{array}{c}
n \\
2
\end{array}
\right) +n=\frac{n(n+1)}{2}$ variables $g_{ij}$.

Now let us consider some more general properties when of the system $g_{ij}%
\widetilde{g}^{jk}=\delta _{i}^{k}$ in the general $n-$dimensional case.
Since the number of equations is greater than the number of variables, from
the $n^{2}$ equations one can select $n^{2}-n$ equations with different
values of $\ i$ and $k$, for which the R. H. S. will be zero. Yet the number
of the chosen equations remains to be greater than the number of variables $%
\left(
\begin{array}{c}
n \\
2
\end{array}
\right) +n$, which is confirmed by the equality
\begin{equation}
(n^{2}-n)-\left[ \left(
\begin{array}{c}
n \\
2
\end{array}
\right) +n\right] =\frac{n(n-3)}{2}\text{ \ \ \ ,}  \tag{2.18}
\end{equation}
fulfilled for $n>3$. \textbf{Only} for the case $n=3$ , as it will be shown
in Appendix C, the number of the equations $n^{2}-n=6$ with a zero R. H. S.
becomes exactly equal to the number of variables $\left(
\begin{array}{c}
n \\
2
\end{array}
\right) +n=\frac{n(n+1)}{2}=\frac{3.\text{ }4}{2}=6$. Therefore for the case
of an arbitrary $n$ from these $n^{2}-n$ equations one can choose $\left[
\left(
\begin{array}{c}
n \\
2
\end{array}
\right) +n\right] $ equations. In each row of the $\left[ \left(
\begin{array}{c}
n \\
2
\end{array}
\right) +n\right] \times \left[ \left(
\begin{array}{c}
n \\
2
\end{array}
\right) +n\right] $ matrix of coefficient functions of this system there
will be only $n$ functions $\widetilde{g}^{jk}$, the rest $\left(
\begin{array}{c}
n \\
2
\end{array}
\right) $ of the \ elements will be zero. The determinant of the matrix will
be equal to zero, a proof of which in the general case of an arbitrary $n$
(and also for $n=3$) is presented in Appendix C. Therefore, \textbf{the
solutions }$g_{ij}$ of this $\left(
\begin{array}{c}
n \\
2
\end{array}
\right) +n$ dimensional homogeneous \ system of equations with a zero
determinant are \textbf{arbitrary. }

Now we are left with $\frac{n(n-3)}{2}$ equations (again with \textbf{\ }$%
\left(
\begin{array}{c}
n \\
2
\end{array}
\right) +n$ variables) with a zero R. H. S. (we shall call it the \textbf{%
first system of equations}) and with another $n$ equations (the \textbf{%
second system})\ with a R. H. S. of each equation, equal to $1$. Since the
number of variables $\left(
\begin{array}{c}
n \\
2
\end{array}
\right) +n$ in the \textbf{first system} is greater than the number $\frac{%
n(n-3)}{2}$ of equations and
\begin{equation}
\left(
\begin{array}{c}
n \\
2
\end{array}
\right) +n-\frac{n(n-3)}{2}=2n\text{ \ \ ,}  \tag{2.19}
\end{equation}
again one can treat as unknown only $\frac{n(n-3)}{2}$ variables and
transfer the rest $2n$ variables in the R. H. S., which will become
different from zero. Analogously, for the \textbf{second system} one can
treat as unknown only $n$ variables and transfer the rest $\left(
\begin{array}{c}
n \\
2
\end{array}
\right) =$ $\frac{n(n-1)}{2}$ variables in the R. H. S. If this R. H. S. is
different from zero and moreover, the determinant of the coefficient
functions $\widetilde{g}^{ij}$ is also different from zero, \textbf{then one
can find unique solutions for these }$n$\textbf{\ variables }$g_{ij}$. Now
comes the most important point of the proof: Since
\begin{equation}
n+\frac{n(n-1)}{2}>2n\text{ \ \ \ \ ,}  \tag{2.20}
\end{equation}
one can take all $n$ of the \ uniquely found solutions plus $n$ more unfixed
(freely varied) variables from the R. H.S. of the second system and
''plunge'' them into the R. H. S. of the first system. Let us remember that
the first system has $\frac{n(n-3)}{2}$ unknown variables, but one may note
that $\left(
\begin{array}{c}
n \\
2
\end{array}
\right) +n-2n=\frac{n(n-3)}{2}$ variables from the second system have not
been transfered in the R. H. S. of the first system. Therefore, one can
choose these $\frac{n(n-3)}{2}$ variables to be the unknown variables for
the first system, and if the determinant of the coefficient functions is
nonzero, an unique solution can be found for them.

\textbf{As a whole , one would have a maximum of }$n+$\textbf{\ }$\frac{%
n(n-3)}{2}=\frac{n(n-1)}{2}$\textbf{\ uniquely fixed variables and the rest
of the variables may be varied freely. }

\subsection*{\protect\bigskip 2.4. ALGEBRAIC \ EQUATIONS \ FOR \ A \ GENERAL
\ CONTRAVARIANT \ METRIC \ TENSOR}

Let us write down the algebraic equations for all admissable
parametrizations of the gravitational Lagrangian for the general case of
contravariant tensor $\widetilde{g}^{ij}:$
\begin{equation*}
\widetilde{g}^{i[k}\widetilde{g}^{l]s}\Gamma _{ik}^{r}g_{rs}+\widetilde{g}%
^{i[k}\widetilde{g}^{l]s}\left( \Gamma _{ik}^{r}g_{rs}\right) _{,l}+
\end{equation*}
\begin{equation}
\widetilde{g}^{ik}\widetilde{g}^{ls}\widetilde{g}^{mr}g_{pr}g_{qs}\left(
\Gamma _{ik}^{q}\Gamma _{lm}^{p}-\Gamma _{il}^{p}\Gamma _{km}^{q}\right) -R=0%
\text{ \ \ \ \ .}  \tag{2.21}
\end{equation}
This equation is again a \textbf{cubic algebraic equation} in \ respect to
the algebraic variety of the variables $\widetilde{g}^{ij}$ and $\widetilde{g%
}_{,k}^{ij}$, and the number of variables in the present case is much
greater than in the previous case for the contravariant tensor $\widetilde{g}%
^{ij}\equiv dX^{i}dX^{j}$ . If the connection is assumed to be the ''tilda
connection''\ (2.11) $\widetilde{\Gamma }_{kl}^{s}\equiv dX^{i}dX^{s}\Gamma
_{i;kl}$, then the same equation can be regarded as a \textbf{fifth - degree
equation} in respect to the contravariant tensor $\widetilde{g}^{ij}$ and
its derivatives $\ \widetilde{g}_{,k}^{ij}$ and at the same time as a
\textbf{fourth - degree algebraic equation} in respect to the covariant
metric tensor $g_{ij}$ and its first and second partial derivatives.
\begin{equation*}
\widetilde{g}^{i[k}\widetilde{g}^{l]s}\widetilde{g}^{rt}\Gamma _{t;ik}\text{
}g_{rs}+\widetilde{g}^{i[k}\widetilde{g}^{l]s}\widetilde{g}_{,l}^{rt}\left(
\Gamma _{t;\text{ }ik}g_{rs}\right) +
\end{equation*}
\begin{equation*}
+\widetilde{g}^{i[k}\widetilde{g}^{l]s}\widetilde{g}^{rt}g_{rs}\left( \Gamma
_{t;\text{ }ik}\right) _{,l}+\widetilde{g}^{i[k}\widetilde{g}^{l]s}%
\widetilde{g}^{rt}\Gamma _{t;ik}\text{ }g_{rs,l}+
\end{equation*}
\begin{equation}
+\widetilde{g}^{ik}\widetilde{g}^{ls}\widetilde{g}^{mr}\widetilde{g}^{qt_{1}}%
\widetilde{g}^{pt_{2}}g_{pr}g_{qs}\left[ \Gamma _{t_{1};ik}\Gamma
_{t_{2;lm}}-\Gamma _{t_{1;mk}}\Gamma _{t_{2;il}}\right] -R=0\text{ \ \ \ \ .}
\tag{2.22}
\end{equation}
Note that the consideration of this equation either as a\textbf{\ fourth-}
or as a \textbf{fifth-} degree equation means that the algebraic equation
will be with \textbf{coefficient functions} and not with \textbf{number
coefficients}, as are the equations in the standardly known \textbf{%
arithmetic theory}. This difficulty can be overcomed if the above equation
is treated \textbf{simultaneously in respect to the algebraic variety of the
covariant and the contravariant variables }- then (2.22)\ is an algebraic
equation with number coefficients . Of course, the algebraic treatment of an
equation of \textbf{nineth degree} is not quite known.

Similarly the Einstein's equations (2.16) can be written as a system of \
\textbf{fifth - degree algebraic equations in respect to the contravariant
variables} and as \textbf{fifth - degree degree equations in respect to the
covariant variables (which is the difference from the previous case). Note
that in such a representation for the general case, when already the
determinant of the components of the contravariant metric is non - degenerate%
}, \textbf{i. e. }$det\parallel \widetilde{g}^{jk}\parallel \neq 0$\textbf{,
the additional equation }$g_{ij}\widetilde{g}^{jk}=\delta _{i}^{k}$\textbf{\
can also be considered together with the Einstein's equations.} \textbf{From
an algebro - geometric point of view, this turns out to be a problem about
the intersection of the Einstein's algebraic equations with the system of }$%
n^{2}$ \ \textbf{(linear)} \textbf{hypersurfaces for the }$\left[ \left(
\begin{array}{c}
n \\
2
\end{array}
\right) +n\right] $\textbf{\ variables of the covariant tensor and }$\left[
\left(
\begin{array}{c}
n \\
2
\end{array}
\right) +n\right] $\textbf{\ variables for the contravariant ones.} If the
corresponding equations are cubic ones, this is an analogue to the well -
known problem in algebraic geometry about the intersection of a cubic
surface with a straight line. In this aspect one can point out the following
three important problems:

1. One can find solutions of the system of Einstein's equations not as
solutions of a system of nonlinear differential equations, but as \textbf{%
elements of an algebraic variety, satisfying the Einstein's algebraic
equations.} The important new moment is that \textbf{this gives an
opportunity to find solutions of \ the Einstein's equations both for the
components of the covariant metric tensor }$g_{ij}$\textbf{\ and for the
contravariant ones }$\widetilde{g}^{jk}$\textbf{. This means that solutions
may exist for which }$g_{ij}\widetilde{g}^{jk}\neq \delta _{i}^{k}$. In
other words, a classification of the solutions of the Einstein's equations
can be performed in an entirely new and nontrivial manner:

a) \textbf{Under a given contravariant tensor, the covariant tensor and its
derivatives have to be found from the algebraic equation}.

b) \textbf{Under a given covariant tensor, the contravariant tensor and its
derivatives have to be found from the corresponding algebraic equation}.

c) \textbf{If} \textbf{both the covariant and the contravariant tensors are
considered to be unknown, then their components have to be found from the
corresponding algebraic equation with number coefficients. This is what is
really meant by ''solutions of \ the Einstein's equations both for the
components of the covariant metric tensor }$g_{ij}$\textbf{\ and for the
contravariant ones }$\widetilde{g}^{jk}$.\textbf{''}

2. The standardly known solutions of the Einstein's equations can be
obtained as an intersection variety of the Einstein's algebraic equations
with the system of linear hypersurfaces $g_{ij}\widetilde{g}^{jk}=\delta
_{i}^{k}$ . However, the strict mathematical proof that such an intersection
of the system of tenth - degree equations (a \ fifth - degree for the
variety of the covariant variables and a fifth - degree for the
contravariant ones) with the system of linear hypersurfaces will give the
standard case is \textbf{still lacking.}

3. The condition for the zero - covariant derivative of the covariant metric
tensor $\nabla _{k}g_{ij}=0$ and of the contravariant metric tensor $\nabla
_{k}\widetilde{g}^{ij}=0$ can be written in the form of the following
\textbf{cubic algebraic equations }in respect to the variables $g_{ij}$, $%
g_{ij,k}$ and $\widetilde{g}^{ls}$ \textbf{:}
\begin{equation}
\nabla _{k}g_{ij}\equiv g_{ij,k}-\widetilde{\Gamma }%
_{k(i}^{l}g_{j)l}=g_{ij,k}-\widetilde{g}^{ls}\Gamma _{s;k(i}g_{j)l}=0
\tag{2.23}
\end{equation}
and
\begin{equation}
0=\nabla _{k}\widetilde{g}^{ij}=\widetilde{g}_{,k}^{ij}+\widetilde{g}^{r(i}%
\widetilde{g}^{j)s}\Gamma _{r;sk}\text{ \ \ \ \ .}  \tag{2.24}
\end{equation}
The first equation (2.23) is linear in respect to $\widetilde{g}^{ls}$ and
quadratic in respect to $g_{ij}$, $g_{ij,k}$, while the second equation
(2.24) is linear in respect to $g_{ij}$, $g_{ij,k}$ and quadratic in respect
to $\widetilde{g}^{ls}$- therefore both equations are cubic in respect to
these variables.

Since the treatment of the above cubic algebraic equations is based on
singling out one variable, let us rewrite equation (2.22) for the effective
parametrization of the gravitational action for the case of diagonal metrics
$g_{\beta \beta }$ and $\widetilde{g}^{\alpha \alpha }$, singling out the
variable $\widetilde{g}^{44}$:
\begin{equation*}
A(\widetilde{g}^{44})^{3}+B_{\alpha }(\widetilde{g}^{44})^{2}\widetilde{g}%
^{\alpha \alpha }+C_{\alpha \alpha }\widetilde{g}^{44}\widetilde{g}^{\alpha
\alpha }+(\Gamma _{44}^{\alpha }g_{\alpha \alpha })\widetilde{g}^{44}%
\widetilde{g}_{,\alpha }^{\alpha \alpha }+
\end{equation*}
\begin{equation}
+D_{\alpha \gamma }\widetilde{g}^{44}\widetilde{g}^{\alpha \alpha }%
\widetilde{g}^{\gamma \gamma }+F_{\alpha \gamma }=0\text{ \ \ \ \ \ ,}
\tag{2.25}
\end{equation}
where the coefficient functions $A$, $B_{\alpha }$, $C_{\alpha \alpha }$, $%
D_{\alpha \gamma }$ and the free term $F_{\alpha \gamma }$ denote the
following expressions:
\begin{equation}
A\equiv g_{4[4}g_{p]p}\Gamma _{44}^{4}\Gamma _{44}^{p}\text{ \ \ ,}
\tag{2.26}
\end{equation}
\begin{equation*}
B_{\alpha }\equiv g_{4[4}g_{\alpha ]\alpha }\Gamma _{44}^{4}\Gamma _{4\gamma
}^{\gamma }+g_{\alpha \alpha }g_{44}(\Gamma _{44}^{4}\Gamma _{\alpha
4}^{\alpha }-\Gamma _{4\alpha }^{4}\Gamma _{44}^{\alpha })+
\end{equation*}
\begin{equation}
+g_{\alpha \alpha }g_{44}\Gamma _{44}^{\alpha }\Gamma _{\alpha
4}^{4}+g_{44}g_{pp}\Gamma _{\alpha \alpha }^{p}\Gamma _{44}^{4}\text{ \ \ \ ,%
}  \tag{2.27}
\end{equation}
\begin{equation}
C_{\alpha \alpha }\equiv \left( \Gamma _{44}^{\alpha }g_{\alpha \alpha
}\right) _{,\alpha }+\left( \Gamma _{\alpha \alpha }^{4}g_{44}\right) _{,4}%
\text{ \ \ \ ,}  \tag{2.28}
\end{equation}
\begin{equation*}
D_{\alpha \gamma }\equiv g_{\alpha \alpha }g_{\gamma \gamma }(\Gamma
_{44}^{\gamma }\Gamma _{\alpha \gamma }^{\alpha }-\Gamma _{4\alpha }^{\gamma
}\Gamma _{4\gamma }^{\alpha })+g_{44}g_{\gamma \gamma }\left[ 2\Gamma
_{\alpha \alpha }^{4}\Gamma _{4\gamma }^{\gamma }+\Gamma _{\alpha \alpha
}^{\gamma }\Gamma _{4\gamma }^{4}\right] +
\end{equation*}
\begin{equation}
+g_{\alpha \lbrack \alpha }g_{\gamma ]\gamma }\Gamma _{44}^{\alpha }\Gamma
_{\alpha \gamma }^{\gamma }+g_{\gamma \lbrack \gamma }g_{4]4}\Gamma _{\alpha
\alpha }^{\gamma }\Gamma _{\gamma 4}^{4}\text{ \ \ \ \ \ \ ,}  \tag{2.29}
\end{equation}
\begin{equation*}
F_{\alpha \gamma }\equiv \widetilde{g}^{\alpha \alpha }\widetilde{g}^{\gamma
\gamma }\widetilde{g}^{\delta _{1}\delta _{1}}g_{\gamma \lbrack \gamma
}g_{\delta _{1}]\delta _{1}}\Gamma _{\alpha \alpha }^{\gamma }\Gamma
_{\gamma \delta _{1}}^{\delta _{1}}+\widetilde{g}^{\alpha \lbrack \alpha }%
\widetilde{g}^{\gamma ]\gamma }[\widetilde{g}^{\delta _{1}\delta
_{1}}g_{\gamma \gamma }g_{\delta _{1}\delta _{1}}\Gamma _{\alpha \alpha
}^{\delta _{1}}\Gamma _{\gamma \delta _{1}}^{\gamma }+
\end{equation*}
\begin{equation}
+\Gamma _{\alpha \alpha }^{r}g_{r\gamma }+\left( \Gamma _{\alpha \alpha
}^{\delta }g_{\delta \delta }\right) _{,\gamma }]\text{ \ \ \ \ \ .}
\tag{2.30}
\end{equation}
In (2.25 - 2.30) the Greek indices run the values $\alpha ,\beta ,\gamma
=1,2,3$, while all the other indices run from $1$ to $4$. Equation (2.25)\
can be considered also as a cubic algebraic equation in respect to $%
\widetilde{g}^{44},\widetilde{g}^{\alpha \alpha }$ and $\widetilde{g}%
_{,\alpha }^{\alpha \alpha }$. It is understood also that the connection
components are known in advance, but if they are not - then the equation
will be no longer a cubic one, but a higher order algebraic equation.

\section{\protect\bigskip \ EMBEDDED \ SEQUENCE \ OF \ ALGEBRAIC \ EQUATIONS
\ AND \ FINDING \ THE \ SOLUTIONS \ OF \ \ THE \ CUBIC \ ALGEBRAIC \ EQUATION%
}

The purpose of the present subsection will be to describe the method for
finding the solution (i. e . the algebraic variety of the differentials $%
dX^{i}$) of the cubic algebraic equation (2.14) (in the limit $d^{2}X^{k}=0$%
). The applied method has been proposed first in [10] but here it will be
developed further and applied in respect to a \textbf{sequence of algebraic
equations with algebraic varieties, which are embedded into the initial one.
This means that if at first the algorithm is applied in respect to the
three-dimensional cubic algebraic equation (2.14) and a solution for }$%
dX^{3} $ (depending on the Weierstrass function and its derivative is
found), then \textbf{the same algorithm will be applied in respect to the
two-dimensional cubic algebraic equation with variables }$dX^{1}$\textbf{\
and }$dX^{2}$\textbf{, and finally to the one-dimensional cubic algebraic
equation of the variable }$dX^{1}$\textbf{\ only. }

The basic knowledge about the parametrization of a cubic algebraic equation
with the Weierstrass function and its derivative are given in almost all
basic textbooks on elliptic functions [9, 11, 32, 33] and many others.
However, the most complete, detailed and exhaustive knowledge about elliptic
functions and automorphic forms is contained in the two two - volume \ books
[34, 35] of \ Felix Klein and Robert Fricke, written more than 100 years
ago. More specific and advanced topics on elliptic curves from a
mathematical point of view such as the group of rational points, cubic
curves over finite fields, families of elliptic curves and torsion points
and etc. are contained in the monographs [36, 37]. A very understandable
exposition of the classical topics on cubic algebraic curves and at the same
time the most contemporary issues such as the Mordell's and Dirichlet's
theorems and $L$ functions, modular forms and theories of Eichler - Shimura
are given in the book of Knapp [38], which can be used for first acquintance
in these topics. A consistent, modern and full exposition of elliptic curves
in the language of modern mathematics is given in the (two consequent)\
monographs of Silverman [39, 40]. A classical and very understandable
exposition of the relation of elliptic curves with modular forms is given in
[41], also in [42]. From a modern standpoint the relation of elliptic curves
with number theory and modular forms is given in the review articles of
Cohen and Don Zagier in [43], also introductory knowledge on hyperelliptic
integrals, compact Riemann surfaces and Abelian varieties are presented in
the review article by Bost also in [43].

The basic and very simple idea about parametrization of a cubic algebraic
equation with the Weierstrass function can be presented as follows: Let us
define the lattice $\Lambda =\{m\omega _{1}+n\omega _{2}\mid m,n\in Z;$ $%
\omega _{1},\omega _{2}\in C,Im\frac{\omega _{1}}{\omega _{2}}>0\}$ and the
mapping $f:$ $C/\Lambda \rightarrow CP^{2}$, which maps the factorized
(along the points of the lattice $\Lambda $) part of the points on the
complex plane into the two \textbf{dimensional complex projective space }$%
CP^{2}$. This means that each point $z$ on the complex plane is mapped into
the point $(x,y)=(\rho (z),\rho ^{^{\prime }}(z))$, where $x$ and $y$ belong
to the \textbf{affine curve }
\begin{equation}
y^{2}=4x^{3}-g_{2}x-g_{3}\text{ \ \ \ ,}  \tag{3.1}
\end{equation}
where the complex numbers $g_{2}$ and $g_{3}$ are the so called \textbf{%
Eisenstein \ series}\textit{\ }$g_{2}=60\sum\limits_{\omega \subset \Gamma }%
\frac{1}{\omega ^{4}}=\sum\limits_{n,m}\frac{1}{(n+m\tau )^{4}};$ $%
g_{3}=140\sum\limits_{\omega \subset \Gamma }\frac{1}{\omega ^{6}}%
=\sum\limits_{n,m}\frac{1}{(n+m\tau )^{6}}$\ \ and $\rho (z)$ denotes the
\textbf{Weierstrass elliptic function} $\rho (z)=\frac{1}{z^{2}}%
+\sum\limits_{\omega }\left[ \frac{1}{(z-\omega )^{2}}-\frac{1}{\omega ^{2}}%
\right] $ and the summation is over the poles in the complex plane. In other
words, the functions $x=\rho (z)$ and $y=\rho ^{^{\prime }}(z)$ are
uniformization functions for the cubic curve \ and it can be proved [9] that
the only cubic algebraic curve (but with number coefficients!) which is
parametrized by the uniformization functions $x=\rho (z)$ and $y=\rho
^{^{\prime }}(z)$ is the above mentioned affine curve. Note that the fact
that the coefficients are numbers is very important, otherwise $\rho (z)$
and $\rho ^{^{\prime }}(z)$ might in principle parametrize a cubic equation
of the general type $y^{2}(z)=a_{3}(z)x^{3}+a_{2}(z)x^{2}+a_{1}(z)x+a_{0}(z)$%
. This problem has also been treated in [10], but still it has numerous
unexplored issues, because it is related to the so called \textbf{%
non-arithmetic theory} of algebraic equations, which has not been yet
investigated.

In the case of the cubic equation (2.14), the aim will be again to bring the
equation to the form (3.1) and afterwards to make equal each of the
coefficient functions to the (numerical) coefficients in (2.1).

In order to provide a more clear description of the developed method, let us
divide it into several steps.

\textbf{Step 1}. The initial cubic algebraic equation (2.14) is presented as
a cubic equation in respect to the variable $dx^{3}$ only
\begin{equation}
A_{3}(dX^{3})^{3}+B_{3}(dX^{3})^{2}+C_{3}(dX^{3})+G^{(2)}(dX^{2},dX^{1},g_{ij},\Gamma _{ij}^{k},R_{ik})\equiv 0%
\text{ \ \ \ \ ,}  \tag{3.2}
\end{equation}
where naturally the coefficient functions $A_{3}$, $B_{3}$ , $C_{3}$ and $%
G^{(2)}$ depend on the variables $dX^{1}$ and $dX^{2}$ of the algebraic
subvariety and on the metric tensor $g_{ij}$, the Christoffel connection $%
\Gamma _{ij}^{k}$ and the Ricci tensor $R_{ij}$:
\begin{equation}
A_{3}\equiv 2p\Gamma _{33}^{r}g_{3r}\text{ \ \ ; \ \ \ \ \ \ \ \ \ \ \ }%
B_{3}\equiv 6p\Gamma _{\alpha 3}^{r}g_{3r}dX^{\alpha }-R_{33}\text{\ \ \ \
,\ }  \tag{3.3 }
\end{equation}
\begin{equation}
C_{3}\equiv -2R_{\alpha 3}dX^{\alpha }+2p(\Gamma _{\alpha \beta
}^{r}g_{3r}+2\Gamma _{3\beta }^{r}g_{\alpha r})dX^{\alpha }dX^{\beta }\text{
\ \ \ .}  \tag{3.4}
\end{equation}
The Greek indices $\alpha ,\beta $ take values $\alpha ,\beta =1,2$ while
the indice $r$ takes values $r=1,2,3$.

\textbf{Step 2}. A linear-fractional transformation
\begin{equation}
dx^{3}=\frac{a_{3}(z)\widetilde{dX}^{3}+b_{3}(z)}{c_{3}(z)\widetilde{dX}%
^{3}+d_{3}(z)}  \tag{3.5 }
\end{equation}
is performed with the purpose of setting up to zero the coefficient
functions in front of the highest (third) degree of $\ \widetilde{dX}^{3}$.
This will be achieved if $G^{(2)}(dX^{2},dX^{1},g_{ij},\Gamma
_{ij}^{k},R_{ik})=-\frac{a_{3}Q}{c_{3}^{3}}$, where
\begin{equation}
Q\equiv A_{3}a_{3}^{2}+C_{3}c_{3}^{2}+B_{3}a_{3}c_{3}+2c_{3}d_{3}C_{3}\text{
\ \ \ \ \ \ , }  \tag{3.6 }
\end{equation}
which gives a cubic algebraic equation in respect to the \textbf{%
two-dimensional algebraic variety} of the variables $dX^{1}$ and $dX^{2}$:
\begin{equation}
p\Gamma _{\gamma (\alpha }^{r}g_{\beta )r}dX^{\gamma }dX^{\alpha }dX^{\beta
}+K_{\alpha \beta }^{(1)}dX^{\alpha }dX^{\beta }+K_{\alpha }^{(2)}dX^{\alpha
}+2p\left( \frac{a_{3}}{c_{3}}\right) ^{3}\Gamma _{33}^{r}g_{3r}=0\text{ \ \
\ }  \tag{3.7 }
\end{equation}
and $K_{\alpha \beta }^{(1)}$ and $K_{\alpha }^{(2)}$ are the corresponding
quantities [10]
\begin{equation}
K_{\alpha \beta }^{(1)}\equiv -R_{\alpha \beta }+2p\frac{a_{3}}{c_{3}}(1+2%
\frac{d_{3}}{c_{3}})(2\Gamma _{\alpha \beta }^{r}g_{3r}+\Gamma _{3\alpha
}^{r}g_{\beta r})\text{ \ \ \ }  \tag{3.8 }
\end{equation}
and
\begin{equation}
K_{\alpha }^{(2)}\equiv 2\frac{a_{3}}{c_{3}}\left[ 3p\frac{a_{3}}{c_{3}}%
\Gamma _{\alpha 3}^{r}g_{3r}-(1+2\frac{d_{3}}{c_{3}})R_{\alpha 3}\right]
\text{ \ \ .}  \tag{3.9 }
\end{equation}
Note that since the linear fractional transformation (with another
coefficient functions) will again be applied in respect to another cubic
equations (further it will be explained how they are obtained), everywhere
in (3.5 - 3.8) the coefficient functions $a_{3}(z)$, $b_{3}(z)$, $c_{3}(z)$
and $d_{3}(z)$ bear the indice ''$3$'' , to distinguish them from the
indices in the other linear-fractional tranformations. In terms of the new
variable $n_{3}=\widetilde{dX}^{3}$ the original cubic equation (2.14)\
acquires the form [10]
\begin{equation}
\widetilde{n}^{2}=\overline{P}_{1}(\widetilde{n})\text{ }m^{3}+\overline{P}%
_{2}(\widetilde{n})\text{ }m^{2}+\overline{P}_{3}(\widetilde{n})\text{ }m+%
\overline{P}_{4}(\widetilde{n})\text{ ,}  \tag{3.10 }
\end{equation}
where $\overline{P}_{1}(\widetilde{n})$ $,\overline{P}_{2}(\widetilde{n}),$ $%
\overline{P}_{3}(\widetilde{n})$ and $\overline{P}_{4}(\widetilde{n})$ are
complicated functions of the ratios $\frac{c_{3}}{d_{3}}$, $\frac{b_{3}}{%
d_{3}}$ and $A_{3},B_{3},C_{3}$ (but not of the ratio $\frac{a_{3}}{d_{3}}$,
which will become evident that is very important). The variable $m$ denotes
the ratio $\frac{a_{3}}{c_{3}}$ and the variable $\widetilde{n}$ is related
through the variable $n$ through the expresssion
\begin{equation}
\widetilde{n}=\sqrt{k_{3}}\sqrt{C_{3}}\left[ n+L_{1}^{(3)}\frac{B_{3}}{C_{3}}%
+L_{2}^{(3)}\right] \text{ \ \ \ \ \ ,}  \tag{3.11}
\end{equation}
where
\begin{equation}
k_{3}\equiv \frac{b_{3}}{d_{3}}\frac{c_{3}}{d_{3}}(\frac{c_{3}}{d_{3}}+2)%
\text{ \ \ \ \ \ ,}  \tag{3.12 }
\end{equation}
\begin{equation}
L_{1}^{(3)}\equiv \frac{1}{2}\frac{\frac{b_{3}}{d_{3}}}{\frac{c_{3}}{d_{3}}+2%
}\text{ \ \ ; \ \ \ }L_{2}^{(3)}\equiv \frac{1}{\frac{c_{3}}{d_{3}}+2}\text{
\ \ .}  \tag{3.13}
\end{equation}

The subscript ''$3$'' in $L_{1}^{(3)}$ and $L_{2}^{(3)}$ means that the
corresponding ratios in the R. H. S. also have the same subscript. Setting
up the coefficient functions $\overline{P}_{1}(\widetilde{n})$ $,\overline{P}%
_{2}(\widetilde{n}),$ $\overline{P}_{3}(\widetilde{n})$ equal to the number
coefficients $4,0,-g_{2},-g_{3}$ respectively, one can now parametrize the
resulting equation
\begin{equation}
\widetilde{n}^{2}=4m^{3}-g_{2}m-g_{3}\text{ }  \tag{3.14 }
\end{equation}
according to the standard prescription
\begin{equation}
\widetilde{n}=\rho ^{^{\prime }}(z)=\frac{d\rho }{dz}\text{ \ \ \ \ \ \ \ \
\ \ \ \ \ \ \ \ \ \ \ \ \ \ \ \ \ \ \ \ \ \ \ \ \ \ \ \ \ \ }\frac{a_{3}}{%
c_{3}}\equiv \text{\ }m=\rho (z)\text{\ \ \ \ .}  \tag{3.15 }
\end{equation}
Taking this into account, representing the linear-fractional transformation
(3.5) as (dividing by $\ c_{3}$)
\begin{equation}
dx^{3}=\frac{\frac{a_{3}}{c_{3}}\widetilde{dX}^{3}+\frac{b_{3}}{c_{3}}}{%
\widetilde{dX}^{3}+\frac{d_{3}}{c_{3}}}  \tag{3.16 }
\end{equation}
and combining expressions (3.11) for $\widetilde{n}$ and (3.16), one can
obtain the final formulae for $dX^{3}$ as a solution of the cubic algebraic
equation
\begin{equation}
dX^{3}=\frac{\frac{b_{3}}{c_{3}}+\frac{\rho (z)\rho ^{^{\prime }}(z)}{\sqrt{%
k_{3}}\sqrt{C_{3}}}-L_{1}^{(3)}\frac{B_{3}}{C_{3}}\rho (z)-L_{2}^{(3)}\rho
(z)}{\frac{d_{3}}{c_{3}}+\frac{\rho ^{^{\prime }}(z)}{\sqrt{k_{3}}\sqrt{C_{3}%
}}-L_{1}^{(3)}\frac{B_{3}}{C_{3}}-L_{2}^{(3)}}\text{ \ \ \ \ \ .}
\tag{3.
17}
\end{equation}
In order to be more precise, it should be mentioned that the identification
of the functions $\overline{P}_{1}(\widetilde{n})$ $,\overline{P}_{2}(%
\widetilde{n}),$ $\overline{P}_{3}(\widetilde{n})$ with the number
coefficients gives some additional equations [10], which in principle have
to be taken into account in the solution for $dx^{3}$. This has been
investigated to a certain extent in [10], and will be continued to be
investigated in the subsequent parts of this paper. It will further be shown
that a cubic equation with coefficient functions, that are quadratic (or
cubic) polynomials can be parametrized in \textbf{two ways}, and it should
be required that the two parametrizations are consistent with one another,
i.e. they would give one and the same result. Here in this paper the main
objective will be to show the dependence of the solutions on the Weierstrass
function and its derivative, which from the given coefficient functions in
Appendix B is evident that is very complicated. Since only the ratios $\frac{%
b}{d}$ and $\frac{c}{d}$ enter these additional relations, and not $\frac{a}{%
c}$ (which is related to the Weierstrass function), they do not affect the
solution in respect to $\rho (z)$ and $\rho ^{^{\prime }}(z)$.

Since $B_{3}$ and $C_{3}$ depend on $dX^{1}$ and $dX^{2}$, the solution
(3.17)\ for $dX^{3}$ shall be called the embedding solution for $dX^{1}$ and
$dX^{2}$.

\textbf{Step 3.} Let us now consider the two-dimensional cubic equation
(3.7). Following the same approach and finding the ''reduced'' cubic
algebraic equation for $dX^{1}$ only, it shall be proved that the solution
for $dX^{2}$ is the embedding solution for $dX^{1}$.

For the purpose, let us again write down eq. (3.7)\ in the form (3.2),
singling out the variable $dX^{2}$:
\begin{equation}
A_{2}(dX^{2})^{3}+B_{2}(dX^{2})^{2}+C_{2}(dX^{2})+G^{(1)}(dX^{1},g_{ij},%
\Gamma _{ij}^{k},R_{ik})\equiv 0\text{ \ \ \ \ ,}  \tag{3.18}
\end{equation}
where the coefficient functions $A_{2},B_{2},C_{2}$ and $G^{(1)}$ are the
following:
\begin{equation}
A_{2}\equiv 2p\Gamma _{22}^{r}g_{2r}\text{ \ \ ; \ \ \ \ \ \ \ \ \ \ \ }%
B_{2}\equiv K_{22}^{(1)}+2p[2\Gamma _{12}^{r}g_{2r}+\Gamma
_{22}^{r}g_{1r}]dX^{1}\text{\ \ \ \ ,\ }  \tag{3.19 }
\end{equation}
\begin{equation}
C_{2}\equiv 2p[\Gamma _{11}^{r}g_{2r}+2\Gamma
_{12}^{r}g_{1r})(dX^{1})^{2}+(K_{12}^{(1)}+K_{21}^{(1)})dX^{1}+K_{2}^{(2)}%
\text{ \ \ \ ,}  \tag{3.20 }
\end{equation}
\begin{equation}
G^{1}\equiv 2p\Gamma
_{11}^{r}g_{1r}(dX^{1})^{3}+K_{11}^{(1)}(dX^{1})^{2}+K_{1}^{(2)}dX^{1}+2p%
\rho ^{3}(z)\Gamma _{33}^{r}g_{3r}\text{ \ \ .}  \tag{3.21 }
\end{equation}
Note that the starting equation (3.7) has the same structure of the first
terms, if one makes the formal substitution $-R_{\alpha \beta }\rightarrow
K_{\alpha \beta }^{(1)}$ in the second terms, but eq. (3.7)\ has two more
additional terms $K_{1}^{(2)}dX^{1}+2p\rho ^{3}(z)\Gamma _{33}^{r}g_{3r}.$
Therefore, one might guess how the coefficient functions will look like just
by taking into account the above substitution and the contributions from the
additional terms. Revealing the general structure of the coefficient
functions might be particularly uiseful in higher dimensions, when one would
have a ''chain'' of cubic algebraic equations, each of which could be
written in the form (3.18) with the prescribed coefficient functions. This
is also an interesting and probably not too easy problem for future
investigation - is it possible to write down the coefficient functions for
arbitrary dimensions. Concretely for the three-dimensional case,
investigated here, $C_{2}$ in (3.20) can be obtained from $C_{3}$ in (3.4),
observing that there will be an additional contribution from the term $%
K_{\alpha }^{(2)}dX^{\alpha }$ for $\alpha =2$. Also, in writing down the
coefficient function (3.2)\ it has been accounted that as a result of the
previous parametrization $\frac{a_{3}}{c_{3}}=\rho (z)$ .

Since eq. (3.18)\ is of the same kind as eq. (3.2), for which we already
wrote down the solution, the expression for $dX^{2}$ will be of the same
kind as in formulae (3.17), but with the corresponding functions $%
A_{2},B_{2},C_{2}$ instead of $A_{3},B_{3},C_{3}$. Taking into account (3.19
- 3.20), the solution for $dX^{2}$ can be written as follows:
\begin{equation}
dX^{2}=\frac{\frac{1}{\sqrt{k_{2}}}\rho (z)\rho ^{^{\prime }}(z)\sqrt{C_{2}}%
+h_{1}(dX^{1})^{2}+h_{2}(dX^{1})+h_{3}}{\frac{1}{\sqrt{k_{2}}}\rho
^{^{\prime }}(z)\sqrt{C_{2}}+l_{1}(dX^{1})^{2}+l_{2}(dX^{1})+l_{3}}\text{ \
\ \ , }  \tag{3.22 }
\end{equation}
where $h_{1},h_{2},h_{3},l_{1},l_{2},l_{3}$ are expressions, depending on $%
\frac{b_{2}}{d_{2}},\frac{d_{2}}{c_{2}},\Gamma _{\alpha \beta }^{r}$ ($%
r=1,2,3$ ; $\alpha ,\beta =1,2$), $g_{\alpha \beta }$, $K_{12}^{(1)}$, $%
K_{21}^{(1)}$ and on the Weierstrass function. They will be presented in
Appendix $B$.

The representation of the solution for $dX^{2}$ in the form (3.22)\ shows
that it is an embedding solution of $dX^{1}$.

\textbf{Step 4.} It remains now to investigate the \textbf{one-dimensional}
cubic algebraic equation
\begin{equation}
A_{1}(dx^{1})^{3}+B_{1}(dx^{1})^{2}+C_{1}(dx^{1})+G^{(0)}(g_{ij},\Gamma
_{ij}^{k},R_{ik})\equiv 0\text{ \ \ \ \ ,}  \tag{3.23 }
\end{equation}
obtained from the two-dimensional cubic algebraic equation (3.18)\ after
applying the linear-fractional transformation
\begin{equation}
dx^{2}=\frac{a_{2}(z)\widetilde{dX}^{2}+b_{2}(z)}{c_{2}(z)\widetilde{dX}%
^{2}+d_{2}(z)}  \tag{3.24 }
\end{equation}
and setting up to zero the coefficient function before the highest (third)\
degree of $(dX^{2})^{3}$. Taking into account that as a result of the
previous parametrization $\frac{a_{2}}{c_{2}}=\rho (z)$ , the coefficient
functions $A_{1},B_{1},C_{1}$and $D_{1}$ are given in a form, not depending
on $dX^{2}$ and $dX^{3}$:
\begin{equation}
A_{1}\equiv 2p\Gamma _{11}^{r}g_{1r}\text{ \ \ \ ,}  \tag{3.25 }
\end{equation}
\begin{equation}
B_{1}\equiv F_{3}\rho (z)+K_{11}^{(1)}=2p(1+2\frac{d_{2}}{c_{2}})[2\Gamma
_{12}^{r}g_{1r}+\Gamma _{11}^{r}g_{2r}]\rho (z)+K_{11}^{(1)}\text{ \ \ ,}
\tag{3.26 }
\end{equation}
\begin{equation*}
C_{1}\equiv F_{1}\rho ^{2}(z)+F_{2}\rho (z)+K_{1}^{(2)}=2p[2\Gamma
_{12}^{r}g_{2r}+\Gamma _{22}^{r}g_{1r}]\rho ^{2}(z)+
\end{equation*}
\begin{equation}
+(1+2\frac{d_{2}}{c_{2}})(K_{12}^{(1)}+K_{21}^{(1)})\rho (z)+K_{1}^{(2)}%
\text{ \ \ \ \ \ ,}  \tag{3.27 }
\end{equation}
\begin{equation}
G^{0}\equiv 2p[\Gamma _{22}^{r}g_{2r}+\Gamma _{33}^{r}g_{3r}]\rho
^{3}(z)+K_{22}^{(1)}\rho ^{2}(z)\text{ \ \ \ \ .}  \tag{3.28 }
\end{equation}
The solution for $dX^{1}$ can again be written in the form (3.17), but with $%
\frac{b_{1}}{c_{1}}$, $\frac{d_{1}}{c_{1}}$, $L_{1}^{(1)}$, $L_{2}^{(1)}$, $%
k_{1}$and $B_{1},C_{1}$ instead of these expressions with the indice ''$3$''.

Taking into account formulaes (3.25 - 3.28)\ for $A_{1},B_{1}$ and $C_{1}$,
the final expression for $dx^{1}$ can be written as
\begin{equation}
dX^{1}=\frac{\frac{1}{\sqrt{k_{1}}}\rho (z)\rho ^{^{\prime }}(z)\sqrt{%
F_{1}\rho ^{2}+F_{2}\rho (z)+K_{1}^{(2)}}+f_{1}\rho ^{3}+f_{2}\rho
^{2}+f_{3}\rho +f_{4}}{\frac{1}{\sqrt{k_{1}}}\rho ^{^{\prime }}(z)\sqrt{%
F_{1}\rho ^{2}(z)+F_{2}\rho (z)+K_{1}^{(2)}}+\widetilde{g}_{1}\rho ^{2}(z)+%
\widetilde{g}_{2}\rho (z)+\widetilde{g}_{3}}\text{ \ \ \ \ \ \ ,}
\tag{3.29
}
\end{equation}
where $F_{1},F_{2},f_{1},f_{2},f_{3},f_{4},\widetilde{g}_{1},\widetilde{g}%
_{2}$ and $\widetilde{g}_{3}$ are functions (also to be given in Appendix $B$%
), depending on $g_{\alpha \beta }$, $\Gamma _{\alpha \beta }^{r}$ ($\alpha
,\beta =1,2$) and on the ratios $\frac{b_{1}}{c_{1}}$, $\frac{b_{1}}{d_{1}}$%
, $\frac{b_{2}}{d_{2}}$, $\frac{d_{1}}{c_{1}}$, $\frac{d_{2}}{c_{2}}$.

\section{\protect\bigskip A \ PROOF \ THAT \ THE \ SOLUTIONS \ $%
dX^{1},dX^{2} $ AND $dX^{3}$ ARE \ NOT \ ELLIPTIC \ FUNCTIONS}

\begin{proposition}
The expressions (3.22) for $dX^{2}$ and (3. 29) for $dX^{1}$\textbf{do not}
represent elliptic functions.
\end{proposition}

Proof: The proof is straightforward and will be based on assuming the
contrary. Let us first assume that $dX^{1}$ is an elliptic function. Then
from standard theory of elliptic functions it follows that $dX^{1}$(being an
elliptic function by assumption) can be represented as
\begin{equation}
dX^{1}=K_{1}(\rho )+\rho ^{^{\prime }}(z)K_{2}(\rho )\text{ \ \ \ \ ,}
\tag{4.1}
\end{equation}
where $K_{1}(\rho )$ and $K_{2}(\rho )$ depend on the Weierstrass function
only. For convenience one may denote the expressions outside the square root
in the nominator and denominator as
\begin{equation}
Z_{1}(\rho )\equiv f_{1}\rho ^{3}(z)+f_{2}\rho ^{2}(z)+f_{3}\rho (z)+f_{4}%
\text{ \ \ \ ,}  \tag{4.2}
\end{equation}
\begin{equation}
Z_{2}(\rho )\equiv g_{1}\rho ^{2}(z)+g_{2}\rho (z)+g_{3}\text{ \ \ \ \ \ \ .}
\tag{4. 3}
\end{equation}
Then, setting up equal the expressions (4.1)\ and (3.29) for $dX^{1}$, one
can express the function $K_{2}(\rho )$ as
\begin{equation}
K_{2}(\rho )=\frac{(1-K_{1}(\rho ))\rho \rho ^{^{\prime }}\sqrt{F_{1}\rho
^{2}+F_{2}\rho +K_{1}^{(2)}}+\sqrt{k_{1}}Z_{1}(\rho )-Z_{2}(\rho )K_{1}(\rho
)}{\rho ^{^{\prime }}\left[ \rho \sqrt{F_{1}\rho ^{2}+F_{2}\rho +K_{1}^{(2)}}%
+\sqrt{k_{1}}Z_{2}(\rho )\right] }\text{ \ \ \ \ .}  \tag{4.4}
\end{equation}
The R. H. S. of the above expression depends on the derivative $\rho
^{^{\prime }}$, while the L.H. S. depends on $\rho $ only. Therefore the
obtained contradiction is due to the initial assumption that $dX^{1}$ is an
elliptic function.

In order to prove that $dX^{2}$ is not an elliptic function, let us first
observe that (3.22) can be solved in respect to $dX^{1}$ as a cubic
algebraic equation by means of the Wiet formulae. For the purpose, first the
variable change
\begin{equation}
dX^{1}=dy^{1}-\frac{B_{1}}{3}  \tag{4.5}
\end{equation}
should be performed, after which equation (3.23) assumes the form
\begin{equation}
\overline{A}_{1}(dy^{1})^{3}+\overline{B}_{1}(dy^{1})+\overline{C}_{1}=0%
\text{ \ \ \ .}  \tag{4. 6}
\end{equation}
The solution of (4.6) is given by
\begin{equation*}
dy^{1}=\sqrt[3]{-\frac{\overline{C}_{1}}{2\overline{A}_{1}}-\sqrt[2]{\frac{1%
}{4}\left( \frac{\overline{C}_{1}}{\overline{A}_{1}}\right) ^{2}+\frac{1}{27}%
\left( \frac{\overline{B}_{1}}{\overline{A}_{1}}\right) ^{3}}}+
\end{equation*}
\begin{equation}
+\sqrt[3]{-\frac{\overline{C}_{1}}{2\overline{A}_{1}}+\sqrt[2]{\frac{1}{4}%
\left( \frac{\overline{C}_{1}}{\overline{A}_{1}}\right) ^{2}+\frac{1}{27}%
\left( \frac{\overline{B}_{1}}{\overline{A}_{1}}\right) ^{3}}}\text{ \ \ \ \
,}  \tag{4. 7}
\end{equation}
where $\overline{A}_{1},\overline{B}_{1}$ and $\overline{C}_{1}$ depend on
the coefficient functions $A_{1},B_{1}$ and $C_{1}$:
\begin{equation}
\overline{A}_{1}=A_{1}\text{ \ \ ; \ \ }\overline{B}_{1}=\frac{A_{1}B_{1}^{2}%
}{3}-\frac{2B_{1}^{2}}{3}+C_{1}\text{ \ \ \ ,}  \tag{4. 8}
\end{equation}
\begin{equation}
\overline{C}_{1}=-\frac{A_{1}B_{1}^{3}}{27}+\frac{B_{1}^{3}}{9}-\frac{%
C_{1}B_{1}}{3}+G^{0}\text{ \ \ .}  \tag{4. 9}
\end{equation}
The important conclusion from the above formulaes (4. 5 - 4. 9)\ is that $%
dX^{1}$ is an \textbf{irrational function}, depending on $A_{1},B_{1},C_{1}$
and therefore only on $g_{\alpha \beta \text{ \ }}$and $\Gamma _{\alpha
\beta }^{r}$ . Therefore
\begin{equation}
dX^{1}=O_{1}(g_{\alpha \beta },\Gamma _{\alpha \beta }^{r})\text{ \ \ .}
\tag{4. 10}
\end{equation}
In the same manner by assuming the contrary, it can be proved that $dX^{2}$
is not an elliptic function.

Now it can be seen that the two equations (3. 22)\ and (3. 29) for $dX^{2}$
and $dX^{1}$represent a complicated relation between $g_{\alpha \beta
},\Gamma _{\alpha \beta }^{r}$ and the Weierstrass function. Unfortunately,
this relation cannot be found explicitely, perhaps for some very simple
choices of the metric. At least, it can be pointed out how it can be found.
For example, from (4. 4) for $dx^{1}=O_{1}(g_{\alpha \beta },\Gamma _{\alpha
\beta }^{r})=K_{1}(\rho )$ and $K_{2}(\rho )=0$ the following relation can
be obtained
\begin{equation}
\int \frac{(1-O_{1})\rho \sqrt{F_{1}\rho ^{2}+F_{2}\rho +K_{1}^{(2)}}}{%
Z_{2}(\rho )O_{1}-\sqrt{k_{1}}Z_{1}(\rho )}d\rho =\int dz\text{ \ \ \ ,}
\tag{4. 11}
\end{equation}
from where, if the integration can be performed, it can be found
\begin{equation}
F(\rho )=z+const.  \tag{4. 12}
\end{equation}

Let us further take the second relation (3. 22) for $dX^{2}$ for the value
(4.10)\ of $dX^{1}=O_{1}(g_{\alpha \beta },\Gamma _{\alpha \beta }^{r}).$The
R. H. S. of (3. 22) can be represented as
\begin{equation}
dX^{2}=\frac{\frac{1}{\sqrt{k_{2}}}\rho \rho ^{^{\prime }}\sqrt{%
C_{2}(O_{1},g_{\alpha \beta },\Gamma _{\alpha \beta }^{r})}%
+K_{1}(O_{1},g_{\alpha \beta },\Gamma _{\alpha \beta }^{r})}{\frac{1}{\sqrt{%
k_{2}}}\rho ^{^{\prime }}\sqrt{C_{2}}+K_{2}(O_{1},g_{\alpha \beta },\Gamma
_{\alpha \beta }^{r})}\text{ \ \ .}  \tag{4. 13}
\end{equation}
At the same time, if expression (3. 29)\ for $dX^{1}$ is substituted into
the formulae (3. 22)\ for $dX^{2}$, one obtains another expression, where
the functions $\overline{C}_{2},\overline{K}_{1}$ and $\overline{K}_{2}$
will depend will depend additionally on a complicated manner on $\rho $ and $%
\rho ^{^{\prime }}$:
\begin{equation}
dX^{2}=\frac{\frac{1}{\sqrt{k_{2}}}\rho \rho ^{^{\prime }}\sqrt{\overline{C}%
_{2}(O_{1},g_{\alpha \beta },\Gamma _{\alpha \beta }^{r},\rho ,\rho
^{^{\prime }})}+\overline{K}_{1}(O_{1},g_{\alpha \beta },\Gamma _{\alpha
\beta }^{r},\rho ,\rho ^{^{\prime }})}{\frac{1}{\sqrt{k_{2}}}\rho ^{^{\prime
}}\sqrt{\overline{C}_{2}}+\overline{K}_{2}(O_{1},g_{\alpha \beta },\Gamma
_{\alpha \beta }^{r},\rho ,\rho ^{^{\prime }})}\text{ \ \ .}  \tag{4. 14}
\end{equation}
Therefore, setting up equal (4.13) and (4. 14), differentiating (4.12) by $z$
in order to get $\rho ^{^{\prime }}=\frac{1}{F^{^{\prime }}(\rho )}$ and
substituting $\rho ^{^{\prime }}$ in the R. H. S. of (4. 14), one obtains an
expression of the kind:
\begin{equation}
G(g_{\alpha \beta },\Gamma _{\alpha \beta }^{r},\rho )=0\text{ \ \ \ \ . }
\tag{4. 15}
\end{equation}

\section{\protect\bigskip COMPLEX \ COORDINATE \ DEPENDENCE \ OF \ THE \
METRIC \ TENSOR \ COMPONENTS \ FROM \ THE \ UNIFORMIZATION \ OF \ A \ CUBIC
\ ALGEBRAIC \ SURFACE}

\bigskip In section 2 it was asserted that the covariant and contravariant
metric tensor components constitute an algebraic variety of the solutions of
the ninth - degree algebraic equation (2.22).

In this Section it will be shown that the solutions (3. 17), (3. 22) and (3.
29) of the cubic algebraic equation (2.14) enable us to express not only the
contravariant metric tensor components through the Weierstrass function and
its derivatives, but the covariant components as well.

Let us write down for convenience the system of equations (3. 17), (3. 22)
and (3. 29) for $dX^{1}$, $dX^{2}$ and $dX^{3}$ as
\begin{equation}
dX^{1}(X^{1},X^{2},X^{3})=F_{1}(g_{ij}(\mathbf{X}),\Gamma _{ij}^{k}(\mathbf{X%
}),\rho (z),\rho ^{^{\prime }}(z))=F_{1}(\mathbf{X},z)\text{ \ \ \ ,}
\tag{5.1}
\end{equation}
\
\begin{equation}
dX^{2}(X^{1},X^{2},X^{3})=F_{2}(g_{ij}(\mathbf{X}),\Gamma _{ij}^{k}(\mathbf{X%
}),\rho (z),\rho ^{^{\prime }}(z))=F_{2}(\mathbf{X},z)\text{ \ \ \ ,}
\tag{5.2}
\end{equation}
\begin{equation}
dX^{3}(X^{1},X^{2},X^{3})=F_{3}(g_{ij}(\mathbf{X}),\Gamma _{ij}^{k}(\mathbf{X%
}),\rho (z),\rho ^{^{\prime }}(z))=F_{3}(\mathbf{X,}z)\text{ \ \ \ ,}
\tag{5.3}
\end{equation}
where the appearence of the complex coordinate $z$ is a natural consequence
of the uniformization procedure, applied in respect to each one of the cubic
equations from the ''embedded'' sequence of equations. Let us put this
statement in a more clear way. \textbf{Expressions (5.1 - 5.3) can be
treated as the uniformization functions for the multivariable cubic
algebraic equation (algebraic surface) (2.14).} In other words, it is
possible to find uniformization functions not only for 'parametrizable''
form $y^{2}=4x^{3}-g_{2}x-g_{3}$ of a two - dimensional cubic algebraic
equation, but also in higher -dimensional case. This opens new possibilities
to investigate multicomponent cubic algebraic equations (surfaces), even in
the framework of the standard arithmetical theory of algebraic equations.

Yet how the appearence of the additional complex coordinate $z$ on the R. H.
S. of (5.1 - 5.3) can be reconciled with the dependence of the differentials
on the L. H. S. only on the generalized coordinates $(X^{1},X^{2},X^{3})$
(and on the initial coordinates $x^{1},x^{2},x^{3}$ because of the mapping $%
X^{i}=X^{i}(x^{1},x^{2},x^{3})$)? The only reasonable assumption will be
that the initial coordinates depend also on the complex coordinate, i.e.
\begin{equation}
X^{1}\equiv X^{1}(x^{1}(z),x^{2}(z),x^{3}(z))=X^{1}(\mathbf{x,}\text{ }z)%
\text{ \ \ \ \ ,}  \tag{5.4}
\end{equation}
\begin{equation}
X^{2}\equiv X^{2}(x^{1}(z),x^{2}(z),x^{3}(z))=X^{2}(\mathbf{x,}\text{ }z)%
\text{ \ \ \ \ ,}  \tag{5.5}
\end{equation}
\begin{equation}
X^{1}\equiv X^{3}(x^{1}(z),x^{2}(z),x^{3}(z))=X^{3}(\mathbf{x,}\text{ }z)%
\text{ \ \ \ \ ,}  \tag{5.6}
\end{equation}

\bigskip and thus the generalized coordinates are defined on a Riemann
surface. Taking into account the important initial assumptions
\begin{equation}
d^{2}X^{1}=0=dF_{1}(\mathbf{X}(z),z)=\frac{dF_{1}}{dz}dz\text{ \ \ ,}
\tag{5.7}
\end{equation}
\begin{equation}
d^{2}X^{2}=0=dF_{2}(\mathbf{X}(z),z)=\frac{dF_{2}}{dz}dz\text{ \ \ ,}
\tag{5.8}
\end{equation}
\begin{equation}
d^{2}X^{3}=0=dF_{3}(\mathbf{X}(z),z)=\frac{dF_{3}}{dz}dz\text{ \ \ ,}
\tag{5.9}
\end{equation}
one easily gets the system of three inhomogeneous linear algebraic equations
in respect to the functions $\frac{\partial X^{1}}{\partial z}$, $\frac{%
\partial X^{2}}{\partial z}$ and $\frac{\partial X^{3}}{\partial z}$ :
\begin{equation}
\frac{\partial F_{1}}{\partial X^{1}}\frac{\partial X^{1}}{\partial z}+\frac{%
\partial F_{1}}{\partial X^{2}}\frac{\partial X^{2}}{\partial z}+\frac{%
\partial F_{1}}{\partial X^{3}}\frac{\partial X^{3}}{\partial z}+\frac{%
\partial F_{1}}{\partial z}=0\text{ \ \ \ ,}  \tag{5.10}
\end{equation}
\begin{equation}
\frac{\partial F_{2}}{\partial X^{1}}\frac{\partial X^{1}}{\partial z}+\frac{%
\partial F_{2}}{\partial X^{2}}\frac{\partial X^{2}}{\partial z}+\frac{%
\partial F_{2}}{\partial X^{3}}\frac{\partial X^{3}}{\partial z}+\frac{%
\partial F_{2}}{\partial z}=0\text{ \ \ \ ,}  \tag{5.11}
\end{equation}
\begin{equation}
\frac{\partial F_{3}}{\partial X^{1}}\frac{\partial X^{1}}{\partial z}+\frac{%
\partial F_{3}}{\partial X^{2}}\frac{\partial X^{2}}{\partial z}+\frac{%
\partial F_{3}}{\partial X^{3}}\frac{\partial X^{3}}{\partial z}+\frac{%
\partial F_{3}}{\partial z}=0\text{ \ \ \ .}  \tag{5.12}
\end{equation}
The solution of this algebraic system ($i,k=1,2,3$)
\begin{equation}
\frac{\partial X^{1}}{\partial z}=G_{1}\left( \frac{\partial F_{i}}{\partial
X^{k}}\right) =G_{1}\left( X^{1},X^{2},X^{3},z\right) \text{ \ \ \ \ \ ,}
\tag{5.13}
\end{equation}
\begin{equation}
\frac{\partial X^{2}}{\partial z}=G_{2}\left( \frac{\partial F_{i}}{\partial
X^{k}}\right) =G_{2}\left( X^{1},X^{2},X^{3},z\right) \text{ \ \ \ \ \ ,}
\tag{5.14}
\end{equation}
\begin{equation}
\frac{\partial X^{3}}{\partial z}=G_{3}\left( \frac{\partial F_{i}}{\partial
X^{k}}\right) =G_{3}\left( X^{1},X^{2},X^{3},z\right) \text{ \ \ \ \ \ }
\tag{5.15}
\end{equation}
represents a system of \textbf{three first - order nonlinear differential
equations. }A solution of this system can always be found in the form
\textbf{\ }
\begin{equation}
X^{1}=X^{1}(z)\text{ \ \ ; \ \ \ }X^{2}=X^{2}(z)\text{ \ \ ; \ \ \ \ }%
X^{3}=X^{3}(z)\text{ \ \ \ \ \ \ \ \ \ .}  \tag{5.16}
\end{equation}
and therefore, the metric tensor components will also depend on the complex
coordinate $z$, i.e. $g_{ij}=g_{ij}(\mathbf{X}(z))$. Note that since the
functions $\frac{\partial F_{i}}{\partial X^{k}}$ in the R. H. S. of (5.13)
- (5.15) depend on the Weierstrass function and its derivatives, it might
seem natural to write that \textbf{\ the solution of the above system of
nonlinear differential equations }$g_{ij}$\textbf{\ will also depend on the
Weierstrass function and its derivatives}
\begin{equation}
g_{ij}=g_{ij}(X^{1}(\rho (z),\rho ^{^{\prime }}(z),X^{2}(\rho (z),\rho
^{^{\prime }}(z),X^{3}(\rho (z),\rho ^{^{\prime }}(z))=g_{ij}(z)\text{ \ \ \
.}  \tag{5.17}
\end{equation}
\textbf{Note however that for the moment we do not have a theorem that the
solution of the system (5.13 - 5.15) will also contain the Weierstrass
function.}

Now let us mention the other equations, which will further be taken into
account.

The first set of equations simply means that the differentials $dF_{1}$, $%
dF_{2}$, $dF_{3}$, equal to the second differentials $d^{2}X^{1}$, $%
d^{2}X^{2}$, $d^{2}X^{3}$ can be taken in respect both to the generalized
coordinates $X^{1}$, $X^{2}$, $X^{3}$ and the initial coordinates $x^{1}$, $%
x^{2}$, $x^{3}$
\begin{equation}
d^{2}X^{1}=dF_{1}(\mathbf{X}(z),z)=dF_{1}(\mathbf{x}(z),z)\text{ \ \ \ \ ,}
\tag{5.18}
\end{equation}
\begin{equation}
d^{2}X^{2}=dF_{2}(\mathbf{X}(z),z)=dF_{2}(\mathbf{x}(z),z)\text{ \ \ \ \ ,}
\tag{5.19}
\end{equation}
\begin{equation}
d^{2}X^{3}=dF_{3}(\mathbf{X}(z),z)=dF_{3}(\mathbf{x}(z),z)\text{ \ \ \ \ .}
\tag{5.20}
\end{equation}
\ For completeness, the formal proof that the second order differentials can
be expressed in different sets of coordinates will be given in Appendix A.
Denoting further $\overset{.}{x}^{1}\equiv \frac{\partial x^{1}}{\partial z}$%
, $\overset{.}{x}^{2}\equiv \frac{\partial x^{2}}{\partial z}$ and $\overset{%
.}{x}^{3}\equiv \frac{\partial x^{3}}{\partial z}$, the above equalities
result again in a system of three inhomogeneous algebraic equations in
respect \ to $\overset{.}{X}^{1}\equiv \frac{\partial X^{1}}{\partial z}$, $%
\overset{.}{X}^{2}\equiv \frac{\partial X^{2}}{\partial z}$ and $\overset{.}{%
X}^{3}\equiv \frac{\partial X^{3}}{\partial z}$
\begin{equation}
\frac{\partial F_{1}}{\partial X^{1}}\frac{\partial X^{1}}{\partial z}+\frac{%
\partial F_{1}}{\partial X^{2}}\frac{\partial X^{2}}{\partial z}+\frac{%
\partial F_{1}}{\partial X^{3}}\frac{\partial X^{3}}{\partial z}=\frac{%
\partial F_{1}}{\partial x^{1}}\overset{.}{x}^{1}+\frac{\partial F_{1}}{%
\partial x^{2}}\overset{.}{x}^{2}+\frac{\partial F_{1}}{\partial x^{3}}%
\overset{.}{x}^{3}\text{ \ \ \ ,}  \tag{5.21}
\end{equation}
\begin{equation}
\frac{\partial F_{2}}{\partial X^{1}}\frac{\partial X^{1}}{\partial z}+\frac{%
\partial F_{2}}{\partial X^{2}}\frac{\partial X^{2}}{\partial z}+\frac{%
\partial F_{2}}{\partial X^{3}}\frac{\partial X^{3}}{\partial z}=\frac{%
\partial F_{2}}{\partial x^{1}}\overset{.}{x}^{1}+\frac{\partial F_{2}}{%
\partial x^{2}}\overset{.}{x}^{2}+\frac{\partial F_{2}}{\partial x^{3}}%
\overset{.}{x}^{3}\text{ \ \ \ ,}  \tag{5.22}
\end{equation}
\begin{equation}
\frac{\partial F_{3}}{\partial X^{1}}\frac{\partial X^{1}}{\partial z}+\frac{%
\partial F_{3}}{\partial X^{2}}\frac{\partial X^{2}}{\partial z}+\frac{%
\partial F_{3}}{\partial X^{3}}\frac{\partial X^{3}}{\partial z}=\frac{%
\partial F_{3}}{\partial x^{1}}\overset{.}{x}^{1}+\frac{\partial F_{3}}{%
\partial x^{2}}\overset{.}{x}^{2}+\frac{\partial F_{3}}{\partial x^{3}}%
\overset{.}{x}^{3}\text{ \ \ \ .}  \tag{5.23}
\end{equation}
Assuming for the moment that we know the functions $\overset{.}{x}^{1}$, $%
\overset{.}{x}^{2}$ and $\overset{.}{x}^{3}$, the solutions of this
algebraic system will give again another system of three first - order
nonlinear differential equations
\begin{equation}
\frac{\partial X^{1}}{\partial z}=H_{1}\left( X^{1},X^{2},X^{3},\text{ }z%
\text{ },\text{ }\overset{.}{x}^{1},\overset{.}{x}^{2},\overset{.}{x}%
^{3}\right) \text{ \ \ \ \ \ ,}  \tag{5.24}
\end{equation}
\begin{equation}
\frac{\partial X^{2}}{\partial z}=H_{2}\left( X^{1},X^{2},X^{3},\text{ }z%
\text{ },\text{ }\overset{.}{x}^{1},\overset{.}{x}^{2},\overset{.}{x}%
^{3}\right) \text{ \ \ \ \ \ ,}  \tag{5.25}
\end{equation}
\begin{equation}
\frac{\partial X^{3}}{\partial z}=H_{3}\left( X^{1},X^{2},X^{3},\text{ }z%
\text{ },\overset{.}{x}^{1},\overset{.}{x}^{2},\overset{.}{x}^{3}\right)
\text{ \ \ \ \ .\ }  \tag{5.26}
\end{equation}
Again, a solution of this system like the one in (5.16) can be obtained but
with account of the dependence on the additional variables $\overset{.}{x}%
^{1}$, $\overset{.}{x}^{2}$ and $\overset{.}{x}^{3}$. Let us also here note
that the solution (5.16) of the nonlinear system of equations (5.13 - 5. 15)
can be assumed to be dependent on some another complex variable $v$
\begin{equation}
X^{1}=X^{1}(z,v)\text{ \ \ \ ; \ \ }X^{1}=X^{1}(z,v)\text{ \ \ \ ; \ \ \ }%
X^{1}=X^{1}(z,v)\text{\ \ \ \ \ .}  \tag{5.27}
\end{equation}
The system of equations (5.18 - 5.20) ($i=1,2,3$)
\begin{equation}
d^{2}X^{i}=dF_{i}(\mathbf{X}(z,v),z)=dF_{1}(\mathbf{x}(z,v),z)\text{ \ \ \
,\ }  \tag{5.28}
\end{equation}
with account of the expressions (5.24 - 5.26) now will be rewritten as
\begin{equation*}
\frac{\partial F_{i}}{\partial X^{1}}\frac{\partial X^{1}}{\partial v}+\frac{%
\partial F_{i}}{\partial X^{2}}\frac{\partial X^{2}}{\partial v}+\frac{%
\partial F_{i}}{\partial X^{3}}\frac{\partial X^{3}}{\partial v}=\frac{%
\partial F_{i}}{\partial x^{1}}\overset{.}{x}^{1}+\frac{\partial F_{i}}{%
\partial x^{2}}\overset{.}{x}^{2}+\frac{\partial F_{i}}{\partial x^{3}}%
\overset{.}{x}^{3}+
\end{equation*}
\ \ \
\begin{equation}
+\frac{\partial F_{i}}{\partial x^{1}}x^{^{\prime }1}+\frac{\partial F_{i}}{%
\partial x^{2}}x^{^{\prime }2}+\frac{\partial F_{i}}{\partial x^{3}}%
x^{^{\prime }3}-\frac{\partial F_{i}}{\partial X^{1}}H_{1}-\frac{\partial
F_{i}}{\partial X^{2}}H_{2}-\frac{\partial F_{i}}{\partial X^{3}}H_{3}\text{
\ \ \ ,}  \tag{5. 29}
\end{equation}
where $x^{^{\prime }1},x^{^{\prime }2},x^{^{\prime }3}$ denote the
derivatives $\frac{\partial x^{1}}{\partial z},\frac{\partial x^{2}}{%
\partial z},\frac{\partial x^{3}}{\partial z}$. The same notation further
will be used in respect to the variables $\frac{\partial X^{1}}{\partial v},%
\frac{\partial X^{2}}{\partial v},\frac{\partial X^{3}}{\partial v}$ .
Similarly to (5. 24) - (5.26), the algebraic solution of this system of
equations can be represented as
\begin{equation}
\frac{\partial X^{i}}{\partial v}=K_{i}\left( \mathbf{X(}z,v\mathbf{),}\text{
}z\text{ },\overset{.}{\mathbf{x}}\text{ },\mathbf{x}^{^{\prime }}\right)
\text{ \ \ \ .}  \tag{5.30}
\end{equation}
Note that instead of (5.28), we could have also written
\begin{equation}
d^{2}X^{i}=dF_{i}(\mathbf{X}(z,v),z)=dF_{1}(\mathbf{x}(z),z,v)\text{ \ \ \
.\ }  \tag{5.31}
\end{equation}

\bigskip Further in section 7 it shall be proved why this would be
incorrect. The complete analysis of the system of equations, when both
system of coordinates depend on the two pair of complex variables $z$ and $v$
will be given in the following sections. For the moment we give just the
general qualitative motivations.

The other set of equations, which will further be used and which relates the
generalized coordinates $X^{i}$ to the initial ones $x^{i}$ is
\begin{equation}
d^{2}X^{i}=0=\frac{\partial ^{2}X^{i}}{\partial x^{k}\partial x^{r}}%
dx^{k}dx^{r}+\frac{\partial X^{i}}{\partial x^{k}}d^{2}x^{k}\text{ \ \ .}
\tag{5.32}
\end{equation}
For the moment we assume that the initial coordinates $x^{k}$ depend only on
the $z$ coordinate, and therefore
\begin{equation}
\frac{\partial ^{2}X^{i}}{\partial x^{k}\partial x^{r}}=\frac{\overset{..}{X}%
^{i}}{\overset{.}{x}^{k}\overset{.}{x}^{r}}-\overset{.}{X}^{i}\frac{\overset{%
..}{x}^{r}}{\overset{.}{x}^{k}\left( \overset{.}{x}^{r}\right) ^{2}}\text{ \
\ .}  \tag{5.33}
\end{equation}

Taking this into account, the system (5.32) in the $n-$dimensional case can
be written as
\begin{equation}
n^{2}\overset{..}{X}^{i}(dz)^{2}-(n-1)\overset{.}{X}^{i}\frac{\overset{..}{x}%
^{r}}{\overset{.}{x}^{r}}(dz)^{2}+n\overset{.}{X}^{i}d^{2}z=0\text{ \ \ \ .}
\tag{5.34}
\end{equation}
Introducing the notation
\begin{equation}
y^{r}=\frac{\partial }{\partial z}\left( \ln \overset{.}{x}^{r}\right) =%
\frac{\overset{..}{x}^{r}}{\overset{.}{x}^{r}}\text{ \ \ \ \ \ ,}  \tag{5.35}
\end{equation}
for the three-dimensional case the system (5.34) can be written as
\begin{equation}
2\overset{.}{X}^{i}(dz)^{2}(y^{1}+y^{2}+y^{3})=9\overset{..}{X}^{i}(dz)^{2}+3%
\overset{.}{X}^{i}d^{2}z\text{ \ \ \ \ .}  \tag{5.36}
\end{equation}
\bigskip\ Dividing the L. H. S. and the R. H. S. of the $i$-th and the $j$%
-th equation of this system, it can easily be obtained
\begin{equation}
\left( dz\right) ^{2}\left( \overset{..}{X}^{i}\overset{.}{X}^{j}-\overset{..%
}{X}^{j}\overset{.}{X}^{i}\right) =0\text{ \ \ \ \ ,}  \tag{5.37}
\end{equation}
which with account of the obvious relation
\begin{equation}
\frac{\partial }{\partial z}\left( \frac{\overset{.}{X}^{i}}{\overset{.}{X}%
^{j}}\right) =\frac{\left( \overset{..}{X}^{i}\overset{.}{X}^{j}-\overset{..%
}{X}^{j}\overset{.}{X}^{i}\right) }{\left( \overset{.}{X}^{j}\right) ^{2}}%
\text{ \ \ \ \ }  \tag{5.38}
\end{equation}
can be written as
\begin{equation}
\left( \overset{.}{X}^{j}\right) ^{2}\frac{\partial }{\partial z}\left(
\frac{\overset{.}{X}^{i}}{\overset{.}{X}^{j}}\right) =0\text{ \ \ \ \ .}
\tag{5.39}
\end{equation}
\ Neglecting the case when $\overset{.}{X}^{j}=0$, the above relation simply
means that $\overset{.}{X}^{2}$ and $\overset{.}{X}^{3}$should be
proportional to $\overset{.}{X}^{1}$%
\begin{equation}
\overset{.}{X}^{2}=C_{2}\overset{.}{X}^{1}\text{ \ ; \ }\overset{.}{X}%
^{3}=C_{3}\overset{.}{X}^{1}\text{ \ \ ,}  \tag{5.40}
\end{equation}
where $C_{2}$ and $C_{3}$ are constants. Indeed, it is easily seen that
(5.40) holds since
\begin{equation}
dX^{1}=\overset{.}{X}^{1}dz=F_{1}\text{ ; \ \ \ }dX^{2}=\overset{.}{X}%
^{2}dz=F_{2}\text{ ; \ \ \ }dX^{3}=\overset{.}{X}^{3}dz=F_{3}  \tag{5.41}
\end{equation}
and consequently
\begin{equation}
C_{2}=\frac{F_{2}}{F_{1}}\text{ \ \ \ \ ; \ \ \ \ }C_{3}=\frac{F_{3}}{F_{1}}%
\text{ \ \ \ .}  \tag{5.42}
\end{equation}

\section{\protect\bigskip FIRST-ORDER \ NONLINEAR \ DIFFERENTIAL \ EQUATIONS
\ FOR \ THE \ COMPLEX \ FUNCTIONS \ $x=x(z)$ AND \ $X=X(z)$}

\bigskip For the purpose, the two systems of algebraic equations (5.10) -
(5.12) and (5.21) - (5.23) will be used. If one substitutes the found
expressions (5.40) for $\overset{.}{X}^{2}$and \ $\overset{.}{X}^{3}$into
the system (5.10) - (5.12), one may treat it as an algebraic system of
equations in respect to the variables $\overset{.}{X}^{1},C_{2}$ and $C_{3}$%
. Introducing the notation
\begin{equation}
\{F_{i},F_{j}\}_{z,X^{k}}\equiv \frac{\partial F_{i}}{\partial z}\frac{%
\partial F_{j}}{\partial X^{k}}-\frac{\partial F_{i}}{\partial X^{k}}\frac{%
\partial F_{j}}{\partial z}\text{ \ \ }  \tag{6.1}
\end{equation}
for the \textbf{''one-dimensional'' Poisson bracket }$\{F_{i},F_{j}%
\}_{z,X^{k}}$ of the coordinates $z,X^{k}$ and also the notation
\begin{equation}
\{F_{1},F_{2},F_{3}\}_{z,\left[ X^{i},X^{j}\right] }\equiv
\{F_{1},F_{2}\}_{z,X^{i}}\{F_{1},F_{3}\}_{z,X^{j}}-\{F_{1},F_{2}\}_{z,X^{j}}%
\{F_{1},F_{3}\}_{z,X^{i}}\text{ \ \ \ ,}  \tag{6.2}
\end{equation}
one can show that the solution of the system of linear algebraic equations
(5.10)-(5.12) in respect to $\overset{.}{X}^{1},C_{2}$ and $C_{3}$
\begin{equation}
\frac{\partial F_{i}}{\partial X^{1}}\overset{.}{X}^{1}+\frac{\partial F_{i}%
}{\partial X^{2}}\overset{.}{X}^{2}+\frac{\partial F_{i}}{\partial X^{3}}%
\overset{.}{X}^{3}+\frac{\partial F_{i}}{\partial z}=0  \tag{6.3}
\end{equation}
can be represented in the following compact form
\begin{equation}
C_{2}=\frac{\{F_{1},F_{2},F_{3}\}_{z,\left[ X^{3},X^{1}\right] }}{%
\{F_{1},F_{2},F_{3}\}_{z,\left[ X^{2},X^{3}\right] }}\text{ \ \ \ ; \ \ \ \
\ }C_{3}=\frac{\{F_{1},F_{2},F_{3}\}_{z,\left[ X^{1},X^{2}\right] }}{%
\{F_{1},F_{2},F_{3}\}_{z,\left[ X^{2},X^{3}\right] }}\text{ \ \ \ ,\ }
\tag{6.4}
\end{equation}
\begin{equation}
\overset{.}{X}^{1}=-\frac{\frac{\partial F_{1}}{\partial z}%
\{F_{1},F_{2},F_{3}\}_{z,\left[ X^{2},X^{3}\right] }}{K_{1}}\text{ \ \ \ \ .}
\tag{6.5}
\end{equation}
In (6.5) the following notation has been introduced for $K_{i}$ ($i=1,2,3$)
\begin{equation}
K_{i}\equiv \frac{\partial F_{i}}{\partial X^{1}}\{F_{1},F_{2},F_{3}\}_{z,%
\left[ X^{2},X^{3}\right] }+\frac{\partial F_{i}}{\partial X^{2}}%
\{F_{1},F_{2},F_{3}\}_{z,\left[ X^{3},X^{1}\right] }+\frac{\partial F_{i}}{%
\partial X^{3}}\{F_{1},F_{2},F_{3}\}_{z,\left[ X^{1},X^{2}\right] }\text{ \
\ \ \ .}  \tag{6.6}
\end{equation}
The usefulness of introducing this notation will soon be understood.

Now let us rewrite the system of equations (5.21) - (5.23) in the form
\begin{equation}
\frac{\partial F_{i}}{\partial x^{1}}\overset{.}{x}^{1}+\frac{\partial F_{i}%
}{\partial x^{2}}\overset{.}{x}^{2}+\frac{\partial F_{i}}{\partial x^{3}}%
\overset{.}{x}^{3}=M_{i}\text{ \ \ \ \ ,}  \tag{6.7}
\end{equation}
where $M_{i}$ will be the notation for
\begin{equation}
M_{i}\equiv \frac{\partial F_{i}}{\partial X^{1}}\overset{.}{X}^{1}+\frac{%
\partial F_{i}}{\partial X^{2}}\overset{.}{X}^{2}+\frac{\partial F_{i}}{%
\partial X^{3}}\overset{.}{X}^{3}\text{ \ \ .}  \tag{6.8}
\end{equation}
Making use of the above formulaes (6.4 - 6.6) and also (5.40), $M_{i}$ can
be calculated to be
\begin{equation}
M_{i}=-\frac{\partial F_{1}}{\partial z}\frac{K_{i}}{K_{1}}\text{ \ \ \ .}
\tag{6.9}
\end{equation}
Further, the solutions of the linear algebraic system of equations (6.7) can
be represented in the form
\begin{equation}
\overset{.}{x}^{i}=S_{1}^{i}M_{1}+S_{2}^{i}M_{2}+S_{3}^{i}M_{3}\text{ \ \ \
\ }  \tag{6.10}
\end{equation}
where the functions $S_{1}^{i},S_{2}^{i}$ and $S_{3}^{i}$ depend on $\frac{%
\partial F_{i}}{\partial x^{k}}$ ($i,k=1,2,3$). Since $M_{1}$, $M_{2}$, $%
M_{3}$ according to (6.9) and (6.6) are proportional to $\frac{\partial F_{1}%
}{\partial z}\frac{\{F_{1},F_{2},F_{3}\}_{z,\left[ X^{k},X^{j}\right] }}{%
K_{1}}$ (where $(k,j)=(2,3)$, $(3,1)$ or $(1,2)$), the resulting solution
(6.10) will be of the kind
\begin{equation}
\overset{.}{x}^{i}=\frac{\overline{S}_{1}^{i}\left( \frac{\partial F_{1}}{%
\partial z}\right) ^{2}+\overline{S}_{2}^{i}\frac{\partial F_{2}}{\partial z}%
\frac{\partial F_{1}}{\partial z}+\overline{S}_{3}^{i}\frac{\partial F_{1}}{%
\partial z}\frac{\partial F_{3}}{\partial z}}{\overline{S}_{4}^{i}\left(
\frac{\partial F_{1}}{\partial z}\right) +\overline{S}_{5}^{i}\frac{\partial
F_{2}}{\partial z}+\overline{S}_{6}^{i}\frac{\partial F_{3}}{\partial z}}%
\text{ \ \ \ \ ,}  \tag{6.11}
\end{equation}
where the functions $\overline{S}_{1}^{i},\overline{S}_{2}^{i},....,%
\overline{S}_{6}^{i}$ depend both on $\frac{\partial F_{i}}{\partial X^{k}}$
and $\frac{\partial F_{i}}{\partial x^{k}}$ and consequently on all the
variables $x^{k},X^{k}$ and $z$. We have used also the following relation,
obtained after simple algebra with account of (6.1) and (6.2)
\begin{equation*}
\{F_{1},F_{2},F_{3}\}_{z,\left[ X^{i},X^{j}\right] }=\left( \frac{\partial
F_{1}}{\partial z}\right) ^{2}\{F_{2},F_{3}\}_{X^{i},X^{j}}+
\end{equation*}
\begin{equation}
+\frac{\partial F_{1}}{\partial z}\frac{\partial F_{2}}{\partial z}%
\{F_{3},F_{1}\}_{X^{i},X^{j}}+\frac{\partial F_{1}}{\partial z}\frac{%
\partial F_{3}}{\partial z}\{F_{1},F_{2}\}_{X^{i},X^{j}}\text{ \ \ .}
\tag{6.12}
\end{equation}
\textbf{Thus we have obtained the system of first order nonlinear
differential equations in respect to the initial coordinates }$%
x^{i}=x^{i}(z) $\textbf{.} An analogous system of nonlinear differential
equations is obtained for $X^{1}=X^{1}(z)$, $X^{2}=X^{2}(z)$ and $%
X^{3}=X^{3}(z)$ - for $X^{1}$ this is equation (6.5), and with aaccount of
(5.40) and expressions (6.4) for $C_{2}$ and $C_{3}$, the corresponding
equations for $X^{2}(z)$ and $X^{3}(z)$ are
\begin{equation}
\overset{.}{X}^{2}=-\frac{\frac{\partial F_{1}}{\partial z}%
\{F_{1},F_{2},F_{3}\}_{z,\left[ X^{3},X^{1}\right] }}{K_{1}}\text{ \ \ ; \ }%
\overset{.}{X}^{1}=-\frac{\frac{\partial F_{1}}{\partial z}%
\{F_{1},F_{2},F_{3}\}_{z,\left[ X^{1},X^{2}\right] }}{K_{1}}\text{ \ .}
\tag{6.13}
\end{equation}
Therefore, if the generalized coordinates $X^{1},X^{2},X^{3}$ are determined
as functions of the complex variable $z$ after solving the system (6.5),
(6.13), the obtained functions $X^{1}=X^{1}(z)$, $X^{2}=X^{2}(z)$ and $%
X^{3}=X^{3}(z)$ can be substituted into the R. H.\ S. of the system (6.11)
for $x^{1},x^{2}$ and $x^{3}$ and the corresponding solutions $%
x^{1}=x^{1}(z) $, $x^{2}=x^{2}(z)$ and $x^{3}=x^{3}(z)$ can be found.
Remember that we started from the assumption that only the generalized
coordinates $X^{1},X^{2},X^{3}$ satisfy the original cubic algebraic
equation and therefore equalities (5.41) are fulfilled. \textbf{%
Nevertheless, in spite of the fact that it had not been assumed that the
initial coordinates satisfy the equations }$dx^{i}=F_{i}$\textbf{, the
corresponding functions }$x^{i}=x^{i}(z)$\textbf{\ is possible to determine
from the system (6.11), the R. H. S. of which also confirms that }$%
dx^{i}\neq F_{i}$\textbf{. }

This conclusion is important since it shows that the two systems of
coordinates should not be treated on an equal footing. This refers of course
to the case of only one complex coordinate.

\section{\protect\bigskip IS IT \ POSSIBLE \ TO \ HAVE \ A \ TWO \ COMPLEX \
COORDINATE \ DEPENDENCE \ OF \ THE \ GENERALIZED \ COORDINATES \ $%
X^{i}=X^{i}\left( \mathbf{x(}z\right) ,z,v)?$ \ }

It will be proved below that such a case is impossible since it leads to an
impossibility to determine the dependance $X^{i}$ on the $v$ coordinate.

Under the above assumption $X^{i}=X^{i}\left( \mathbf{x(}z\right) ,z,v)$,
the first set of three equations
\begin{equation}
dX^{i}=\frac{\partial X^{i}}{\partial x^{1}}dx^{1}+\frac{\partial X^{i}}{%
\partial x^{2}}dx^{2}+\frac{\partial X^{i}}{\partial x^{3}}dx^{3}\text{ \ \ }
\tag{7.1}
\end{equation}
can be represented as
\begin{equation}
F_{i}=\overset{.}{X}^{i}\frac{\partial z}{\partial x^{1}}\overset{.}{x}%
^{1}dz+\overset{.}{X}^{i}\frac{\partial z}{\partial x^{2}}\overset{.}{x}%
^{2}dz+\overset{.}{X}^{i}\frac{\partial z}{\partial x^{3}}\overset{.}{x}%
^{3}dz=3\overset{.}{X}^{i}dz\text{ \ \ \ ,}  \tag{7.2}
\end{equation}
so again relations (5.40) - (5.42) $\ \overset{.}{X}^{2}=\frac{F_{2}}{F_{1}}%
\overset{.}{X}^{1}$, \ $\overset{.}{X}^{3}=\frac{F_{3}}{F_{1}}\overset{.}{X}%
^{1}$will hold.

The second set of equations
\begin{equation}
d^{2}X^{i}=dF_{i}(X,z)=dF_{i}(X(z,v),z)=dF_{i}(z,v)\text{ \ \ \ \ }
\tag{7.3}
\end{equation}
will express the equality of the differentials, expressed in terms of the
two different sets of coordinates $(X,z)$ and $(z,v)$
\begin{equation*}
d^{2}X^{i}=\frac{\partial F_{i}}{\partial X^{1}}dX^{1}+\frac{\partial F_{i}}{%
\partial X^{2}}dX^{2}+\frac{\partial F_{i}}{\partial X^{3}}dX^{3}+\frac{%
\partial F_{i}}{\partial z}dz=
\end{equation*}
\begin{equation*}
=\left[ \frac{\partial F_{i}}{\partial X^{1}}\overset{.}{X}^{1}+\frac{%
\partial F_{i}}{\partial X^{2}}\overset{.}{X}^{2}+\frac{\partial F_{i}}{%
\partial X^{3}}\overset{.}{X}^{3}+\frac{\partial F_{i}}{\partial z}\right]
dz+
\end{equation*}
\begin{equation}
+\left[ \frac{\partial F_{i}}{\partial X^{1}}dX^{^{\prime }1}+\frac{\partial
F_{i}}{\partial X^{2}}dX^{^{\prime }2}+\frac{\partial F_{i}}{\partial X^{3}}%
dX^{^{\prime }3}\right] dv\text{ \ \ \ .}  \tag{7.4}
\end{equation}
Taking into account that according to (5.42) $dX_{1}=F_{1}$, $dX_{2}=F_{2}$
and $dX_{3}=F_{3}$ and also the expressed from (5.42) differential
\begin{equation}
dz=\frac{1}{3}\frac{F_{1}}{\overset{.}{X}^{1}}\text{ \ \ ,}  \tag{7.5}
\end{equation}
one can obtain for (7.4)
\begin{equation*}
\left[ \frac{\partial F_{i}}{\partial X^{1}}dX^{^{\prime }1}+\frac{\partial
F_{i}}{\partial X^{2}}dX^{^{\prime }2}+\frac{\partial F_{i}}{\partial X^{3}}%
dX^{^{\prime }3}\right] dv=
\end{equation*}
\begin{equation}
=\frac{2}{3}\left[ \frac{\partial F_{i}}{\partial X^{1}}F_{1}+\frac{\partial
F_{i}}{\partial X^{2}}F_{2}+\frac{\partial F_{i}}{\partial X^{3}}F_{3}\right]
\text{ \ \ \ \ \ .}  \tag{7.6}
\end{equation}
Dividing the L. H. S. and the R. H. S. for different values of the indice $%
i=1,2,3$, one can obtain the following system of linear homogeneous
algebraic equations in respect to $X^{^{\prime }1},X^{^{\prime }2}$ and $%
X^{^{\prime }3}$
\begin{equation*}
\left( \frac{\partial F_{1}}{\partial X^{1}}Q_{2}-\frac{\partial F_{2}}{%
\partial X^{1}}Q_{1}\right) X^{^{\prime }1}+\left( \frac{\partial F_{1}}{%
\partial X^{2}}Q_{2}-\frac{\partial F_{2}}{\partial X^{2}}Q_{1}\right)
X^{^{\prime }2}+
\end{equation*}
\begin{equation}
+\left( \frac{\partial F_{1}}{\partial X^{3}}Q_{2}-\frac{\partial F_{2}}{%
\partial X^{3}}Q_{1}\right) X^{^{\prime }3}=0\text{ \ \ ,}  \tag{7.7}
\end{equation}
\begin{equation*}
\left( \frac{\partial F_{1}}{\partial X^{1}}Q_{3}-\frac{\partial F_{3}}{%
\partial X^{1}}Q_{1}\right) X^{^{\prime }1}+\left( \frac{\partial F_{1}}{%
\partial X^{2}}Q_{3}-\frac{\partial F_{3}}{\partial X^{2}}Q_{1}\right)
X^{^{\prime }2}+
\end{equation*}
\begin{equation}
+\left( \frac{\partial F_{1}}{\partial X^{3}}Q_{3}-\frac{\partial F_{3}}{%
\partial X^{3}}Q_{1}\right) X^{^{\prime }3}=0\text{ \ \ ,}  \tag{7.8}
\end{equation}
\begin{equation*}
\left( \frac{\partial F_{2}}{\partial X^{1}}Q_{3}-\frac{\partial F_{3}}{%
\partial X^{1}}Q_{2}\right) X^{^{\prime }1}+\left( \frac{\partial F_{2}}{%
\partial X^{2}}Q_{3}-\frac{\partial F_{3}}{\partial X^{2}}Q_{2}\right)
X^{^{\prime }2}+
\end{equation*}
\begin{equation}
+\left( \frac{\partial F_{2}}{\partial X^{3}}Q_{3}-\frac{\partial F_{2}}{%
\partial X^{3}}Q_{2}\right) X^{^{\prime }3}=0\text{ \ \ ,}  \tag{7.9}
\end{equation}
where $Q_{i}$ ($i=1,2,3$) denotes the expression
\begin{equation}
Q_{i}\equiv \frac{\partial F_{i}}{\partial X_{1}}F_{1}+\frac{\partial F_{i}}{%
\partial X^{2}}F_{2}+\frac{\partial F_{i}}{\partial X^{3}}F_{3}\text{ \ \ .}
\tag{7.10}
\end{equation}
Note that for the moment we have not yet used the equations $%
d^{2}X^{i}=dF_{i}=0$, from where $Q_{i}=0$. Then the system of equations
(7.7) - (7.9) would be identically satisfied for all $X^{^{\prime
}1},X^{^{\prime }2}$ and $X^{^{\prime }3}$ and it would be impossible to
express them as solutions of the system. But even without making use of the
equations $d^{2}X^{i}=dF_{i}=0$, the consistency (or inconsistensy) of the
system (7.7) - (7.9) is a necessary condition for the consistency (or
inconsistency) of the assumption about $X^{i}=X^{i}\left( \mathbf{x(}%
z\right) ,z,v)$.

Making use of the notation (6.1), the determinant of the system can be
written as
\begin{equation}
\begin{vmatrix}
\dsum\limits_{l_{1}\neq 1}F_{l_{1}}\{F_{1},F_{2}\}_{1,l_{1}} &
\dsum\limits_{l_{2}\neq 2}F_{l_{2}}\{F_{1},F_{2}\}_{2,l_{2}} &
\dsum\limits_{l_{3}\neq 3}F_{l_{3}}\{F_{1},F_{3}\}_{3,l_{3}} \\
\dsum\limits_{m_{1}\neq 1}F_{m_{1}}\{F_{1},F_{3}\}_{1,m_{1}} &
\dsum\limits_{m_{2}\neq 2}F_{m_{2}}\{F_{1},F_{3}\}_{2,m_{2}} &
\dsum\limits_{m_{3}\neq 3}F_{m_{3}}\{F_{1},F_{3}\}_{3,m_{3}} \\
\dsum\limits_{n_{1}\neq 1}F_{n_{1}}\{F_{2},F_{3}\}_{1,n_{1}} &
\dsum\limits_{n_{2}\neq 2}F_{n_{2}}\{F_{2},F_{3}\}_{2,n_{2}} &
\dsum\limits_{n_{3}\neq 3}F_{n_{3}}\{F_{2},F_{3}\}_{3,n_{3}}
\end{vmatrix}
\text{ \ \ \ \ ,}  \tag{7.11}
\end{equation}
where instead of $\{F_{i},F_{j}\}_{X^{k},X^{n_{k}}}$ we have written only $%
\{F_{i},F_{j}\}_{k,n_{k}}$.

The determinant is equal to
\begin{equation*}
\dsum\limits_{l_{k},m_{k},n_{k\text{ (}l_{k},m_{k},n_{k}\neq k\text{)}%
}}F_{l_{1}}F_{m_{2}}F_{n_{3}}\{F_{1},F_{2}\}_{1,l_{l1}}\{F_{1},F_{3}%
\}_{2,m_{2}}\{F_{2},F_{3}\}_{3,n_{3}}-
\end{equation*}
\begin{equation*}
-F_{l_{1}}F_{m_{3}}F_{n_{2}}\{F_{1},F_{2}\}_{1,l_{l1}}\{F_{1},F_{3}%
\}_{3,m_{3}}\{F_{2},F_{3}\}_{2,n_{2}}-
\end{equation*}
\begin{equation*}
-F_{l_{1}}F_{m1}F_{l_{2}}\{F_{1},F_{2}\}_{1,m_{l1}}\{F_{1},F_{2}\}_{2,l_{2}}%
\{F_{2},F_{3}\}_{3,n_{3}}+
\end{equation*}
\begin{equation*}
+F_{m_{1}}F_{l_{3}}F_{n_{2}}\{F_{1},F_{2}\}_{1,m_{l1}}\{F_{1},F_{2}%
\}_{3,l_{3}}\{F_{2},F_{3}\}_{2,n_{2}}
\end{equation*}
\begin{equation*}
+F_{n_{1}}F_{l_{2}}F_{m_{3}}\{F_{2},F_{3}\}_{1,n_{l1}}\{F_{1},F_{2}%
\}_{2,l_{2}}\{F_{1},F_{3}\}_{3,m_{3}}-
\end{equation*}
\begin{equation}
-F_{n_{1}}F_{l_{3}}F_{n_{2}}\{F_{2},F_{3}\}_{1,n_{1}}\{F_{1},F_{2}%
\}_{3,l_{3}}\{F_{1},F_{3}\}_{2,m_{2}}\text{ \ \ \ .}  \tag{7.12}
\end{equation}
It is seen from the above expression that the first and the third terms
cancel but the other terms remain. Indeed, the explicite calculation of the
determinant (7.11) gives the non-zero expression
\begin{equation}
\frac{\partial F_{2}}{\partial X^{1}}\{F_{1},F_{3}\}_{2,3}+\frac{\partial
F_{3}}{\partial X^{1}}\{F_{1},F_{2}\}_{1,2}\text{ \ \ .}  \tag{7.13}
\end{equation}
Since the determinant is non-zero, the system of linear \textbf{homogeneous}
algebraic equations does not have a solution and consequently the assumption
that $X^{i}=X^{i}\left( \mathbf{x(}z\right) ,z,v)$ turns out to be i\textbf{%
ncorrect}.

\section{\protect\bigskip COMPLEX \ STRUCTURE \ \ $X^{i}=X^{i}\left( \mathbf{%
x(}z,v\right) ,z) $ $\ $OF \ \ THE \ GENERALIZED \ COORDINATES \ AND \ OF \
THE \ METRIC \ TENSOR \ COMPONENTS}

\textbf{\bigskip }Now it shall be proved that the parametrization (5.1) -
(5.3) of the initially given cubic algebraic curve (surface) can be extended
to a parametrization in terms of a pair of \ complex coordinates $(z,v)$ and
thus a complex structure can be introduced. Of particular interest in view
of possible physical applications to theories with extra dimensions and
relation to $ADS$ \ theories, which will be discussed in the Conlusion, will
be the case of $v=\overline{z}$, when a pair of holomorphic -
-antiholomorphic variables can be introduced.

In principle a manifold may admit a complex structure [44], if it can be
covered with opened sets $U$ $_{1},V_{1},U_{2},V_{2}.....$, such that in any
intersection $U_{i}\cap V_{i}$ the associated transformations $z^{k^{\prime
}}=z^{k^{\prime }}(z_{i},v_{i})$ are complex (analytical) functions.

The investigated problem may be formulated as follows. Let (again) the
system of equations (5.1) - (5.3) is given, subjected to the additional
constraining equation $d^{2}X^{i}=0$. \textbf{Then the parametrization (5.1)
- (5.3) of the initially given cubic algebraic surface can \ be extended to
a parametrization by means of a pair of complex coordinates }$(z,v)$\textbf{%
\ in the following way}
\begin{equation}
dX^{i}(\mathbf{X})=F_{i}(\mathbf{X}(\mathbf{x}(z,v)),z)\text{ \ \ .}
\tag{8.1}
\end{equation}
Therefore, it should be proved that the same system of equations,
investigated in the previous sections, is not contradictable under the
assumption $X^{i}=X^{i}\left( \mathbf{x(}z,v\right) ,z)$.

The first set of equations to be used is similar to (7.3), but this time
expressing the equality of the differentials
\begin{equation}
dF_{i}(\mathbf{X}(z,v),z)\text{ }=dF_{i}(\mathbf{x}(z,v),z)\text{ \ \ \ \ ,}
\tag{8.2}
\end{equation}
written in terms of the coordinates $(\mathbf{X,}z)$ and $(\mathbf{x},z)$
\begin{equation*}
\left[ \frac{\partial F_{i}}{\partial X^{1}}\overset{.}{X}^{1}+\frac{%
\partial F_{i}}{\partial X^{2}}\overset{.}{X}^{2}+\frac{\partial F_{i}}{%
\partial X^{3}}\overset{.}{X}^{3}+\frac{\partial F_{i}}{\partial z}\right]
dz+
\end{equation*}
\begin{equation*}
+\left[ \frac{\partial F_{i}}{\partial X^{1}}X^{^{\prime }1}+\frac{\partial
F_{i}}{\partial X^{2}}X^{^{\prime }2}+\frac{\partial F_{i}}{\partial X^{3}}%
X^{^{\prime }3}\right] dv=
\end{equation*}
\begin{equation*}
=\left[ \frac{\partial F_{i}}{\partial x^{1}}\overset{.}{x}^{1}+\frac{%
\partial F_{i}}{\partial x^{2}}\overset{.}{x}^{2}+\frac{\partial F_{i}}{%
\partial x^{3}}\overset{.}{x}^{3}+\frac{\partial F_{i}}{\partial z}\right]
dz+
\end{equation*}
\begin{equation}
+\left[ \frac{\partial F_{i}}{\partial x^{1}}x^{^{\prime }1}+\frac{\partial
F_{i}}{\partial x^{2}}x^{^{\prime }2}+\frac{\partial F_{i}}{\partial x^{3}}%
x^{^{\prime }3}\right] dv\text{ \ \ \ \ .}  \tag{8.3}
\end{equation}
The second set of equations takes into account the fact that the second
differential $d^{2}X^{i}$ is zero, or equivalently
\begin{equation}
d^{2}X^{i}=dF_{i}(\mathbf{X}(z,v),z)=0\text{ \ \ \ \ \ ,}  \tag{8.4}
\end{equation}
where $dF_{i}(\mathbf{X}(z,v),z)$ is given by the L. H. S. of equation (8.3).

The third set of equations is
\begin{equation}
dX^{i}=F^{i}=\frac{\partial X_{i}}{\partial z}dz+\frac{\partial X_{i}}{%
\partial v}dv\text{ \ \ \ \ .}  \tag{8.5}
\end{equation}
Let us now introduce the notations
\begin{equation}
M_{i}(X,z)\equiv \frac{\partial F_{i}}{\partial X^{k}}\overset{.}{X}^{k}%
\text{ \ \ ; \ \ \ \ }M_{i}(x,z)\equiv \frac{\partial F_{i}}{\partial x^{k}}%
\overset{.}{x}^{k}\text{ \ \ ,}  \tag{8.6}
\end{equation}
\begin{equation}
M_{i}(X,v)\equiv \frac{\partial F_{i}}{\partial X^{k}}X^{^{\prime }k}\text{
\ \ ; \ \ \ \ }M_{i}(x,v)\equiv \frac{\partial F_{i}}{\partial x^{k}}%
x^{^{\prime }k}\text{ \ \ ,}  \tag{8.7}
\end{equation}
which will allow us to write down the the first and the second set of
equations (7.3) - (7.4) in the following compact form
\begin{equation}
\left[ M_{i}(X,z)-M_{i}(x,z)\right] dz+\left[ M_{i}(X,v)-M_{i}(x,v)\right]
dv=0\text{ \ \ ,}  \tag{8.8}
\end{equation}
\
\begin{equation}
\left[ M_{i}(X,z)+\frac{\partial F_{i}}{\partial z}\right] dz+M_{i}(X,v)dv=0%
\text{ \ \ .}  \tag{8.9}
\end{equation}
Expressing $\frac{\partial X^{i}}{\partial v}dv$ from (8.5), it can easily
be proved that
\begin{equation}
M_{i}(X,v)dv=\frac{\partial F_{i}}{\partial X^{k}}X^{^{\prime }k}dv=-\frac{%
\partial F_{i}}{\partial X^{k}}\overset{.}{X}^{k}dz+\frac{\partial F_{i}}{%
\partial X^{k}}F_{k}\text{ \ \ \ \ ,}  \tag{8.10}
\end{equation}
where the last term is zero due to the fulfillment of the second set of
equations (8.5). Consequently, from (8.10) it follows
\begin{equation}
M_{i}(X,z)dz+M_{i}(X,v)dv=dF_{i}(z,v)=dF_{i}(\mathbf{X}(z,v),z)=0\text{ \ \
\ .}  \tag{8.11}
\end{equation}
Additionally, if (8.11) is substracted from (8.8) and (8.9), one easily
obtains
\begin{equation}
M_{i}(x,z)dz+M_{i}(x,v)dv=dF_{i}(z,v)=dF_{i}(\mathbf{x}(z,v),z)=0\text{ \ \
\ ,}  \tag{8.12}
\end{equation}
\begin{equation}
\frac{\partial F_{i}}{\partial z}dz=0\text{ \ \ \ }\Rightarrow \text{ \ \ }%
\frac{\partial F_{i}}{\partial v}dv=0\text{ \ \ \ \ .}  \tag{8.13}
\end{equation}
In other words, if the differential $dF_{i}(\mathbf{X}(z,v),z)$ is zero in
terms of the coordinates $(\mathbf{X,}z\mathbf{)}$, then it necessarily
should be zero in the coordinates $(\mathbf{x},z)$. But in the spirit of the
discussion at the end of section 6, this does not mean that if $dX^{i}=F_{i}
$, then the same should hold also for the initial coordinates $x^{i}$, i.e. $%
dx^{i}\neq F_{i}$. Indeed, we can find
\begin{equation*}
M_{i}(x,z)\equiv \frac{\partial F_{i}}{\partial x^{k}}\overset{.}{x}^{k}=%
\frac{\partial F_{i}}{\partial X^{l}}\frac{\partial X^{l}}{\partial x^{k}}%
\overset{.}{x}^{k}=
\end{equation*}
\begin{equation}
=\frac{\partial F_{i}}{\partial X^{l}}\left[ \frac{\overset{.}{X}^{l}}{%
\overset{.}{x}^{k}}+\frac{X^{^{\prime }l}}{x^{^{\prime }k}}\right] \overset{.%
}{x}^{k}=M_{i}(X,z)+M_{i}(X,v)\frac{\overset{.}{x}^{k}}{x^{^{\prime }k}}%
\text{ \ \ \ \ .}  \tag{8.14}
\end{equation}
Similarly
\begin{equation}
M_{i}(x,v)=M_{i}(X,v)+M_{i}(X,z)\frac{x^{^{\prime }k}}{\overset{.}{x}^{k}}%
\text{ \ \ .}  \tag{8.15}
\end{equation}
If the above two expressions are substituted into (8.12), and (8.11) is
taken into account, one can obtain
\begin{equation}
M_{i}(X,v)\frac{\overset{.}{x}^{k}}{x^{^{\prime }k}}dz+M_{i}(X,z)\frac{%
x^{^{\prime }k}}{\overset{.}{x}^{k}}dv=0\text{ \ \ \ .}  \tag{8.16}
\end{equation}
Additionally, we have
\begin{equation}
M_{i}(X,v)=M_{i}(X,z)\frac{x^{^{\prime }k}}{\overset{.}{x}^{k}}\text{ \ \ \
; \ \ \ \ \ }dv=-\frac{M_{i}(x,z)}{M_{i}(x,v)}dz\text{ \ \ \ \ .}  \tag{8.17}
\end{equation}
Therefore, the following equation in partial derivatives in respect to $%
F_{i}=F_{i}(x^{l})$ can be derived
\begin{equation}
\frac{\partial F_{i}}{\partial x^{l}}x^{^{\prime }l}\frac{\overset{.}{x}^{k}%
}{x^{^{\prime }k}}\frac{x^{^{\prime }m}}{\overset{.}{x}^{m}}-\frac{\partial
F_{i}}{\partial x^{l}}\overset{.}{x}^{l}\frac{x^{^{\prime }k}}{\overset{.}{x}%
^{k}}=0\text{ \ \ \ \ .}  \tag{8.18}
\end{equation}
As partial case, for an arbitrary $\frac{\partial F_{i}}{\partial x^{l}}$ ($%
l=l_{1}$ is fixed; there is an independent summation along the indices $k$
and $m$), the above equation holds when the following complicated nonlinear
equation in partial derivatives is satisfied
\begin{equation}
x^{^{\prime }l_{1}}\frac{\overset{.}{x}^{k}}{x^{^{\prime }k}}\frac{%
x^{^{\prime }m}}{\overset{.}{x}^{m}}-\overset{.}{x}^{l_{1}}\frac{x^{^{\prime
}k}}{\overset{.}{x}^{k}}=0\text{ \ \ \ \ .}  \tag{8.19}
\end{equation}
\ Note that the equation takes a different form if one assumes that
\begin{equation}
dx^{i}=F_{i}\text{ \ \ \ }\Rightarrow \text{ \ }dv=\frac{F_{i}}{x^{^{\prime
}i}}-\frac{\overset{.}{x}^{i}}{x^{^{\prime }i}}dz\text{ \ \ \ .}  \tag{8.20}
\end{equation}
Then instead of (8.19), the obtained equation will be of the form
\begin{equation}
\frac{\overset{.}{x}^{k}}{x^{^{\prime }k}}\left( \frac{x^{^{\prime }m}}{%
\overset{.}{x}^{m}}\right) ^{2}-\frac{x^{^{\prime }k}}{\overset{.}{x}^{k}}=0%
\text{ \ \ \ \ .}  \tag{8.21}
\end{equation}
In fact, even a stronger statement may be proved, clearly showing that from $%
dF_{i}(X,z)=dF_{i}(x,z)=0$ and $dX_{i}=F_{i}$ it does not follow that $%
dx^{i}=F_{i}$. If expression (8.17) for $M_{i}(X,v)=M_{i}(X,z)\frac{%
x^{^{\prime }k}}{\overset{.}{x}^{k}}$ is substituted into (8.11), one
obtains
\begin{equation}
M_{i}(X,z)\left[ dz+\frac{x^{^{\prime }k}}{\overset{.}{x}^{k}}dv\right] =0
\tag{8.22}
\end{equation}
and since $M_{i}(X,z)\neq 0$ (and if $\frac{\partial F_{i}}{\partial X^{l}}%
\neq 0$), it follows
\begin{equation}
dx^{k}=\overset{.}{x}^{k}dz+x^{^{\prime }k}dv=0\text{ \ \ \ .}  \tag{8.23}
\end{equation}

\section{\protect\bigskip ANALYSIS \ OF \ THE \ FOURTH \ AND \ THE \ FIFTH \
SET \ OF \ EQUATIONS \ FOR \ THE \ PREVIOUS \ CASE \ $X^{i}=X^{i}(\mathbf{x}%
(z,v),z)$}

\bigskip The \textbf{fourth set} of equations, which will be considered is
\begin{equation*}
dX^{k}=\frac{\partial X^{k}}{\partial x^{1}}dx^{1}+\frac{\partial X^{k}}{%
\partial x^{2}}dx^{2}+\frac{\partial X^{k}}{\partial x^{3}}dx^{3}=
\end{equation*}
\begin{equation}
=3\overset{.}{X}^{k}dz+X^{^{\prime }k}\frac{\overset{.}{x}^{m}}{x^{^{\prime
}m}}dz+\overset{.}{X}^{k}\frac{x^{^{\prime }m}}{\overset{.}{x}^{m}}%
dv+X^{^{\prime }k}dv\text{ \ \ \ \ .}  \tag{9.1}
\end{equation}
If multiplied by $\frac{\partial F_{i}}{\partial X^{k}}dzdv$ and also
relation (8.11) $M_{i}(X,z)dz+M_{i}(X,v)dv=0$ is taken into account, the
fourth set of equations can be written as
\begin{equation}
\frac{\partial F_{i}}{\partial X^{k}}F_{k}dv=M_{i}(X,z)dz\left[ \frac{%
x^{^{\prime }m}}{\overset{.}{x}^{m}}\left( dv\right) ^{2}-\frac{\overset{.}{x%
}^{m}}{x^{^{\prime }m}}\left( dz\right) ^{2}\right] \text{ \ \ \ \ .}
\tag{9.2}
\end{equation}
The \textbf{fifth set} of equations is
\begin{equation}
d^{2}X^{k}=0=\frac{\partial ^{2}X^{k}}{\partial x^{m}\partial x^{n}}%
dx^{m}dx^{n}+\frac{\partial X^{k}}{\partial x^{m}}d^{2}x^{m}\text{ \ \ ,}
\tag{9.3}
\end{equation}
where
\begin{equation}
\frac{\partial ^{2}X^{k}}{\partial x^{m}\partial x^{n}}=\frac{1}{2}\left[
\frac{\partial }{\partial z}\left( \frac{\partial X^{k}}{\partial x^{m}}%
\right) \frac{\partial z}{\partial x^{n}}+\frac{\partial }{\partial v}\left(
\frac{\partial X^{k}}{\partial x^{m}}\right) \frac{\partial v}{\partial x^{n}%
}\right] =  \tag{9.4}
\end{equation}
\begin{equation*}
=\frac{1}{2}\{\frac{\overset{..}{X}^{k}}{\overset{.}{x}^{m}\overset{.}{x}^{n}%
}+\frac{X^{^{\prime \prime }k}}{x^{^{\prime }m}x^{^{\prime }n}}+\overset{.}{X%
}^{^{\prime }k}\left[ \frac{1}{\overset{.}{x}^{m}x^{^{\prime }n}}+\frac{1}{%
\overset{.}{x}^{n}x^{^{\prime }m}}\right] -
\end{equation*}
\begin{equation}
-\overset{.}{X}^{k}\left[ \frac{\overset{.}{x}^{^{\prime }m}}{x^{^{\prime }n}%
}+\frac{\overset{..}{x}^{m}}{\overset{.}{x}^{n}}\right] -\frac{X^{^{\prime
}k}}{\left( x^{^{\prime }m}\right) ^{2}}\left[ \frac{\overset{.}{x}%
^{^{\prime }m}}{\overset{.}{x}^{n}}+\frac{x^{^{\prime \prime }m}}{%
x^{^{\prime }n}}\right] \}\text{ \ \ \ \ ,}  \tag{9.5}
\end{equation}
\begin{equation}
\frac{\partial X^{k}}{\partial x^{m}}=\frac{\overset{.}{X}^{k}}{\overset{.}{x%
}^{m}}+\frac{X^{^{\prime }k}}{x^{^{\prime }m}}\text{ \ \ \ \ ,}  \tag{9.6 }
\end{equation}
\begin{equation}
dx^{m}dx^{n}=\overset{.}{x}^{m}\overset{.}{x}^{n}(dz)^{2}+(x^{^{\prime }m}%
\overset{.}{x}^{n}+\text{ }\overset{.}{x}^{m}x^{^{\prime
}n})dzdv+x^{^{\prime }m}x^{^{\prime }n}(dv)^{2}\text{ \ \ \ ,}  \tag{9.7 }
\end{equation}
\begin{equation}
d^{2}x^{m}=\overset{..}{x}^{m}(dz)^{2}+x^{^{\prime \prime }m}(dv)^{2}+2%
\overset{.}{x}^{^{\prime }m}dxdv+\overset{.}{x}^{m}d^{2}z+x^{^{\prime
}m}d^{2}v\text{ \ \ .}  \tag{9.8 }
\end{equation}
Our goal further will be to show whether the fifthe equation (9.3)
constitutes a separate equation or whethetr it follows from the preceeding
four ones.

For the purpose, let us multiply both sides of the fifth equation by $\frac{%
\partial F_{i}}{\partial X^{k}}$ and see which are the terms, containing the
second differentials $d^{2}z$ and $d^{2}v$
\begin{equation*}
\left( \frac{\partial F_{i}}{\partial X^{k}}\frac{\overset{.}{X}^{k}}{%
\overset{.}{x}^{m}}+\frac{\partial F_{i}}{\partial X^{k}}\frac{X^{^{\prime
}k}}{x^{^{\prime }m}}\right) \left( \overset{.}{x}^{m}d^{2}z+x^{^{\prime
}m}d^{2}v\right) =
\end{equation*}
\begin{equation}
=M_{i}(x,z)d^{2}z+M_{i}(x,v)d^{2}v\text{ \ \ .}  \tag{9.9 }
\end{equation}
In (9.9) we have used relations (8.14) and (8.15) for $M_{i}(x,z)$ and $%
M_{i}(x,v)$. But we may note that the obtained term in (9.9) can be found
from the relation (8.12) $M_{i}(x,z)dz+M_{i}(x,v)dv=0$, if it is
differentiated by $z$ and $v$ and the resulting equations are summed up.
Therefore
\begin{equation*}
M_{i}(x,z)d^{2}z+M_{i}(x,v)d^{2}v=-M_{i}(x,z)(dz)^{2}-M_{i}(x,v)(dv)^{2}-
\end{equation*}
\begin{equation}
-(\overset{.}{M}_{i}(x,v)+M_{i}^{^{\prime }}(x,z))dxdv\text{ \ \ \ . }
\tag{9.10 }
\end{equation}
The derivatives $\overset{.}{M}_{i}(x,v)$ and $M_{i}^{^{\prime }}(x,z)$ can
be found also from the already used expressions (8.14) and (8.15)
\begin{equation*}
M_{i}^{^{\prime }}(x,z)=M_{i}^{^{\prime }}(X,z)+\frac{\overset{.}{x}^{m}}{%
x^{^{\prime }m}}M_{i}^{^{\prime }}(X,v)+
\end{equation*}
\begin{equation}
+M_{i}(X,v)\frac{\left[ \overset{.}{x}^{^{\prime }m}x^{^{\prime }m}-\overset{%
.}{x}^{m}x^{^{\prime \prime }m}\right] }{\left( x^{^{\prime }m}\right) ^{2}}%
\text{ \ \ \ \ ,}  \tag{9.11 }
\end{equation}
\begin{equation*}
\overset{.}{M}_{i}(x,v)=\overset{.}{M}_{i}(X,v)+\frac{x^{^{\prime }m}}{%
\overset{.}{x}^{m}}\overset{.}{M}_{i}(X,z)+
\end{equation*}
\begin{equation}
+M_{i}(X,z)\frac{\left[ \overset{.}{x}^{^{\prime }m}\overset{.}{x}%
^{m}-x^{^{\prime }m}\overset{.}{x}^{m}\right] }{\left( x^{^{\prime
}m}\right) ^{2}}\text{ \ \ \ \ .}  \tag{9.12  }
\end{equation}

\bigskip Making use of all the expressions (9.9) - (9.12), the following
expression for the fifth equation (9.3), multiplied by $\frac{\partial F_{i}%
}{\partial X^{k}}$, can be obtained:
\begin{equation*}
(dz)^{2}[-2M_{i}(X,z)+M_{i}(X,z)\frac{\overset{..}{x}^{m}}{\overset{.}{x}^{m}%
}+\frac{\partial F_{i}}{\partial X^{k}}\frac{\partial ^{2}X^{k}}{\partial
x^{m}\partial x^{n}}\overset{.}{x}^{m}\overset{.}{x}^{n}]+
\end{equation*}
\begin{equation*}
+(dv)^{2}[-2\frac{x^{^{\prime }m}}{\overset{.}{x}^{m}}M_{i}(X,z)+M_{i}(X,z)%
\frac{x^{^{\prime \prime }m}}{\overset{.}{x}^{m}}+\frac{\partial F_{i}}{%
\partial X^{k}}\frac{\partial ^{2}X^{k}}{\partial x^{m}\partial x^{n}}%
x^{^{\prime }m}x^{^{\prime }n}]+
\end{equation*}
\begin{equation*}
+dzdv[2M_{i}(X,z)\frac{\overset{.}{x}^{^{\prime }m}}{\overset{.}{x}^{m}}+2%
\frac{\partial F_{i}}{\partial X^{k}}\frac{\partial ^{2}X^{k}}{\partial
x^{m}\partial x^{n}}\overset{.}{x}^{m}x^{^{\prime }n}-2M_{i}^{^{\prime
}}(X,z)-
\end{equation*}
\begin{equation}
-2\overset{.}{M_{i}}(X,z)\frac{x^{^{\prime }m}}{\overset{.}{x}^{m}}%
+2M_{i}(X,z)\frac{\left[ x^{^{\prime }m}\overset{..}{x}^{m}-\overset{.}{x}%
^{^{\prime }m}\overset{.}{x}^{m}\right] }{\left( \overset{.}{x}^{m}\right)
^{2}}]=0\text{ \ \ .}  \tag{9.13 }
\end{equation}
The last two terms $\frac{\partial F_{i}}{\partial X^{k}}\frac{\partial
^{2}X^{k}}{\partial x^{m}\partial x^{n}}\overset{.}{x}^{m}\overset{.}{x}^{n}$%
and $\frac{\partial F_{i}}{\partial X^{k}}\frac{\partial ^{2}X^{k}}{\partial
x^{m}\partial x^{n}}x^{^{\prime }m}x^{^{\prime }n}$ in the first two square
brackets can be found as follows: First, the derivatives $\overset{.}{M}%
_{i}(X,v)$ and $M_{i}^{^{\prime }}(X,v)$ can be expressed from the relation
(8.17) $M_{i}(X,v)=M_{i}(X,z)\frac{x^{^{\prime }k}}{\overset{.}{x}^{k}}$:
\begin{equation}
\overset{.}{M_{i}}(X,v)=\overset{.}{M_{i}}(X,z)\frac{x^{^{\prime }m}}{%
\overset{.}{x}^{m}}+M_{i}(X,z)\frac{\left[ \overset{.}{x}^{^{\prime }m}%
\overset{.}{x}^{m}-x^{^{\prime }m}\overset{..}{x}^{m}\right] }{\left(
\overset{.}{x}^{m}\right) ^{2}}\text{ \ \ \ ,}  \tag{9.14 }
\end{equation}
\begin{equation}
M_{i}^{^{\prime }}(X,v)=M_{i}^{^{\prime }}(X,z)\frac{x^{^{\prime }m}}{%
\overset{.}{x}^{m}}+M_{i}(X,z)\frac{\left[ x^{^{\prime \prime }m}\overset{.}{%
x}^{m}-x^{^{\prime }m}\overset{.}{x}^{^{\prime }m}\right] }{\left( \overset{.%
}{x}^{m}\right) ^{2}}\text{ \ \ \ .}  \tag{9.15 }
\end{equation}
But on the other hand, the same derivatives can be found by using the
defining expressions (8.6 - 8.7)
\begin{equation*}
\overset{.}{M}_{i}(X,v)=\frac{\partial ^{2}F_{i}}{\partial X^{k}\partial
X^{l}}\frac{\partial X^{l}}{\partial x^{m}}\frac{\partial X^{k}}{\partial
x^{n}}\overset{.}{x}^{m}x^{^{\prime }n}+\frac{\partial F_{i}}{\partial X^{k}}%
\frac{\partial ^{2}X^{k}}{\partial x^{m}\partial x^{n}}\overset{.}{x}%
^{n}x^{^{\prime }m}+
\end{equation*}
\begin{equation}
+M_{i}(X,z)\frac{x^{^{\prime }m}\overset{.}{x}^{n}}{\overset{.}{x}^{m}}%
+M_{i}(X,z)\overset{.}{x}^{^{\prime }n}\text{ \ \ ,}  \tag{9.16 }
\end{equation}
\begin{equation*}
M_{i}^{^{\prime }}(X,v)=\frac{\partial ^{2}F_{i}}{\partial X^{k}\partial
X^{l}}\frac{\partial X^{l}}{\partial x^{m}}\frac{\partial X^{k}}{\partial
x^{n}}x^{^{\prime }m}x^{^{\prime }n}+\frac{\partial F_{i}}{\partial X^{k}}%
\frac{\partial ^{2}X^{k}}{\partial x^{m}\partial x^{n}}x^{^{\prime
}n}x^{^{\prime }m}+
\end{equation*}
\begin{equation}
+\frac{\partial F_{i}}{\partial X^{k}}\frac{\partial X^{k}}{\partial x^{n}}%
\frac{x^{^{\prime }m}\overset{.}{x}^{^{\prime }n}}{\overset{.}{x}^{m}}+\frac{%
\partial F_{i}}{\partial X^{k}}\frac{\partial X^{k}}{\partial x^{m}}%
x^{^{\prime \prime }n}\text{ \ \ .}  \tag{9.17 }
\end{equation}
Therefore, the desired expressions can be found by setting up formulae
(9.14) equal to (9.15) and also formulae (9.15) equal to (9.17). It can
easily be derived how eq. (9.13) will transform, but unfortunately, this
would not result in any simplification of the equation in respect to the
generalized coordinates $X^{i}$.

It remained only to show whether the fourth equation (8.2) is independent
from the preceeding ones and thus can be treated separately or if it follows
naturally from these equations. For the purpose, let us take the
differential of (8.2) and use equations (8.10). After some lengthy, but
straightforward calculations it can be obtained
\begin{equation*}
(dz)(dv)^{2}[\overset{.}{M_{i}}(X,z)\frac{x^{^{\prime }m}}{\overset{.}{x}^{m}%
}+2\frac{x^{^{\prime }m}}{\overset{.}{x}^{m}}M_{i}^{^{\prime
}}(X,z)-M_{i}(X,z)\frac{\left( \overset{.}{x}^{^{\prime }m}\overset{.}{x}%
^{m}-x^{^{\prime }m}\overset{..}{x}^{m}\right) }{\left( \overset{.}{x}%
^{m}\right) ^{2}}+
\end{equation*}
\begin{equation*}
+M_{i}(X,z)\frac{2\left( \overset{.}{x}^{^{\prime }m}\overset{.}{x}%
^{m}-x^{^{\prime }m}\overset{..}{x}^{m}+x^{^{\prime \prime }m}\overset{.}{%
\overset{.}{x}^{m}-\overset{.}{x}^{^{\prime }m}x^{^{\prime }m}}\right) }{%
\left( \overset{.}{x}^{m}\right) ^{2}}]+
\end{equation*}
\begin{equation*}
+(dz)^{2}(dv)[M_{i}^{^{\prime }}(X,z)\frac{\overset{.}{x}^{m}}{x^{^{\prime
}m}}+2\overset{.}{M}_{i}(X,z)+M_{i}(X,z)\frac{\left( \overset{.}{x}%
^{^{\prime }m}x^{^{\prime }m}-\overset{.}{x}^{m}x^{^{\prime \prime
}m}\right) }{\left( x^{^{\prime }m}\right) ^{2}}]+
\end{equation*}
\begin{equation*}
+(dz)^{3}[\overset{.}{M_{i}}(X,z)\frac{\overset{.}{x}^{m}}{x^{^{\prime }m}}%
+M_{i}(X,z)\frac{\left( \overset{..}{x}^{m}x^{^{\prime }m}-\overset{.}{x}^{m}%
\overset{.}{x}^{^{\prime }m}\right) }{\left( x^{^{\prime }m}\right) ^{2}}]+
\end{equation*}
\begin{equation}
+(dv)^{3}[M_{i}^{^{\prime }}(X,z)\frac{x^{^{\prime }m}}{\overset{.}{x}^{m}}%
+M_{i}(X,z)\frac{\left( x^{^{\prime \prime }m}\overset{.}{x}^{m}-x^{^{\prime
}m}\overset{.}{x}^{^{\prime }m}\right) }{\left( \overset{.}{x}^{m}\right)
^{2}}]=0\text{ \ \ \ \ \ .}  \tag{9.18 }
\end{equation}
This is an equation both for the initial coordinates $x^{i}$ and for the
generalized ones $X^{i}$. The equation would have been only for the initial
coordinates if the sum of all the terms with $\overset{.}{M}_{i}(X,z)$ and $%
M_{i}^{^{\prime }}(X,z)$ is sero. However, it can be found by using relation
(8.23)\ $dx^{m}=0$, that the sum of these terms is equal to
\begin{equation*}
-2M_{i}^{^{\prime }}(X,z)\left[ dz(dv)^{2}+(dz)^{2}dv\right] =
\end{equation*}
\begin{equation}
=-2M_{i}^{^{\prime }}(X,z)(dz)^{3}\left[ \left( \frac{\overset{.}{x}^{m}}{%
x^{^{\prime }m}}\right) ^{2}-\frac{\overset{.}{x}^{m}}{x^{^{\prime }m}}%
\right] \text{ \ \ .}  \tag{9.19 }
\end{equation}
Also, evidently the fourth equation (8.18) is different from the fifth one
(9.13).

It can easily be seen that if the initial coordinates are known, equation
(9.18) can be treated as an differential equation in respect to $M_{i}(X,z)$.

\section{\protect\bigskip DISCUSSION - OBTAINED \ RESULTS \ AND \ POSSIBLE \
APPLICATIONS \ IN \ THEORIES \ WITH \ EXTRA \ DIMENSIONS}

\subsection*{\protect\bigskip 10.1. SUMMARY \ OF \ THE \ OBTAINED \ RESULTS}

\bigskip In this paper we continued the investigation of cubic algebraic
equations in gravity theory, which has been initiated in the previous paper
[10].

It has been demonstrated that there is a wide variety of algebraic equations
of third, fourth, fifth, ninth and tenth order. Their derivation is based on
two important initial assumptions:

1. The covariant and contravariant metric components are treated
independently, which is a natural approach in the framework of affine
geometry [15 - 18].

2. Under the above assumption, the gravitational Lagrangian (or Ricci
tensor) should remain the same as in the standard gravitational theory with
inverse contravariant metric tensor components.

It has been proved in Appendix A that if the contravariant metric tensor
components are assumed to be represented as a factorized product of the
components of the vector field $dX=(dX^{1},dX^{2},.....,dX^{n})$, lying in
the tangent space of the given manifold, i.e. $\widetilde{g}%
^{ij}=dX^{i}dX^{j}$, then the new connection $\widetilde{\Gamma }_{ij}^{k}=%
\frac{1}{2}dX^{k}dX^{s}(g_{js,i}+g_{is,j}-g_{ij,s})$ has again an affine
connection transformation property.

The proposed approach allows to treat the Einstein's equations as algebraic
equations, and \textbf{thus to search for separate classes of solutions for
the covariant and contravariant metric tensor components.} The mathematical
approach, based again on the application of the linear-fractional
transformation, will be developed in the third part of this paper. It can be
supposed also that the existence of such separate classes of solutions might
have some interesting and unexplored until now physical consequences. Some
of them will be demonstrated in reference to theories with extra dimensions,
but no doubt the physical applications are much more numerous.

Since the derived solutions should be of a most general type (for example -
non-zero covariant derivative of the metric tensor), the "transition" to the
standard Einsteinian theory \ of gravity can be performed by investigating
the intersection with the corresponding algebraic equations. For example,
the zero covariant derivative of the metric tensor represents again a cubic
algebraic equation, while the condition for the existence of an
contravariant inverse metric tensor can be considered as a quadratic
algebraic equation $g_{ij}dX^{j}dX^{k}=\delta _{i}^{k}$ in respect to the
tangent space vector components $dX^{i}$. On the contrary, if $dX^{i}$ are
considered to be known, then the covariant components $g_{ij}$ satisfy a
\textbf{system of linear operator algebraic equations} (since the unknown
variables $g_{ij}$ form a matrix and not a vector-column, as in the usual
case of linear algebraic systems). This system is well defined
irrespectively of the fact that $\det \parallel dX^{i}dX^{j}\parallel =0$
and moreover, in Appendix C it has been shown how such a system \textbf{in
the general }$N-$\textbf{dimensional case can be transformed to a system of
linear algebraic equations.} For that purpose, an original approach has been
consistently developed for the first time, \textbf{based on the block
structure method. }In a purely algebraic aspect, further investigation is
needed to establish how the type of solutions of the predetermined linear
algebraic system is related to the matrix $dX^{i}dX^{j}$, characterizing the
initially given operator system of equations.

These are more or less the "conceptual" issues (excluding of course the
developed method in respect to the system of linear operator equations) for
further investigation, raised in this first part of the present paper.

\textbf{The most important result in this paper is related to the
possibility to find the uniformization functions for a multicomponent cubic
algebraic surface. } More concretely, the derived in [10] s.c. ''cubic
algebraic equation (2.14) of reparametrizational invariance of the
gravitational Lagrangian'' is investigated by performing consequently the
linear - fractional transformations (3.5) and (3.24) in respect to the
variables $dX^{3}$ and $dX^{2}$. It is important to stress that (2.14) is a
\textbf{multi - variable cubic algebraic equation} (i. e. algebraic
surface), which for the three - dimensional case can be represented as
\begin{equation*}
2p\Gamma _{11}^{r}g_{1r}(dX^{1})^{3}+2p\Gamma
_{22}^{r}g_{2r}(dX^{2})^{3}+2p\Gamma _{33}^{r}g_{3r}(dX^{3})^{3}+
\end{equation*}
\begin{equation*}
+6p\Gamma _{13}^{r}g_{3r}dX^{1}(dX^{3})^{2}+6p\Gamma
_{23}^{r}g_{3r}(dX^{2})(dX^{3})^{2}+2p(\Gamma _{22}^{r}g_{3r}+
\end{equation*}
\begin{equation*}
+2\Gamma _{32}^{r}g_{2r})(dX^{2})^{2}(dX^{3})+2p(\Gamma
_{11}^{r}g_{3r}+2\Gamma _{13}^{r}g_{1r})(dX^{1})^{2}(dX^{3})+
\end{equation*}
\begin{equation*}
+2p(2\Gamma _{12}^{r}g_{2r}+\Gamma
_{22}^{r}g_{1r})(dX^{2})^{2}(dX^{1})+2p(\Gamma _{11}^{r}g_{2r}+
\end{equation*}
\begin{equation*}
+2\Gamma _{12}^{r}g_{1r})(dX^{1})^{2}(dX^{2})+4p(\Gamma
_{12}^{r}g_{3r}+\Gamma _{3(1}^{r}g_{2)r})dX^{1}dX^{2}dX^{3}-
\end{equation*}
\begin{equation*}
-2R_{12}dX^{1}dX^{2}-2R_{13}dX^{1}dX^{3}-2R_{23}dX^{2}dX^{3}-
\end{equation*}
\begin{equation}
-R_{11}(dX^{1})^{2}-R_{22}(dX^{2})^{2}-R_{33}(dX^{3})^{2}=0  \tag{10.1}
\end{equation}
At the same time, all we know from standard algebraic geometry [9] is how to
parametrize a plane (i. e. a two - dimensional) cubic algebraic equation of
the form (3.14) $\widetilde{n}^{2}=4m^{3}-g_{2}m-g_{3}$. Therefore, the
advantage of applying the linear- fractional transformations (3.5) and
(3.23) is that by adjusting their coefficient functions (so that the highest
- third degree in the transformation equation will vanish), the following
sequence of plane cubic algebraic equations is fulfilled (the analogue of
eq.(65) in [10]):
\begin{equation}
P_{1}^{(3)}(n_{(3)})m_{(3)}^{3}+P_{2}^{(3)}(n_{(3)})m_{(3)}^{2}+P_{3}^{(3)}(n_{(3)})m_{(3)}+P_{(4)}^{(3)}=0%
\text{ \ \ \ \ \ ,}  \tag{10.2}
\end{equation}
\begin{equation}
P_{1}^{(2)}(n_{(2)})m_{(2)}^{3}+P_{2}^{(2)}(n_{(2)})m_{(2)}^{2}+P_{3}^{(2)}(n_{(2)})m_{(2)}+P_{(4)}^{(2)}=0%
\text{ \ \ \ \ \ ,}  \tag{10.3}
\end{equation}
\begin{equation}
P_{1}^{(1)}(n_{(1)})m_{(1)}^{3}+P_{2}^{(1)}(n_{(1)})m_{(1)}^{2}+P_{3}^{(1)}(n_{(1)})m_{(1)}+P_{(4)}^{(1)}=0%
\text{ \ \ \ \ \ ,}  \tag{10.4}
\end{equation}
where $m_{(3)}$, $m_{(2)}$, $m_{(1)}$ denote the ratios $\frac{a_{3}}{c_{3}}$%
, $\frac{a_{2}}{c_{2}}$, $\frac{a_{1}}{c_{1}}$ in the corresponding linear -
fractional transformations and $n_{(3)}$, $n_{(2)}$, $n_{(1)}$ are the
''new'' variables $\widetilde{dX}^{3}$, $\widetilde{dX}^{2}$, $\widetilde{dX}%
^{1}$. The sequence of plane cubic algebraic equations (10.2 - 10.4) should
be understood as follows: the first one (10.2) holds if the second one
(10.3) is fulfilled; the second one (10.3) holds if the third one (10.4) is
fulfilled. Of course, in the case of $n$ variables (i. e. $n$ component
cubic algebraic equation) the generalization is straightforward. Further,
since each one of the above plane cubic curves can be transformed to the
algebraic equation ($i=1,2,3$)
\begin{equation}
\widetilde{n}_{(i)}^{2}=\overline{P}_{1}^{(i)}(\widetilde{n}%
_{(i)})m_{(i)}^{3}+\overline{P}_{2}^{(i)}(\widetilde{n}_{(i)})m_{(i)}^{2}+%
\overline{P}_{3}^{(i)}(\widetilde{n}_{(i)})m_{(i)}+\overline{P}_{4}^{(i)}(%
\widetilde{n}_{(i)})  \tag{10.5}
\end{equation}
and subsequently to its parametrizable form, one obtains the solutions (5.1
- 5.3) of the initial multicomponent cubic algebraic equation. \textbf{%
Consequently, these expressions are nothing else but the uniformization
functions for the variables }$dX^{1}$\textbf{, }$dX^{2}$\textbf{, }$dX^{3}$%
\textbf{\ of the cubic equation.} This is so because in Sections 5 - 9 it
has been shown that from the expressions (5.1 - 5.3) a system of first -
order nonlinear differential equations is obtained, for which always a
solution $X^{1}=X^{1}(z)$, $X^{1}=X^{1}(z)$, $X^{1}=X^{1}(z)$ exists. Thus
the dependence on the generalized coordinates $X^{1}$\textbf{, }$X^{2}$%
\textbf{, }$X^{3}$ in the uniformization functions (5.1 - 5.3) dissappears
and only the dependence on the complex coordinate $z$ remains, as it should
be for uniformization functions.

Moreover, the initial assumption $dX^{i}=0$ for obtaining the solutions (5.1
- 5.3) allows us to derive a system of nonlinear differential equations also
for the initial variables $x^{1}$, $x^{2}$, $x^{3}$ and thus the
corresponding solutions $x^{1}=x^{1}(z)$, $x^{2}=x^{2}(z)$, $x^{3}=x^{3}(z)$
in principle can be found. This analysis has been performed in section 6. In
fact, it can easily be guessed that if we have the solutions $X^{1}=X^{1}(z)$%
, $X^{2}=X^{2}(z)$, $X^{3}=X^{3}(z)$ and the additional condition $%
d^{2}X^{i}=0$ (which in fact relates the generalized and the initial sets of
coordinates), then the solutions $x^{1}=x^{1}(z)$, $x^{2}=x^{2}(z)$, $%
x^{3}=x^{3}(z)$ should also be "coordinated" with the previous ones. Indeed,
this is evident from the dependence of the functions $\overline{S}_{1}^{(i)}$%
,$\overline{S}_{2}^{(i)}$,....,$\overline{S}_{6}^{(i)} $ in the system
(6.11) for $\frac{dx^{i}}{dz}$ both on the functions $\frac{\partial F_{i}}{%
\partial X^{k}}$ and $\frac{\partial F_{i}}{\partial x^{k}}$ , i.e. on both
system of coordinates. Of particular importance is the conclusion at the end
of Sect. VI that the two sets of coordinates $X^{1}$, $X^{2}$, $X^{3}$ and $%
\ x^{1}$, $x^{2}$, $x^{3}$ should not be treated on an equal footing. This
means that if $dX^{1}$, $dX^{2}$, $dX^{3}$ satisfy the originally derived
cubic algebraic equation, then it is not necessary to assume this for $\
dx^{1}$, $dx^{2}$, $dx^{3}$.

Much more interesting is the other investigated case in Sections 7 - 9,
where a pair of complex coordinates $z,v$ has been introduced and thus
through the generalized coordinates $X^{1}=X^{1}(z,v)$, $X^{2}=X^{2}(z,v)$, $%
X^{3}=X^{3}(z,v)$ a complex structure of the metric tensor components is
introduced. For the investigated case under the assumption $d^{2}X^{i}=0$
there is only one way for introducing this complex structure - namely,
through the dependence of the initial coordinates on $z$ and $v$, i. e. $%
X^{i}=X^{i}(\mathbf{x}(z,v),z)$. Otherwise, if some other possibility is
assumed, for example $X^{i}=X^{i}(\mathbf{x}(z),z,v)$, then, as proved in
section 7, the obtained system of equations is contradictory. Therefore, it
remains to investigate the full system of equations for the only allowed
case $X^{i}=X^{i}(\mathbf{x}(z,v),z)$, which has been performed in Sections
8 and 9. Remarkably, a nonlinear differential equation is obtained only for
the initial coordinates. However, no such an equation only for the
generalized coordinates can be obtained - the derived equation depends in a
complicated manner on both system of coordinates. Since the existence of
these noncontradictory systems of equations confirms that a complex
structure can be introduced, one may express the line element $ds^{2}=g_{ij}(%
\mathbf{X})dX^{i}dX^{j}$ as
\begin{equation}
ds^{2}=\widetilde{g}_{zz}(z,v)(dz)^{2}+\widetilde{g}_{zv}(z,v)dzdv+%
\widetilde{g}_{vv}(z,v)(dv)^{2}\text{ \ \ \ \ ,}  \tag{10.6}
\end{equation}
where
\begin{equation}
\widetilde{g}_{zz}(z,v)\equiv g_{ij}(\mathbf{X}(z,v))\overset{.}{X}^{i}%
\overset{.}{X}^{j}\text{ \ \ ; \ \ }\widetilde{g}_{vv}(z,v)\equiv g_{ij}(%
\mathbf{X}(z,v))X^{^{\prime }i}X^{^{\prime }j}\text{\ \ \ \ ,}  \tag{10.7}
\end{equation}
\begin{equation}
\widetilde{g}_{zv}(z,v)\equiv g_{ij}(\mathbf{X}(z,v))\left[ \overset{.}{X}%
^{i}X^{^{\prime }j}+X^{^{\prime }i}\overset{.}{X}^{j}\right] \text{ \ \ .}
\tag{10.8}
\end{equation}
This result will be of particular importance in reference to possible
physical applications.

\subsection*{\protect\bigskip 10.2. SOME \ POSSIBLE \ PHYSICAL \
APPLICATIONS \ IN \ GRAVITATIONAL \ THEORIES \ WITH \ EXTRA \ DIMENSIONS}

\subsubsection*{\protect\bigskip 10.2.1. FUNDAMENTAL \ PARALELLOGRAM \ ON \
THE \ COMPLEX \ PLANE, ORBIFOLD \ COMPACTIFICATION \ AND \ PERIODIC \
IDENTIFICATION}

\bigskip As it is well - known, a class of two - dimensional metrics exists
[45]
\begin{equation}
ds^{2}=R^{2}\frac{(a^{2}-v^{2})du^{2}+2uvdudv+(a^{2}-u^{2})dv^{2}}{%
(a^{2}-u^{2}-v^{2})^{2}}\text{ \ \ \ ,}  \tag{10.9}
\end{equation}
representing the linear element of a unit surface in the Lobachevsky space
with a constant negative curvature $-\frac{1}{R^{2}}$. Performing the
transformations
\begin{equation}
\frac{a^{2}-u\text{ }u_{0}-vv_{0}}{\sqrt{a^{2}-u^{2}-v^{2}}}=ae^{-\frac{\rho
}{R}}\text{ \ \ \ \ \ ;\ \ \ }\frac{u_{0}v-uv_{0}}{a^{2}-u\text{ }%
u_{0}-vv_{0}}=\frac{\sigma }{R}\text{ \ \ \ \ ,}  \tag{10.10}
\end{equation}
the above metric (10.9) can be rewritten as
\begin{equation}
ds^{2}=d\rho ^{2}+e^{-\frac{2\rho }{R}}d\sigma ^{2}\text{ \ \ \ \ ,}
\tag{10.11}
\end{equation}
which turns out to be the same as the metric
\begin{equation}
ds^{2}=e^{-2kr_{-}\Phi }\eta _{\mu \nu }dx^{\mu }dx^{\nu }+r_{c}^{2}d\Phi
^{2}\text{ \ \ \ \ \ \ ,}  \tag{10.12}
\end{equation}
extensively used in the first version of the Randall - Sundrum model [[46].
In (10.12) $\eta _{\mu \nu }$ is the flat Minkowski metric, $0\leq \Phi \leq
\pi $ and the extra dimension is a finite interval, whose size is set by the
compactification radius $r_{c}$. A nice and effective generalization of this
model implies that the SM (Standard Model) particles and forces with the
exception of gravity are confined to a 4 - dimensional subspace, but within
a $(4+n)$- dimensional spacetime.

The first problem in reference to Lobachevsky geometries and theories with
extra dimensions is the following one: \textbf{starting from the }$5-$%
\textbf{dimensional metric (10.12) in its relatively simple form and
applying the developed in this paper approach, to find its equivalent form
in the two-dimensional coordinates }$(z,v)$\textbf{. }Obviously, a new class
of Lobachevsky metric, depending on the Weierstrass function and its
derivative can be obtained. Remember also that the complex coordinate $z$ is
related to the Weierstrass function and as mentioned, it is defined on the
lattice $\Lambda =\{m\omega _{1}+n\omega _{2}\mid m,n\in Z;$ $\omega
_{1},\omega _{2}\in C,Im\frac{\omega _{1}}{\omega _{2}}>0\}$ on the two -
dimensional projective plane $CP^{2}$. Then let us define the complex
uniformization coordinate $z$ as $z=\pi r_{c}(\cos \Phi +i\sin \Phi )$ and $%
0\leq \Phi \leq \pi $ is the periodic coordinate. Under the transformation $%
\Phi =arctg\frac{z}{r_{c}}$, the metric (10.12) will transform as
\begin{equation}
ds^{2}=e^{-2kr_{-}\frac{r_{c}}{\sqrt{z^{2}+r_{c}^{2}}}}\eta _{\mu \nu
}dx^{\mu }dx^{\nu }+\frac{r_{c}^{4}}{\sqrt{z^{2}+r_{c}^{2}}}dz^{2}\text{ \ \
\ \ \ .}  \tag{10.13}
\end{equation}
Now the advantage of such a formulation is clear: the nice properties of the
Weierstrass function and its derivative
\begin{equation}
\rho ^{^{\prime }}(z+\omega _{i})=\rho ^{^{\prime }}(z)\text{ \ \ \ ; \ \ \
\ }\rho (\pi r_{c})=\rho (-\pi r_{c})\text{ \ \ \ \ \ }  \tag{10.14}
\end{equation}
exactly matches the requirement for orbifold identification of the points $%
+\pi r_{c}$ and $-\pi r_{c}$. \textbf{In other words, by making the above
transformation for the periodical coordinate }$\Phi $\textbf{\ of the
additional extra dimension [46, 47], a periodical identification is achieved
of the identical points under orbifold compactification with a fundamental
domain of lenght }$2\pi r_{c}$\textbf{\ with the lattice points of the
fundamental paralellogram on the complex plane}. A general overview of
orbifold compactifications is presented in [48]. However, one should not
expect also a periodic identification $X_{\mu }\rightarrow X_{\mu }+2\pi
r_{c}$ for the generalized coordinates of the flat Minkowski space, since
there is no strictly proved theorem that the solutions $X^{1}=X^{1}(z,v)$, $%
X^{2}=X^{2}(z,v)$, $X^{3}=X^{3}(z,v)$ and $X^{4}=X^{4}(z,v)$ of the system
of equations (3.11), (3.22) and (3.29) (more exactly, the equivalent to it
system, written for the five - dimensional case of the metric (10.12))
should depend on $z$ only through the Weierstrass function and its
derivative.

\subsubsection*{\protect\bigskip 10.2.2. FACTORIZATION \ AND \ NON -
FACTORIZATION \ OF \ THE \ VOLUME \ ELEMENT \ }

Obtaining some estimates for the fundamental lenght $2\pi r_{c}$ would be
interesting, since for $d$ additional compactified dimensions, each one of
radius $r_{i}$, the fundamental (Planck) scale of gravity is related to the
gravity scale in the $(4+d)$- dimensional space as [49, 50, 51]
\begin{equation}
M_{pl}^{2}=M_{fund}^{2+d}r_{i}^{d}=M_{fund}^{2+d}\text{ }V^{d}\text{ \ \ \ .}
\tag{10.15}
\end{equation}

Obtaining an estimate of the volume of the extradimensiona space is
important, since by taking a large volume the large discrepancy between the
Planck scale of $10^{19}$\bigskip GeV and the electroweak scale of $100$ GeV
can be diminished and thus the hierarchy problem can be solved. For a
derivation of the relation (10. 15) between the gravity scales on the base
of dimensional analysis of the higher - dimensional Einstein - Hilbert
action, one may use the excellent review article [52]. In such a case, the
metric (10.1) should be generalized to the $(d+4)$- dimensional metric of
the Lobachevsky space. Naturally, the most simple case [52] is of a flat
extradimensional space, when $V^{d}=(2\pi r$ $)^{d}$ and also a flat $4D$
Minkowski \ metric. However, it would be much more interesting to consider
an $(d+4)$ - dimensional $ADS$ (Lobachevsky) space with a constant negative
curvature, whose volume element is given by [53]
\begin{equation}
dV_{n}=\frac{c_{n}dx_{1}dx_{2}...dx_{n}}{(c^{2}-x_{\alpha }x_{\alpha })^{%
\frac{(n+1)}{2}}}\text{ \ \ \ }  \tag{10.16}
\end{equation}
and can be found by splitting up the $n$- dimensional volume by means of $%
(n-1)$- dimensional hyperplanes, perpendicular to the coordinate axis.
Details can be found again in the monograph [53]. For example, the five- and
four- dimensional volume elements are calculated to be
\begin{equation}
V_{5}=\frac{1}{12}\pi ^{2}c^{6}(sh\frac{4r}{c}-8sh\frac{2r}{c}+12\frac{r}{c})%
\text{ \ \ \ ,}  \tag{10.17}
\end{equation}
\begin{equation}
V_{4}=\pi c^{3}(sh\frac{2r}{c}-\frac{2r}{c})\text{ \ \ \ \ \ ,}  \tag{10.18}
\end{equation}
where $r$ denotes the natural (euclidean) length and $c=\frac{1}{k}$ is the
unit length parameter for the Lobachevsky space, which enters not only the
expressions (10.17 - 10.18), but also the exponential factor $%
e^{-2kr_{-}\Phi }$. This factor should be present also in the $(d+4)$-
dimensional analogue of the metric (10.12). Unfortunately, in most of the
existing papers on theories with extra dimensions, the origin and meaning of
the parameter $k$ is not clarified, due to which we point out this here.

\bigskip Since in the limit $c\rightarrow \infty $ the usual Euclidean
geometry is recovered [54], then the above formulaes would give the volumes
of the five- and of the four- dimensional (Euclidean) spheres respectively
\begin{equation}
V_{5}=\frac{8}{15}\pi ^{2}r^{5}\text{ \ \ \ ; \ \ \ }V_{4}=\frac{4}{3}\pi
r^{3}(1+\frac{1}{5}\frac{r^{2}}{c^{2}}+...)\text{ }=\text{\ }\frac{4}{3}\pi
r^{3}\text{\ \ \ .}  \tag{10.19}
\end{equation}
The volumes of the $n$- dimensional (Euclidean) spheres for $n=2\lambda $
and $n=2\lambda +1$ [53]
\begin{equation}
V_{2\lambda }=\frac{\pi ^{\lambda }}{\lambda !}r^{2\lambda }\text{ \ \ \ \ ;
\ \ \ \ }V_{2\lambda +1}=\frac{2^{\lambda +1}\pi ^{\lambda }}{(2\lambda
+1)(2\lambda -1)....3.1}r^{2\lambda +1}\text{ }  \tag{10.20}
\end{equation}
can also be derived in the limit $c\rightarrow \infty $ from the (recurrent)
formulae for the $n$- dimensional hyperbolic volume
\begin{equation}
V_{n}=\frac{2\pi c^{2}}{(n-1)}\left[ \frac{P_{n-4}}{(n-2)}c^{n-2}sh^{n-2}%
\frac{r}{c}ch\frac{r}{c}-V_{n-2}\right] \text{ \ \ \ ,}  \tag{10.21}
\end{equation}
where
\begin{equation}
P_{2\lambda }=\frac{2\pi ^{\lambda +1}}{\lambda !}\text{ \ \ \ \ ; \ \ }%
P_{2\lambda +1}=\frac{2^{\lambda +2}\pi ^{\lambda +1}}{(2\lambda
+1)(2\lambda -1)(2\lambda -3)....3.1}\text{\ \ \ \ .}  \tag{10.22}
\end{equation}
Therefore, only in the flat (Euclidean) $(4+d)$- dimensional space, which is
a product of a $4$- dimensional Minkowski space ($\mu ,\nu =1,2,..,4$) and a
flat $d$- (extra)dimensional space [52]
\begin{equation}
ds^{2}=\eta _{\mu \nu }dX^{\mu }dX^{\nu }-r^{2}d\Omega _{(d)}^{2}\text{ \ \
\ ,}  \tag{10.23}
\end{equation}
one can factorize the volume element in the $(4+d)$ Einstein - Hilbert
action (assuming also that $R^{(4+d)}=R^{(4)}$) as
\begin{equation}
S_{4+d}=-M_{\ast }^{d+2}\int d^{4+d}X\sqrt{g^{(4+d)}}R^{(4+d)}=-M_{\ast
}^{(d+2)}\int d\Omega _{(d)}r^{d}\int d^{4}X\sqrt{g^{(4)}}R^{(4)}\text{ \ \ .%
}  \tag{10.24}
\end{equation}
\textbf{It is clear however from expressions (10.17-18), (10.21) that in the
general case of a multidimensional non-euclidean (Lobachevsky) space such a
factorization of the volume element is impossible. Even in the limit of
small ratios }$\frac{r}{c}$\textbf{, possible corrections to the volume
element (see f.(10. 19)) have to be taken into account and therefore, the
non-euclidean geometry would ''induce'' correction terms in the relation
between the gravitational couplings.}

But again in the flat euclidean case, from (10.24) in [52] the relation
(10.15) $M_{pl}^{2}=M_{\ast }^{d+2}V_{(d)}$ between the gravitational
couplings has been correctly found. Remarkably, the same matching relation
has been found in [47] on the base of a different approach, namely the $%
(4+d) $- dimensional Gauss law
\begin{equation}
\oint\limits_{surf.C}F\text{ }dS=S_{(3+d)}G_{N(4+d)}\times Mass\text{ \ \ \ ,%
}  \tag{10.25}
\end{equation}
where it is important to point out that $S_{(3+d)}\equiv \frac{2\pi ^{\frac{%
3+d}{2}}}{\Gamma (\frac{3+d}{2})}$ is the surface area of the euclidean (!)
unit sphere in $(3+d)$- dimensions. So again, provided that the Newton's
force law in the $(4+d)$- dimensional space is
\begin{equation}
F_{(4+d)}(r)=G_{N(4+d)}\frac{m_{1}m_{2}}{r^{n+2}}\text{ \ \ \ \ \ ,}
\tag{10.26}
\end{equation}
the s. c. ''matching relation'' has been recovered. This really proves that
in a $(4+d)$- dimensional flat space the correct Newton's law is indeed
given by expression (10.26). It should be stressed that there is a perfect
matching between all results due to the initial assumption about the flat
(Euclidean) metric.

\subsubsection*{\protect\bigskip 10.2.3. COORDINATE \ TRANSFORMATIONS \ WITH
\ THE \ LOBACHEVSKY \ CONSTANT \ IN \ "WARP" \ TYPE \ OF \ METRICS}

Now let us turn to the other frequently used case [55, 56] of a $5D$- metric
with an exponentially suppressed ''warp'' factor in front of the flat $4D$
Minkowski \ metric ($0\leq y\leq \pi r_{c}$)
\begin{equation}
ds^{2}=e^{-2\widetilde{k}\mid y\mid }\eta _{\mu \nu }dX^{\mu }dX^{\nu
}+dy^{2}  \tag{10.27}
\end{equation}
and for the moment it shall be assumed that $\widetilde{k}$ is different
from the constant $k=\frac{1}{c}$. The five - dimensional effective action
can be factorized as [55]
\begin{equation}
S_{eff}=\int d^{4}X\dint\limits_{0}^{\pi r_{-}}dy2M^{3}r_{c}e^{-2\widetilde{k%
}\mid y\mid }\sqrt{g}R\text{ \ \ \ }  \tag{10.28}
\end{equation}
and the metric (10.27) is chosen so that the $5$- dimensional Ricci scalar
curvature is equal to the $4$- dimensional Minkowski one. From (10.28), the
matching relation between the gravitational couplings is obtained to be [55]
\begin{equation}
M_{pl}^{2}=2M^{3}\dint\limits_{0}^{\pi r_{-}}dye^{-2\widetilde{k}\mid y\mid
}=\frac{M^{3}}{\widetilde{k}}(1-e^{-2\widetilde{k}r_{-}\pi })\ \ \ .
\tag{10.29}
\end{equation}
\ \ $\ \ \ \ $If one gives up the assumption that the four- dimensional
scalar curvature is equal to the five- dimensional one, but just fixes a
given scalar curvature and keeps the exponential prefactor in the $4D$
Minkowski part of the metric, the approach of the cubic algebraic equation
in this paper might be used to find the metric component in front of $dy^{2}$%
, which will be nothing else, but the contravariant metric tensor component $%
\widetilde{g}^{55}=dX^{5}dX^{5}$. For this special case, it is not
obligatory to find the solution in terms of elliptic functions or to express
it in complex coordinates - as already mentioned, parametrization with the
Weierstrass function is just one of the ways for solving a cubic algebraic
equation.

Let us note that the result of the integration in (10.28 - 29) will not be
coordinate independent. In other words, if we map the $4D$ Minkowski part of
the metric with an exponentially ''damped'' prefactor into a flat Minkowski $%
4D$ metric without the exponential prefactor, this would result in the
appearence of an exponentially growing prefactor in front of the $dy^{2}$
part of the metric (10.27). To illustrate this, let us use a coordinate
transformation, similar to the one, used in [57]
\begin{equation}
X^{1}=a\text{ }ch\frac{\rho }{c}\text{ \ \ ; \ \ \ }X^{2}=b\text{ }sh\frac{%
\rho }{c}\sin \Theta \cos \varphi \text{ \ \ \ ,}  \tag{10.30}
\end{equation}
\begin{equation}
X^{3}=b\text{ }sh\frac{\rho }{c}\sin \Theta \sin \varphi \text{ \ \ ; \ \ \ }%
X^{4}=b\text{ }sh\frac{\rho }{c}\cos \Theta \text{ \ \ \ .\ }  \tag{10.31}
\end{equation}
The signature of the Minkowski space is $(+,-,-,-)$, i.e. $\eta _{\mu \nu
}=(+1,-1,-1,-1)$, $c$ and $r$ are the Lobachevsky constant (natural distance
unit in the Lobachevsky space) and the euclidean distance respectively and $%
\rho $ is the distance in the Lobachevsky space, related to the euclidean
distance $r$ by the formulae
\begin{equation}
r=c\text{ }sh\frac{\rho }{c}\text{ \ \ .}  \tag{10.32}
\end{equation}
The constants $a$ and $b$ are to be chosen so that a flat Minkowski metric
without any prefactor is obtained. In fact, if $\widetilde{k}=\frac{1}{c}$,
the metric (10.27) will be exactly the one, known from Lobachevsky geometry.
Now we shall establish the physical meaning of the relation $\widetilde{k}=%
\frac{1}{c}$, but in reference to theories with extra dimensions. An
elementary introduction into Lobachevsky geometry can be found in [58] and a
more comprehensive and detailed exposition - in [59].

For the choice $a=b=ce^{\widetilde{k}\mid y\mid }$ and after applying the
transformations (10.30 - 31), the metric (10.27) can be rewritten as
\begin{equation}
ds^{2}=-d\rho ^{2}-c^{2}sh^{2}\frac{\rho }{c}d\Theta ^{2}-c^{2}sh^{2}\frac{%
\rho }{c}\sin ^{2}\Theta d\varphi ^{2}+(c\widetilde{k})^{2}e^{2\widetilde{k}%
\mid y\mid }dy^{2}\text{ \ \ \ .}  \tag{10.33}
\end{equation}
Obviously, the first three terms give the unit lenght element in the
Lobachevsky space, which in the limit $c\rightarrow \infty $ (then from
(10.32 \ $\rho \rightarrow r$) gives the usual euclidean lenght element $%
dr^{2}+r^{2}(d\Theta ^{2}+\sin ^{2}\Theta d\varphi ^{2})$ in spherical
coordinates $(r,\Theta ,\varphi )$. Most interesting is the last term in
(10.33) - it goes to infinity if $\mid y\mid \rightarrow \infty $ and in the
limit $c\rightarrow \infty $ (when $c\neq \frac{1}{\widetilde{k}}$), but it
tends to $1$ in the limit $c\rightarrow \infty $ and when $\widetilde{k}=%
\frac{1}{c}$, which is physically reasonable because a flat euclidean
geometry is obtained, as it should be. Thus the exponential increase of the
''extra - dimensional'' distance, when $c\neq \frac{1}{\widetilde{k}}$, can
be regarded as an effect of the non-euclidean nature of space-time. Indeed,
it is physically unacceptable to take the limit $\widetilde{k}=\frac{1}{c}%
\rightarrow 0$, because if the five - dimensional Planck mass is assumed to
be not very far from the electroweak scale $M_{W}\approx TeV$, then a fine-
tuning $\widetilde{k}$ $r_{c}\approx 50$ is needed [49].

Of course, the same approach may be applied to more complicated models with
an arbitrary ''warp'' exponential factor and $(D-4)$ compact non-flat extra-
dimensional spacetime [60, 61]
\begin{equation}
ds^{2}=g_{ab}(\mathbf{X})dX^{a}dX^{b}=2e^{2A(y)}\eta _{\mu \nu }dX^{\mu
}dX^{\nu }+h_{ij}(y)dy^{i}dy^{j}\text{ \ \ \ ,}  \tag{10.34}
\end{equation}
where $(a,b)=1,2,...D$; $(\mu ,\nu )=1,..,4$; $(i,j)=5,...,D$. The
transformations (10.30 - 31) can again be used (with $a=b=\widetilde{k}$ $%
e^{-A(y)}$) and an expression for the warp factor $A(y)$ can be found so
that the metric tensor components $h_{ij}$ of the extra- dimensional space
are left unchanged. In principle, the motivation for different warp factors
comes from $M$- theory (see [62] for a recent review), where for example
three warp factors might be allowed: one for the $2+1$ dimensional spacetime
and two for the internal manifold.

\subsubsection*{\protect\bigskip 10.2.4. ALGEBRAIC $\ $EQUATIONS $\ $IN $\
4D $ \ SCHWARZSCHILD \ BLACK \ HOLES \ IN \ HIGHER \ DIMENSIONAL \ BRANE \
WORLDS}

\bigskip Now suppose that the metric (10.27) does contain a flat Minkowski $%
4D$ space, but a $4D$ black hole instead
\begin{equation}
ds^{2}=e^{-2\widetilde{k}\mid y\mid }g_{\mu \nu }dX^{\mu }dX^{\nu }+dy^{2}%
\text{ \ \ ,}  \tag{10.35}
\end{equation}
where
\begin{equation*}
g_{\mu \nu }dX^{\mu }dX^{\nu }=-(1-\frac{2M}{r})dt^{2}+(1-\frac{2M}{r}%
)^{-1}dr^{2}+
\end{equation*}
\begin{equation}
+r^{2}(d\Theta ^{2}+\sin ^{2}\Theta d\varphi ^{2})\text{ \ \ \ .}
\tag{10.36}
\end{equation}
For such a model [63, 64], if a negative tension brane is introduced at a
distance $y=l<\infty $, the five dimensional BH singularity will have a
finite size and a black tube will extend into the bulk, thus interpolating
between the two black holes. In analogy with the previous considerations, an
interesting problem is to find whether there exists a transformation of the
type
\begin{equation}
X^{1}\equiv t=e^{\widetilde{k}\mid y\mid }g_{1}(\rho ,\widetilde{\Theta },%
\widetilde{\varphi })\text{ }ch\frac{\rho }{c}\text{ \ \ ; \ \ \ }%
X^{2}\equiv r=e^{\widetilde{k}\mid y\mid }g_{2}(\rho ,\widetilde{\Theta },%
\widetilde{\varphi })\text{ }sh\frac{\rho }{c}\sin \widetilde{\Theta }\cos
\widetilde{\varphi }\text{ \ \ \ ,}  \tag{10.37}
\end{equation}
\begin{equation}
X^{3}\equiv \Theta =e^{\widetilde{k}\mid y\mid }g_{3}(\rho ,\widetilde{%
\Theta },\widetilde{\varphi })\text{ }sh\frac{\rho }{c}\sin \widetilde{%
\Theta }\sin \widetilde{\varphi }\text{ \ \ ; \ \ \ }X^{4}\equiv \varphi =e^{%
\widetilde{k}\mid y\mid }\text{ }g_{4}(\rho ,\widetilde{\Theta },\widetilde{%
\varphi })sh\frac{\rho }{c}\cos \widetilde{\Theta }\text{ \ \ \ ,\ }
\tag{10.38}
\end{equation}
so that in terms of the new coordinates again the known hyperbolic spherical
element (without any exponential prefactor) will be obtained plus a term in
front of the extra- dimensional coordinate $y$. \textbf{If the functions }$%
g_{i}$\textbf{\ include also the variable }$\widetilde{t}$\textbf{, i.e. }$%
g_{i}=g_{1}(\rho ,\widetilde{t},\widetilde{\Theta },\widetilde{\varphi })$%
\textbf{, it will be interesting also to see does a transformation exist, so
that the black hole in terms of the coordinates }$(\rho ,\widetilde{t},%
\widetilde{\Theta },\widetilde{\varphi })$\textbf{\ in the limit }$%
c\rightarrow \infty $\textbf{\ will give again the known Schwarzschild black
hole element (10.36)? }

A possible application of the formalism in this paper is related to the
Riemann scalar curvature invariant $R_{ABCD}R^{ABCD}$ [63], which for the
background metric (10.36) and using the conventional contravariant metric
components $g^{ij}$ is calculated to be [63]
\begin{equation}
R_{ABCD}R^{ABCD}=40k^{4}+\frac{48M^{2}e^{4k\mid y\mid }}{r^{6}}\text{ \ \ \ .%
}  \tag{10.39}
\end{equation}
This expression contains an important physical information - it diverges at
the black hole singularity at $r=0$ and also at the $ADS$ horizon at $\mid
y\mid \rightarrow \infty $. The elimination of this singulariry (i. e.
giving it a finite size) is the main motivation for introducing the second,
negative tension brane at a distance $y=L$. But even in the case of a single
brane configuration the presence of a singularity is essential since there
might be a non- vanishing energy flow into the bulk singularities, which is
not desirable. Such an energy flow will exist if the limit [63]
\begin{equation}
\lim_{\mid y\mid \rightarrow \infty }\sqrt{-g}J_{(\mu )}^{y}=\lim_{\mid
y\mid \rightarrow \infty }\sqrt{-g}T^{yN}K_{N}^{(\mu )}  \tag{10.40}
\end{equation}
is non- zero, where $J_{(\mu )}^{y}$ and $T^{yN}$ are the current and the
energy- momentum tensor \ of a massless scalar field and $K_{N}^{(\mu
)}=e^{2A}\delta _{M}^{\mu }$ ($\mu =t,\Theta ,\varphi $) is the Killing
vector for the BH metric (10.35 - 36).

Therefore, it is essential to check whether the presence of the singularity
in (10.39) and of the zero energy flow in (10.40) will be confirmed if the
same scalar curvature $R$ will be obtained by contracting the Riemann tensor
with another contravariant metric tensor field $\widetilde{g}^{ij}$ such
that
\begin{equation}
R=g^{AC}g^{BD}R_{ABCD}=\widetilde{g}^{AC}\widetilde{g}^{BD}\widetilde{R}%
_{ABCD}  \tag{10.41}
\end{equation}
where $\widetilde{R}_{ABCD}$, following the notation and approach in Sect.
II, is the \textbf{modified Riemann tensor }
\begin{equation*}
\widetilde{R}_{ABCD}\equiv \frac{1}{2}%
(g_{AD,BC}+g_{BC,AD}-g_{AC,BD}-g_{BD,AC})+g_{np}(\widetilde{\Gamma }_{BC}^{n}%
\widetilde{\Gamma }_{AD}^{p}-\widetilde{\Gamma }_{BD}^{n}\widetilde{\Gamma }%
_{AC}^{p})=
\end{equation*}
\begin{equation}
=\frac{1}{2}(....)+g_{np}g_{rs}g_{qt}\widetilde{g}^{ns}\widetilde{g}%
^{pt}(\Gamma _{BC}^{n}\Gamma _{AD}^{p}-\Gamma _{BD}^{n}\Gamma _{AC}^{p})%
\text{ \ \ \ .}  \tag{10.42}
\end{equation}
If the scalar curvature $R$, the connection $\Gamma _{AB}^{C}$ are
calculated from the initially given metric, equation (10.41) can be treated
as an \textbf{fourth- order} algebraic equation in respect to the components
$\widetilde{g}^{AB}$ and as an \textbf{eight- order }algebraic equation in
respect to the variables $dX^{A}$, if again the factorization $\widetilde{g}%
^{AB}=dX^{A}dX^{B}$ is used. This example clearly shows the necessity to go
beyond the assumption about the contravariant metric factorization. But on
the other hand, even if the factorization assumption is used, the same
scalar curvature can be obtained by contracting the (modified) Ricci tensor
with the contravariant metric tensor $\widetilde{g}^{AB}$, i. e. $R=%
\widetilde{g}^{AB}R_{AB}$, which was in fact the cubic algebraic equation,
investigated in the previous sections.

But there is also one more way for obtaining the scalar curvature $R$ - by
assuming that the following algebraic equation with the usual Riemann tensor
components holds
\begin{equation}
R=\widetilde{g}^{AC}\widetilde{g}^{BD}R_{ABCD}\text{ \ \ .}  \tag{10.43}
\end{equation}
Fortunately, this equation is second order in respect to $\widetilde{g}^{AB}$
and fourth order in respect to $dX^{A}$ and moreover, it \textbf{does no}t
contain any derivatives of the components $\widetilde{g}^{AB}$ and $dX^{A}$,
unlike the investigated cubic algebraic equation, which contains such
derivatives and consequently a special approach is needed for solving such
an equation.

Let us now assume that in the framework of the factorization assumption,
\textbf{both} equations (10.41) and (10.43) are fulfilled. Then the
fulfillment of these equations is a necessary condition for the preservation
of the scalar curvature invariant because
\begin{equation}
R_{ABCD}R^{ABCD}=R_{ABCD}\widetilde{g}^{Ai}\widetilde{g}^{Bj}\widetilde{g}%
^{Ck}\widetilde{g}^{Dl}\widetilde{R}_{ijkl}=  \tag{10.44}
\end{equation}
\begin{equation}
=\left( R_{ABCD}dX^{A}dX^{B}dX^{C}dX^{D}\right) \left( \widetilde{R}%
_{ijkl}dX^{i}dX^{j}dX^{k}dX^{l}\right) =  \tag{10.45}
\end{equation}
\begin{equation}
=\left( R_{ABCD}\widetilde{g}^{AC}\widetilde{g}^{BD}\right) \left(
\widetilde{R}_{ijkl}\widetilde{g}^{ik}\widetilde{g}^{jl}\right) =R^{2}\text{
\ \ \ \ .}  \tag{10.46}
\end{equation}
Motivated by the necessity to investigate lower degree algebraic equations,
one may take equation (10.43) and also the equation
\begin{equation}
\widetilde{g}^{AC}\widetilde{g}^{BD}\widetilde{R}_{ABCD}-\widetilde{g}^{AC}%
\widetilde{g}^{BD}R_{ABCD}=0\text{ \ \ \ \ .}  \tag{10.47}
\end{equation}
A subclass of solutions of this equation will be represented by the
algebraic equation
\begin{equation}
\widetilde{g}^{BD}\widetilde{R}_{ABCD}-\widetilde{g}^{BD}R_{ABCD}=0\text{ \
\ \ \ }  \tag{10.48}
\end{equation}
(cubic in respect to $\widetilde{g}^{AB}$ and of sixth order in respect to $%
dX^{A}$) and another, more restricted class of solutions - by the equation
\begin{equation}
\widetilde{R}_{ABCD}-R_{ABCD}=0\text{ \ \ \ \ ,}  \tag{10.49}
\end{equation}
which is quadratic in $\widetilde{g}^{AB}$ and quartic in respect to $dX^{A}$%
. \textbf{Therefore, even in such a complicated case, the investigation of
the intersection varieties of the two quartic equations (10.43) and (10.49),
written respectively as (again, it shall be used that }$\widetilde{\Gamma }%
_{ij}^{k}=dX^{k}dX^{s}g_{rs}\Gamma _{ij}^{r}$\textbf{) }
\begin{equation}
dX^{A}dX^{B}dX^{C}dX^{D}R_{ABCD}-R=0  \tag{10.50}
\end{equation}
\textbf{\ and }
\begin{equation*}
g_{np}g_{rs}g_{qt}(\Gamma _{BC}^{r}\Gamma _{AD}^{q}-\Gamma _{BD}^{r}\Gamma
_{AC}^{q})dX^{n}dX^{s}dX^{p}dX^{t}-
\end{equation*}
\begin{equation}
-g_{np}(\Gamma _{BC}^{n}\Gamma _{AD}^{p}-\Gamma _{BD}^{n}\Gamma _{AC}^{p})=0%
\text{ \ \ \ \ ,}  \tag{10.51}
\end{equation}
\textbf{\ may give some solutions for the contravariant metric tensor
components }$\widetilde{g}^{AB}=dX^{A}dX^{B}$\textbf{, which will preserve
both the scalar curvature and the scalar curvature invariant.} Respectively,
if only the scalar curvature $R$ is to be preserved, one may find the
solutions of the algebraic equation (10.43) and then substitute them in the
expression for the scalar curvature invariant $R_{ABCD}R^{ABCD}$.

It is clear also that if one takes only equation (10.41) $R=\widetilde{g}%
^{AC}\widetilde{g}^{BD}\widetilde{R}_{ABCD}$ and not equation (10.43), from
(10.44- 46) one may obtain not the equality $R_{ABCD}R^{ABCD}=R^{2}$, but an
fourth- order algebraic equation in respect to $dX^{A}$ for the preservation
of the scalar curvature invariant
\begin{equation}
R.R_{ABCD}dX^{A}dX^{B}dX^{C}dX^{D}-R_{ABCD}R^{ABCD}=0\text{ \ \ \ \ .}
\tag{10.52}
\end{equation}
But this is not the only possibility. One may take also \textbf{only }%
equation (10.43) $R=\widetilde{g}^{AC}\widetilde{g}^{BD}R_{ABCD}$\textbf{\ }%
and disregard equation (10.41). Then the resulting algebraic equation from
(10.44 - 46) will be
\begin{equation*}
\frac{1}{2}(g_{AD,BC}+g_{BC,AD}-g_{AC,BD}-g_{BD,AC})dX^{A}dX^{B}dX^{C}dX^{D}+
\end{equation*}
\begin{equation*}
+g_{np}g_{rs}g_{qt}(\Gamma _{BC}^{r}\Gamma _{AD}^{q}-\Gamma _{BD}^{r}\Gamma
_{AC}^{q})dX^{A}dX^{B}dX^{C}dX^{D}dX^{n}dX^{s}dX^{p}dX^{t}-
\end{equation*}
\begin{equation}
-R_{ABCD}R^{ABCD}=0\text{ \ \ .}  \tag{10.53}
\end{equation}
This equation is of eight order and due to the presence of the last scalar
curvature invariant term it is impossible to find subclasses of solutions of
(lower - order) algebraic equations, as in the case of eq. (10.47).

\subsubsection*{\protect\bigskip 10.2.5. COMPACTIFICATION \ RADIUS \ AND \
SCALAR \ FIELD \ EQUATION $\ $IN $\ 4D$ \ SCHWARZSCHILD \ BLACK \ HOLES \ IN
\ HIGHER \ DIMENSIONAL \ BRANE \ WORLDS}

\textbf{\bigskip }In theories with extra dimensions, for example $(4+n)$-
dimensional Schwarzschild black hole [64, 65, 66, 67]
\begin{equation}
ds^{2}=-h(r)dt^{2}+h^{-1}(r)dr^{2}+r^{2}d\Omega _{n+2}^{2}  \tag{10.54}
\end{equation}
with
\begin{equation}
h(r)=1-\left( \frac{r_{H}}{r}\right) ^{n+1}  \tag{10.55}
\end{equation}
($r_{H}$- \ the horizon radius) it is important to distinguish between
distances $r\ll R_{1}$ ($R_{1}$- the compactification radius), when the BH
is a $(4+n)$- dimensional one, and distances $r\gg R_{1}$, when the BH
metric goes over to the usual four dimensional Schwarzschild metric
\begin{equation}
ds^{2}=-(1-\frac{2M}{M_{Hr}^{2}})dt^{2}+(1-\frac{2M}{M_{Hr}^{2}}%
)^{-1}dr^{2}+r^{2}d\Omega ^{2}\text{ \ \ .}  \tag{10.56}
\end{equation}
However, when solving the scalar wave equation $g^{IJ}\Phi _{I;J}=0$, there
is no way to introduce the scale factor $R_{1}$ in the solution of the
scalar equation - in such a case the scalar field behaviour can be compared
in the transition from one limit to another.

The use of the more general contravariant tensor $\widetilde{g}^{ij}$ gives
the opportunity to introduce such a scale factor. Let us first note that
\begin{equation}
g_{AB}\widetilde{g}^{BC}=l_{A}^{C}(\mathbf{x})\text{ \ \ \ }\Rightarrow
\text{ \ }\widetilde{g}^{BC}=l_{D}^{B}g^{DC}\text{ \ \ \ \ ,}  \tag{10.57}
\end{equation}
where $A,B,C,D$ concretely in this case will denote the $(4+n)$- dimensional
indices, $\mu ,\nu $- only the four- dimensional indices and $i,j,k$ denote
the indices of the additional $n$- dimensional space. Then the metric can be
represented as
\begin{equation}
ds^{2}=g_{AB}dX^{A}dX^{B}=g_{\mu \nu }dX^{\mu }dX^{\nu
}+\sum_{i=5}^{n+4}l_{i}^{i}=ds_{(4)}^{2}+nR_{1}  \tag{10.58}
\end{equation}
where it has been assumed that $l_{i}^{i}=R_{1}$ for all $i$. Consequently,
some of the components $\widetilde{g}^{jB}$ of the contravariant metric
tensor can be expressed as
\begin{equation}
\widetilde{g}^{jB}=l_{\nu }^{j}g^{\nu B}+l_{i}^{j}g^{iB}=l_{\nu }^{j}g^{\nu
B}+R_{1}g^{jB}+\underset{i\neq j}{l_{i}^{j}g^{iB}}  \tag{10.59}
\end{equation}
and evidently the solutions of the scalar wave equation will depend on the
compactification radius $R_{1}$.

\subsubsection*{\protect\bigskip 10.2. 6. A \ COMPLIMENTARY \ PROPOSAL \ FOR
\ HIGGS \ MASS \ GENERATION \ IN \ THEORIES \ WITH \ TWO \ THREE - BRANES}

\bigskip Closely related to the above problem about the contravariant metric
tensor components as coupling constants is the problem about \textbf{Higgs
mass generation} in theories with two branes [46, 56] - the so called
\textbf{''hidden''} and \textbf{''visible}'' branes at the orbifold fixed
points $\Phi =0$ and $\Phi =\pi $ (the metric is again (10.10)). These three
branes couple to the four dimensional components $G_{\mu \nu }$ of the bulk
metric as [46]
\begin{equation}
g_{\mu \nu }^{vis}(X^{\mu })=G_{\mu \nu }(X^{\mu },\Phi =\pi )\text{ \ \ \ \
; \ \ \ \ }g_{\mu \nu }^{hid}(X^{\mu })=G_{\mu \nu }(X^{\mu },\Phi =0)\text{%
\ }  \tag{10.60}
\end{equation}
and the action includes the gravity part plus the action the action for the
visible and hidden branes and also the part of the action for the
fundamental Higgs field
\begin{equation}
S_{vis}=\dint d^{4}X\sqrt{-g}_{vis}\left[ g_{vis}^{\mu \nu }D_{\mu
}H^{+}D_{\nu }H-\lambda \left( \mid H\mid ^{2}-v_{0}^{2}\right) ^{2}\right]
\text{ \ \ \ ,}  \tag{10.61}
\end{equation}
where $v_{0}$ is the vacuum expectation value (VEV) for the Higgs field $H$,
$\lambda $ is a coupling constant [46]. Similar coupling of the
contravariant metric tensor components to a gauge field can be found also in
radion cosmology theories [68]. Since $g_{\mu \nu }^{vis}=e^{-2kr_{-}\pi
}g_{\mu \nu }$, it is believed that by a proper normalization of the fields
one can determine the physical masses. In particular, if the Higgs field
wave function is normalized as $H\rightarrow e^{kr_{-}\pi }H$, then
\begin{equation}
S_{vis}=\int d^{4}X\sqrt{-g}\left[ g^{\mu \nu }D_{\mu }H^{+}D_{\nu
}H-\lambda \left( \mid H\mid ^{2}-e^{-2kr_{-}\pi }v_{0}^{2}\right) ^{2}%
\right] \text{ \ \ . }  \tag{10.62}
\end{equation}
\textbf{Therefore, since }$v=e^{-2kr_{-}\pi }v_{0}$\textbf{, any mass }$%
m_{0} $\textbf{\ on the visible three- brane in the fundamental higher-
dimensional theory will correspond to a physical mass }
\begin{equation}
m=e^{-kr_{-}\pi }m_{0}\text{ \ \ \ ,}  \tag{10.63}
\end{equation}
\textbf{\ ''measured'' with the metric }$g^{\mu \nu }$\textbf{\ in the
effective Einstein- Hilbert action. If }$kr_{c}\approx 50$\textbf{\ (i.e. }$%
e^{kr_{-}\pi }\approx 10^{15}$\textbf{), this is the physical mechanism that
is supposed to produce TeV physical mass scales from mass parameters around
the Planck scale }$\approx 10^{19}$\textbf{\ GeV. }

\textbf{In the context of the developed approach in this paper, now it shall
be shown that the above physical mechanism of generation of TeV mass scales
may turn out to be more complicated and diverse. Namely, for a given scalar
curvature, there will be a multitude of contravariant metric tensors, thus
suggesting that the possibilities for the mass scales will be much more in
number.}

Following the two- dimensional approach, the contravariant metric tensor
components $\widetilde{g}^{\mu \nu }$ can be written as
\begin{equation}
\widetilde{g}^{\mu \nu }=dX^{\mu }dX^{\nu }=F_{\mu }(\mathbf{X}(z,v),\Phi
(z,v),z)F_{\nu }(\mathbf{X}(z,v),\Phi (z,v),z)\text{ \ \ \ ,}  \tag{10.64}
\end{equation}
from where the (contravariant) metric on the visible brane can be expressed
as
\begin{equation}
\widetilde{g}_{vis}^{\mu \nu }=L_{2}(z,v)\widetilde{g}^{\mu \nu }\text{ \ \ ,%
}  \tag{10.65}
\end{equation}
where
\begin{equation}
L_{2}(z,v)\equiv \frac{F_{\mu }(\mathbf{X}(z,v),\Phi (z,v)=\pi ,z)F_{\nu }(%
\mathbf{X}(z,v),\Phi (z,v)=\pi ,z)}{F_{\mu }(\mathbf{X}(z,v),\Phi
(z,v),z)F_{\nu }(\mathbf{X}(z,v),\Phi (z,v),z)}\text{ \ \ .}  \tag{10.66}
\end{equation}
\textbf{\ }Formulaes (10.65 - 66) have been derived as a ratio of the
''visible'' and the usual contravariant metric components for each fixed
indices $(\mu ,\nu )=(\mu _{1},\nu _{1})$ and for the moment, \textbf{%
without assuming} that the points on the complex plane, for which \textbf{\ }%
$\Phi (z_{0},v_{0})=\pi $, are known. Further it shall be shown how the
calculation will be modified if these points are assumed to be known.

The transition from the four- dimensional variables $%
d^{4}X=dX_{1}dX_{2}dX_{3}dX_{4}$ to the two- dimensional complex variables $%
(z,v)$ can be performed by using the formulae
\begin{equation}
d^{4}X=\sum_{1\leq i_{1}<i_{k}\leq 4}\det
\begin{Vmatrix}
\frac{\partial X_{i_{1}}}{\partial z} & \frac{\partial X_{i_{k}}}{\partial v}
\\
\frac{\partial X_{i_{k}}}{\partial z} & \frac{\partial X_{i_{k}}}{\partial v}
\end{Vmatrix}
dz\wedge dv=L_{3}(z,v)dz\wedge dv\text{ \ \ \ ,}  \tag{10.67}
\end{equation}
but since we are interested in rescaling only the Higgs field and the
contravariant metric as
\begin{equation}
H\rightarrow \widetilde{H}f\text{ \ \ ;\ \ \ \ }v_{0}\rightarrow \widetilde{v%
}_{0}\text{ \ \ \ \ \ }(f-a\text{ }function)\text{ \ \ \ ,}  \tag{10.68}
\end{equation}
the change of\textbf{\ }variables in the volume integration is not necessary
to be taken into account. Next it is necessary to find how the volume
element $\sqrt{-g}_{vis}$ of the visible brane can be expressed through the
volume element $\sqrt{-g}$ in terms of the metric (10.10). It can easily be
calculated that
\begin{equation}
\sqrt{-g}=\sqrt{K_{1}(\Phi ,\frac{\partial \Phi }{\partial z},\frac{\partial
\Phi }{\partial v},\frac{\partial X^{\mu }}{\partial z},\frac{\partial
X^{\mu }}{\partial z})+e^{-4kr_{-}\Phi }K_{2}(\frac{\partial X^{\mu }}{%
\partial z},\frac{\partial X^{\mu }}{\partial v}})\text{ \ \ \ ,}
\tag{10.69}
\end{equation}
where
\begin{equation*}
K_{1}\equiv r_{c}^{2}e^{-2kr_{-}\Phi }[(\frac{\partial \Phi }{\partial z}%
)^{2}(\frac{\partial X^{1}}{\partial v})^{2}+(\frac{\partial \Phi }{\partial
v})^{2}(\frac{\partial X^{1}}{\partial z})^{2}-(\frac{\partial \Phi }{%
\partial z})^{2}\sum_{i=2}^{4}(\frac{\partial X^{i}}{\partial v})^{2}-
\end{equation*}
\begin{equation*}
-(\frac{\partial \Phi }{\partial v})^{2}\sum_{i=2}^{4}(\frac{\partial X^{i}}{%
\partial z})^{2}]+3r_{c}^{4}(\frac{\partial \Phi }{\partial z})^{2}(\frac{%
\partial \Phi }{\partial v})^{2}-8r_{c}^{2}\frac{\partial \Phi }{\partial z}%
\frac{\partial \Phi }{\partial v}\frac{\partial X^{1}}{\partial z}\frac{%
\partial X^{1}}{\partial v}e^{-2kr_{-}\Phi }+
\end{equation*}
\begin{equation}
+8r_{c}^{2}e^{-2kr_{-}\Phi }\frac{\partial \Phi }{\partial z}\frac{\partial
\Phi }{\partial v}\sum_{i=2}^{4}\frac{\partial X^{i}}{\partial z}\frac{%
\partial X^{i}}{\partial v}\text{ \ \ \ ,}  \tag{10.70}
\end{equation}
\begin{equation*}
K_{2}\equiv 8\frac{\partial X^{1}}{\partial z}\frac{\partial X^{1}}{\partial
v}\sum_{i=2}^{4}\frac{\partial X^{i}}{\partial z}\frac{\partial X^{i}}{%
\partial v}-\left( \frac{\partial X^{1}}{\partial z}\right)
^{2}\sum_{i=2}^{4}\left( \frac{\partial X^{i}}{\partial v}\right)
^{2}-\left( \frac{\partial X^{1}}{\partial v}\right)
^{2}\sum_{i=2}^{4}\left( \frac{\partial X^{i}}{\partial z}\right) ^{2}-
\end{equation*}
\begin{equation*}
-3\left( \frac{\partial X^{1}}{\partial z}\right) ^{2}\left( \frac{\partial
X^{1}}{\partial v}\right) ^{2}+\left( \sum_{i=2}^{4}\frac{\partial X^{i}}{%
\partial z}\right) ^{2}\left( \sum_{i=2}^{4}\frac{\partial X^{i}}{\partial z}%
\right) ^{2}-
\end{equation*}
\begin{equation}
-4\left( \sum_{i=2}^{4}\frac{\partial X^{i}}{\partial z}\frac{\partial X^{i}%
}{\partial v}\right) ^{2}\text{ \ \ \ .}  \tag{10.71}
\end{equation}
Setting up $\Phi (z,v)=\pi $ (note that then $K_{1}=\pi $), one obtains
\begin{equation}
\sqrt{-g}_{vis}=L_{1}(\Phi ,X^{\mu },\sqrt{-g})\sqrt{-g}\text{ \ \ \ \ ,}
\tag{10.72}
\end{equation}
where
\begin{equation}
L_{1}(\Phi ,X^{\mu },\sqrt{-g})\equiv \frac{e^{-2kr_{-}\pi }}{%
e^{-2kr_{-}\Phi }}\sqrt{1-\frac{K_{1}(\Phi ,X^{\mu })}{\left( \sqrt{-g}%
\right) ^{2}}\text{ \ \ .}}  \tag{10.73}
\end{equation}
Unlike the previously discussed in [46] case, when the ''visible'' volume
element is represented as a product of some factor (constant), multiplying
the volume element $\sqrt{-g}$ (i.e. $\sqrt{-g}_{vis}=e^{4kr_{-}\pi }\sqrt{-g%
}$), the present case might seem to be quite different, since the function $%
L_{1}$ depends again on $\sqrt{-g}$. However, it shall be proved below that
by requiring the action of the ''visible'' brane to remain unchanged after
the rescaling (10.68), still such a possibility will exist, but in a more
general form. Indeed, after the rescaling (10.68) $H\rightarrow \widetilde{H}%
f$ \ \ ;\ \ $v_{0}\rightarrow \widetilde{v}_{0}$ the action (10.61) becomes
(written in the two - dimensional coordinates $(z,v)$)
\begin{equation*}
S_{vis}=\dint dzdv\sqrt{-g}L_{3}L_{1}\left[ g^{\mu \nu }L_{2}\text{ }%
f^{2}D_{\mu }\widetilde{H}^{+}D_{\nu }\widetilde{H}-\lambda f^{4}\left( \mid
\widetilde{H}\mid ^{2}-\widetilde{v}_{0}^{2}\right) ^{2}\right] +\ \ \
\end{equation*}
\begin{equation}
+\int dzdv\sqrt{-g}\text{ }L_{3}L_{add}\text{ \ \ ,}  \tag{10.74}
\end{equation}
where
\begin{equation*}
L_{add}\equiv L_{2}g^{\mu \nu }[\mid \widetilde{H}\mid ^{2}A_{\nu }\partial
_{\mu }\mid f\mid ^{2}+\mid \widetilde{H}\mid ^{2}A_{\nu }\partial _{\mu
}f^{+}\partial _{\nu }f+
\end{equation*}
\begin{equation}
+\widetilde{H}^{+}f\text{ }\partial _{\mu }f^{+}\partial _{\nu }\widetilde{H}%
+\widetilde{H}\text{ }f\text{ }^{+}\partial _{\mu }f\text{ }\partial _{\nu }%
\widetilde{H}^{+}\text{ }  \tag{10.75}
\end{equation}
and the covariant derivative $D_{\mu }$ is expressed as $D_{\mu }=\partial
_{\mu }+A_{\mu }$. Clearly, the visible brane actions before and after the
rescaling will remain unchanged if
\begin{equation}
L_{1}L_{2}f^{2}=1\text{ \ \ \ ; \ \ \ \ \ \ }L_{1}f^{4}=1  \tag{10.76}
\end{equation}
and
\begin{equation}
L_{add}=0\text{ \ \ \ \ \ .}  \tag{10.77}
\end{equation}
The first two relations (10.76) give
\begin{equation}
f=\pm (L_{2})^{\frac{1}{2}}=\pm (L_{1})^{-\frac{1}{6}}\text{ \ \ \ \ ,}
\tag{10.78}
\end{equation}
which can be rewritten as
\begin{equation}
\frac{1}{L_{2}^{3}}=\frac{e^{-2kr_{-}\pi }}{e^{-2kr_{-}\Phi }}\sqrt{1-\frac{%
K_{1}(\Phi ,X^{\mu })}{(\sqrt{-g})^{2}}\text{ }}\text{ \ \ \ \ \ ,}
\tag{10.79}
\end{equation}
from where the function $K_{1}(\Phi ,X^{\mu })$ can be expressed and
substituted into expression (10.69) for $\sqrt{-g}$. Thus one obtains
\begin{equation}
\sqrt{-g}=L_{2}^{3}e^{-2kr_{-}\pi }\sqrt{K_{2}(X^{\mu })}\text{ \ \ \ .}
\tag{10.80}
\end{equation}
From (10.69) for $\Phi (z,v)=\pi $ one can easily derive
\begin{equation}
\sqrt{-g}_{vis}=\sqrt{e^{-4kr_{-}\pi }}.\sqrt{K_{2}(X^{\mu })}=\frac{1}{%
L_{2}^{3}}\sqrt{-g}\text{ \ \ .}  \tag{10.81}
\end{equation}
Therefore, even in the more general case of contravariant metric tensor,
different from the inverse one, there is a relation similar to $\sqrt{-g}%
_{vis}=e^{-4kr_{-}\pi }\sqrt{-g}$, but with the function $\frac{1}{L_{2}^{3}}
$, multiplying the volume element. Let us remind that for performing the
calculation it was sufficient to know the function $\Phi (z,v)$ as a
solution of the system of nonlinear differential equations, but not the
points $(z_{0}^{(l)},v_{0}^{(l)})$, at which $\Phi
(z=z_{0}^{(l)},v=v_{0}^{(l)})=\pi $. Consequently, in the final result
(10.81) one \textbf{cannot set up}
\begin{equation}
\sqrt{-g}_{vis}\left( \mathbf{X(}z=z_{0}^{(l)},v=v_{0}^{(l)}),\Phi =\pi
\right) =\frac{1}{L_{2}^{3}(z=z_{0}^{(l)},v=v_{0}^{(l)},\Phi =\pi )}\sqrt{-g}%
\text{ \ \ \ .}  \tag{10.82}
\end{equation}
Then \textbf{to any mass }$m_{0}$\textbf{\ on the visible three- brane would
correspond a single physical mass, ''measured'' with the metric }$g^{\mu \nu
}$
\begin{equation}
m^{(l)}=m_{0}f=m_{0}\sqrt[4]{L_{2}^{(l)}(z=z_{0}^{(l)},v=v_{0}^{(l)},\Phi
=\pi )\text{ }}\text{ \ \ \ ,}  \tag{10.83}
\end{equation}
i. e. there is no degeneracy of masses. The corresponding additional
condition (10.77) $L_{add}=0$ can be written as
\begin{equation*}
L_{add}=f^{2}\text{ }\partial _{\mu }\ln f\text{ }[2\mid \widetilde{H}\mid
^{2}A_{\nu }+2\widetilde{H}^{+}+\widetilde{H}^{2}\text{ \ }\partial _{\nu
}\left( \frac{\widetilde{H}^{+}}{\widetilde{H}}\right) +
\end{equation*}
\begin{equation}
+2\mid \widetilde{H}\mid ^{2}\partial _{\nu }(\ln f)\text{ }-\widetilde{H}%
^{2}\partial _{\nu }(\ln f)]=0\text{ \ \ \ ,}  \tag{10.84}
\end{equation}
from where the trivial case $f=const$ is obtained from $\partial _{\mu }\ln
f=0$.

Let us now see how the above approach will change if the points $%
(z_{0}^{(l)},v_{0}^{(l)})$ on the complex plane, at which the equation $\Phi
(z=z_{0}^{(l)},v=v_{0}^{(l)})=\pi $ holds, are considered to be known. The
function $L_{2}(z,v)$ in the ratio of $g_{vis}^{\mu \nu }$ and $g^{\mu \nu }$
will be different and will be denoted as $\widetilde{L}_{2}(z,v)$
\begin{equation}
\widetilde{L}_{2}(z,v)\equiv \frac{F_{\mu }(\mathbf{X}%
(z_{0}^{(l)},v_{0}^{(l)}),\Phi =\pi ,z_{0}^{(l)})F_{\nu }(\mathbf{X}%
(z_{0}^{(l)},v_{0}^{(l)}),\Phi =\pi ,z_{0}^{(l)})}{F_{\mu }(\mathbf{X}%
(z,v),\Phi (z,v),z)F_{\nu }(\mathbf{X}(z,v),\Phi (z,v),z)}\text{ \ \ .}
\tag{10.85}
\end{equation}
Also from formulae (10.68) for $\Phi =\pi $ and for all points $%
(z,v)=(z_{0}^{(l)},v_{0}^{(l)})$ one can obtain
\begin{equation}
\sqrt{-g}_{vis}=\sqrt{-g}e^{-2kr_{-}\pi }\sqrt{\frac{K_{2}^{0}(X^{\mu
}(z_{0}^{(1)},v_{0}^{(1)})}{K_{1}+e^{-4kr_{-}\Phi }K_{2}(X^{\mu }(z,v))}}=%
\widetilde{L}_{1}(z,v)\text{ \ \ \ ,}  \tag{10.86}
\end{equation}
which evidently is different from expression (10.59). Consequently, for this
case instead of (10.79) one receives
\begin{equation}
\frac{1}{\widetilde{L}_{2}^{6}}=e^{-4kr_{-}\pi }\frac{K_{2}^{0}(X^{\mu
}(z_{0}^{(1)},v_{0}^{(1)})}{K_{1}+e^{-4kr_{-}\Phi }K_{2}(X^{\mu }(z,v))}%
\text{ \ \ ,}  \tag{10.87}
\end{equation}
from where the function $K_{1}$ can be expressed and substituted into
expression (10.86) for $\sqrt{-g}_{vis}$. Taking into account again equality
(10.68) for $\sqrt{-g}$, one obtains
\begin{equation}
\sqrt{-g}_{vis}=\sqrt{-g}\frac{1}{\widetilde{L}_{2}^{3}}=\frac{\sqrt{%
K_{2}^{0}}}{e^{2kr_{-}\pi }}\text{ \ \ \ .}  \tag{10.88}
\end{equation}
Therefore, the volume element of the "visible" brane is a constant, while
the real volume element is $\widetilde{L}_{2}^{3}$ times the volume of the
"visible" brane.

In this case, to any mass $m_{0}$ on the "visible" brane would correspond $l$
in number physical masses, determined by the formulae
\begin{equation}
m^{(l)}=m_{0}f^{(l)}=m_{0}\sqrt{\widetilde{L}_{2}^{(l)}(z,v)\text{ }}\text{
\ \ \ ,}  \tag{10.89}
\end{equation}
where the function $\widetilde{L}_{2}(z,v)$ is given by (10.85). Therefore,
there will be a degeneracy of masses.

\subsubsection*{\protect\bigskip 10.2.7. TENSOR \ LENGTH \ SCALE, RESCALING
\ AND \ COMPACTIFICATION \ IN \ THE \ LOW \ ENERGY \ ACTION \ OF \ TYPE \ I
\ TEN - DIMENSIONAL \ STRING \ THEORY}

\bigskip Our next \ example of possible application of theories with
covariant and contravariant metric tensors is related to the low - energy
action of type I string theory in ten dimensions [47, 70, 71, 72]
\begin{equation}
S=\int d^{10}x\left( \frac{m_{s}^{8}}{(2\pi )^{7}\lambda ^{2}}R+\frac{1}{4}%
\frac{m_{s}^{6}}{(2\pi )^{7}\lambda }F^{2}+...\right) \text{ }=\int
d^{4}xV_{6}(......)\text{\ \ \ \ ,}  \tag{10.90}
\end{equation}
where $\lambda \sim \exp (\Phi )$ is the string coupling (as remarked in
[72], in the first term the coupling is $\lambda ^{2}$, because it is
generated by an world - sheet path integral on an sphere and the coupling $%
\lambda $ in the second term - by an world - sheet path integral on the
disc), $m_{s}$ is the string scale, which we can identify with $m_{grav.}$.
Compactifying to $4$ dimensions on a manifold of volume $V_{6}$, one can
identify the resulting coefficients in front of the $R$ and $\frac{1}{4}%
F^{2} $ terms with $M_{(4)}^{2}$ and $\frac{1}{g_{4}^{2}}$, from where one
obtains [47]
\begin{equation}
M_{(4)}^{2}=\frac{(2\pi )^{7}}{V_{6}m_{s}^{4}g_{4}^{2}}\text{ \ \ \ ; \ \ \
\ \ }\lambda =\frac{g_{4}^{2}V_{6}m_{s}^{6}}{(2\pi )^{7}}\text{ \ \ \ .}
\tag{10.91}
\end{equation}
The physical meaning of the performed identification is that since the
length scale $\sqrt{\alpha ^{^{\prime }}}$ of string theory, the volume $V$
of the (Calabi - Yau) manifold and the expectation value of the dilaton
field cannot be determined experimentally, \textbf{they can be adjusted in
such a way so that to give the desired values of the Newton's constant, the
GUT (Grand Unified Theory) scale }$M_{GUT}$\textbf{\ and the GUT\ coupling
constant [72].} It should be stressed that in the weakly coupled heterotic
string theory (when there are no different string couplings $\lambda \sim
\exp (2\Phi )$ and $\lambda \sim \exp (\Phi )$, but just one), the obtained
bound on the Newton's constant [72] $G_{N}\geq \frac{\alpha ^{\frac{4}{3}%
_{GUT}}}{M_{GUT}^{2}}$ is too large, but in the same paper [72] it was
remarked that \textbf{\textquotedblright the problem might be ameliorated by
considering an anisotropic Calabi - Yau with a scale }$\sqrt{\alpha
^{^{\prime }}}$\textbf{\ in }$d$\textbf{\ directions and }$\frac{1}{M_{GUT}}$%
\textbf{\ in }$(6-d)$\textbf{\ directions\textquotedblright . }

Now we shall propose, in the spirit of the affine geometry approach, how
such a different metric scale on the given manifold can be introduced by
defining more general contravariant tensors. The key idea is that the
contraction of the covariant metric tensor $g_{ij}$ with the contravariant
one $\widetilde{g}^{jk}=dX^{j}dX^{k}$ gives exactly (when $i=k$) the length
interval [10]
\begin{equation}
l=ds^{2}=g_{ij}dX^{j}dX^{i}\text{ \ \ .}  \tag{10.92}
\end{equation}
Naturally, for $i\neq k$ the contraction will give a tensor function $%
l_{i}^{k}=g_{ij}dX^{j}dX^{k}$, which can be interpreted as a
\textquotedblright tensor\textquotedblright\ length scale for the different
directions. In the spirit of the remark in [72], one can take for example
\begin{equation}
l_{i}^{k}=g_{ij}dX^{j}dX^{k}=L_{1}\delta _{i}^{k}\text{ \ for \ }%
i,j,k=1,....,d\text{ \ \ \ \ ,}  \tag{10.93a}
\end{equation}
\begin{equation}
l_{a}^{b}=g_{ac}dX^{c}dX^{b}=L_{2}\delta _{a}^{b}\text{ \ for \ }%
a,b,c=1,....,6-d\text{ \ \ \ \ .}  \tag{10.93b}
\end{equation}
For simplicity and as a starting point, further we shall assume that for all
indices $i,j,k...$
\begin{equation}
l_{i}^{k}=l\text{ }\delta _{i}^{k}\text{ \ \ .}  \tag{10.94}
\end{equation}
In fact, this will be fulfilled if we assume that the contravariant metric
tensor components $\widetilde{g}^{ij}$ are proportional to the usual inverse
contravariant metric tensor $g^{ij}$ with a function of proportionality $l(%
\mathbf{x})$, i.e. $\widetilde{g}^{ij}=l(\mathbf{x})g^{ij}$ (it will be
called a \textquotedblright conformal\textquotedblright\ rescaling). Also,
further in the next parts of this paper it will be shown that in the more
complicated case of different functions of proportionality for the different
components, the tensor length scale can be calculated uniquely too. Further
we shall call the function $l(\mathbf{x})$ \textquotedblright a length scale
function\textquotedblright .

\textbf{Our next purpose will be to prove that if one imposes the
requirement for invariance of the low - energy type I string action (10.90)
under the "conformal" rescaling, i.e. }
\begin{equation*}
S=\int d^{10}x\left( \frac{m_{s}^{8}}{(2\pi )^{7}\lambda ^{2}}\widetilde{R}+%
\frac{1}{4}\frac{m_{s}^{6}}{(2\pi )^{7}\lambda }\widetilde{F}^{2}\right)
=\int d^{4}xV_{6}\left( ....\right) =
\end{equation*}
\begin{equation}
=\int d^{4}x\left( M_{(4)}^{2}R+\frac{1}{4}\frac{1}{g_{4}^{2}}F^{2}\right)
\text{ \ \ \ \ ,}  \tag{10.95}
\end{equation}
\ \ t\textbf{hen the length scale }$l(x)$\textbf{\ will be possible to be
determined from a differential equqtion in partial derivatives. }In other
words, unlike the previously described in [47, 70, 71, 72] case, when the
coefficients in front of $R$ and $F^{2}$ \textbf{before} and \textbf{after}
the compactification are identified, here we shall propose another approach
to the same problem. Concretely, first a rescaling of the contravariant
metric components shall be performed, and \textbf{after that} the
compactification shall be realized, resulting again in the R.H.S. of the
standard $4D$ action (10.90).

However, in principle another approach is also possible. One may start from
the "unrescaled" ten - dimensional action (10.90), then perform a
compactification to the four - dimensional manifold and \textbf{afterwards }%
a transition to the usual "unrescaled" scalar quantities $R$ and $F^{2}.$
Thus it is required that the "unrescaled" ten - dimensional effective action
(10.90) (i.e. the L. H. S. of (10.90)) is equivalent to the four -
dimensional effective action after compactification, but in terms of the
rescaled quantities $\widetilde{R}$ and $\widetilde{F}^{2}$ in the R.H.S of
(10.90). This can be expressed as follows \
\begin{equation*}
S=\int d^{10}x\left( \frac{m_{s}^{8}}{(2\pi )^{7}\lambda ^{2}}R+\frac{1}{4}%
\frac{m_{s}^{6}}{(2\pi )^{7}\lambda }F^{2}\right) =\int d^{4}xV_{6}\left(
....\right) =
\end{equation*}
\begin{equation}
=\int d^{4}x\left( M_{(4)}^{2}\widetilde{R}+\frac{1}{4}\frac{1}{g_{4}^{2}}%
\widetilde{F}^{2}\right) \text{ \ \ \ \ .}  \tag{10.96}
\end{equation}
In the next subsections both cases shall be investigated, deriving the
corresponding (quasilinear) differential equations in partial derivatives
and moreover, finding concrete solutions of these equations for the special
case of the metric (10.35) of a flat $4D$ Minkowski space, embedded in a
five - dimensional $ADS$ space of constant negative curvature. It will be
shown also that for a definite scale factor $h(y)=\beta y^{n}$ ($\beta $ is
a constant) in front of the extra -coordinate $y$ in the metric, the derived
differential equations are still solvable, in spite of the fact that the
five - dimensional space is no longer of a constant negative curvature.
Besides the opportunity to extend the results to such spaces of non-constant
curvature, there is one more reason for the necessity to investigate such
quasilinear differential equations for concrete cases - in the next parts of
this paper examples will be given, when such equations cannot be explicitely
solved. \textbf{But evidently, some special kinds of metrics like (10.35)
will allow the solution of these equations and consequently the
determination of the scale length function }$l(x)$\textbf{\ in terms of all
the important parameters in the low - energy type I string thery action. If
for certain metrics this is possible , then it will turn out to be possible
to test whether there will be deviations from the standardly known
gravitational theory with }$l=1$\textbf{, if the electromagnetic coupling
constant }$g_{4}$\textbf{, the }$4D $\textbf{\ Planck constant }$M_{(4)}$%
\textbf{, the string scale }$m_{s}$\textbf{\ and the string coupling }$%
\lambda $\textbf{\ are known, presumably from future experiments or
cosmological data}. Even if one assumes that there no deviations from the
standard theory with $l=1$, the obtained solutions will allow to find some
new relations between the above mentioned parameters. It should be stressed
that this refers to the solutions of these equations, otherwise the obtained
differential equations in the limit of $l=1 $ will result in the simple
algebraic relations (10.91), already found in the literature.

One may also require the equivalence of the two approaches, expressed
mathematically by (10.95) and (10.96), although for the moment it is not
known whether there is some physical reason for this equivalence.

\subsubsection*{\protect\bigskip 10.2.8. ALGEBRAIC \ RELATION \ AND A \
QUASILINEAR \ DIFFERENTIAL \ EQUATION \ IN \ PARTIAL \ DERIVATIVES \ FROM \
THE \ "RESCALED + COMPACTIFIED" \ LOW - ENERGY TYPE I \ STRING \ ACTION}

\bigskip In order to rewrite the ''rescaled+compactified'' string action
(10.95), let us first define the ''rescaled'' square of the electromagnetic
field strength as
\begin{equation*}
\widetilde{F}^{2}=\widetilde{F}_{AB}\widetilde{F}^{AB}=F_{AB}\widetilde{g}%
^{AM}\widetilde{g}^{BN}F_{MN}=
\end{equation*}
\begin{equation}
=l^{2}F_{AB}g^{AM}g^{BN}F_{MN}=l^{2}F^{2}\text{ \ \ \ .}  \tag{10.97}
\end{equation}
Using the formulaes for the Riemann tensor and for the rescaled affine
connection
\begin{equation}
\widetilde{\Gamma }_{AC}^{D}=\widetilde{g}^{DG}g_{GF}\Gamma
_{AC}^{F}=l\Gamma _{AC}^{D}\text{ \ \ ,}  \tag{10.98}
\end{equation}
the rescaled scalar gravitational curvature $\widetilde{R}$ can be written
as
\begin{equation*}
\widetilde{R}=\widetilde{g}^{DG}\widetilde{g}_{GF}\widetilde{R}_{ABCD}=\frac{%
1}{2}l^{2}g^{AC}g^{BD}(g_{AD,BC}+g_{BC,AD}-g_{AC,BD}-
\end{equation*}
\begin{equation}
-g_{BD,AC})+l^{4}g^{AC}g^{BD}g_{FG}(\Gamma _{CB}^{F}\Gamma _{AD}^{G}-\Gamma
_{DB}^{F}\Gamma _{AC}^{G})=  \tag{10.99}
\end{equation}
\begin{equation}
=l^{4}R-\frac{1}{2}%
l^{2}(l^{2}-1)g^{AC}g^{BD}(g_{AD,BC}+g_{BC,AD}-g_{AC,BD}-g_{BD,AC})\text{ \
\ \ \ .}  \tag{10.100}
\end{equation}
Substituting the above expressions (10.97)\ and (10.100) for $\widetilde{F}%
^{2}$ and $\widetilde{R}$ into the L. H. S. of the low - energy string
action (10.95) and setting up equal the corresponding coefficients in front
of the $\frac{1}{4}F^{2}$ term in the L. H. S. and the R. H. S. of (10.95),
one can derive
\begin{equation}
\lambda =\frac{g_{4}m_{s}^{6}V_{6}}{(2\pi )^{7}}l^{2}\text{ \ \ \ .}
\tag{10.101}
\end{equation}
This is almost the same expression as in (10.91), but now corrected wih the
function of proportionality $l(\mathbf{x})$. The string coupling $\lambda $
is thus a non - local physical quantity, depending on the space - time
coordinates.

Next, after the elimination of the terms with $\frac{1}{4}F^{2}$ on both
sides of (10.95)\ and substituting the found formulae for $\lambda $ into
the resulting expression on both sides of (10.95), one derives the algebraic
relation
\begin{equation}
\left[ \frac{(2\pi )^{7}}{V_{6}m_{s}^{4}g_{4}^{4}}-M_{(4)}^{2}\right] R=%
\frac{(2\pi )^{7}(l^{2}-1)}{2m_{s}^{4}V_{6}l^{2}g_{4}^{2}}g^{AC}g^{BD}(....)%
\text{ \ \ .}  \tag{10.102}
\end{equation}
For brevity, the brackets $(....)$ will denote the term in (10.100)\ with
the second derivatives of the metric tensor. For $l=1$, as expected, we
obtain the usual relation for $M_{(4)}^{2}$ as in (10.91). Therefore,
physically any possible deviations from relation (10.91) can be attributed
to the appearence of the new length scale $l(\mathbf{x})$. Let us introduce
the notation
\begin{equation}
\beta \equiv \left[ \frac{(2\pi )^{7}}{V_{6}m_{s}^{4}g_{4}^{4}}-M_{(4)}^{2}%
\right] m_{s}^{4}V_{6}\frac{2}{(2\pi )^{7}}  \tag{10.103}
\end{equation}
and assume that the deviation from the relation $M_{(4)}^{2}=\frac{(2\pi
)^{7}}{V_{6}m_{s}^{4}g_{4}^{2}}$ is small, i.e. $\beta \ll 1$. Moreover, the
number $(2\pi )^{7}$ in the denominator of (10.103) is great, so one can
expect that $\beta $ is really a small quantity, in spite of the fact that
the string scale $m_{s}$ (remember that in [47] it was set up $\sim 1$) and
the compactification volume $V_{6}$ are not known. Then the lenght scale $%
l(x)$ can be expressed from the algebraic relation (10.102) as
\begin{equation}
l^{2}=\frac{1}{1-\beta \frac{R}{g^{AC}g^{BD}(...)}}\approx 1+\beta \frac{R}{%
g^{AC}g^{BD}(...)}\text{ \ \ \ .}  \tag{10.104}
\end{equation}
Consequently the deviation from the \textquotedblright
standard\textquotedblright\ length scale $l=1$ in the case of a
gravitational theory with $l\neq 1$ in the case of small $\beta $ shall be
proportional to the ratio $\frac{R}{g^{AC}g^{BD}(...)}$. In the concrete
example of an $4D$ Minkowski space, embedded in a $5D$ $ADS$ space of
constant negative curvature, this ratio will be
\begin{equation}
\frac{R}{g^{AC}g^{BD}(...)}=\frac{(-8k^{2})}{(-32k^{2})}=\frac{1}{4}\text{ \
}  \tag{10.105}
\end{equation}
and therefore, this constant factor will not affect the smallness of the
number $\beta $.

Let us now derive the differential equation in partial derivatives, starting
from the second representation of the \textquotedblright
rescaled\textquotedblright\ scalar gravitational curvature $\widetilde{R}$ $%
\ $by means of the \textquotedblright rescaled\textquotedblright\ Ricci
tensor $\widetilde{R}_{ij}$
\begin{equation}
\widetilde{R}=\widetilde{g}^{AB}\widetilde{R}_{AB}=lg^{AB}\left[ \frac{%
\partial \widetilde{\Gamma }_{AB}^{C}}{\partial x^{C}}-\frac{\partial
\widetilde{\Gamma }_{AC}^{C}}{\partial x^{B}}+\widetilde{\Gamma }_{AB}^{C}%
\widetilde{\Gamma }_{CD}^{D}-\widetilde{\Gamma }_{AC}^{D}\widetilde{\Gamma }%
_{BD}^{C}\right] \text{ \ \ \ .}  \tag{10.106}
\end{equation}
It can easily be found that the rescaled gravitational curvature is
expressed through the usual one as
\begin{equation*}
\widetilde{R}=\widetilde{g}^{AB}\widetilde{R}_{AB}=l^{2}R+l^{2}(l-1)g^{AB}%
\left( \Gamma _{AB}^{C}\Gamma _{CD}^{D}-\Gamma _{AC}^{D}\Gamma
_{BD}^{C}\right) +
\end{equation*}
\begin{equation}
+l\text{ }\frac{\partial l}{\partial x^{C}}g^{AB}\Gamma _{AB}^{C}-l\frac{%
\partial l}{\partial x^{B}}g^{AB}\Gamma _{AC}^{C}\text{ \ \ \ .}
\tag{10.107}
\end{equation}
Again, this expression and also (10.97) for $\widetilde{F}^{2}$ are
substituted into the L. H. S. of the action (10.95) and the corresponding
coefficients in front of the term $\frac{1}{4}F^{2}$ in the L. H. S. and the
R. H. S. of (10.91)\ are set up equal. Thus one obtains
\begin{equation}
\lambda ^{2}=\frac{g_{4}^{4}m_{s}^{12}V_{6}l^{4}}{(2\pi )^{14}}\text{ \ \ .}
\tag{10.108}
\end{equation}
Substituting this expression into the resulting one on both sides of
(10.91), we receive the following equation in partial derivatives in respect
to the scale function $l(x)$:
\begin{equation*}
\left[ \frac{(2\pi )^{7}}{m_{s}^{4}V_{6}g_{4}^{4}l^{2}}-M_{4}^{2}\right] R+%
\frac{(2\pi )^{7}}{m_{s}^{4}V_{6}g_{4}^{4}}\frac{(l-1)}{l^{2}}g^{AB}\left(
\Gamma _{AB}^{C}\Gamma _{CD}^{D}-\Gamma _{AC}^{D}\Gamma _{BD}^{C}\right) +
\end{equation*}
\begin{equation}
+\frac{(2\pi )^{7}}{m_{s}^{4}V_{6}g_{4}^{4}}\frac{1}{l^{3}}\left[ \frac{%
\partial l}{\partial x^{C}}g^{AB}\Gamma _{AB}^{C}-\frac{\partial l}{\partial
x^{B}}g^{AB}\Gamma _{AC}^{C}\right] =0\text{ \ \ .}  \tag{10.109}
\end{equation}
Note that for $l=1$ (the known gravitational theory) we obtain again
expression (10.91)\ for $M_{4}^{2}$. It turns out that for $l(x)$ we have
both the algebraic relation (10.102)\ and the above differential equation
(10.109).

Now there are two possibilities:\

First possibility: The ten - dimensional space - time is represented as a
factorized product of the compactification manifold $V_{6}$ and the
remaining four - dimensional spacetime $K^{(4)}$, i.e. $V_{6}\times K^{(4)}$%
. Then the coordinates of $V_{6}$ are independent from the coordinates of $%
K^{(4)}$ and consequently the derivatives of $l$, if calculated from the
found algebraic relation \ (10.104), will depend not on the compactification
volume, but on the ratio $\frac{R}{g^{AC}g^{BD}(...)}$ :
\begin{equation}
\frac{\partial l}{\partial x^{B}}=\frac{\beta }{2}l^{3}\frac{\partial }{%
\partial x^{B}}\left( \frac{R}{g^{AC}g^{BD}(...)}\right) \text{ \ .}
\tag{10.110}
\end{equation}
If the above expression and also formulae (10.104) for $l$ are substituted
into the differential equation (10.109), then an algebraic relation in
respect to $V_{6}$ is obtained, depending on the parameters in the initial
string action (without the string coupling constant $\lambda $).

Second possibility: The ten - dimensional spacetime cannot be represented as
factorized product and therefore the compactification volume $V_{6}$ depends
on the coordinates of thee four - dimensional spacetime. Then an additional
term $\frac{1}{2}\left( \frac{R}{g^{AC}g^{BD}(...)}\right) l^{3}\frac{%
\partial \beta }{\partial x^{B}}$ has to be added to the derivative
expression for $\frac{\partial l}{\partial x^{B}}$ in (10.110). Substituting
into (10.109), a nonlinear differential equation in partial derivatives will
be obtained in respect to the compactification volume.

\subsubsection*{\protect\bigskip 10.2.9. (ANOTHER) \ ALGEBRAIC \ RELATION \
AND A \ QUASILINEAR \ DIFFERENTIAL \ EQUATION \ FROM \ THE \
"COMPACTIFIED+RESCALED" \ LOW \ ENERGY \ TYPE \ I \ STRING \ THEORY \ ACTION}

\bigskip This time we start from the action (10.96) and substitute the
expressions for the "unrescaled" scalar quantities $F^{2}$ and $R$
\begin{equation}
F^{2}=\frac{1}{l^{2}}\widetilde{F}^{2}\text{ \ \ \ ; \ \ \ \ }R=\frac{1}{%
l^{4}}\widetilde{R}+\frac{(l^{2}-1)}{2l^{2}}g^{AC}g^{BD}(...)  \tag{10.111}
\end{equation}
into the L. H. S. of (10.96).

Following the method, described in the previous section, we find for $%
\lambda $
\begin{equation}
\lambda =\frac{g_{4}m_{s}^{6}V_{6}}{(2\pi )^{7}l^{2}}\text{ \ \ \ ,}
\tag{10.112}
\end{equation}
which in respect to the function $l(x)$ can be considered as the "dual" one,
if compared with (10.101). However, in comparison with (10.102), the
obtained algebraic relation will be different
\begin{equation}
\left[ \frac{(2\pi )^{7}}{V_{6}m_{s}^{4}g_{4}^{4}}-M_{(4)}^{2}\right]
\widetilde{R}+\frac{(2\pi )^{7}l^{2}(l^{2}-1)}{2m_{s}^{4}V_{6}g_{4}^{4}}%
g^{AC}g^{BD}(....)=0\text{ \ \ .}  \tag{10.113}
\end{equation}
If again expression (10.100) for $\widetilde{R}$ is used, the algebraic
relation (10.113) can be rewritten as
\begin{equation}
\frac{1}{2}l^{2}(l^{2}-1)g^{AC}g^{BD}(....)=\frac{P^{2}R}{(P-NR)^{2}}+l^{4}R%
\text{ \ \ \ \ \ ,}  \tag{10.114}
\end{equation}
where $P$ and $N$ denote the expressions
\begin{equation}
P\equiv \frac{(2\pi )^{7}}{2m_{s}^{4}V_{6}g_{4}^{4}}g^{AC}g^{BD}(...)\text{
\ \ \ ; \ \ \ \ }N\equiv \frac{(2\pi )^{7}}{m_{s}^{4}V_{6}g_{4}^{4}}%
-M_{4}^{2}\text{ \ \ .}  \tag{10.115}
\end{equation}
In the same way, starting from the second representation (10.107) of the
gravitational Lagrangian for $R$ in terms of $\ \widetilde{R}$ \ and again
making use of formulae (10.107) for $\widetilde{R}$, one can obtain the
\textbf{second quasilinear equation in partial derivatives }
\begin{equation*}
\left[ \frac{(2\pi )^{7}}{m_{s}^{4}V_{6}g_{4}^{4}}l^{6}-M_{4}^{2}l^{4}\right]
R-\left[ \frac{(2\pi )^{7}l^{2}}{m_{s}^{4}V_{6}g_{4}^{4}}-M_{4}^{2}\right]
\frac{l^{2}(l^{2}-1)}{2}g^{AC}g^{BD}(...)-
\end{equation*}
\begin{equation*}
-\frac{(2\pi )^{7}l^{4}(l-1)}{m_{s}^{4}V_{6}g_{4}^{4}}g^{AB}\left( \Gamma
_{AB}^{C}\Gamma _{CD}^{D}-\Gamma _{AC}^{D}\Gamma _{BD}^{C}\right) -
\end{equation*}
\begin{equation}
-\frac{(2\pi )^{7}l^{3}}{m_{s}^{4}V_{6}g_{4}^{4}}\left( \frac{\partial l}{%
\partial x^{C}}g^{AB}\Gamma _{AB}^{C}-\frac{\partial l}{\partial x^{B}}%
g^{AB}\Gamma _{AC}^{C}\right) =0\text{ \ \ \ }.  \tag{10.116}
\end{equation}
Substituting the algebraic relation (10.114) into the second term of
(10.116), the differential equation is obtained in a simpler form
\begin{equation*}
\frac{(2\pi )^{7}l^{3}}{m_{s}^{4}V_{6}g_{4}^{4}}\left( \frac{\partial l}{%
\partial x^{C}}g^{AB}\Gamma _{AB}^{C}-\frac{\partial l}{\partial x^{B}}%
g^{AB}\Gamma _{AC}^{C}\right) +\frac{RP^{2}}{l^{3}(P-NR)^{2}}\left[ \frac{%
(2\pi )^{7}l^{2}}{m_{s}^{4}V_{6}g_{4}^{4}}-M_{4}^{2}\right] +
\end{equation*}
\begin{equation}
+\frac{(2\pi )^{7}l(l-1)}{m_{s}^{4}V_{6}g_{4}^{4}}g^{AB}\left( \Gamma
_{AB}^{C}\Gamma _{CD}^{D}-\Gamma _{AC}^{D}\Gamma _{BD}^{C}\right) =0\text{ \
\ \ \ . }  \tag{10.117}
\end{equation}
This (second) differential equation evidently is different from the first
one (10.109) and in this aspect an interesting conclusion can be made.
Suppose that the two differential equations (10.109) and (10.117)
simultaneously hold, which means that it does not matter whether we perform
"rescaling + compactification" or "compactification +rescaling" in the low
energy type I string theory action. Then, if the initial term with the
derivatives in (10.117) is expressed and substituted into the first
differential equation (10.109), then the square of the length scale function
$l^{2}$ can be found as a solution of the following algebraic equation
\begin{equation}
M_{4}^{2}l^{6}-\frac{(2\pi )^{7}}{m_{s}^{4}V_{6}g_{4}^{4}}l^{4}+\frac{(2\pi
)^{7}}{m_{s}^{4}V_{6}g_{4}^{4}}\frac{P^{2}}{(P-NR)^{2}}l^{2}-M_{4}^{2}\frac{%
P^{2}}{(P-NR)^{2}}=0\text{ \ \ .}  \tag{10.118}
\end{equation}

\textbf{\bigskip }For the moment, this expression shall be used. Since the
functions in the third and the fourth term depend on the function $l$, it
would become clear later on, that the equation is a cubic one.

Both the above mentioned approaches of " rescaling + compactification " and
"compactification + rescaling" would be consistent in the case of a non -
imaginary Lobachevsky space, if the function $l(\mathbf{x})$ is a real one
and not a complex one, i.e. the roots of the above equation should not be
imaginary functions and there should be \textbf{at least} one root, which is
a real function. This may lead additionallly to some restrictions on the
parameters in the initial string action. Also, for $l=1$ the above equation
can be written as
\begin{equation}
\left[ M_{4}^{2}-\frac{(2\pi )^{7}}{m_{s}^{4}V_{6}g_{4}^{4}}\right] \left[
\frac{P^{2}}{(P-NR)^{2}}+1\right] =0\text{ \ \ .}  \tag{10.119}
\end{equation}

\bigskip There is no other relation from this equation besides the known one
$M_{4}^{2}-\frac{(2\pi )^{7}}{m_{s}^{4}V_{6}g_{4}^{4}}=0$, since the
nominator of the second term can be written as
\begin{equation}
2\left[ \left( \frac{P}{N}\right) ^{2}-\left( \frac{P}{N}\right) R+\frac{1}{2%
}R^{2}\right] =\frac{1}{2}\left[ \left( \frac{P}{N}-\frac{R}{2}\right) ^{2}+%
\frac{R^{2}}{4}\right] \text{ \ \ \ \ }  \tag{10.120}
\end{equation}
and evidently this term is positive and different from zero.

However, if solutions of the two quasilinear differential equations are
found, then some new relations may be written. It should be kept in mind
that these solutions are found by means of the characteristic system of
equations, and the general solutions depend on the solutions of the
characteristic system.

\subsubsection*{\protect\bigskip 10.2.10. ALGEBRAIC \ INEQUALITIES \ FOR \
THE \ PARAMETERS \ IN \ THE \ LOW - ENERGY \ TYPE \ I STRING \ THEORY \
ACTION \ \ }

\bigskip Taking into account expressions (10.115) for $P$ and $N$,
eq.(10.118) after dividing by $Q^{2}M_{4}^{2}m_{s}^{4}V_{6}g_{4}^{2}$ can be
written in the form of the following cubic algebraic equation in respect to
the variable $l_{1}=l^{2}$
\begin{equation}
l_{1}^{3}+a_{1}l_{1}^{2}+a_{2}l_{1}+a_{3}=0\text{ \ \ \ \ ,}  \tag{10.121}
\end{equation}
where $Q$, $a_{1}$, $a_{2}$ and $a_{3}$ denote the expressions
\begin{equation}
Q\equiv \frac{g^{AC}g_{BD}(2\pi )^{7}g_{4}^{4}(....)}{\left[
g^{AC}g^{BD}(...)(2\pi )^{7}g_{4}^{4}-2R\left( (2\pi
)^{7}-M_{4}^{2}V_{6}m_{s}^{4}g_{4}^{4}\right) \right] }\text{ \ \ \ ,}
\tag{10.122}
\end{equation}
\begin{equation}
a_{1}\equiv -\frac{(2\pi )^{7}}{M_{4}^{2}m_{s}^{4}V_{6}g_{4}^{2}}\text{ \ \
\ ; \ \ \ \ }a_{2}\equiv \frac{(2\pi )^{7}}{%
M_{4}^{2}m_{s}^{4}V_{6}g_{4}^{2}Q^{2}}\text{ \ \ ; \ \ }a_{3}\equiv -\frac{%
g_{4}^{2}}{Q^{2}}\text{ \ \ .}  \tag{10.123}
\end{equation}
After the variable change $l_{1}=x-\frac{a_{1}}{3}$ equation (10.121) is
brought to the form
\begin{equation}
x^{3}+ax+b=0\text{ \ \ , }  \tag{10.124}
\end{equation}
where $a$ and $b$ are the expressions
\begin{equation}
a\equiv a_{2}-\frac{a_{1}^{2}}{3}\text{ \ \ ; \ \ \ }b\equiv 2\frac{a_{1}^{3}%
}{27}-\frac{a_{1}a_{2}}{3}+a_{3}\text{ \ \ \ .}  \tag{10.125}
\end{equation}
The roots of the cubic equation (10.124) are given by the formulae [9]
\begin{equation}
x=\sqrt[3]{p}-\frac{a}{3\sqrt[3]{p}}\text{ \ \ \ ,}  \tag{10.126}
\end{equation}
where $p$ denotes the expression
\begin{equation}
p\equiv -\frac{b}{2}\pm \sqrt{\frac{b^{2}}{4}+\frac{a^{3}}{27}}\text{ \ \ \ .%
}  \tag{10.127}
\end{equation}
The roots of the cubic equation will not depend on the $+$ $\ $or $-$ sign
in front of the square in the above expression.

It may be noted that if the expression for $\frac{b^{2}}{4}+\frac{a^{3}}{27}$
is negative, then the corresponding roots $x_{1}$, $x_{2}$, $x_{3}$ and the
length function $l(x)$ will be imaginary. From a physical point of view,
this would be unacceptable, but with one exception - in the imaginary
Lobachevsky space [73], which is realized by all the straight lines outside
the absolute cone (on which the scalar product is zero, i.e. $[x,x]=0$), the
length may may take imaginary values in the interval $[0,\frac{\pi \text{ }i%
}{2k}]$ ($k$ is the Lobachevsky constant). Further we shall assume that $%
l(x) $ is a real function, but in principle it is interesting that the sign
of the inequalities, relating the paeameters in the string action, may
change, if the spacetime is an imaginary Lobachevsky one.

The expression for $\frac{b^{2}}{4}+\frac{a^{3}}{27}$ can be written as
\begin{equation*}
\frac{b^{2}}{4}+\frac{a^{3}}{27}=\frac{1}{Q^{6}d^{6}}[\frac{1}{27}%
g_{4}^{6}d^{3}-\frac{1}{4.27}g_{4}^{8}d^{2}Q^{2}-\frac{2}{27^{2}}%
g_{4}^{12}Q^{6}+
\end{equation*}
\begin{equation}
+\frac{1}{4}g_{4}^{2}Q^{2}-\frac{1}{6}g_{4}^{6}Q^{2}d^{4}+\frac{1}{27}%
g_{4}^{8}Q^{4}d^{3}]\text{ \ \ \ ,}  \tag{10.128}
\end{equation}
where $d$ is the introduced notation for
\begin{equation}
d\equiv \frac{M_{4}^{2}V_{6}m_{s}^{4}g_{4}^{4}}{\left( 2\pi \right) ^{7}}%
\text{ \ \ \ .}  \tag{10.129}
\end{equation}
It is difficult to check when expression (10.128) will be non - negative,
since $Q$ depends also on $d$ and a higher - degree polynomial in respect to
$d$ will be obtained. However, it may be noted that since
\begin{equation}
l^{2}=l_{1}=x-\frac{a_{1}}{3}>0\text{ \ \ }  \tag{10.130}
\end{equation}
and since $a_{1}$ is a non - complex quantity, then all the roots $x_{1}$, $%
x_{2}$, $x_{3}$ are real. Therefore from the Wiet formulae
\begin{equation}
-a=x_{1}+x_{2}+x_{3}>a_{1}  \tag{10.131}
\end{equation}
and with account of the expressions for $a$ and $a_{1}$ an equality can be
obtained in respect to $d$
\begin{equation}
\frac{1}{3}g_{4}^{2}>W(W+2)d^{4}-2W(W+1)d^{3}+W^{2}d^{2}\text{ \ \ \ \ \ ,}
\tag{10.132}
\end{equation}
where $W$ is the notation for
\begin{equation}
W\equiv \frac{2R}{g^{AB}g^{CD}(...)g_{4}^{4}}\text{ \ \ \ .}  \tag{10.133}
\end{equation}
The last (third) inequality with the parameters of the low - energy type I
string theory action can be obtained from the restriction (10.130) $x>\frac{%
a_{1}}{3}$ for the roots of the cubic equation and expression (10.126) for $%
x $
\begin{equation}
\frac{3\sqrt[3]{p^{2}}-a}{3\sqrt[3]{p}}>\frac{a_{1}}{3}\text{ \ \ \ .}
\tag{10.134}
\end{equation}
Denoting
\begin{equation}
q_{1}=\sqrt[3]{p^{2}}\text{ \ \ ,}  \tag{10.135}
\end{equation}
the above inequality can be rewritten as
\begin{equation}
9q_{1}^{2}-(a_{1}^{2}+6a)q_{1}+a^{2}>0\text{ \ \ \ \ \ \ ,}  \tag{10.136}
\end{equation}
which is satisfied for
\begin{equation}
p^{2}=\frac{b^{2}}{2}+\frac{a^{3}}{27}-b\sqrt{\frac{b^{2}}{2}+\frac{a^{3}}{27%
}}>\left[ \frac{a_{1}+6a}{18}+\frac{a_{1}}{18}\sqrt{a_{1}^{2}+12a}\right]
^{3}  \tag{10.137}
\end{equation}
or for
\begin{equation}
p^{2}=\frac{b^{2}}{2}+\frac{a^{3}}{27}-b\sqrt{\frac{b^{2}}{2}+\frac{a^{3}}{27%
}}<\left[ \frac{a_{1}+6a}{18}-\frac{a_{1}}{18}\sqrt{a_{1}^{2}+12a}\right]
^{3}\text{ \ \ .}  \tag{10.138}
\end{equation}

\subsubsection*{\protect\bigskip 10.2.11 SOLUTIONS \ OF \ THE \ FIRST \
QUASILINEAR \ DIFFERENTIAL \ EQUATION \ (10.109) \ FOR \ THE \ CASE \ OF \ A
\ FLAT \ $4D$ \ MINKOWSKI \ METRIC, \ EMBEDDED \ IN \ A \ FIVE - DIMENSIONAL
SPACE (OF \ CONSTANT \ NEGATIVE \ OR \ NON - CONSTANT \ CURVATURE)\ }

\bigskip The purpose will be to show that the differential equation in
partial derivatives (10.109) will be solvable for the previously considered
case of the metric (10.12), written now as
\begin{equation}
ds^{2}=e^{-2k\epsilon y}\eta _{\mu \nu }dx^{\mu }dx^{\nu }+h(y)dy^{2}\text{
\ \ \ \ \ \ ,}  \tag{10.139}
\end{equation}
$\eta _{\mu \nu }=(+,-,-,-)$ with $h(y)=1$ and $\epsilon =\pm 1$. Moreover,
the equation will be solvable also for the case of a power - like dependence
of the scale factor $h(y)=\gamma y^{n}$ ($\gamma =const$), when the five -
dimensional scalar curvature is no longer a constant one. In principle, it
is necessary to know for what kind of metrics quasilinear differential
equations of the type (10.109) admit exact analytical solutions, since it
may be shown that for more complicated metrics (for example, when the
embedded four - dimensional spacetime is a Schwarzschild Black hole with an
warp factor), such analytical solutions cannot be found in the sense that
some algebraic relations can be found but the unknown function cannot be
expressed from them. This will be shown also in the next parts of this paper.

As usual, the Greek indices $\mu ,\nu ,\alpha ,\beta $ will run from $1$ to $%
4$ and the extra - dimensional metric tensor component is $h(y)\equiv g_{55}$%
. The big letters $A,B,C...$ will denote the coordinates of the five -
dimensional spacetime.

The corresponding affine connection components are
\begin{equation}
\Gamma _{\mu \nu }^{\alpha }=\Gamma _{5\nu }^{\alpha }=0\text{ \ \ \ ; \ \ }%
\Gamma _{\mu \nu }^{5}=\frac{k\varepsilon }{h}\eta _{\mu \nu
}e^{-2k\varepsilon y}\text{ \ \ \ ,}  \tag{10.140}
\end{equation}
\begin{equation}
\Gamma _{\mu 5}^{5}=0\text{ \ ;\ \ }\Gamma _{55}^{5}=\frac{1}{2}\frac{%
h^{^{\prime }}(y)}{h(y)}\text{ \ \ ; \ \ }\Gamma _{55}^{\alpha }=\frac{1}{2}%
e^{2k\varepsilon y}\eta _{\alpha \alpha }h^{^{\prime }}(y)\text{ \ \ \ .}
\tag{10.141}
\end{equation}
The expressions for the scalar curvatureb$R$ and for $g^{AB}(\Gamma
_{AB,C}^{C}-\Gamma _{AC,B}^{C})$ are
\begin{equation}
R=-\frac{8k^{2}}{h}-4k\epsilon \frac{h^{^{\prime }}}{h^{2}}\text{ }%
=g^{AB}(\Gamma _{AB,C}^{C}-\Gamma _{AC,B}^{C})\text{\ \ ,}  \tag{10.142}
\end{equation}
from where, taking the difference of the two expressions, it is found that $%
g^{AB}(\Gamma _{AB}^{C}\Gamma _{CD}^{D}-\Gamma _{AC}^{D}\Gamma _{BD}^{C})=0$
and the differential equation (10.109) is written as
\begin{equation*}
\frac{(2\pi )^{7}}{m_{s}^{4}V_{5}g_{5}^{5}}\frac{e^{2k\epsilon y}h^{^{\prime
}}}{2h}\left[ \frac{\partial l}{\partial x^{1}}-\frac{\partial l}{\partial
x^{2}}-\frac{\partial l}{\partial x^{3}}-\frac{\partial l}{\partial x^{4}}%
\right] +
\end{equation*}
\begin{equation}
+\frac{(2\pi )^{7}4k\epsilon }{m_{s}^{4}V_{5}g_{5}^{4}h}\frac{\partial l}{%
\partial y}=Cl-Dl^{3}\text{ \ ,}  \tag{10.143}
\end{equation}
where $C$ and $D$ denote the expressions
\begin{equation}
C\equiv \frac{(2\pi )^{7}4k(2kh+\epsilon h^{^{\prime }})}{%
m_{s}^{4}V_{5}g_{5}^{4}h^{2}}\text{ \ \ ; \ \ }D\equiv \frac{%
M_{5}^{2}4k(2kh+\epsilon h^{^{\prime }})}{h^{2}}\text{ \ \ \ .}  \tag{10.144}
\end{equation}
The characteristic system of equations for the equation (10.143) is
\begin{equation}
\frac{dl}{Cl-Dl^{3}}=\frac{\varepsilon m_{s}^{4}V_{5}g_{5}^{4}h}{(2\pi
)^{7}4k}dy=d\sigma \text{ \ \ \ ,}  \tag{10.145}
\end{equation}
\begin{equation}
\frac{2m_{s}^{4}V_{5}g_{5}^{4}e^{-2k\varepsilon y}}{(2\pi )^{7}(\ln
h)^{^{\prime }}}dx^{1}=-\frac{2m_{s}^{4}V_{5}g_{5}^{4}e^{-2k\varepsilon y}}{%
(2\pi )^{7}(\ln h)^{^{\prime }}}dx^{i}=d\sigma \text{ \ \ ,}  \tag{10.146}
\end{equation}
where the indice $i=2,3,4$ and $\sigma $ is some parameter. The solution of
the first characteristic equation for the $y$ and $l$ variables is
\begin{equation}
\frac{\varepsilon _{1}l}{\left[ \varepsilon _{2}(l^{2}-\alpha _{1}^{2})%
\right] ^{\frac{1}{2}}}=C_{1}(x_{1},x_{i})e^{2k\varepsilon _{3}y}h\text{ \ \
\ \ ,}  \tag{10.147}
\end{equation}
where
\begin{equation}
\alpha _{1}=\sqrt{\frac{C}{D}}\text{ \ \ \ \ }  \tag{10.148}
\end{equation}
and $\varepsilon _{1}$, $\varepsilon _{2}$, $\varepsilon _{3}$ take values $%
\pm 1$ independently one from another.

In order to find the function $C_{1}(x_{1},x_{i})$, the obtained solution
(10.147) should be differentiated by $x_{1}$ and the characteristic
equations for $\frac{\partial l}{\partial y}$ and $\frac{\partial y}{%
\partial x_{1}}$ have to be used. As a result, the function $%
C_{1}(x_{1},x_{i})$ is found as a solution of the following simple
differential equation
\begin{equation}
M=S\frac{\partial C_{1}(x_{1},x_{i})}{\partial x_{1}}+TC_{1}(x_{1},x_{i})%
\text{ \ \ \ \ ,}  \tag{10.149}
\end{equation}
where the functions $M$, $S$ and $T$ are defined as follows
\begin{equation}
M\equiv -\frac{\varepsilon _{1}(2l^{2}-\alpha
_{1}^{2})l8kM_{5}^{2}(2kh+\varepsilon _{3}h^{^{\prime
}})m_{s}^{4}V_{5}g_{5}^{4}}{\left[ \varepsilon _{2}(l^{2}-\alpha _{1}^{2})%
\right] ^{\frac{1}{2}}(2\pi )^{7}hh^{^{\prime }}e^{2k\varepsilon _{3}y}}%
\text{ \ for \ }\varepsilon _{1}\varepsilon _{2}=-1\text{ \ ,}  \tag{10.150}
\end{equation}
\begin{equation}
M\equiv \frac{\varepsilon _{1}l\alpha _{1}^{2}8kM_{5}^{2}(2kh+\varepsilon
_{3}h^{^{\prime }})m_{s}^{4}V_{5}g_{5}^{4}}{\left[ \varepsilon
_{2}(l^{2}-\alpha _{1}^{2})\right] ^{\frac{1}{2}}(2\pi )^{7}hh^{^{\prime
}}e^{2k\varepsilon _{3}y}}\text{ \ for \ \ }\varepsilon _{1}\varepsilon
_{2}=+1\text{ \ \ \ \ ,}  \tag{10.151}
\end{equation}
\begin{equation}
S\equiv he^{2k\varepsilon _{3}y}\text{ \ \ ; \ \ \ }T\equiv \frac{%
8k(2k+\varepsilon _{3}h^{^{\prime }})}{h^{^{\prime }}}\text{ \ \ \ \ .}
\tag{10.152}
\end{equation}
The solution of the differential equation (10.149) can be written as
\begin{equation}
C_{1}(x_{1},x_{i})=\frac{M}{T}-\varepsilon _{4}\frac{C_{2}(x_{i})}{T}%
e^{-\int \frac{T}{S}dx_{1}}\text{ \ \ .}  \tag{10.153}
\end{equation}
Again, the obtained solution (10. 147) can be differentiated in respect to $%
x_{i}$ and taking into account from the characteristic equations that
\begin{equation}
\frac{\partial l}{\partial x_{i}}=-\frac{\partial l}{\partial x_{1}}\text{ \
\ \ ; \ \ \ }\frac{\partial y}{\partial x_{i}}=-\frac{\partial y}{\partial
x_{1}}\text{ \ \ \ ,}  \tag{10.154}
\end{equation}
the solution of the corresponding equation (10.149) with $\widetilde{M}=-M$,
$\widetilde{T}=-T$, $\widetilde{S}=S$ can be represented as
\begin{equation}
C_{1}(x_{1},x_{i})=\frac{M}{T}+\varepsilon _{4}\frac{C_{3}(x_{1})}{T}e^{\int
\frac{T}{S}dx_{i}}\text{ \ \ .}  \tag{10.155}
\end{equation}
Substracting the two relations (10.153) and (10.155), the following relation
between the functions $C_{2}(x_{i})$ and $C_{3}(x_{1})$ can be derived
\begin{equation}
C_{2}(x_{i})=e^{\int \frac{T}{S}dx_{1}}C_{3}(x_{1})e^{\int \frac{T}{S}dx_{i}}%
\text{ \ \ \ \ .}  \tag{10.156}
\end{equation}
Now let us differentiate relation (10.155) in respect to $x_{1}$ and make
use of (10.156) and its derivative in respect to $x_{1}$. Then the following
differential equation is derived in respect to the function $C_{3}(x_{1})$
\begin{equation}
\frac{\partial C_{3}(x_{1})}{\partial x_{1}}-C_{3}(x_{1})\frac{T}{S}=0\text{
\ \ \ \ ,}  \tag{10.157}
\end{equation}
from where
\begin{equation}
C_{3}(x_{1})=const.\text{ }e^{^{\int \frac{T}{S}dx_{1}}}\text{ \ \ .}
\tag{10.158}
\end{equation}
Therefore from (10.155)
\begin{equation}
C_{1}(x_{1},x_{i})=\frac{M+\varepsilon _{4}}{T}  \tag{10.159}
\end{equation}
and substituting into (10.147), a final expression for $l$ can be found (for
the case $\varepsilon _{1}\varepsilon _{2}=+1$)
\begin{equation}
l^{2}=\frac{\varepsilon _{2}\alpha _{1}^{2}e^{4\varepsilon _{3}ky}h^{2}}{%
\varepsilon _{2}h^{2}e^{4\varepsilon _{3}ky}-\frac{T^{2}}{(M+\varepsilon
_{4})^{2}}}\text{ \ \ \ .}  \tag{10.160}
\end{equation}
The solution for the other case $\varepsilon _{1}\varepsilon _{2}=-1$ can be
found analogously.

The general solution of the quasilinear differential equation in partial
derivativatives will be not only expression (10.160), but also any function $%
V$, depending on the first integrals $K_{1},K_{2},K_{3}$,....,$K_{6}$ of the
characteristic system of equations [74]
\begin{equation}
V=V(K_{1},K_{2},K_{3},K_{4},K_{5},K_{6})\text{ \ \ \ \ .}  \tag{10.161}
\end{equation}
Now let us find a solution of the characteristic equation for the $x_{i}$
and $y$ variables for the case of the function $h(y)=\gamma y^{n}$. The
equation can be written as
\begin{equation}
e^{2k\varepsilon _{4}y}\text{ }y^{n-1}=-\varepsilon _{4}\frac{8k}{n\gamma }%
dx^{i}\text{ \ \ ,}  \tag{10.162}
\end{equation}
from where $x^{i}$ can be expressed as
\begin{equation}
x^{i}=-\varepsilon _{4}\frac{n\gamma }{8k}I_{n-1}(k,y)\text{ \ }
\tag{10.163}
\end{equation}
and $I_{n-1}(k,y)$ denotes the integral
\begin{equation}
I_{n-1}(k,y)=\int e^{2k\varepsilon _{4}y}y^{n-1}dy\text{ \ \ .}  \tag{10.164}
\end{equation}
This integral can be exactly calculated (see Appendix D). Note that because
of the complicated expression for the integral $I_{n-1}(k,y)$, $y$ cannot be
expressed as a function of the $x_{i}$ and the $x_{1}$ coordinate. Also, the
solvability of the quasilinear differential equation is determined mostly by
the presence of the embedded flat $4D$ Minkowski spacetime. Therefore, it
may be expected that there might be another functions $h(y)$, for which
exact analytical solution may be found.

\subsubsection*{\protect\bigskip 10.2.12 SOLUTIONS \ OF \ THE \ SECOND \
QUASILINEAR \ DIFFERENTIAL \ EQUATION \ (10.116) \ FOR \ THE \ CASE \ OF \ A
\ FLAT \ $4D$ \ MINKOWSKI \ METRIC, \ EMBEDDED \ IN \ A \ FIVE - DIMENSIONAL
SPACE \ }

\bigskip The same approach, developed in the previous subsection, shall be
applied in respect to the second quasilinear differential equation in
partial derivatives (10.116). The aim will be to show that the analytical \
solution will be different, compared to the first one for the differential
equation (10.109).

The differential equation (10.116) for the case of the metric (10.139) can
be written as
\begin{equation}
-D\frac{\partial l}{\partial x^{1}}+D\left[ \frac{\partial l}{\partial x^{2}}%
+\frac{\partial l}{\partial x^{3}}+\frac{\partial l}{\partial x^{4}}\right]
+E\frac{\partial l}{\partial y}+Al^{4}+Bl^{2}+C=0\text{ \ \ \ ,}
\tag{10.165}
\end{equation}
where $A,B,C,D,E$ denote the expressions
\begin{equation}
A\equiv \frac{(2\pi )^{7}4k(2kh-\varepsilon h^{^{\prime }})}{%
h^{2}m_{s}^{4}V_{5}g_{s}^{4}}\text{ \ \ ; \ \ }C\equiv \frac{16k^{2}M_{5}^{2}%
}{h}\text{ \ \ \ \ ,}  \tag{10.166}
\end{equation}
\begin{equation}
B\equiv \frac{16k^{2}(2\pi )^{7}}{hm_{s}^{4}g_{5}^{4}V_{5}}+\frac{4kM_{5}^{2}%
}{h^{2}}(\varepsilon h^{^{\prime }}-2kh)\text{ \ \ \ \ ,}  \tag{10.167}
\end{equation}
\begin{equation}
D\equiv -\frac{(2\pi )^{7}le^{2k\varepsilon y}h^{^{\prime }}}{%
m_{s}^{4}V_{5}g_{5}^{4}2h}\text{ \ \ \ ; \ \ \ }E\equiv -\frac{4k\varepsilon
(2\pi )^{7}l}{m_{s}^{4}V_{5}g_{5}^{4}h}\text{ \ \ \ .}  \tag{10.168}
\end{equation}
The characteristic system of equations is
\begin{equation}
\frac{dx_{1}}{D}=-\frac{dx_{i}}{D}=-\frac{dy}{E}=\frac{dl}{Al^{4}+Bl^{2}+C}%
\text{ \ \ \ .}  \tag{10.169}
\end{equation}
The characteristic equation for the $y$ and $l$ variables can be written as
\begin{equation}
d\left[ \ln \left( \varepsilon _{1}\left( \frac{s-G}{s+G}\right) \right) %
\right] =2AG\frac{\varepsilon _{4}hm_{s}^{4}V_{5}g_{5}^{4}}{4k(2\pi )^{7}}dy%
\text{ \ \ \ ,}  \tag{10.170}
\end{equation}
where
\begin{equation}
G\equiv \frac{B^{2}}{4A^{2}}-\frac{C}{A}=\frac{M_{5}\sqrt{(F-\frac{k}{2}%
)^{2}+k^{2}\left( 2(2\pi )^{7}-\frac{1}{4}\right) }}{\sqrt{\pi }(2\pi
)^{3}\mid 2k-\varepsilon _{4}\frac{h^{^{\prime }}}{h}\mid }\text{ \ \ }
\tag{10.171}
\end{equation}
is a \textbf{non - negative function }and
\begin{equation}
s\equiv l^{2}+\frac{B}{2A}\text{ \ \ ; \ \ \ }F\equiv
m_{s}^{4}V_{5}g_{5}^{4}(2k-\varepsilon _{4}\frac{h^{^{\prime }}}{h})\text{ \
\ \ \ .}  \tag{10.172}
\end{equation}
The integration of the characteristic equation (10.170) results in the
expression
\begin{equation}
l^{2}=-\frac{B}{2A}+G\frac{(1+\varepsilon _{1}D_{1}(x_{1},x_{i})e^{Z\text{ }%
\widetilde{J}(y)})}{(1-\varepsilon _{1}D_{1}(x_{1},x_{i})e^{Z\text{ }%
\widetilde{J}(y)}}\text{ \ \ \ \ \ ,}  \tag{10.173}
\end{equation}
where $Z$ is the expression
\begin{equation}
Z\equiv \frac{2M_{5}\varepsilon _{4}m_{s}^{2}g_{5}^{2}\sqrt{V_{5}}}{(2\pi
)^{3}}  \tag{10.174}
\end{equation}
and $\widetilde{J}(y)$ is the integral
\begin{equation}
\widetilde{J}(y)\equiv \int \frac{\sqrt{K_{1}y^{2}+K_{2}y+1}}{y}dy\text{ \ \
\ .}  \tag{10.175}
\end{equation}
The functions $K_{1}$ and $K_{2}$ are the following
\begin{equation}
K_{1}\equiv \frac{k^{2}\left[ 2(2\pi )^{7}-\frac{1}{4}\right] }{%
m_{s}^{8}V_{5}^{2}g_{5}^{8}}+k^{2}\left( 2-\frac{1}{2m_{s}^{4}V_{5}g_{5}^{4}}%
\right) ^{2}\text{ \ \ ,}  \tag{10.176}
\end{equation}
\begin{equation}
K_{2}\equiv -2\varepsilon _{4}\left( 2k-\frac{k}{2m_{s}^{4}V_{5}g_{5}^{4}}%
\right) \text{ \ \ \ .}  \tag{10.177}
\end{equation}
It is important to note that the free term under the square in the integral $%
\widetilde{J}(y)$ (10.175) is positive (it is $+1$) and the function $K_{1}$
is also positive. The analytical solution of integrals of the type (10.175)
depends on the sign of the function $K_{1}$ and of the free term, which in
the present case are both positive. The explicite solution can be found in
the book of Timofeev [75]:
\begin{equation*}
\widetilde{J}(y)\equiv \sqrt{K_{1}y^{2}+K_{2}y+1}+\frac{K_{2}}{2\sqrt{K_{1}}}%
\ln (K_{1}y+\frac{K_{2}}{2}+
\end{equation*}
\begin{equation}
+\sqrt{K_{1}}\sqrt{K_{1}y^{2}+K_{2}y+1})-\ln \left( \frac{1+\frac{K_{2}}{2}y+%
\sqrt{K_{1}y^{2}+K_{2}y+1}}{y}\right) \text{ \ \ \ \ \ .}  \tag{10.178}
\end{equation}
Now it remains to determine the function $D_{1}(x_{1},x_{i})$ in expression
(10.173) for $l^{2}$. For the purpose, let us denote the under - integral
expression in (10.175) by $J(y)$, differentiate both sides of (10.173) by $%
x_{1}$ and take into account the expressions for $\frac{\partial l}{\partial
x_{1}}$ and $\frac{\partial y}{\partial x_{1}}$ from the characteristic
system of equations. After rearranging the terms and denoting by $V$%
\begin{equation}
V\equiv \varepsilon _{1}\frac{-\frac{2l(Al^{4}+Bl^{2}+C)}{E}+\frac{\partial
}{\partial y}\left( \frac{B}{2A}\right) -\frac{\partial Q}{\partial y}\frac{%
(l^{2}+\frac{B}{2A})}{G}}{G+l^{2}+\frac{B}{2A}}\text{ \ \ \ \ ,}
\tag{10.179}
\end{equation}
the following differential equation can be obtained for the function $%
D_{1}(y)$
\begin{equation}
\frac{\partial D_{1}}{\partial y}+D_{1}(ZJ(y)+V)-Ve^{-Z\widetilde{J}(y)}=0%
\text{ \ \ \ .}  \tag{10.180}
\end{equation}
It may seem strange at first glance that the function $D_{1}$ depends on the
$y$ coordinate, while in (10.173) it was assumed that $%
D_{1}=D_{1}(x_{1},x_{i})$. In fact, from the characteristic equations
(10.169) for the $y$ and $x_{1}$ variables it follows
\begin{equation}
\frac{dy}{e^{2k\varepsilon _{4}y}h^{^{\prime }}}=-\frac{\varepsilon _{4}}{8k}%
dX_{1}\text{ \ \ .}  \tag{10.181a}
\end{equation}
For the concrete expression for the function $h(y)=\gamma y^{n}$, the
coordinates $x_{1}$ and $x_{i}$ can be expressed as
\begin{equation}
x_{1}=-\varepsilon _{4}\frac{8k}{n\gamma }I(-k,1-n)+const.\text{ \ ; \ }%
x_{i}=-x_{1}  \tag{10.181b}
\end{equation}
and therefore it is reasonable to consider that $%
D_{1}=D_{1}(x_{1}(y),x_{i}(y))=D_{1}(y)$. Note however that due to the
complicated structure of the integral $I(-k,1-n)$, it is impossible to
express $y$ as a function of $x_{1}$ (or $x_{i}$).

As in the previous case, the general solution of the equation depends on all
the first integrals of the characteristic system of equations.

\subsubsection*{\protect\bigskip 10.2.13 LENGTH \ FUNCTION \ $l(x)$ \ FROM \
THE \ CONSTANCY \ OF \ THE \ SCALAR \ CURVATURE $\ R$ \ UNDER \ "RESCALINGS"
\ OF \ THE \ CONTRAVARIANT \ METRIC \ TENSOR \ FOR \ THE \ CASE \ OF \ A \
FLAT \ $4D$ \ MINKOWSKI \ METRIC, \ EMBEDDED \ IN A \ $5D$ \ SPACETIME. \ }

\bigskip Now we shall find solutions of the corresponding differential
equation in partial derivatives, when the second representation of the
scalar curvature $\widetilde{R}$ \ (10.107) is equal to the initial scalar
curvature $R$ (i.e. $\widetilde{R}=R$). The obtained differential equation
under this identification is
\begin{equation*}
l^{3}\left[ R-\frac{1}{2}g^{AC}g^{BD}(...)\right] +l^{2}\left[
-R+g^{AB}(\Gamma _{AB,C}^{C}-\Gamma _{AC,B}^{C})\right] +
\end{equation*}
\begin{equation*}
+l\left[ \frac{1}{2}g^{AC}g^{BD}(...)-g^{AB}(\Gamma _{AB,C}^{C}-\Gamma
_{AC,B}^{C})\right] +
\end{equation*}
\begin{equation}
+\frac{\partial l}{\partial x^{B}}g^{AB}\Gamma _{AC}^{C}-\frac{\partial l}{%
\partial x^{C}}g^{AB}\Gamma _{AB}^{C}=0\text{ \ \ \ \ .}  \tag{10.182.}
\end{equation}
The expression in the small brackets is the same as in (10.99), i.e. $%
(...)\equiv (g_{AD,BC}+g_{BC,AD}-g_{AC,BD}-g_{BD,AC})$. The equation
(10.182) for the case of the metric (10.107) with the affine connection
components (10.139) acquires the form
\begin{equation*}
\varepsilon \frac{\partial l}{\partial y}+\frac{h^{^{\prime }}}{8k}%
e^{2k\varepsilon y}\left( \frac{\partial l}{\partial x_{1}}-\frac{\partial l%
}{\partial x_{2}}-\frac{\partial l}{\partial x_{3}}-\frac{\partial l}{%
\partial x_{4}}\right) =
\end{equation*}
\begin{equation}
=(2k-\varepsilon \frac{h^{^{\prime }}}{h})(l^{3}-l)\text{ \ \ .}
\tag{10.183.}
\end{equation}
The characteristic system of equations is
\begin{equation}
\frac{dl}{(2k-\varepsilon \frac{h^{^{\prime }}}{h})l(l^{2}-1)}=\varepsilon
dy=  \tag{10.184.}
\end{equation}
\begin{equation}
=\frac{dx_{1}}{h^{^{\prime }}}8ke^{-2k\varepsilon y}=-\frac{dx_{i}}{%
h^{^{\prime }}}8ke^{-2k\varepsilon y}\text{ \ \ .}  \tag{10.185.}
\end{equation}
The solutions of the characteristic system for the $x_{1}$ and $y$ variables
are correspondingly
\begin{equation}
x_{1}=C_{4}(x_{i},l)+\varepsilon _{2}\frac{e^{2k\varepsilon _{1}y}h}{8k}-%
\frac{1}{4}\int he^{2k\varepsilon _{1}y}dy\text{ \ \ ,}  \tag{10.186.}
\end{equation}
\begin{equation}
l^{2}=\frac{h^{2}}{h^{2}-D_{2}(x_{1},x_{i})e^{4k\varepsilon _{1}y}}\text{ \
\ \ .}  \tag{10.187.}
\end{equation}
Unfortunately, (10.187) cannot be considered as an expression for $l$, since
from (10.186) it is obvious that $D_{2}(x_{1},x_{i})$ also depends on the
function $l$. Also, it should be understood that $D_{2}$ depends on all the
variables $x_{i}$, $i=2,3,4$.

Let us differentiate both sides of (10.187) by $x_{1}$
\begin{equation*}
2l\frac{\partial l}{\partial x_{1}}=\frac{2lh^{^{\prime }}}{h}\frac{\partial
y}{\partial x_{1}}-\frac{l^{4}}{h^{2}}(2hh^{^{\prime }}\frac{\partial y}{%
\partial x_{1}}-
\end{equation*}
\begin{equation}
-2D_{2}\frac{\partial D_{2}}{\partial x_{1}}e^{4k\varepsilon
_{1}y}-D_{2}4k\varepsilon _{1}e^{4k\varepsilon _{1}y}\frac{\partial y}{%
\partial x_{1}})\text{ \ \ .}  \tag{10.188.}
\end{equation}
If the same operation is applied also in respect to the $x_{i}$ coordinate
and the derived equation is summed up with (10.188) with account also of $%
\frac{\partial l}{\partial x_{1}}=-\frac{\partial l}{\partial x_{i}}$, $%
\frac{\partial y}{\partial x_{1}}=-\frac{\partial y}{\partial x_{i}}$, then
it can be obtained
\begin{equation}
\frac{2l^{4}}{h^{2}}e^{4k\varepsilon _{1}y}D_{2}\left( \frac{\partial D_{2}}{%
\partial x_{1}}+\frac{\partial D_{2}}{\partial x_{i}}\right) =0\text{ \ \ \
\ .}  \tag{10.189.}
\end{equation}
The equality is satisfied also for $D_{2}\equiv 0$, which evidently
corresponds to the standard case $l=1$ in gravity theory. The other case,
when the equality is fulfilled, is $\frac{\partial D_{2}}{\partial x_{1}}=-%
\frac{\partial D_{2}}{\partial x_{i}}$.

Now let us rewrite equation (10.188) with account of the expressions for $%
\frac{\partial l}{\partial x_{1}}$ and $\frac{\partial y}{\partial x_{1}}$
from the characteristic system of equations (10.184 - 10.185). Then the
following nonlinear differential equation in respect to the function $%
D_{2}^{2}$ is derived
\begin{equation*}
\frac{\partial D_{2}^{2}}{\partial y}-4k\varepsilon _{1}(1+\frac{4}{h}%
)D_{2}^{2}-
\end{equation*}
\begin{equation}
-2\varepsilon _{1}\varepsilon _{5}\frac{h^{^{\prime }}}{h}(2k-\varepsilon
_{1}\frac{h^{^{\prime }}}{h})e^{-4k\varepsilon
_{1}y}(h^{2}-D_{2}^{2}e^{4k\varepsilon _{1}y})^{\frac{3}{2}}=0\text{ \ \ \ \
.}  \tag{10.190.}
\end{equation}
Finding the function $D_{2}=D_{2}(y)$ as a solution of this equation, from
(10.187) $l^{2}$ can also be found as a function of the extra - coordinate $%
y $, i.e. $l=l(y)$. But then, differentiating the solution for $x_{1}$
(10.186) by $y$, the function $C_{4}(x_{i},l)$ can be found as a solution of
the following differential equation
\begin{equation}
\frac{\partial C_{4}(x_{i},l)}{\partial y}=F_{1}\text{ \ ,}  \tag{10.191.}
\end{equation}
where the function $F_{1}$ is determined as
\begin{equation}
F_{1}\equiv e^{2k\varepsilon _{1}y}[\frac{\varepsilon
_{2}l(l^{2}-1)h^{^{\prime }}}{8k}-\frac{\varepsilon _{1}\varepsilon _{2}h}{4}%
-\frac{\varepsilon _{2}h^{^{\prime }}}{8k}+\frac{h}{4}]  \tag{10.192.}
\end{equation}
and $l$ has to be substituted with expression (10.187), in which $D_{2}$ is
determined as a solution of the differential equation (10.190). The
representation in the form (10.191) is particularly convenient for the case $%
h(y)=\gamma y^{n}$, when the difficulty will be only in calculaing the
integral along $y$ in the first term of (10.192). The advantage of the
representation (10.191) will become evident if we differentiate the solution
(10.186) by $l$, obtaining thus the differential equation
\begin{equation}
E_{1}(y)=\frac{\partial C_{4}(x_{i},l)}{\partial l}+\frac{E_{2}(y)}{%
l(l^{2}-1)}\text{ \ \ ,}  \tag{10.193.}
\end{equation}
where
\begin{equation}
E_{1}(y)\equiv \frac{hh^{^{\prime }}e^{2k\varepsilon _{1}y}}{%
8k(2kh-\varepsilon _{2}h^{^{\prime }})}\text{ \ \ \ ,}  \tag{10.194.}
\end{equation}
\begin{equation}
E_{2}(y)\equiv \frac{\varepsilon _{2}he^{2k\varepsilon _{1}y}}{%
(2kh-\varepsilon _{2}h^{^{\prime }}}\left[ \left( \varepsilon
_{1}\varepsilon _{2}-1\right) +\frac{\varepsilon _{2}h^{^{\prime }}}{8k}%
\right] \text{ \ \ .}  \tag{10.195.}
\end{equation}
Now it is important to stress that the second representation (10.193) is
\textbf{inconvenient} to use for the case $h(y)=\gamma y^{n}$. The reason is
that the integration is along the $l$ coordinate, which means that the $y$
coordinate in $E_{1}(y)$ and $E_{2}(y)$ has to be expressed from expression
(10.187) as a function of $l$. However, in view of the extremely complicated
expression, this is not possible. Instead, the differential equation
(10.193) will be very helpful for the case $h(y)\equiv 1$, which is
frequently encountered in most of the papers on theories with extra
dimensions. Indeed, then the nonlinear differential equation (10.190) is of
a particularly simple form:
\begin{equation}
\frac{\partial D_{2}^{2}}{\partial y}-20k\varepsilon D_{2}^{2}=0\text{ \ \ ,}
\tag{10.196.}
\end{equation}
from where with the help of (10.187)
\begin{equation}
l^{2}=\frac{1}{1-const.e^{24k\varepsilon _{1}y}}\text{ \ \ \ .}
\tag{10.197.}
\end{equation}
Note one interesting property of the obtained solution, already mentioned in
the Introduction - \textbf{when }$\varepsilon _{1}=-1$\textbf{\ and }$y$%
\textbf{\ tends to infinity, the known case in gravity theory }$l^{2}=1$%
\textbf{\ is recovered.} Since we received this solution for a partial case,
it would be interesting to check whether this happens also for an arbitrary
function $h(y)$. But this is more complicated since it is necessary to find
the solution of the nonlinear differential equation (10.190).

Now $y$ can be expressed easily and the resulting differential equation
(10.193) with $E_{1}(y)=0$ can be rewritten as
\begin{equation}
\frac{\partial C_{4}(x_{i},l)}{\partial l}+\frac{\varepsilon
_{2}(\varepsilon _{1}\varepsilon _{2}-1)e^{\frac{1}{12}}}{const.8kl^{3}}=0%
\text{ \ \ \ .}  \tag{10.198.}
\end{equation}
The solution of the equation can be represented as
\begin{equation}
C_{4}(x_{i},l)=-\frac{\varepsilon _{2}e^{\frac{1}{12}}(1-const\text{ }%
e^{24k\varepsilon _{1}y})}{8k\text{ }const\text{.}}\widetilde{C}_{4}(x_{i})%
\text{ \ \ \ \ .}  \tag{10.199.}
\end{equation}
The unknown function $\widetilde{C}_{4}(x_{i})$ can be found if expression
(10.186) for $x_{1}$ is differentiated in respect to $x_{i}$. Unfortunately,
the resulting formulae will contain the expressions for $\frac{\partial y}{%
\partial x_{i}}$ and $\frac{\partial l}{\partial x_{i}}$, which are singular
when $h^{^{\prime }}(y)=0.$But if we multiply by $\frac{\partial x_{i}}{%
\partial y}$, the following differential equation will be obtained
\begin{equation*}
\frac{\partial x_{1}}{\partial y}=\frac{\varepsilon _{1}(\varepsilon
_{2}-\varepsilon _{1})}{4}e^{2k\varepsilon _{1}y}+\frac{\varepsilon _{2}e^{%
\frac{1}{12}}}{8k\text{ }const\text{ }l^{3}}\widetilde{C}_{4}(x_{i})\frac{%
\partial l}{\partial y}-
\end{equation*}
\begin{equation}
-\frac{\varepsilon _{2}e^{\frac{1}{12}}}{8k\text{ }const\text{ }l^{2}}\frac{%
\partial \widetilde{C}_{4}(x_{i})}{\partial y}\text{ \ \ \ \ ,}
\tag{10.200
}
\end{equation}
which is no longer singular in the limit $h^{^{\prime }}(y)=0$, because the
expressions for $\frac{\partial x_{1}}{\partial y}$ and $\frac{\partial l}{%
\partial y}$ are
\begin{equation}
\frac{\partial x_{1}}{\partial y}=0\text{ \ \ ; \ \ }\frac{\partial l}{%
\partial y}=\varepsilon _{2}2kl(l^{2}-1)\text{ \ \ \ .}  \tag{10.201 }
\end{equation}
Taking into account these formulaes, the solution of the differential
equation (10.200) in respect to the function $\widetilde{C}_{4}(x_{i})$ can
be found in the form
\begin{equation}
\widetilde{C}_{4}(x_{i})=const_{2}\mid 1-const\text{ }e^{24k\varepsilon
_{1}y}\mid ^{-\frac{\varepsilon _{1}}{12}}\text{ \ .}  \tag{10.202 }
\end{equation}
Substituting into (10.199), the final expressions for the function $%
C_{4}(x_{i},l)$ can be found and also for the coordinate $x_{1}$, which for $%
\varepsilon _{1}=\varepsilon _{2}$ and $h(y)=1$ is simply
\begin{equation*}
x_{1}=C_{4}(x_{i}(y),l(y))=
\end{equation*}
\begin{equation}
=-\frac{\varepsilon _{2}\text{ }const_{2}\text{ }e^{\frac{1}{12}}}{8k\text{ }%
const\text{ }}\frac{\left( 1-const\text{ }e^{24k\varepsilon _{1}y}\right) }{%
\mid 1-const\text{ }e^{24k\varepsilon _{1}y}\mid ^{\frac{\varepsilon _{1}}{12%
}}}\text{ \ \ \ \ \ .}  \tag{10.203 }
\end{equation}

\bigskip

\section*{\protect\bigskip Acknowledgments}

The author is grateful to Prof. N. A. Chernikov, Dr. L. K. Alexandrov, Dr.
D. M. Mladenov (BLTP, JINR, Dubna \& DESY, Germany), St. Mishev (BLTP, JINR,
Dubna), Dr. N. S. Shavokhina and especially to Prof. V. V. Nesterenko, Dr.
O. Santillan (BLTP, JINR, Dubna) and to Prof. Sawa Manoff (INRNE, Bulgarian
Academy of Sciences, Sofia) for valuable comments, discussions and critical
remarks. It is a pleasure to thank all the participants in the seminar of
the Theoretical Physics Department at the INRNE, Sofia, where some parts of
the present work had been reported. \

This paper is written in memory of \ Prof. S. S. Manoff (1943 - 27.05.2005)
- a specialist in classical gravitational theory and physics and a true
friend to all his colleagues everywhere. \

The author is grateful also to Dr. A. Zorin (LNP, JINR) and to J. Yanev
(BLTP, JINR) for various helpful advises ; to Dr. G. V. Kraniotis (Texas
A\&M University, USA) for sending to me his published paper (ref. [8]) and
to Dr.V. Gvaramadze (Sternberg Astronomical Institute, MSU, Moscow) and his
family for their moral support and encouragement.

\section{\protect\bigskip APPENDIX\ \ A: SOME \ PROPERTIES \ OF \ THE \
NEWLY \ INTRODUCED \ CONNECTION $\widetilde{\Gamma }_{ij}^{k}=\frac{1}{2}%
dX^{k}dX^{l}(g_{jl,i}+g_{il,j}-g_{ij,l})$.\ }

\subsection*{\protect\bigskip A1: \ FIRST \ AND \ SECOND \ DIFFERENTIALS}

The first and the second differentials $dX^{k}=dX^{k}(x^{1},x^{2},...,x^{n})$
and $d^{2}X^{i}=d^{2}X^{k}(x^{1},x^{2},...,x^{n})$ are functions, in
principle, of the initial coordinates $x^{1},x^{2},...,x^{n}$. However, it
is important to note that since the mapping $%
X^{i}=X^{i}(x^{1},x^{2},...,x^{n})$ is unique and therefore $\det \parallel
\frac{\partial X^{k}}{\partial x^{l}}\parallel \neq 0$, the inverse mapping $%
x^{k}=x^{k}(X^{1},X^{2},...,X^{n})$ can also be defined. Therefore we may
also write
\begin{equation}
d^{2}X^{i}=d^{2}X^{i}(x^{1},x^{2},...,x^{n})=d^{2}X^{i}(X^{1},X^{2},...,X^{n})=
\tag{A1}
\end{equation}
\begin{equation}
=\sum\limits_{r=1}^{n}\frac{\partial (dX^{k})}{\partial X^{r}}dX^{r}\text{ \
\ .}  \tag{A2}
\end{equation}
Our next purpose will be to express the second differential, starting from
the R. H. S. of the above expression, through the initial coordinates and
thus really to prove that the second differential is one and the same in all
system of coordinates. Let us transform the R. H. S. of (A1) in the
following way
\begin{equation}
d^{2}X^{i}(X^{1},X^{2},...,X^{n})=\sum\limits_{s=1}^{n}\frac{\partial }{%
\partial X^{s}}\left[ \sum\limits_{r=1}^{n}\frac{\partial (dX^{k})}{\partial
X^{r}}dX^{r}\right] \sum\limits_{f=1}^{n}\frac{\partial X^{s}}{\partial x^{t}%
}dx^{t}=  \tag{A3}
\end{equation}
\begin{equation}
=\sum\limits_{r,s,f=1}^{n}\frac{\partial }{\partial x^{p}}\left[ \frac{%
\partial X^{k}}{\partial x^{r}}dx^{r}\right] \frac{\partial x^{p}}{\partial
X^{s}}\frac{\partial X^{s}}{\partial x^{t}}dx^{t}=  \tag{A4}
\end{equation}
\begin{equation}
=\sum\limits_{r,t}\frac{\partial ^{2}X^{k}}{\partial x^{t}\partial x^{r}}%
dx^{r}dx^{t}+\sum\limits_{r}\frac{\partial X^{k}}{\partial x^{r}}d^{2}x^{r}%
\text{ \ \ \ \ ,}  \tag{A5}
\end{equation}
where due to the existence of the inverse transformation $%
x^{k}=x^{k}(X^{1},X^{2},...,X^{n})$ it has been used that
\begin{equation}
\frac{\partial x^{p}}{\partial X^{s}}\frac{\partial X^{s}}{\partial x^{t}}%
=\delta _{t}^{p}\text{ \ \ .}  \tag{A6}
\end{equation}
As expected and as it can easily be checked, the expression (A5) is the same
as if one starts with the second differential, expressed in terms of the
initial coordinates $x^{1},x^{2},...,x^{n}$. This precludes the proof that
\begin{equation}
d^{2}X^{k}(X^{1},X^{2},...,X^{n})=d^{2}X^{i}(x^{1},x^{2},...,x^{n})=\sum%
\limits_{r=1}^{n}\frac{\partial (dX^{k})}{\partial x^{r}}dx^{r}\text{ \ \ .}
\tag{A7}
\end{equation}

\subsection*{\protect\bigskip A2: \ A \ PROOF \ OF \ THE \ AFFINE \
TRANSFORMATION \ LAW \ FOR \ THE \ CONNECTION \ $\widetilde{\Gamma }%
_{ij}^{k} $}

Next we proceed with the proof that the defined in (2.11) (and in the
preceeding paper [10]) connection
\begin{equation}
\widetilde{\Gamma }_{ij}^{k}=\frac{1}{2}%
dX^{k}dX^{l}(g_{jl,i}+g_{il,j}-g_{ij,l})=dX^{k}dX^{r}g_{sr}(X)\Gamma
_{ij}^{s}(X)  \tag{A8}
\end{equation}
has the transformation property of an affine connection under the coordinate
transformations $X^{i}=X^{i}(x^{1},x^{2},...,x^{n})$. The connections $%
\widetilde{\Gamma }_{ij}^{k}$ and $\Gamma _{ij}^{k}$ at the spacetime point $%
\ \mathbf{X}=(X^{1},X^{2},...,X^{n})$ will be denoted as $\widetilde{\Gamma }%
_{ij}^{k^{\prime }}(\mathbf{X})$ and $\Gamma _{ij}^{k^{\prime }}(\mathbf{X})$%
, while the same connections at the initial coordinate points $\mathbf{x}%
=(x^{1},x^{2},...,x^{n})$ will be denoted simply as $\widetilde{\Gamma }%
_{ij}^{k}(\mathbf{x})$ and $\Gamma _{ij}^{k}(\mathbf{x})$ (the same for the
notation $g_{sr}^{^{\prime }}(\mathbf{X})$).

From the defining equation (A8) for $\widetilde{\Gamma }_{ij}^{k}$, the
tensor transformation property for $g_{ij}^{^{\prime }}(\mathbf{X})$
\begin{equation}
g_{ij}^{^{\prime }}(\mathbf{X})=\frac{\partial x^{k}}{\partial X^{i}}\frac{%
\partial x^{l}}{\partial X^{j}}g_{kl}(\mathbf{x})\text{ \ \ ,}  \tag{A9}
\end{equation}
the affine transformation law for the "usual" connection $\Gamma _{ij}^{k}$
\begin{equation}
\Gamma _{ij}^{k^{\prime }}(\mathbf{X})=\Gamma _{np}^{m}(\mathbf{x})\frac{%
\partial X^{k}}{\partial x^{m}}\frac{\partial x^{n}}{\partial X^{i}}\frac{%
\partial x^{p}}{\partial X^{j}}+\frac{\partial ^{2}x^{m}}{\partial
X^{i}\partial X^{j}}\frac{\partial X^{k}}{\partial x^{m}}  \tag{A10}
\end{equation}
and from the expressions for the differentials $dX^{k}$ and $dX^{r}$ we may
write down
\begin{equation}
\widetilde{\Gamma }_{ij}^{k^{\prime }}(\mathbf{X})==dX^{k}(\mathbf{X})dX^{r}(%
\mathbf{X})g_{sr}^{^{\prime }}(\mathbf{X})\Gamma _{ij}^{s^{\prime }}(\mathbf{%
X})=  \tag{A11}
\end{equation}
\begin{equation}
=\Gamma _{np}^{m}(\mathbf{x})\frac{\partial X^{k}}{\partial x^{\alpha }}%
\frac{\partial x^{n}}{\partial X^{i}}\frac{\partial x^{p}}{\partial X^{j}}%
g_{m\beta }(\mathbf{x})dx^{\alpha }dx^{\beta }+\frac{\partial ^{2}x^{m}}{%
\partial X^{i}\partial X^{j}}\frac{\partial X^{k}}{\partial x^{\alpha }}%
g_{m\beta }(\mathbf{x})dx^{\alpha }dx^{\beta }\text{ \ \ .}  \tag{A12}
\end{equation}
Again, the existence of the inverse transformation and of relation (A6) has
been taken into account.

On the other hand, if $\widetilde{\Gamma }_{ij}^{k}(\mathbf{X})$ is an
affine connection, then it should satisfy the affine connection
transformation law (A10)
\begin{equation}
\widetilde{\Gamma }_{ij}^{k^{\prime }}(\mathbf{X})=\widetilde{\Gamma }%
_{np}^{m}(\mathbf{x})\frac{\partial X^{k}}{\partial x^{m}}\frac{\partial
x^{n}}{\partial X^{i}}\frac{\partial x^{p}}{\partial X^{j}}+\frac{\partial
^{2}x^{m}}{\partial X^{i}\partial X^{j}}\frac{\partial X^{k}}{\partial x^{m}}%
\text{ \ \ \ .}  \tag{A13}
\end{equation}
Making use of the defining equation (A8) (but in terms of the initial
coordinates $x^{1},x^{2},....,x^{n}$), the above expression can be written
also as
\begin{equation}
\widetilde{\Gamma }_{ij}^{k^{\prime }}(\mathbf{X})=\Gamma _{np}^{m}(\mathbf{x%
})\frac{\partial X^{k}}{\partial x^{\alpha }}\frac{\partial x^{n}}{\partial
X^{i}}\frac{\partial x^{p}}{\partial X^{j}}g_{m\beta }(\mathbf{x})dx^{\alpha
}dx^{\beta }+\frac{\partial ^{2}x^{\alpha }}{\partial X^{i}\partial X^{j}}%
\frac{\partial X^{k}}{\partial x^{\alpha }}\text{ \ \ \ .}  \tag{A14}
\end{equation}
Clearly, if $\widetilde{\Gamma }_{ij}^{k^{\prime }}(\mathbf{X})$ is an
affine connection, from the R. H. S. of \ (A12) and (A14) it would follow
that the following relation has to be satisfied
\begin{equation}
dx^{\alpha }dx^{\beta }g_{m\beta }(\mathbf{x})\frac{\partial ^{2}x^{m}}{%
\partial X^{i}\partial X^{j}}\frac{\partial X^{k}}{\partial x^{\alpha }}-%
\frac{\partial ^{2}x^{\alpha }}{\partial X^{i}\partial X^{j}}\frac{\partial
X^{k}}{\partial x^{\alpha }}=0\text{ \ \ .}  \tag{A15}
\end{equation}
Let us first prove that the second term is equal to zero. We \ have
\begin{equation*}
\frac{\partial X^{k}}{\partial x^{\alpha }}\frac{\partial ^{2}x^{\alpha }}{%
\partial X^{i}\partial X^{j}}=\frac{\partial }{\partial X^{i}}\left[ \frac{%
\partial X^{k}}{\partial x^{\alpha }}\frac{\partial x^{\alpha }}{\partial
X^{j}}\right] -\frac{\partial x^{\alpha }}{\partial X^{j}}\frac{\partial }{%
\partial x^{\alpha }}\left[ \frac{\partial X^{k}}{\partial X^{i}}\right] =
\end{equation*}
\begin{equation}
=\frac{\partial }{\partial X^{i}}\delta _{j}^{k}-\frac{\partial x^{\alpha }}{%
\partial X^{j}}\frac{\partial }{\partial x^{\alpha }}\delta _{i}^{k}=0\text{
\ .}  \tag{A16}
\end{equation}
Making use of the tensor transformation property (A9), the first term in
(A15) can be transformed as follows
\begin{equation}
dx^{\alpha }dx^{\beta }g_{m\beta }(\mathbf{x})\frac{\partial ^{2}x^{m}}{%
\partial X^{i}\partial X^{j}}\frac{\partial X^{k}}{\partial x^{\alpha }}=
\tag{A17}
\end{equation}
\begin{equation}
=\frac{\partial x^{\alpha }}{\partial X^{r}}\frac{\partial x^{\beta }}{%
\partial X^{s}}dX^{r}dX^{s}\frac{\partial X^{p}}{\partial x^{m}}\frac{%
\partial X^{q}}{\partial x^{\beta }}\frac{\partial X^{k}}{\partial x^{\alpha
}}\frac{\partial ^{2}x^{m}}{\partial X^{i}\partial X^{j}}=  \tag{A18}
\end{equation}
\begin{equation}
=dX^{r}dX^{s}\delta _{r}^{k}\delta _{s}^{q}\frac{\partial X^{p}}{\partial
x^{m}}\frac{\partial ^{2}x^{m}}{\partial X^{i}\partial X^{j}}\text{ \ \ \ .}
\tag{A19}
\end{equation}
But the above expression has already been proved to be equal to zero.
Therefore, equation (A15) is satisfied and consequently, $\widetilde{\Gamma }%
_{ij}^{k}$ has an affine connection transformation property.

\subsection*{\protect\bigskip A3: \ THE \ CONNECTION \ $\widetilde{\Gamma }%
_{ij}^{k}$ \ AS \ AN EQUIAFFINE \ CONNECTION}

\bigskip We have to prove that the connection $\widetilde{\Gamma }_{ij}^{k}$
for $j=k$ can be represented in the form of a gradient of a scalar quantity,
i. e.
\begin{equation}
\widetilde{\Gamma }_{ij}^{k}=\partial _{i}lne\text{ \ \ .}  \tag{A20.}
\end{equation}
From the defining formulae (A8), it follows
\begin{equation*}
\widetilde{\Gamma }_{ij}^{k}=\frac{1}{2}\left[ dX^{k}dX^{s}g_{ks}\right]
_{,i}-\frac{1}{2}g_{is}\left[ dX^{k}dX^{s}\right] _{,i}+
\end{equation*}
\begin{equation*}
+\frac{1}{2}\left[ dX^{k}dX^{s}g_{is}\right] _{,k}-\frac{1}{2}g_{is}\left[
dX^{k}dX^{s}\right] _{,k}-
\end{equation*}
\begin{equation}
-\frac{1}{2}\left[ dX^{k}dX^{s}g_{ik}\right] _{,s}-\frac{1}{2}g_{ik}\left[
dX^{k}dX^{s}\right] _{,s}\text{ \ \ .}  \tag{A21. }
\end{equation}
The last four terms cancel, so therefore from the first two terms it is
evident that in the approximation ($dX^{i})_{,k}=0$ the connection $%
\widetilde{\Gamma }_{ij}^{k}$ is indeed an equiaffine one, since one can set
up
\begin{equation}
lne\equiv \frac{1}{2}dX^{k}dX^{s}g_{ks}  \tag{A22}
\end{equation}
and so $e$ will be fully determined and the defining equality (A20) then
will be fulfilled.

The more complicated and interesting task is to prove that even in the case (%
$dX^{i})_{,k}\neq 0$, the connection $\widetilde{\Gamma }_{ij}^{k}$ will
again be an equiaffine one. For the purpose, note that
\begin{equation}
\widetilde{\Gamma }_{ik}^{k}=\frac{1}{2}dX^{s}dX^{k}g_{ks,i}=\frac{1}{2}%
dX^{k}dX^{s}g_{r(s}\Gamma _{k)i}^{r}=W_{i}\text{ \ }  \tag{A23.}
\end{equation}
and consequently $\widetilde{\Gamma }_{ik}^{k}$ will be an equiaffine
connection if the scalar quantity $e$ can be determined as a solution of the
differential equation
\begin{equation}
\partial _{i}lne=W_{i}\text{ \ }  \tag{A24.}
\end{equation}
as
\begin{equation}
e=g(X_{1},X_{2},..,X_{i-1},X_{i+1},..,X_{n})e^{\int
W_{i}(X_{1},.....,X_{n})dX^{i}}\text{ \ .}  \tag{A25.}
\end{equation}
Note that the function $g$ depends on all variables $%
X_{1},X_{2},..,X_{i-1},X_{i+1},..,X_{n}$ with the exception of $X_{i}$,
while the function $W_{i}$ depends on all the variables, including also $%
X_{i}$.

Unfortunately, the proof at this stage will be incomplete, since $e$ will
depend on the choice of the variable $X_{i}$, which should not happen with a
scalar quantity.Consequently, it should be proved that the function $%
g(X_{1},X_{2},..,X_{i-1},X_{i+1},..,X_{n})$ can be determined in a proper
way so (that for every choice of $W_{i}$ the expression (A25) for $e$ would
be a scalar quantity. Until we have not proved it, we shall denote the L.H.
S. of (A25) with $e^{(i)}$.

Let us differentiate both sides of (A25) for $e\equiv e^{(i)}$ and $e\equiv
e^{(j)}$ by $X^{j}$ and $X^{i}$ respectively ($i\neq j$). We shall write
down only the first equation, since the second one is obtained from the
first after a change of the indices $i\Longleftrightarrow $ $j$.
\begin{equation*}
\frac{\partial e^{(i)}}{\partial X^{j}}=\frac{\partial \ln
g(X_{1},X_{2},.,X_{i-1},X_{i+1},.,X_{n})}{\partial X^{j}}e^{(i)}+
\end{equation*}
\begin{equation}
+g(X_{1},X_{2},..,X_{i-1},X_{i+1},..,X_{n})e^{\int \frac{\partial
W_{i}(X_{1},.....,X_{n})}{\partial X^{j}}dX^{i}}\text{ \ \ \ .}  \tag{A26}
\end{equation}
Now differentiate again the derived equation (A26) for $\frac{\partial
e^{(i)}}{\partial X^{j}}$ by $X^{i}$ and the other equation for $\frac{%
\partial e^{(j)}}{\partial X^{i}}$ by $X^{j}$. Taking into account also that
$\frac{\partial e^{(i)}}{\partial X^{i}}=e^{(i)}W_{i}$,the result for the
first equation will be
\begin{equation*}
\frac{\partial ^{2}e^{(i)}}{\partial X^{j}\partial X^{i}}=\frac{\partial \ln
g(X_{1},X_{2},.,X_{i-1},X_{i+1},.,X_{n})}{\partial X^{j}}e^{(i)}W_{i}+
\end{equation*}
\begin{equation}
+g(X_{1},X_{2},..,X_{i-1},X_{i+1},..,X_{n})\frac{\partial W_{i}}{\partial
X^{j}}e^{\int \frac{\partial W_{i}(X_{1},.....,X_{n})}{\partial X^{j}}dX^{i}}%
\text{ \ \ \ .}  \tag{A27}
\end{equation}
In respect to the second term in (A27), again the equality (A26) may be
applied and after that a summation along the indices $i$ and $j$ can be
defined. Equation (A27) acquires the form
\begin{equation*}
\sum_{i,j}\frac{\partial }{\partial X^{j}}\left( \frac{\partial e^{(i)}}{%
\partial X^{i}}\right) =grad\left[ \ln
g(X_{1},X_{2},.,X_{i-1},X_{i+1},.,X_{n})\right] (\mathbf{e}\text{ }.\mathbf{W%
})+
\end{equation*}
\begin{equation}
+\sum_{i,j}\left[ \frac{\partial W_{i}}{\partial X^{j}}\frac{\partial e^{(i)}%
}{\partial X^{j}}-\frac{\partial \widetilde{W}_{ij}}{\partial X^{j}}.e^{(i)}%
\right] +(\mathbf{e}\text{ }.\mathbf{W})\bigtriangleup \ln
g(X_{1},X_{2},.,X_{i-1},X_{i+1},.,X_{n})\text{ \ \ \ \ ,}  \tag{A28}
\end{equation}
where $(\mathbf{e}$ $.\mathbf{W})$ denotes a scalar product and the
following notations have been introduced

\begin{equation}
\bigtriangleup \ln g(....)\equiv \sum_{j}\frac{\partial ^{2}}{\partial X^{j2}%
}\bigtriangleup \ln g(.....)\text{ \ \ \ ,}  \tag{A29.}
\end{equation}
\begin{equation}
\widetilde{W}_{ij}\equiv W_{i}\frac{\partial \ln
g(X_{1},X_{2},.,X_{i-1},X_{i+1},.,X_{n})}{\partial X^{j}}\text{ \ \ \ .}
\tag{A30}
\end{equation}
Again, the second equation will be the same as (A28), but with $%
i\Leftrightarrow j$. Substracting the two equations and taking into account
that
\begin{equation}
graddive=\sum_{i,j}\frac{\partial }{\partial X^{j}}\left( \frac{\partial
e^{(i)}}{\partial X^{i}}\right) =\sum_{i,j}\frac{\partial }{\partial X^{i}}%
\left( \frac{\partial e^{(i)}}{\partial X^{j}}\right) \text{ \ \ \ ,}
\tag{A31.}
\end{equation}
one can derive
\begin{equation*}
\sum_{i,j}\left[ \frac{\partial W_{i}}{\partial X^{j}}\frac{\partial e^{(i)}%
}{\partial X^{i}}-\frac{\partial W_{j}}{\partial X^{i}}\left( \frac{\partial
e^{(j)}}{\partial X^{i}}\right) \right] -
\end{equation*}
\begin{equation}
-\sum_{i,j}\left[ \frac{\partial \widetilde{W}_{ij}}{\partial X^{j}}e^{(i)}-%
\frac{\partial \widetilde{W}_{ji}}{\partial X^{i}}e^{(j)}\right] =0\text{ \
\ \ \ .}  \tag{A32. }
\end{equation}
But it can be written also
\begin{equation*}
\sum_{i,j}\left[ \frac{\partial W_{i}}{\partial X^{j}}\frac{\partial e^{(i)}%
}{\partial X^{i}}-\frac{\partial \widetilde{W}_{ij}}{\partial X^{j}}e^{(i)}%
\right] =
\end{equation*}
\begin{equation}
=\sum_{i,j}\left[ \frac{\partial W_{i}}{\partial X^{j}}\frac{\partial e^{(i)}%
}{\partial X^{i}}-e^{(i)}\frac{\partial W_{i}}{\partial X^{j}}\frac{\partial
\ln g}{\partial X^{j}}-(\mathbf{e}\text{ .}\mathbf{W})\bigtriangleup \ln g%
\right] \text{ \ \ .}  \tag{A33.}
\end{equation}
Taking the above expression into account, equation (A32) acquires the form
\begin{equation*}
\sum_{i,j}\left[ \frac{\partial W_{i}}{\partial X^{j}}\frac{\partial e^{(i)}%
}{\partial X^{i}}-\frac{\partial W_{i}}{\partial X^{j}}\frac{\partial e^{(i)}%
}{\partial X^{i}}\right] +
\end{equation*}
\begin{equation}
+\sum_{i,j}\left[ e^{(j)}\frac{\partial W_{j}}{\partial X^{i}}\frac{\partial
\ln g}{\partial X^{i}}-e^{(i)}\frac{\partial W_{i}}{\partial X^{j}}\frac{%
\partial \ln g}{\partial X^{j}}\right] =0\text{ \ \ \ .}  \tag{A34.}
\end{equation}
Further we shall assume that each term in the sum is zero, i.e. the equation
is fulfilled for each $i$ and $j$. Substituting expressions (A25) for $e^{(i)%
\text{ }}$and $e^{(j)}$ and (A26) for $\frac{\partial e^{(i)}}{\partial X^{j}%
}$ and $\frac{\partial e^{(j)}}{\partial X^{i}}$, equation (A34) can be
rewritten in the following simple form:
\begin{equation*}
\frac{\partial W_{i}}{\partial X^{j}}g(X_{1},..,X_{i-1},X_{i+1},..,X_{n})e^{%
\int \frac{\partial W_{i}}{\partial X^{j}}dX^{i}}=
\end{equation*}
\begin{equation}
=\frac{\partial W_{j}}{\partial X^{i}}%
g(X_{1},..,X_{j-1},X_{j+1},..,X_{n})e^{\int \frac{\partial Wj}{\partial Xi}%
dX^{j}}\text{ \ \ .}  \tag{A35.}
\end{equation}
Differentiating this expression by $X^{i}$ and making use again of (A25) and
(A26), the following simple differential equation can be obtained:
\begin{equation*}
W_{j,i}\frac{\partial \ln g(X_{1},.,X_{j-1,}X_{j+1},..,X_{n})}{\partial X^{i}%
}+(W_{j,ii}-W_{i,j}W_{j,i}-
\end{equation*}
\begin{equation}
-\frac{W_{i,ji}W_{j,i}}{W_{i,j}})+W_{j,i}e^{\int \left[ \frac{\partial
^{2}W_{j}}{\partial X^{i2}}-\frac{\partial W_{j}}{\partial X^{i}}\right]
dX^{j}}=0\text{ \ \ .}  \tag{A36.}
\end{equation}
The first case, when this equation will be satisfied will be
\begin{equation}
W_{j,i}=\frac{\partial W_{j}}{\partial X^{i}}=0\text{ \ \ }\Longrightarrow
W_{j}=f(X_{1},..,X_{i-1},X_{i+1},..,X_{n})\text{ \ \ .}  \tag{A37.}
\end{equation}
Since this will be fulfilled for \textbf{every} $i$, then $W_{j}$ should be
a constant, which of course is a very rare and special case.

The second, more realistic case is when the function $g$ is a solution of
the differential equation (A36) for every $i$ and $j$ ($i\neq j$):
\begin{equation}
g(X_{1},.,X_{j-1,}X_{j+1},..,X_{n})=F(X_{1},.,X_{i-1,}X_{i+1},..,X_{n})e^{%
\int \widetilde{Q}(X_{1},...,X_{n})dX^{i}}\text{ \ \ ,}  \tag{A38.}
\end{equation}
where
\begin{equation}
\widetilde{Q}(X_{1},...,X_{n})\equiv \left( W_{i,j}+\frac{W_{i,ji}}{W_{i,j}}-%
\frac{W_{j,ii}}{W_{j,i}}\right) -e^{\int \left( \frac{\partial ^{2}W_{j}}{%
\partial X^{i2}}-\frac{\partial W_{j}}{\partial X^{i}}\right) dX^{j}}\text{ .%
}  \tag{A39.}
\end{equation}
Since the function $g(X_{1},.,X_{j-1,}X_{j+1},..,X_{n})$ on the L. H. S. of
(A38) does not depend on the variable $X_{j}$, then for each $j$ the unknown
function $F(X_{1},.,X_{i-1,}X_{i+1},..,X_{n})$ can be obtained after
differentiating both sides of (A38) by $X^{j}$. Thus the function $F$ is a
solution of the following differential equation
\begin{equation}
0=\frac{\partial F(X_{1},.,X_{i-1,}X_{i+1},..,X_{n})}{\partial X^{j}}e^{\int
\widetilde{Q}dX^{i}}+F(X_{1},.,X_{i-1,}X_{i+1},..,X_{n})e^{\int \frac{%
\partial \widetilde{Q}}{\partial X^{j}}dX^{i}}\text{ \ \ .}  \tag{A40.}
\end{equation}
This precludes the proof that the function $g$ in (A25) can be determined in
such a way that $e^{(i)}$ would be indeed a scalar quantity and therefore $%
e\equiv e^{(i)}$. Throughout the whole proof, we assumed that $W_{i}$,
determined by (A23), is a vector. This of course should be proved in the
same way, in which it was proved that the connection $\widetilde{\Gamma }%
_{ij}^{k}$ has affine transformation properties.

\section{\protect\bigskip APPENDIX \ B: \ SOME \ COEFFICIENT \ FUNCTIONS \
IN \ THE \ FINAL \ SOLUTIONS \ FOR \ $dX^{1}$, $dX^{2}$, $dX^{3}$ \ IN
SECTION \ III}

\bigskip The functions $h_{1},h_{2},h_{3}$ (depending on the Weierstrass
function $\rho (z)$) and the functions $l_{1},l_{2},l_{2}$ (not depending on
$\rho (z)$) in the expression (3.21) for the solution $dX^{2}$ of the cubic
algebraic equation are
\begin{equation}
h_{1}\equiv 2p\left( \frac{b_{2}}{c_{2}}-2\frac{b_{2}}{d_{2}}L_{1}^{(2)}\rho
(z)\right) \left( 2\Gamma _{12}^{r}g_{1r}+\Gamma _{11}^{r}g_{2r}\right)
\text{ \ \ \ \ ,}  \tag{B1.}
\end{equation}
\begin{equation*}
h_{2}\equiv \left( \frac{b_{2}}{c_{2}}-2\frac{b_{2}}{d_{2}}L_{1}^{(2)}\rho
(z)\right) \left( K_{12}^{(1)}+K_{21}^{(1)}\right) -
\end{equation*}
\begin{equation}
-2pL_{1}^{(2)}\rho (z)\left( 2\Gamma _{12}^{r}g_{2r}+\Gamma
_{22}^{r}g_{1r}\right) \text{ \ \ \ \ \ ,}  \tag{B2.}
\end{equation}
\begin{equation}
h_{3}\equiv \left( \frac{b_{2}}{c_{2}}-2\frac{b_{2}}{d_{2}}L_{1}^{(2)}\rho
(z)\right) K_{2}^{(2)}-L_{1}^{(2)}\rho (z)K_{22}^{(1)}\text{ \ \ \ ,}
\tag{B3.}
\end{equation}
\begin{equation}
l_{1}\equiv 2p\left( \frac{d_{2}}{c_{2}}-2\frac{b_{2}}{d_{2}}%
L_{1}^{(2)}\right) \left( 2\Gamma _{12}^{r}g_{1r}+\Gamma
_{11}^{r}g_{2r}\right) \text{ \ \ \ \ ,}  \tag{B4.}
\end{equation}
\begin{equation*}
l_{2}\equiv \left( \frac{d_{2}}{c_{2}}-2\frac{b_{2}}{d_{2}}%
L_{1}^{(2)}\right) \left( K_{12}^{(1)}+K_{21}^{(1)}\right) -
\end{equation*}
\begin{equation}
-2pL_{1}^{(2)}\left( 2\Gamma _{12}^{r}g_{2r}+\Gamma _{22}^{r}g_{1r}\right)
\text{ \ \ \ \ \ ,}  \tag{B5.}
\end{equation}
\begin{equation}
l_{3}\equiv \left( \frac{d_{2}}{c_{2}}-2\frac{b_{2}}{d_{2}}%
L_{1}^{(2)}\right) K_{2}^{(2)}-L_{1}^{(2)}K_{22}^{(1)}\text{ \ \ \ .}
\tag{B6.}
\end{equation}
The functions $F_{1},F_{2},F_{3},f_{1},f_{2},f_{3},f_{4},\widetilde{g}_{1},%
\widetilde{g}_{2},\widetilde{g}_{3}$ in the solution for $dX^{1}$ are the
following
\begin{equation}
F_{1}\equiv 2p\left( 2\Gamma _{12}^{r}g_{2r}+\Gamma _{22}^{r}g_{1r}\right)
\text{ \ \ ; \ }F_{2}\equiv \left( 1+2\frac{d_{2}}{c_{2}}\right) \left(
K_{12}^{(1)}+K_{21}^{(1)}\right) \text{ \ \ ,}  \tag{B7.}
\end{equation}
\begin{equation}
F_{3}\equiv 2p\left( 1+2\frac{d_{2}}{c_{2}}\right) \left( 2\Gamma
_{12}^{r}g_{1r}+\Gamma _{11}^{r}g_{2r}\right) \text{ \ \ \ \ ,}  \tag{B8.}
\end{equation}
\begin{equation}
f_{1}\equiv -2\frac{b_{1}}{d_{1}}L_{1}^{(1)}F_{1}\text{ \ \ ; \ \ }%
f_{3}\equiv \frac{b_{1}}{c_{1}}F_{2}-L_{1}^{(1)}\text{ \ ,}  \tag{B9.}
\end{equation}
\begin{equation}
f_{2}\equiv \frac{b_{1}}{c_{1}}F_{1}-L_{1}^{(1)}F_{3}-2\frac{b_{1}}{d_{1}}%
L_{1}^{(1)}F_{2}\text{ \ \ ,}  \tag{B10.}
\end{equation}
\begin{equation}
\widetilde{g}_{1}\equiv \left( \frac{d_{1}}{c_{1}}-2\frac{b_{1}}{d_{1}}%
\right) F_{1}\text{ \ \ ,}  \tag{B11.}
\end{equation}
\begin{equation}
\widetilde{g}_{2}\equiv \left( \frac{d_{1}}{c_{1}}-2\frac{b_{1}}{d_{1}}%
\right) F_{2}-L_{1}^{(1)}F_{3}\text{ \ \ ,}  \tag{B12.}
\end{equation}
\begin{equation}
\widetilde{g}_{3}\equiv \left( \frac{d_{1}}{c_{1}}-2\frac{b_{1}}{d_{1}}%
\right) K_{1}^{(2)}-L_{1}^{(1)}K_{11}^{(1)}\text{ \ \ \ .}  \tag{B13.}
\end{equation}

\section{\protect\bigskip\ APPENDIX C: BLOCK STRUCTURE METHOD \ FOR SOLVING
\ THE SYSTEM OF EQUATIONS $\ g_{ij}g^{jk}=\protect\delta _{i}^{k} $ \ IN \ \
THE \ GENERAL \ N - DIMENSIONAL \ CASE.}

\subsection*{\protect\bigskip C1. BLOCK STRUCTURE METHOD FOR \ THE \ PARTIAL
$n=3$ \ CASE}

\bigskip The purpose of the present section will be to develop a method for
solving the system of equations $g_{ij}g^{jk}=\delta _{i}^{k}$ for the case $%
n=3$ . Since in principle the system for $n=3$ can be solved in an
elementary manner, the aim will be not to find another more convenient
method, but rather than that find a method, which can further be generalized
to higher dimensions.

Let us first write down the system of six equations for different indices $i$
and $k:$
\begin{equation}
g_{11}g^{12}+g_{12}g^{22}+g_{13}g^{32}=0\text{ \ \ ,}  \tag{C1. }
\end{equation}

\begin{equation}
g_{21}g^{11}+g_{22}g^{21}+g_{23}g^{31}=0\text{ \ \ ,}  \tag{C2. }
\end{equation}
\begin{equation}
g_{11}g^{13}+g_{12}g^{23}+g_{13}g^{33}=0\text{ \ ,}  \tag{C3. }
\end{equation}
\begin{equation}
g_{31}g^{11}+g_{32}g^{21}+g_{33}g^{31}=0\text{ \ ,}  \tag{C4. }
\end{equation}
\begin{equation}
g_{21}g^{13}+g_{22}g^{23}+g_{23}g^{33}=0\text{ \ ,}  \tag{C5. }
\end{equation}
\begin{equation}
g_{31}g^{12}+g_{32}g^{22}+g_{33}g^{32}=0\text{ \ .}  \tag{C6. }
\end{equation}
In matrix notations and for the unknown variables $%
g_{11},g_{22},g_{33},g_{12},g_{23}$ and $g_{13}$ the system can be written
as
\begin{equation}
AX=0\text{ \ ,}  \tag{C7. }
\end{equation}
where $A$ is the $6\times 6$ dimensional matrix
\begin{equation}
A\equiv \left(
\begin{array}{cccccc}
g^{12} & 0 & 0 & g^{22} & 0 & g^{32} \\
0 & g^{21} & 0 & g^{11} & g^{31} & 0 \\
g^{13} & 0 & 0 & g^{23} & 0 & g^{33} \\
0 & 0 & g^{31} & 0 & g^{21} & g^{11} \\
0 & g^{23} & 0 & g^{13} & g^{33} & 0 \\
0 & 0 & g^{32} & 0 & g^{22} & g^{12}
\end{array}
\right)  \tag{C8. }
\end{equation}
and the transponed vector $X^{T}$ is

\begin{equation}
X^{T}\equiv (g_{11},g_{22},g_{33},g_{12},g_{23},g_{13})\text{ \ .}
\tag{C9.
}
\end{equation}
By a direct calculation it can easily be checked that $detA=0$ and therefore
the homogeneous sysytem of equations (C7) has arbitrary solutions.

Now let us write down the rest of the system of equations for the case $i=k$%
, which in matrix notations reads as
\begin{equation}
BX=\overline{1}^{T}\text{ ,}  \tag{C10.}
\end{equation}
where $\overline{1}^{T}=(1,1,1)$ and $B$ is the $6\times 3$ matrix
\begin{equation}
B\equiv \left(
\begin{array}{cccccc}
g^{11} & 0 & 0 & g^{12} & 0 & g^{13} \\
0 & g^{22} & 0 & g^{12} & g^{23} & 0 \\
0 & 0 & g^{33} & 0 & g^{23} & g^{13}
\end{array}
\right) \text{ \ \ \ .}  \tag{C11. }
\end{equation}
Since the system (C10) can be written either as
\begin{equation}
B_{1}X=B_{2}\text{ \ \ \ \ \ \ }or\text{ \ \ \ \ \ \ }B_{2}X=B_{1}\text{ \ \
\ ,}  \tag{C12. }
\end{equation}
\begin{equation}
B_{1}\equiv \left(
\begin{array}{ccc}
g^{11} & 0 & 0 \\
0 & g^{22} & 0 \\
0 & 0 & g^{33}
\end{array}
\right) \text{ \ \ \ ;\ \ \ \ \ }B_{2}\equiv \left(
\begin{array}{ccc}
g^{12} & 0 & g^{13} \\
g^{12} & g^{23} & 0 \\
0 & g^{23} & g^{13}
\end{array}
\right)  \tag{C13. }
\end{equation}
and $detB_{1}\neq 0$; $detB_{2}\neq 0$, it is clear that one can choose
freely either the three components $\left( g_{11},g_{22},g_{33}\right) $ or $%
\left( g_{12},g_{23},g_{13}\right) $ and after that \textbf{fix the other
three components }by finding the unique solutions of the system (C10). Also,
it is important to note that \textbf{the above conclusions do not change if
the contravariant metric tensor is chosen in the form of the factorized
product }$g^{ij}=dX^{i}dX^{j}$\textbf{.}

Now it is worth noting that the entire system of equations $%
g_{ij}g^{jk}=\delta _{i}^{k}$ $\ $\ for $n=3$ can be written as
\begin{equation}
\widetilde{A}X=E\text{ \ \ \ ,}  \tag{C13. }
\end{equation}
where $E^{T}$ is the transponed $9-$ vector:
\begin{equation}
E^{T}=(1,0,0,0,1,1,0,0,1)\text{ \ \ \ \ .}  \tag{C14. }
\end{equation}
The $6\times 9$ matrix $\widetilde{A}$ has the following interesting block
structure:
\begin{equation}
\widetilde{A}\equiv \left(
\begin{array}{cc}
P_{1} & Q_{1} \\
P_{2} & Q_{2} \\
P_{3} & Q_{3}
\end{array}
\right) \text{ \ \ \ ,}  \tag{C15. }
\end{equation}
where ($s=1,2,3$) the matrices $P_{s}$ and $Q_{s}$ are the following:
\begin{equation}
P_{s}\equiv \left(
\begin{array}{ccc}
g^{s1} & g^{s2} & g^{s3} \\
0 & g^{s1} & 0 \\
0 & 0 & g^{s1}
\end{array}
\right) \text{ \ \ \ \ ;\ \ \ \ \ }Q_{s}\equiv \left(
\begin{array}{ccc}
0 & 0 & 0 \\
g^{2s} & g^{3s} & 0 \\
0 & g^{2s} & g^{3s}
\end{array}
\right) \text{ \ \ .}  \tag{C16. }
\end{equation}
In order to find the solution $X=\widetilde{A}^{-1}E$ of the system (C13),
one has to find the inverse matrix $\widetilde{A}^{-1}$. For the case of
\textbf{quadratic} matrices with the block structure
\begin{equation}
M\equiv \left(
\begin{array}{cc}
A & B \\
C & D
\end{array}
\right) \text{ \ ,}  \tag{C17. }
\end{equation}
where $A,B,C$ and $D$ are $n\times n,q\times n,n\times q$ and $q\times q$
matrices correspondingly, the so called \textbf{Frobenius formulae} [20] for
finding the inverse matrix $M^{-1}$ is valid
\begin{equation}
M^{-1}=\left(
\begin{array}{cc}
A^{-1}+A^{-1}BH^{-1}CA^{-1} & -A^{-1}BH^{-1} \\
-H^{-1}CA^{-1} & H^{-1}
\end{array}
\right) \text{ \ \ \ ,}  \tag{C18. }
\end{equation}
where $H$ is the matrix:
\begin{equation}
H\equiv D-CA^{-1}B\text{ \ .}  \tag{C19. }
\end{equation}
In the present case, the \textbf{Frobenius formulae cannot be applied to the
block matrix (C15), since it is not a quadratic one.}However, if $X$ is a
solution of the system (C13), then it is a solution also of the equation $%
\widetilde{A}^{T}\widetilde{A}X=E\widetilde{A}^{T}$, where the $6\times 9$
matrix $\widetilde{A}$, multiplied to the left with its transponed one,
gives already the quadratic $9\times 9$ matrix $\widetilde{A}^{T}\widetilde{A%
}$. Further it shall be demonstrated how the Frobenius inversion formulae
can be applied twice in respect to $\widetilde{A}^{T}\widetilde{A}$.

The matrix $\widetilde{A}^{T}\widetilde{A}$ can be calculated to be the
following block matrix:
\begin{equation}
\widetilde{A}^{T}\widetilde{A}=\left(
\begin{array}{cc}
\sum\limits_{i=1}^{3}P_{i}^{T}P_{i} & \sum\limits_{j=1}^{3}P_{j}^{T}Q_{j} \\
\sum\limits_{k=1}^{3}Q_{k}^{T}P_{k} & \sum\limits_{l=1}^{3}Q_{l}^{T}Q_{l}
\end{array}
\right) \text{ \ }  \tag{C20. }
\end{equation}
and $P_{i},Q_{j}$ are the corresponding matrices (C15) and their transponed
ones. The block matrices in (C20) are found to be the following $3\times 3$
matrices, which shall further be identified with the corresponding block -
matrices $A,B,C$ and $D$ in (C17):
\begin{equation}
A\equiv \sum\limits_{i=1}^{3}P_{i}^{T}P_{i}=\left(
\begin{array}{ccc}
\sum\limits_{i=1}^{3}(g^{i1})^{2} & \sum\limits_{i=1}^{3}g^{i1}g^{i2} &
\sum\limits_{i=1}^{3}g^{i1}g^{i3} \\
\sum\limits_{i=1}^{3}g^{i1}g^{i2} & \sum\limits_{i=1}^{3}\left[
(g^{i1})^{2}+(g^{i2})^{2}\right] & \sum\limits_{i=1}^{3}g^{i2}g^{i3} \\
\sum\limits_{i=1}^{3}g^{i1}g^{i3} & \sum\limits_{i=1}^{3}g^{i2}g^{i3} &
\sum\limits_{i=1}^{3}\left[ (g^{i1})^{2}+(g^{i3})^{2}\right]
\end{array}
\right) \text{ \ \ \ \ \ ,}  \tag{C21. }
\end{equation}
\begin{equation}
B\equiv
\sum\limits_{j=1}^{3}P_{j}^{T}Q_{j}=\sum\limits_{j=1}^{3}g^{j1}\left(
\begin{array}{ccc}
0 & 0 & 0 \\
g^{2j} & g^{3j} & 0 \\
0 & g^{j2} & g^{j3}
\end{array}
\right) \text{ \ \ \ ,}  \tag{C22. }
\end{equation}
\begin{equation}
C\equiv
\sum\limits_{k=1}^{3}Q_{k}^{T}P_{k}=\sum\limits_{k=1}^{3}g^{k1}\left(
\begin{array}{ccc}
0 & g^{k2} & 0 \\
0 & g^{k3} & g^{k2} \\
0 & 0 & g^{k3}
\end{array}
\right) \text{ \ \ \ \ \ ,}  \tag{C23. }
\end{equation}
\begin{equation}
D\equiv \sum\limits_{l=1}^{3}Q_{l}^{T}Q_{l}=\left(
\begin{array}{ccc}
\sum\limits_{l=1}^{3}(g^{2l})^{2} & \sum\limits_{l=1}^{3}g^{2l}g^{3l} & 0 \\
\sum\limits_{l=1}^{3}g^{2l}g^{3l} & \sum\limits_{l=1}^{3}\left[
(g^{21})^{2}+(g^{3l})^{2}\right] & \sum\limits_{l=1}^{3}g^{2l}g^{3l} \\
0 & \sum\limits_{l=1}^{3}g^{2l}g^{3l} & \sum\limits_{i=1}^{3}(g^{3l})^{2}
\end{array}
\right) \text{ \ \ \ .}  \tag{C24. }
\end{equation}
Note that the diagonal block - matrices $A$ and $D$ have non - zero
determinants (even if \ $g^{ij}=dX^{i}dX^{j}$), while the non - diagonal
block - matrices $B$ and $D$ have zero - determinants. However, in order to
apply the Frobenius formulae for inverting the matrix (C20) it is sufficient
to have as invertible only the matrix $A$ (and of course the matrix $H$).

\subsection*{\protect\bigskip C2. MODIFICATION OF THE BLOCK STRUCTURE OF THE
MATRIX A}

The above presented method has nevertheless the following shortcomings:

1. It deals with an rectangular $p\times q$ matrix $A$ for a system of
equations with $p=\left(
\begin{array}{c}
n \\
2
\end{array}
\right) +n$ variables and $q=n^{2}$ equations. At the same time it would
have been much better to deal with a quadratic matrix at the beginning.

2. The block - matrix $A$ contains two types of matrices $P_{s}$ and $Q_{s},$%
while it would be more convenient to have just one type of an elementary
''constituent'' $E_{k}^{(i)}$ with a definite structure, where the indice $i$
denotes the corresponding column in the block matrice $A$ (i.e. the number
of the column, containing block - matrices) and the indice $k$ - the
corresponding row of block - matrices.

3. The matrix $Q_{s},$ given by the formulae (C15) has a zero determinant.

To avoid these shortcomings, let us define an extended $9-$ dimensional
vector $Y$, whose transponed one is
\begin{equation}
Y^{T}\equiv (g_{11},g_{12},g_{13},g_{21},g_{22},g_{23},g_{31},g_{32},g_{33})%
\text{ \ \ , }  \tag{C26. }
\end{equation}

where the elements $g_{21},g_{32}$ and $g_{31}$ formally shall be considered
unknown, although they are equal to their symmetric counterparts $%
g_{12},g_{23}$ and $g_{13}.$ Then the system of equations can be written as
\begin{equation}
MY=\overline{1}\text{ \ \ \ ,}  \tag{C27. }
\end{equation}
where $\overline{1}^{T}$ is the transponed $9$ -dimensional vector
\begin{equation}
\overline{1}^{T}\equiv (1,0,0,0,1,0,0,0,1)  \tag{C28. }
\end{equation}
and the $9\times 9$ matrix $M$ \ has the following block structure
\begin{equation}
M\equiv \left(
\begin{array}{ccc}
E_{1}^{(1)} & E_{1}^{(2)} & E_{1}^{(3)} \\
E_{2}^{(1)} & E_{2}^{(2)} & E_{2}^{(3)} \\
E_{3}^{(1)} & E_{3}^{(2)} & E_{3}^{(3)}
\end{array}
\right) \text{ \ \ \ \ . }  \tag{C29. }
\end{equation}
The elementary $3\times 3$ block matrices $E_{k}^{(1)}$, $E_{k}^{(2)}$ and $%
E_{k}^{(3)}$ ($k=1,2,3$ denotes the number of the row) in each column are
the following
\begin{equation}
E_{k}^{(1)}\equiv \frac{1}{2}\left(
\begin{array}{ccc}
2g^{k1} & g^{k2} & g^{k3} \\
0 & g^{k1} & 0 \\
0 & 0 & g^{k1}
\end{array}
\right) \text{ \ \ \ \ ,}  \tag{C30. }
\end{equation}
\begin{equation}
E_{k}^{(2)}\equiv \frac{1}{2}\left(
\begin{array}{ccc}
g^{2k} & 0 & 0 \\
g^{1k} & 2g^{2k} & g^{3k} \\
0 & 0 & g^{2k}
\end{array}
\right) \text{ \ \ \ \ ,}  \tag{C31. }
\end{equation}
\begin{equation}
E_{k}^{(3)}\equiv \frac{1}{2}\left(
\begin{array}{ccc}
g^{3k} & 0 & 0 \\
0 & g^{3k} & 0 \\
g^{1k} & g^{2k} & g^{3k}
\end{array}
\right) \text{ \ \ \ .}  \tag{C32. }
\end{equation}
From (C30 - C32) it is seen that depending on the indice $k$ of the
elementary matrices row and of the indice $(s)$ for the elementary matrices
column, the matrix $E_{k}^{(s)}$ for the $n-$ dimensional case can be
written as
\begin{equation}
E_{k}^{(s)}\equiv \frac{1}{(n-1)}\left(
\begin{array}{cccccc}
g^{ks} & 0... & .... & ... & 0 & 0 \\
0 & g^{ks} & 0.. & ... & 0 & 0 \\
.... & .... & ... & .... & .... & ... \\
g^{k1} & g^{k2}. & .. & (n-1)g^{ks}.. & ... & g^{kn} \\
.... & ... & ... & ............ & .... & .... \\
0 & 0 & 0 & ............. & 0 & g^{ks}
\end{array}
\right) \text{ \ \ \ .}  \tag{C33. }
\end{equation}
In other words, on the main diagonal of the matrix $E_{k}^{(s)}$ the
elements $g^{ks}$ are situated, on the $s-$th row - the elements $\left(
\begin{array}{cccccc}
g^{k1}, & g^{k2},. & .. & (n-1)g^{ks}.. & ... & ,g^{kn}
\end{array}
\right) $ with an element $(n-1)g^{ks}$ on the $s-$th row and $s-$th column$%
. $This block structure (with some slight modifications) shall be obtained
also for the $n-$ dimensional case, and thus the $3-$ dimensional case
really helps to make the corresponding generalization for the $n-$
dimensional case.

Another advantage of the block - matrice representation (C29) is that it
gives a possibility to apply \textbf{twice} the Frobenius formulae.
Correspondingly, in the $n-$ dimensional case the Frobenius formulae will be
applied $(n-1)$ times.

For the representation (C29) this can be performed in the following two
steps:

\textbf{Step 1}. Apply the Frobenius formulae to the sub - matrix $%
\widetilde{A}$ of the matrix (C29), where $\widetilde{A}$ is the matrix of
the first two columns and first two rows
\begin{equation}
\widetilde{A}\equiv \left(
\begin{array}{cc}
E_{1}^{(1)} & E_{1}^{(2)} \\
E_{2}^{(1)} & E_{2}^{(2)}
\end{array}
\right)  \tag{C34. }
\end{equation}
The inverse matrix $\widetilde{A}^{-1}$ will be given by the formulae (C18)
with the following identifications
\begin{equation}
A\equiv E_{1}^{(1)}\text{; \ \ \ \ \ }B\equiv E_{1}^{(2)}\text{ \ ; \ \ \ }%
C\equiv E_{2}^{(1)}\text{ \ ; \ \ \ }D\equiv E^{(2)}\text{ \ \ .}
\tag{C35.
}
\end{equation}
\textbf{Step 2}. Denote by $\widetilde{B},\widetilde{C}$ and $\widetilde{D}$
the following $3\times 6,6\times 3$ and $3\times 3$ matrices:
\begin{equation}
\widetilde{B}\equiv \left(
\begin{array}{c}
E_{1}^{(3)} \\
E_{2}^{(3)}
\end{array}
\right) \text{; \ \ \ \ \ \ \ \ }\widetilde{C}\equiv \left(
\begin{array}{cc}
E_{3}^{(1)} & E_{3}^{(2)}
\end{array}
\right) \text{ \ \ ; \ \ \ \ \ \ \ \ }\widetilde{D}\equiv E_{3}^{(3)}\text{
\ \ \ \ .}  \tag{C36. }
\end{equation}

Therefore the matrix $M$ is written as
\begin{equation}
M\equiv \left(
\begin{array}{cc}
\widetilde{A} & \widetilde{B} \\
\widetilde{C} & \widetilde{D}
\end{array}
\right) \text{ \ }  \tag{C37. }
\end{equation}
and the Frobenius formulae is applied again to the block - matrix $M$,
resulting in the inverse matrix
\begin{equation}
M^{-1}=\left(
\begin{array}{cc}
\widetilde{A}^{-1}+\widetilde{A}^{-1}\widetilde{B}\widetilde{H}^{-1}%
\widetilde{C}\widetilde{A}^{-1} & -\widetilde{A}^{-1}\widetilde{B}\widetilde{%
H}^{-1} \\
-\widetilde{H}^{-1}\widetilde{C}\widetilde{A}^{-1} & \widetilde{H}^{-1}
\end{array}
\right) \text{ \ \ \ ,}  \tag{C38. }
\end{equation}
where $\widetilde{H}$ is the matrix:
\begin{equation}
\widetilde{H}\equiv \widetilde{D}-\widetilde{C}\widetilde{A}^{-1}\widetilde{B%
}\text{ \ \ \ }  \tag{C39. }
\end{equation}
and $\widetilde{A}^{-1}$ is the calculated inverse matrix from step 1.

Now let us find the matrix $\widetilde{A}^{-1}\widetilde{B}$ in terms of the
matrices $E_{k}^{(s)}:$
\begin{equation}
\widetilde{A}^{-1}\widetilde{B}=\left(
\begin{array}{cc}
E_{1}^{(1)-1}+E_{1}^{(1)-1}E_{1}^{(2)}H_{1}^{-1}E_{2}^{(1)}E_{1}^{(1)-1} &
-E_{1}^{(1)}E_{1}^{(2)}H_{1}^{-1} \\
-H_{1}^{-1}E_{2}^{(1)}E_{1}^{(1)-1} & H_{1}^{-1}
\end{array}
\right) \left(
\begin{array}{c}
E_{1}^{(3)} \\
E_{2}^{(3)}
\end{array}
\right) \text{ \ \ \ ,}  \tag{C40. }
\end{equation}
\bigskip\ where $H_{1}$ is given by
\begin{equation}
H_{1}\equiv E_{2}^{(2)}-E_{2}^{(1)}E_{1}^{(1)-1}E_{1}^{(2)}  \tag{C41. }
\end{equation}
and is obviously different from $\widetilde{H}$ in (C39). Finally $%
\widetilde{A}^{-1}\widetilde{B}$ is calculated to be the following matrix -
column:
\begin{equation}
\widetilde{A}^{-1}\widetilde{B}=\left(
\begin{array}{c}
E_{2}^{(1)-1}E_{2}^{(2)}E_{1}^{(2)-1}E_{1}^{(3)}-E_{1}^{(1)-1}E_{1}^{(2)}X^{-1}E_{1}^{(1)}E_{2}^{(1)-1}E_{2}^{(3)}
\\
-X^{-1}E_{1}^{(3)}+X^{-1}E_{1}^{(1)}E_{2}^{(1)-1}E_{2}^{(3)}
\end{array}
\right)  \tag{C42. }
\end{equation}
with
\begin{equation}
X\equiv E_{1}^{(1)}E_{2}^{(1)-1}E_{2}^{(2)}-E_{1}^{(2)}\text{ \ \ \ \ \ \ \ .%
}  \tag{C43. }
\end{equation}
Note the obvious advantage of the method, in which all the sub - matrices
are invertible in comparison with the \ method in the previous sub-section
(C1), when only the diagonal submatrices had been invertible. For example,
in the present case it had been used that block element $%
E_{1}^{(1)}E_{1}^{(2)}H_{1}^{-1}$ in (C40) can be represented as
\begin{equation}
-E_{1}^{(1)}E_{1}^{(2)}H_{1}^{-1}=-(H_{1}E_{1}^{(2)-1}E_{1}^{(1)})^{-1}\text{
\ \ \ , }  \tag{C44. }
\end{equation}
which obviously cannot be applied in \ the previous case.

Further the expression for $\widetilde{H\text{ }}$ \ can be found, which
participates in the formulae for $M^{-1}$ :
\begin{equation*}
\widetilde{H}\equiv \widetilde{D}-\widetilde{C}\widetilde{A}^{-1}\widetilde{B%
}=E_{3}^{(3)}-E_{3}^{(1)}\{E_{2}^{(1)-1}E_{2}^{(2)}E_{1}^{(2)-1}E_{1}^{(3)}-
\end{equation*}
\begin{equation}
-E_{1}^{(1)-1}E_{1}^{(2)}X^{-1}E_{1}^{(1)}E_{2}^{(1)-1}E_{2}^{(3)}%
\}-E_{3}^{(2)}X^{-1}\{-E_{1}^{(3)}+E_{1}^{(1)}E_{2}^{(1)-1}E_{2}^{(3)}\}%
\text{ \ \ \ .}  \tag{C45. }
\end{equation}
In order to calculate $M^{-1}$ according to (C38), one needs to find also
the expression for $\widetilde{C}\widetilde{A}^{-1}:$
\begin{equation}
\widetilde{C}\widetilde{A}^{-1}=\left(
\begin{array}{c}
E_{3}^{(1)}E_{1}^{(1)-1}-E_{3}^{(1)}E_{1}^{(1)-1}E_{1}^{(2)}X^{-1}-E_{3}^{(2)}X^{-1}
\\
\lbrack
E_{3}^{(2)}-E_{3}^{(1)}E_{1}^{(1)-1}E_{1}^{(2)}](X^{-1}E_{1}^{(1)}E_{2}^{(1)-1}
\end{array}
\right) \text{ \ \ \ .}  \tag{C46. }
\end{equation}

The entire calculation for $M^{-1}$ is rather cumbersome and wil not be
presented. Instead of this, let us denote by $M^{(k)}$ the matrix, obtained
after taking the first $k$ matrix - rows and $k$ matrix - columns in the $n-$
dimensional block matrix $M^{(n)}:$%
\begin{equation}
M^{(n)}\equiv \left(
\begin{array}{cccc}
E_{1}^{(1)} & E_{1}^{(2)} & ............. & E_{1}^{(n)} \\
E_{2}^{(1)} & E_{2}^{(2)} & .............. & E_{2}^{(n)} \\
......... & ...... & .............. & ...... \\
E_{n}^{(1)} & E_{n}^{(2)} & ............... & E_{n}^{(n)}
\end{array}
\right) \text{ \ \ \ .}  \tag{C47. }
\end{equation}
Then one begins first with the calculation of $E_{1}^{(1)-1},$then with $%
M^{(2)}$ (which is in fact the caculated inverse matrix $\widetilde{A}^{-1}$
in (C40), then with $M^{(3)}$ and so on. Each subsequent inverse matrix $%
M^{(k+1)-1}$ is calculated by applying the Frobenius formulae to the
following block matrix:
\begin{equation}
M^{(k+1)}\equiv \left(
\begin{array}{ccc}
M^{(k)} & . & E_{1}^{(k+1)} \\
& . & E_{k}^{(k+1)} \\
E_{k+1}^{(1)}...... & E_{k+1}^{(k)} & E_{k+1}^{(k+1)}
\end{array}
\right) \text{ \ .}  \tag{C48. }
\end{equation}

\subsection*{\protect\bigskip C3. BLOCK \ MATRIX \ STRUCTURE \ IN \ THE \
N-DIMENSIONAL \ CASE}

Following the same algorithm as in the preceeding subsections, we shall try
to find the block structure of the system of equations $g_{ij}g^{jk}=\delta
_{i}^{k}$ in the $n-$dimensional case. For different values of $k$ the
system can be written as
\begin{equation*}
g^{11}g_{1i}+g^{12}g_{2i}+..........+g^{1n}g_{ni}=\delta _{i}^{1}
\end{equation*}
\begin{equation*}
g^{21}g_{1i}+g^{22}g_{2i}+.........+g^{2n}g_{ni}=\delta _{i}^{2}
\end{equation*}
\begin{equation*}
..............................................................
\end{equation*}
\begin{equation}
g^{n1}g_{1i}+g^{n2}g_{2i}+..............+g^{nn}g_{ni}=\delta _{i}^{n}
\tag{C49. }
\end{equation}
In order to understand the structure of the matrix, it would be useful to
write the above system of equations for different values of $i$.

For $i=1$ the system (C49) is
\begin{equation*}
g^{11}g_{11}+g^{12}g_{21}+..........+g^{1n}g_{ni}=1
\end{equation*}
\begin{equation*}
g^{21}g_{11}+g^{22}g_{21}+.........+g^{2n}g_{n1}=0
\end{equation*}
\begin{equation*}
..............................................................
\end{equation*}
\begin{equation}
g^{n1}g_{11}+g^{n2}g_{21}+..............+g^{nn}g_{n1}=0\text{ \ \ \ ,}
\tag{C50. }
\end{equation}
or it can be written as
\begin{equation}
AX_{1}=\left(
\begin{array}{c}
1 \\
0 \\
.. \\
0
\end{array}
\right) \text{ \ \ \ ,}  \tag{C51.}
\end{equation}
where the matrix $A$ is
\begin{equation}
A\equiv \left(
\begin{array}{cccc}
g^{11} & g^{12} & ........... & g^{1n} \\
g^{21} & g^{22} & ........... & g^{2n} \\
..... & ..... & .......... & ..... \\
g^{n1} & g^{n2} & ......... & g^{nn}
\end{array}
\right) \text{ \ \ \ }  \tag{C52. }
\end{equation}
and $X_{1}^{T}$ is the transponed vector
\begin{equation}
X_{1}^{T}\equiv (g_{11},g_{21},........,g_{n1})\text{ \ \ \ \ .}  \tag{C53.}
\end{equation}
For $i=2$ the system (C49) can be written as
\begin{equation}
AX_{2}=\left(
\begin{array}{c}
0 \\
1 \\
.. \\
0
\end{array}
\right) \text{ \ \ ,}  \tag{C54. }
\end{equation}
where $X_{2}^{T}$ is the transponed vector
\begin{equation}
X_{2}^{T}\equiv (g_{12},g_{22},...........,g_{2n})\text{ \ \ \ \ .}
\tag{C55.}
\end{equation}
For $i=k$ the system will be $AX_{k}=\left(
\begin{array}{c}
0 \\
.. \\
1 \\
...
\end{array}
\right) $, where $"1"$ is on the $k-$th place and the vector $X_{k}^{T}$ is
\begin{equation}
X_{k}^{T}\equiv (g_{k1},g_{k2},........,g_{kn})\text{ \ \ \ \ .}  \tag{C56.}
\end{equation}
The corresponding vectors $X_{1},X_{2},......,X_{k}$ represent the rows of t
he symmetric matrix $N$
\begin{equation}
N\equiv \left(
\begin{array}{cccc}
g_{11} & g_{12} & ...... & g_{1n} \\
g_{21} & g_{22} & ..... & g_{2n} \\
.... & .... & .... & ... \\
g_{n1} & g_{n2} & .... & g_{nn}
\end{array}
\right) \text{ \ \ \ \ ,}  \tag{C57. }
\end{equation}
in which the unknown variables are in the lower triangular (half) part of
the matrix (denoted by $N^{tr}$).

Let us construct a $\frac{n(n+1)}{2}\times n^{2}$ dimensional matrix $B$,
which will multiply a $\frac{n(n+1)}{2}$ dimensional vector $Y$, formed by
joining all the consequent rows of the triangular matrix $\ N^{tr}$
\begin{equation}
Y\equiv
(g_{11},g_{12},.......,g_{1n},g_{22},g_{23},......,g_{2n},......,g_{n1},g_{n2},.....,g_{nn})%
\text{ \ \ .}  \tag{C58.}
\end{equation}
Correspondingly the matrix $B$ will have the \textbf{following block
triangular structure}:
\begin{equation}
B\equiv \left(
\begin{array}{cccc}
B_{11} & 0 & ...... & 0 \\
B_{21} & B_{22} & ..... & 0 \\
.... & .... & .... & ... \\
B_{n1} & B_{n2} & .... & B_{nn}
\end{array}
\right) \text{ \ \ \ ,}  \tag{C59. }
\end{equation}
where each of the block matrices $B_{kk}$ on the diagonal is an $n\times
(n-k+1)$ (i.e. $(n-k+1)$ columns and $n$ rows) dimensional matrix, obtained
from the matrix $A$ by removing the first $(k-1)$ columns. For example, $%
B_{11}\equiv A$, but
\begin{equation}
B_{22}\equiv \left(
\begin{array}{cccc}
g^{12} & g^{13} & ...... & g^{1n} \\
g^{22} & g^{23} & ..... & g^{2n} \\
.... & .... & .... & ... \\
g^{n2} & g^{n3} & .... & g^{nn}
\end{array}
\right) \text{ \ \ \ \ .}  \tag{C60. }
\end{equation}
The corresponding matrix $B_{kk}$ will be
\begin{equation}
B_{kk}\equiv \left(
\begin{array}{cccc}
g^{1k} & g^{1,(k+1)} & .... & g^{1,n} \\
g^{2k} & g^{2,(k+1)} & .... & g^{2,n} \\
... & ........ & .... & ..... \\
g^{nk} & g^{n,(k+1)} & ... & g^{nn}
\end{array}
\right) \text{ \ \ \ \ \ .}  \tag{C61. }
\end{equation}
The block - matrices $B_{sk}$ ($s>k)$ are $n\times (n-k+1)$ dimensional ones
with just one nonzero column (the $s-k+1$ column) with the elements $%
g^{k1},g^{k2},.......$
\begin{equation}
B_{sk}\equiv \left(
\begin{array}{ccccc}
0 & ... & g^{k1} & ... & 0 \\
0 & .... & g^{k2} & .... & 0 \\
... & .... & ... & ... & ... \\
0 & ... & g^{k,(n-1)} & .... & 0 \\
0 & ... & g^{k,n} & .... & 0
\end{array}
\right) \text{ \ \ .}  \tag{C62. }
\end{equation}
Let us now have a look at the complex of neighbouring block matrices around
the main block diagonal
\begin{equation}
K^{(k)}\equiv \left(
\begin{array}{cc}
B_{k-1,k-1} & B_{k-1,k} \\
B_{k,k-1} & B_{kk}
\end{array}
\right) \text{ \ \ \ .}  \tag{C63.}
\end{equation}
For illustration of the block matrix multiplication and in order to derive
some useful formulaes, let us calculate $K^{(p)T}K^{(k)}$, which will be
equal to
\begin{equation}
\left(
\begin{array}{cc}
B_{p-1,p-1}^{T}B_{k-1,k-1}+B_{p,p-1}^{T}B_{k,k-1} & B_{p,p-1}^{T}B_{k,k} \\
B_{pp}^{T}B_{k,k-1} & B_{pp}^{T}B_{kk}
\end{array}
\right) \text{ \ \ \ \ . }  \tag{C64. }
\end{equation}
The corresponding terms in the above matrix are:
\begin{equation*}
B_{p-1,p-1}^{T}B_{k-1,k-1}=
\end{equation*}
\begin{equation*}
=\left(
\begin{array}{cccc}
g^{1,(p-1)} & g^{2,(p-1)} & .... & g^{n,(p-1)} \\
g^{1,p} & g^{2,p} & .... & g^{n,p} \\
... & ........ & .... & ..... \\
g^{1,n} & g^{2,n} & ... & g^{nn}
\end{array}
\right) \left(
\begin{array}{cccc}
g^{1,(k-1)} & g^{1,k} & .... & g^{1,n} \\
g^{2,(k-1)} & g^{2,k} & .... & g^{2,n} \\
... & ........ & .... & ..... \\
g^{n,k-1} & g^{n,k} & ... & g^{nn}
\end{array}
\right) =
\end{equation*}
\begin{equation}
=\left(
\begin{array}{cccc}
P_{11} & P_{12} & ... & P_{n-k+2} \\
.... & .... & ... & ..... \\
.. & .... & P_{sr.....} & ...... \\
P_{n-p+2,1} & P_{n-p+2,2} & ...... & P_{n-p+2,n-k+2}
\end{array}
\right) \text{ \ ,}  \tag{C65. }
\end{equation}
where all the elements of the matrix are nonzero and the element $P_{sr}$ on
the $s-$th row and on \ the $r-$th column is equal to $P_{sr}\equiv
\sum\limits_{i=1}^{n}g^{i,p+s-2}g^{i,k+r-2}$,
\begin{equation*}
B_{p,p-1}^{T}B_{k,k-1}\equiv \left(
\begin{array}{cccc}
0 & 0 & .... & 0 \\
g^{p-1,1} & g^{p-1,2} & ... & g^{p-1,n} \\
...... & ...... & .... & ..... \\
0 & 0 & .... & 0
\end{array}
\right) \left(
\begin{array}{cccc}
0 & g^{k-1,1} & ..... & 0 \\
0 & g^{k-1,2} & ... & 0 \\
..... & ...... & .... & ..... \\
0 & g^{k-1,n} & ... & 0
\end{array}
\right) =
\end{equation*}
\begin{equation}
=\left(
\begin{array}{cccc}
0 & 0 & .. & 0 \\
0 & \sum\limits_{i=1}^{n}g^{(p-1),i}g^{(k-1),i} & ... & 0 \\
..... & ...... & .... & ... \\
0 & 0 & ..... & 0
\end{array}
\right) \text{ \ \ \ \ , }  \tag{C66. }
\end{equation}
where $B_{p,p-1}^{T}$ and $B_{k,k-1}$ are $(n-p+2)\times n$ $\ $and $\
n\times (n-k+2)$ matrices and the resulting $n\times n$ matrix has only one
nonzero element $\sum\limits_{i=1}^{n}g^{(p-1),i}g^{(k-1),i}$ \textbf{on the
second row and on second column}. Next let us calculate the matrix $%
B_{pp}^{T}B_{k,k-1}$, which is a product of \ the \ $(n-p+1)\times n$ matrix
$B_{pp}^{T}$ and the $\ n\times (n-k+2)$ matrix $B_{k,k-1}$ with the only
nonzero second column:
\begin{equation*}
B_{pp}^{T}B_{k,k-1}\equiv
\end{equation*}
\begin{equation*}
\equiv \left(
\begin{array}{ccccc}
g^{1,p} & g^{2,p} & .. & .. & g^{n,p} \\
.... & .... & .. & ... & .... \\
g^{1,p+r-1} & g^{2,p+r-1} & ... & ... & g^{n,p+r-1} \\
.... & ... & .. & .. & .. \\
g^{1,n} & g^{2,n} & .. & .... & g^{nn}
\end{array}
\right) \left(
\begin{array}{ccccc}
0 & g^{k-1,1} & ... & 0 & 0 \\
0 & g^{k-1,2} & ... & 0 & 0 \\
... & .... & .... & .. & ... \\
0 & g^{k-1,n-1} & ... & 0 & 0 \\
0 & g^{k-1,n} & .. & 0 & 0
\end{array}
\right) =
\end{equation*}
\begin{equation}
=\left(
\begin{array}{cccc}
0 & F_{12} & ... & 0 \\
0 & F_{22} & ... & 0 \\
.... & .. & ... & ... \\
0 & F_{n-p+1,2} & ... & 0
\end{array}
\right) \text{ \ \ .}  \tag{C67. }
\end{equation}

\bigskip The resulting matrix $B_{pp}^{T}B_{k,k-1}$ has a dimension $%
(n-p+1)\times (n-k+2)$ with the only nonzero second column with an element
on the $r-$th row and on the $2-$nd \ column $F_{r2}\equiv
\sum\limits_{i=1}^{n}g^{i,(p+r-1)}g^{k-1,i}$.

Next let us find the matrix $B_{p,p-1}^{T}B_{k,k}$, which is a product of \
the $(n-p+2)\times n$ \ matrix $B_{p,p-1}^{T}$ and the $n\times (n-k+1)$
matrix $\ B_{k,k}:$
\begin{equation*}
B_{p,p-1}^{T}B_{k,k}=
\end{equation*}
\begin{equation*}
=\left(
\begin{array}{cccc}
0 & 0 & ... & 0 \\
g^{(p-1),1} & g^{(p-1),2} & ... & g^{(p-1),n} \\
... & ... & ... & .... \\
0 & 0 & .. & 0
\end{array}
\right) \left(
\begin{array}{ccccc}
g^{1,k} & .. & g^{1,(k+s-1)} & .. & g^{1,n} \\
g^{2,k} & .. & g^{2,(k+s-1)} & .. & g^{2,n} \\
.. & . & .. & .. & .. \\
... & .. & ... & ... & .. \\
g^{kk} & g^{k,(k+1)} & ... & .. & g^{kn}
\end{array}
\right) =
\end{equation*}
\begin{equation}
=\left(
\begin{array}{ccccc}
0 & .. & 0 & .. & 0 \\
\sum\limits_{i=1}^{n}g^{(p-1),i}g^{i,k} & .. & \sum%
\limits_{i=1}^{n}g^{(p-1),i}g^{i,(k+s-1)} & .. & \sum%
\limits_{i=1}^{n}g^{(p-1),i}g^{i,n} \\
.. & ... & .... & ... & .. \\
0 & .. & 0 & .. & 0 \\
0 & .. & 0 & .. & 0
\end{array}
\right) \text{ \ \ \ \ ,}  \tag{C68. }
\end{equation}
where in the last formulae we have used that the obtained matrix has $%
(n-k+1) $ columns and therefore the indice $k+s-1$ in the expression for the
element $\sum\limits_{i=1}^{n}g^{(p-1),i}g^{i,(k+s-1)}$ on the $2-$nd $\ $%
row and $s- $th column ranges from $k$ $\ $to $k+s-1=k+(n-k+1)-1=n$.

\bigskip It remains only to calculate the matrix $B_{pp}^{T}B_{kk}$, but it
is the same as \ (C65), this time with an element
\begin{equation}
P_{sr}\equiv \sum\limits_{i=1}^{n}g^{i,p+s-1}g^{i,k+r-1}  \tag{C69.}
\end{equation}

on the $s-$th row and on the $r-$th column.

Using the above developed techniques for matrix multiplication, let us
calculate the $\frac{n(n+1)}{2}\times \frac{n(n+1)}{2}$ matrix $B^{T}B$
(recall - $B$ is an $n^{2}\times \frac{n(n+1)}{2}$ matrix and $B^{T}$ is an $%
\frac{n(n+1)}{2}\times n^{2}$ matrix), which is the $n-$dimensional analogue
of the matrix (C19). Using (C63), one has
\begin{equation}
B^{T}B=\left(
\begin{array}{cccc}
\sum\limits_{i=1}^{n}B_{i1}^{T}B_{i1} & \sum\limits_{i=2}^{n}B_{i1}^{T}B_{i2}
& ... & \sum\limits_{i=n}^{n}B_{i1}^{T}B_{in} \\
\sum\limits_{i=2}^{n}B_{i2}^{T}B_{i1} & \sum\limits_{i=2}^{n}B_{i2}^{T}B_{i2}
& ... & \sum\limits_{i=n}^{n}B_{i2}^{T}B_{ni} \\
.... & .. & ... & ... \\
\sum\limits_{i=n}^{n}B_{in}^{T}B_{i1} & .. & ... & \sum%
\limits_{i=n}^{n}B_{in}^{T}B_{in}
\end{array}
\right) \text{ \ \ \ \ \ \ .}  \tag{C70. }
\end{equation}
The above matrix contains \textbf{three types of elements}: \

\textbf{1-st type. Elements below the block diagonal of the type }$%
\sum\limits_{\alpha =j}^{n}B_{\alpha j}^{T}B_{\alpha k}$\textbf{\ with }$k<j$%
\textbf{.} Carrying out the matrix multiplication and for the moment not
taking the summation over $\alpha ,$ we find
\begin{equation}
B_{\alpha j}^{T}B_{\alpha k}=\left(
\begin{array}{ccccc}
0 & 0 & .. & 0 & 0 \\
.. & .. & .. & .. & .. \\
0 & .. & \sum\limits_{i=1}^{n}g^{ji}g^{ki} & .. & 0 \\
0 & .. & .. & .. & .0 \\
0 & 0 & .. & .. & 0
\end{array}
\right) \text{ \ \ \ \ ,}  \tag{C71. }
\end{equation}
where the only nonzero element $\sum\limits_{i=1}^{n}g^{ji}g^{ki}$ in the
matrix is on the $(\alpha -j+1)$ row and on the $(\alpha -k+1)$ column and
the matrices $B_{\alpha j}^{T}$ and $B_{\alpha k}$ are of dimensions $%
(n-j+1)\times n$ and $n\times (n-k+1)$ correspondingly.

Since the first term in the sum $\sum\limits_{\alpha =j}^{n}B_{\alpha
j}^{T}B_{\alpha k}$ will be $B_{\alpha \alpha }^{T}B_{\alpha k}$, let us
find it, performing the same kind of matrix multiplication as in (C67):
\begin{equation}
B_{\alpha \alpha }^{T}B_{\alpha k}=\left(
\begin{array}{ccccc}
0 & 0 & \sum\limits_{i=1}^{n}g^{i\alpha }g^{ki} & 0 & 0 \\
.. & .. & .. & .. & .. \\
0 & .. & \sum\limits_{i=1}^{n}g^{i,\alpha +r-1}g^{k,i} & .. & 0 \\
0 & .. & .. & .. & .0 \\
0 & 0 & \sum\limits_{i=1}^{n}g^{i,n}g^{k,i} & .. & 0
\end{array}
\right) \text{ \ \ \ \ ,}  \tag{C72. }
\end{equation}
where the only nonzero column is the $(\alpha -k+1)$ one and the element in
this column and on the $r-$th row is $\sum\limits_{i=1}^{n}g^{i,\alpha
+r-1}g^{k,i}.$ The matrices $B_{\alpha \alpha }^{T}$ and $B_{\alpha k}$ are
of dimensions $(n-\alpha +1)\times n$ and $\ n\times (n-k+1)$
correspondingly and the resulting matrix $B_{\alpha \alpha }^{T}B_{\alpha k}$
is $(n-\alpha +1)\times (n-k+1)$ dimensional.

\textbf{2-nd type. Elements above the block diagonal of the type }$%
\sum\limits_{\alpha =k}^{n}B_{\alpha j}^{T}B_{\alpha k}$\textbf{\ with }$k>j$%
\textbf{\ [}The summation indice $\alpha $ takes at first the value of that
indice ($k$ or $j)$, which is greater than the other\textbf{]. }The first
term in the above sum is $B_{\alpha j}^{T}B_{\alpha \alpha }$, which
similarly to the expression (C68) can \ be found to be
\begin{equation*}
B_{\alpha j}^{T}B_{\alpha \alpha }=
\end{equation*}
\begin{equation*}
=\left(
\begin{array}{ccccc}
0 & 0 & .. & .. & 0 \\
.. & .. & .. & .. & .. \\
g^{j1} & g^{j2} & .. & . & g^{jn} \\
.. & .. & .. & .. & .. \\
0 & 0 & .. & 0 & 0
\end{array}
\right) \left(
\begin{array}{cccc}
g^{1\alpha } & g^{2\alpha } & .. & g^{n\alpha } \\
g^{1,(\alpha +1)} & g^{2,(\alpha +1)} & . & g^{n,(\alpha +1)} \\
... & ... & .. & .. \\
g^{1,n} & g^{2,n} & ... & g^{n,n}
\end{array}
\right) =
\end{equation*}
\begin{equation}
=\left(
\begin{array}{ccccc}
0 & 0 & .. & 0 & 0 \\
.. & .. & .. & .. & .. \\
\sum\limits_{i=1}^{n}g^{j,i}g^{1,(\alpha +i-1)} & ... & \sum%
\limits_{i=1}^{n}g^{j,i}g^{p,(\alpha +i-1)} & .. & \sum%
\limits_{i=1}^{n}g^{j,i}g^{n,(\alpha +i-1)} \\
.. & .. & ... & .. & .. \\
0 & 0 & ... & 0 & 0
\end{array}
\right) \text{ \ \ \ .}  \tag{C73. }
\end{equation}
The above $(n-j+1)\times (n-\alpha +1)$ matrix $B_{\alpha j}^{T}B_{\alpha
\alpha }$ contains a nonzero $(\alpha -j+1)$ row with an element in the $p-$%
th column, equal to
\begin{equation}
K_{\alpha -j+1,p}=\sum\limits_{i=1}^{n}g^{j,i}g^{p,(\alpha +i-1)}\text{ \ \
\ \ .}  \tag{C73b.}
\end{equation}

\textbf{3-rd type. Elements situated on the block diagonal of the type }
\begin{equation}
\sum\limits_{\alpha =k}^{n}B_{\alpha k}^{T}B_{\alpha k}=B_{\alpha \alpha
}^{T}B_{\alpha \alpha }+\sum\limits_{\alpha =k+1}^{n}B_{\alpha
k}^{T}B_{\alpha k}\text{ \ .}  \tag{C74. }
\end{equation}
Similarly to the calculation of (C69), the first term in (C74) ($B_{\alpha
\alpha }^{T}$ is an $(n-\alpha +1)\times n$ \ matrix ; $B_{\alpha \alpha }$
is an $n\times (n-\alpha +1)$ matrix) is found to be the following $%
(n-\alpha +1)\times (n-\alpha +1)$ matrix (also, $\alpha =k)$ :
\begin{equation}
B_{\alpha \alpha }^{T}B_{\alpha \alpha }=\left(
\begin{array}{cccc}
N_{11} & N_{12} & ... & N_{1,(n-\alpha +1)} \\
N_{21} & N_{22} & ... & N_{2,(n-\alpha +1)} \\
... & ... & ... & .. \\
N_{n-\alpha +1,1} & ... & ... & N_{n-\alpha +1,n-\alpha +1}
\end{array}
\right) \text{ \ \ ,}  \tag{C75. }
\end{equation}
where
\begin{equation}
N_{pq}=\sum\limits_{i=1}^{n}g^{i,\alpha +p-1}g^{i,\alpha +q-1}\text{ \ \ .}
\tag{C76. }
\end{equation}
The second term in (C74) is in fact the $(n-k+1)\times (n-k+1)$ matrix (C71)
for $j=k.$ The summation over $\alpha $ from $\alpha =k+1$ to $n$ will give
a diagonal $(n-k+1)\times (n-k+1)$ matrix with an element $G_{\alpha
-k+1,\alpha -k+1}=\sum\limits_{i=1}^{n}(g^{k,i})^{2}$ on the $(\alpha -k+1)$
row and on the $\alpha -k+1)$ column, which will be situated on the main (%
\textbf{block)} diagonal from $\alpha =k+1$ to $n$. Since for $\alpha =k+1$
we have $\alpha -k+1=k+1-k+1=2$ and for $\alpha =n$ \ we have also $\alpha
-k+1=n-k+1$, this means that the matrix $\sum\limits_{\alpha
=k+1}^{n}B_{\alpha k}^{T}B_{\alpha k}$ will have the following structure:

\begin{equation}
\sum\limits_{\alpha =k+1}^{n}B_{\alpha k}^{T}B_{\alpha k}=\left(
\begin{array}{ccccc}
0 & 0 & .. & 0 & 0 \\
0 & \sum\limits_{i=1}^{n}(g^{k,i})^{2} & .. & 0 & 0 \\
0 & 0 & \sum\limits_{i=1}^{n}(g^{k,i})^{2} & .. & 0 \\
... & ... & .... & .... & ... \\
0 & 0 & .... & 0 & \sum\limits_{i=1}^{n}(g^{k,i})^{2}
\end{array}
\right) \text{ \ \ \ \ .}  \tag{C77. }
\end{equation}
Therefore, summing up the matrices (C75) and (C77), one obtains \textbf{the
general structure of the matrices (C74) }$\sum\limits_{\alpha
=k}^{n}B_{\alpha k}^{T}B_{\alpha k}$\textbf{\ on the block diagonal:}
\begin{equation}
\sum\limits_{\alpha =k}^{n}B_{\alpha k}^{T}B_{\alpha k}=\left(
\begin{array}{cccc}
N_{11} & N_{12} & ... & N_{1,(n-\alpha +1)} \\
N_{21} & N_{22}+G_{22} & ... & N_{2,(n-\alpha +1)} \\
... & ... & ... & .. \\
N_{n-\alpha +1,1} & ... & ... & N_{n-\alpha +1,n-\alpha +1}+G_{22}
\end{array}
\right) \text{ \ \ ,}  \tag{C78. }
\end{equation}
where $N_{pq}$ and $G_{22}$ are given by expressions (C76) and $%
G_{2,2}=\sum\limits_{i=1}^{n}(g^{k,i})^{2}$ respectively. Consequently (for $%
p\geq 2$)
\begin{equation}
N_{pp}+G_{22}=\sum\limits_{i=1}^{n}\left[
g^{i,(k+p-1)}g^{i,(k+p-1)}+(g^{k,i})^{2}\right] \text{ \ \ \ \ \ \ .}
\tag{C79. }
\end{equation}

Let us find now the structure of the \textbf{off - diagonal block matrices},
situated \textbf{below the diagonal in the block matrice (C70)}. Each such
an element can be decomposed as
\begin{equation}
\sum\limits_{\alpha =j(j>k)}B_{\alpha j}^{T}B_{\alpha k}=B_{\alpha \alpha
}^{T}B_{\alpha k}+\sum\limits_{\alpha =j+1}^{n}B_{\alpha j}^{T}B_{\alpha k}%
\text{ \ \ \ .}  \tag{C80. }
\end{equation}
The first term in (C80) is the already known $(n-\alpha +1)\times (n-\alpha
+1)$ matrix (C72) with the only nonzero $(j-k+1)$ column. The second term is
the sum from $\alpha =j+1$ to $\alpha =n$ of the $(n-\alpha +1)\times
(n-k+1) $ matrices $B_{\alpha j}^{T}B_{\alpha k}$, already calculated in
(C71) and having the only nonzero element $\sum\limits_{i=1}^{n}g^{ji}g^{ki}$
on the $(\alpha -j+1)$ row and on the $(\alpha -k+1)$ column. The summation
over $\alpha $ from $\alpha =j+1$ to $\alpha =n$ means that in the sum $%
\sum\limits_{\alpha =j+1}^{n}B_{\alpha j}^{T}B_{\alpha k}$ the element $%
\sum\limits_{i=1}^{n}g^{ji}g^{ki}$ will appear beginning from the row $%
\alpha -j+1=2$ (for $\alpha =j+1$) up to the row $\alpha -j+1=n-j+1$ (for $%
\alpha =n$), which in fact is the \textbf{last row. }Correspondingly, the
same element will appear beginning from the column $\alpha -k+1=j-k+2$ (for $%
\alpha =j+1)$ and ending at the column $\alpha -k+1=n-k+1$ (at $\alpha =n$),
which is \textbf{the last column}. In other words, the summation of the
matrices $B_{\alpha j}^{T}B_{\alpha k},$containing one element, effectively
results in a matrix, filled up from the second row to the end and from the $%
(j-k+2)$ column to the end:
\begin{equation*}
\sum\limits_{\alpha =j+1}^{n}B_{\alpha j}^{T}B_{\alpha
k}=\sum\limits_{\alpha =j+1}^{n}\left(
\begin{array}{ccccc}
0 & .. & 0 & .. & 0 \\
... & .. & .... & ... & ... \\
0 & .... & \sum\limits_{i=1}^{n}g^{ji}g^{ki} & .... & 0 \\
... & ... & ... & ... & ... \\
0 & .. & 0 & ... & 0
\end{array}
\right) =
\end{equation*}
\begin{equation}
=\left(
\begin{array}{ccccc}
0 & ... & \text{column }(j-k+2)... & 0 & 0 \\
0 & ... & \sum\limits_{i=1}^{n}g^{ji}g^{ki} & .. & \sum%
\limits_{i=1}^{n}g^{ji}g^{ki} \\
.. & .. & .. & .. & .. \\
0 & .. & \sum\limits_{i=1}^{n}g^{ji}g^{ki} & .. & \sum%
\limits_{i=1}^{n}g^{ji}g^{ki} \\
0 & .. & \sum\limits_{i=1}^{n}g^{ji}g^{ki} & .. & \sum%
\limits_{i=1}^{n}g^{ji}g^{ki}
\end{array}
\right) \text{ \ \ .}  \tag{C81. }
\end{equation}
Now recall that the matrix $B_{\alpha \alpha }^{T}B_{\alpha k}$ (C72) had a
nonzero $(j-k+1)$ column, so therefore the structure of the whole matrix $%
\sum\limits_{\alpha =j}^{n}B_{\alpha j}^{T}B_{\alpha k}$ ($j>k$) \textbf{%
below the diagonal }is similar to (C81), but with the additional $(j-k+1)$
column:
\begin{equation}
\left(
\begin{array}{cccccc}
0 & 0.......... & \sum\limits_{i=1}^{n}g^{i,\alpha }g^{k,i} & 0 & .. & 0 \\
0 & 0...... & \sum\limits_{i=1}^{n}g^{i,\alpha +1}g^{k,i} &
\sum\limits_{i=1}^{n}g^{ji}g^{ki} & ... & \sum\limits_{i=1}^{n}g^{ji}g^{ki}
\\
... & ... & ..... & ... & ... & ... \\
0 & 0........ & \sum\limits_{i=1}^{n}g^{i,\alpha +r-1}g^{k,i} &
\sum\limits_{i=1}^{n}g^{ji}g^{ki} & .. & \sum\limits_{i=1}^{n}g^{ji}g^{ki}
\\
.. & ..... & .. & .. & ... & .. \\
0 & 0..... & \sum\limits_{i=1}^{n}g^{i,n}g^{k,i} & \sum%
\limits_{i=1}^{n}g^{ji}g^{ki} & .. & \sum\limits_{i=1}^{n}g^{ji}g^{ki}
\end{array}
\right) \text{ \ \ .}  \tag{C82a. }
\end{equation}
Since the structure of this matrix is important, let us write it in a more
precise way:
\begin{equation}
\sum\limits_{\alpha =j}^{n}B_{\alpha j}^{T}B_{\alpha k}\equiv \left(
\begin{array}{ccc}
M_{1} & M_{2} & M_{3} \\
M_{4} & M_{5} & M_{6}
\end{array}
\right) \text{ \ \ \ \ ,}  \tag{C82b.}
\end{equation}
where $M_{1},M_{3}$ and $M_{4}$ are zero matrices of dimensions $1\times
(j-k)$, $1\times l$ and $(n-j)\times (j-k)$ respectively and the number $l$
is equal to $l=n-k+1-(j-k+1)+1=n-j+1$. The matrix $M_{2}$ is $1\times 1$ and
contains the only element $\sum\limits_{i=1}^{n}g^{i,\alpha }g^{k,i}$, $%
M_{5} $ is $(n-j)\times 1$ dimensional and is therefore the column
\begin{equation}
M_{5}\equiv \left(
\begin{array}{c}
\sum\limits_{i=1}^{n}g^{i,\alpha +1}g^{k,i} \\
... \\
\sum\limits_{i=1}^{n}g^{i,\alpha +r-1}g^{k,i} \\
.... \\
\sum\limits_{i=1}^{n}g^{i,n}g^{k,i}
\end{array}
\right)  \tag{C82c.}
\end{equation}

and $M_{6}$ is a $(n-j)\times l$ matrix, in which each element is equal to $%
\sum\limits_{i=1}^{n}g^{ji}g^{ki}$.

\ \ \ In a completely analogous way the elements \textbf{above the block
diagonal} in (C70) can be found. These elements $\sum\limits_{\alpha
=k}^{n}B_{\alpha j}^{T}B_{\alpha k}$ ($k>j$) can be decomposed as
\begin{equation}
\sum\limits_{\alpha =k(k>j)}B_{\alpha j}^{T}B_{\alpha
k}=B_{kj}^{T}B_{kk}+\sum\limits_{\alpha =k+1}^{n}B_{\alpha j}^{T}B_{\alpha k}%
\text{ \ \ \ .}  \tag{C83. }
\end{equation}
This formulae is similar to (C80) for the below - diagonal elements, but
here in (C83) we have the matrix $B_{kj}^{T}B_{kk}$ instead of the matrix $%
B_{\alpha \alpha }^{T}B_{\alpha k}$ and the summation over the indice $%
\alpha $ in the second term is from $\alpha =k+1$ to $\alpha =n$ instead of $%
\alpha =j+1$ to $\alpha =n$ in (C80). The first term in (C83) $%
B_{kj}^{T}B_{kk}$ is the $(n-j+1)\times (n-k+1)$ matrix (C73) \ (with $%
\alpha =k$) with the only nonzero $(k-j+1)$ row with the elements
\begin{equation}
\sum\limits_{i=1}^{n}g^{ji}g^{1,(k+i-1)}\text{ \ \ ; ....}%
\sum\limits_{i=1}^{n}g^{ji}g^{p,(k+i-1)}\text{;.....}\sum%
\limits_{i=1}^{n}g^{ji}g^{n,(k+i-1)}\text{\ \ \ }  \tag{C84. }
\end{equation}

The second term in (C83) is again the sum of the matrices (C71). The only
nonzero element $\sum\limits_{i=1}^{n}g^{ji}g^{ki}$ will now appear in the
final sum from the $\alpha -j+1=k-j+2$ $\ $row \ ($\alpha =k+1$) until the $%
k-j+1=n-j+1$ ($\alpha =n$) row, which is the \textbf{last one}. Also, the
same element will fill up the columns from $\alpha -k+1=2$ $\ $($\alpha =k+1$%
) to the column $\alpha -k+1=n-k+1$ ($\alpha =n$), which is also the \textbf{%
last one}. Therefore, summing up the two terms in (C83), one obtains the
following matrix for the \textbf{above - diagonal block terms, }in which the
$(k-j+1)$ row is filled up with the elements $\sum%
\limits_{i=1}^{n}g^{j,i}g^{p,(\alpha +i-1)}$ and from the next $(k-j+2)$ row
to \ the end and from the second column to the end column the matrix is
filled up with the other element $\sum\limits_{i=1}^{n}g^{ji}g^{ki}:$
\begin{equation}
\left(
\begin{array}{cccccc}
0 & 0.......... & 0.. & 0.. & .. & 0 \\
.. & ...... & ... & ... & ... & .... \\
\sum\limits_{i=1}^{n}g^{j,i}g^{1,(\alpha +i-1)} & \sum%
\limits_{i=1}^{n}g^{j,i}g^{2,(\alpha +i-1)} & ........\sum%
\limits_{i=1}^{n}g^{j,i}g^{p,(\alpha +i-1)} & ... & ... &
\sum\limits_{i=1}^{n}g^{j,i}g^{n,(\alpha +i-1)} \\
0 & \sum\limits_{i=1}^{n}g^{ji}g^{ki} & ..... & \sum%
\limits_{i=1}^{n}g^{ji}g^{ki} & .. & \sum\limits_{i=1}^{n}g^{ji}g^{ki} \\
.. & ..... & .. & .. & ... & .. \\
0 & \sum\limits_{i=1}^{n}g^{ji}g^{ki} & \sum\limits_{i=1}^{n}g^{i,n}g^{k,i}
& \sum\limits_{i=1}^{n}g^{ji}g^{ki} & .. & \sum\limits_{i=1}^{n}g^{ji}g^{ki}
\end{array}
\right) \text{ \ \ .}  \tag{C85. }
\end{equation}

\bigskip Let us now summarize the obtained results for the $n-$dimensional
case. The (predetermined) system of equations $g_{ij}g^{jk}=\delta _{i}^{k}$
was represented as $BY=\overline{1}$ , where \ $\overline{1}$ is an $n^{2}$
dimensional vector, whose transponed one is defined as $\overline{1}^{T}=(%
\overline{1}^{T1},\overline{1}^{T2},.............,\overline{1}^{Tn})$ and
the corresponding transponed $n-$ dimensional vectors $\overline{1}^{T1},%
\overline{1}^{T2},.............,\overline{1}^{Tn}$ are defined as follows: $%
\overline{1}^{T1}=(1,0,........,0),\overline{1}^{T2}=(0,1,0,........0)$ and
the $k-$th transponed vector $\overline{1}^{Tk}=(0,0,.....,1,0,...0)$
contains the number $1$ on the $k-$th place. The $n^{2}\times \frac{n(n+1)}{2%
}$ matrix $B$ in terms of the elementary block matrices $B_{ij}$ and the $%
\frac{n(n+1)}{2}$ dimensional vector $Y$ were defined by formulaes (C59 -
C62) and (C58) respectively. In order to solve the system, we multiplied it
to the left with the transponed matrix $B^{T}$ and thus the solution for the
vector $Y$ in matrix notations can be found as $Y=(B^{T}B)^{-1}B^{T}%
\overline{1}$, where the expression for $B^{T}\overline{1}$ can easily be
found to be
\begin{equation}
B^{T}\overline{1}=\left(
\begin{array}{cccc}
B_{11}^{T} & B_{21}^{T} & ...... & B_{n1}^{T} \\
0 & B_{22}^{T} & .... & B_{n2}^{T} \\
... & ... & ... & .. \\
0 & 0 & .. & B_{nn}^{T}
\end{array}
\right) \left(
\begin{array}{c}
\overline{1}^{1} \\
\overline{1}^{2} \\
... \\
\overline{1}^{m}
\end{array}
\right) =\left(
\begin{array}{c}
\sum\limits_{i=1}^{n}B_{i1}^{T}\overline{1}^{i} \\
\sum\limits_{i=1}^{n}B_{i2}^{T}\overline{1}^{i} \\
... \\
\sum\limits_{i=1}^{n}B_{im}^{T}\overline{1}^{i}
\end{array}
\right) \text{ \ \ \ \ .}  \tag{C86. }
\end{equation}
The expressions for the elements of the vector in the R. H. S. can also be
found, but this will not be performed here and will be left for the
interested reader. Let us remind again \ that each ''element'' of this
vector is in itself an $n-$dimensional vector and $m=\frac{n(n+1)}{2}$ gives
the number of the ''block'' elements of this vector - column. The number $m$
should be an integer number, but this requirement can be fulfilled and this
will be commented in the next subsection.

\textbf{The main result of this subsection are contained in the expressions
(C78), (C82a) and (C85) for the elements of the matrix }$B^{T}B$\textbf{\ on
the block diagonal, below the block diagonal and above the block diagonal
respectively. In such a way the detailed structure of the matrtix }$B^{T}B$%
\textbf{\ is known in terms of the elementary constituent block matrices }$%
B_{ij}$\textbf{. \ }The structure of the matrix $B^{T}B$ is important for
the following reasons:\

1. If one chooses the contravariant metric tensor in the form of a
factorized product $\widetilde{g}^{ij}=dX^{i}dX^{j}$, then one can answer
the question whether it is possible the matrix $B^{T}B$ to have a lower rank
and some common multiplier. But from the above expressions and since in each
''cell'' of the elementary block matrices one has a summation of the kind $%
\sum\limits_{i=1}^{n}g^{i,n}g^{k,i}$, it is evident that such a common
multiplier does not exist and therefore the rank of the matrix $B^{T}B$ (and
therefore of $B$) cannot be lowered due to the above made choice of $g^{ij}$.

2. The quadratic structure of \ the \ matrix $B^{T}B$ gives an opportunity \
to apply the Frobenius formulae for finding the inverse matrix $%
(B^{T}B)^{-1} $ in the sense, in which this was discussed at the end of the
previous subsection. However, the application of the Frobenius foemulaes
presumes that the inverse matrix of the upper left ''element'' $%
\sum\limits_{i=1}^{n}B_{i1}^{T}B_{i1}$ in the matrix (C70) exits (which
probably is the case due to the expression (C78) for the block diagonal
elements), and the subsequent application of the Frobenius formulae to the $%
2\times 2$ block matrix ''complex'' in (C70) will also give an invertible
matrix, then also with the $3\times 3$ complex and so on. Probably it is
interesting to derive a recurrent formulae for $(B^{T}B)^{-1}$, depending on
the elementary constituent block matrices. In any case, if they are
invertible matrices, this shall turn out to be important, because it will be
shown further how the matrix $B$ may be ''divided'' into $n\times n$ block
matrices, which \textbf{might not \ }be invertible.

\subsection*{\protect\bigskip C4. BLOCK MATRIX REPRESENTATION OF THE
HOMOGENEOUS SYSTEM \ OF \ EQUATIONS WITH \ A \ ZERO \ R. H. S. \ }

Earlier it was shown that for the case $n=3$ the sub - system of equations $%
g_{ij}g^{jk}=0$ with a zero R. H. S. has a determinant of coefficient
(functions), equal to zero. \textbf{The question which naturally arises is
whether this property is valid only for the }$n=3$\textbf{\ case and is it
valid for the }$n-$\textbf{\ dimensional case.}

Below it shall be proved that this can be done for the general case.\
Namely, it shall be established that from the $n^{2}-n$ equations with a
zero R. H. S. there may be chosen a sub-system of $\left(
\begin{array}{c}
n \\
2
\end{array}
\right) +n=\frac{n(n+1)}{2}$ equations with a determinant of coefficients,
equal to zero. It shall be stressed that the proof will be that \textbf{such
a system exists} (i.e. can be chosen) and \textbf{not that all other choices
of the subsystem of equations will also satisfy this requirement.} In fact,
investigating under other choices of the sub - system of equations with a
zero R. H. S. this property will \ be preserved represents an interesting
problem for further research.

For the purpose, let us again use the block matrix representation (C59).
Then, in order to obtain the system of equations with a zero R. H. S. , one
has to remove the first row (for $i=k=1$) from the system (C50) of the first
$n$ equations, then the second row from the second system of $n$ equations
and so on, one has to remove the $k-$th row from the $k-$th system of $n$
equations. Further, in order to receive again a block matrix structure with
(elementary) submatrices with $n$ rows, one should \textbf{add the first row}
\textbf{of the second system} of $n$ equations as the \textbf{last row of
the first system} of equations. In effect, since the \textbf{first two rows
of the second system have been removed}, one should add t\textbf{he first
two rows of the third system} of $n$ equations as \textbf{the last two rows
of the second system}. Therefore, since also the third row in the third
system ( for $i=k=3$) has been removed, it has a total of three rows
removed, and subsequently \textbf{three last rows} have to be added from the
fourth system. Continuing in the same manner, from the $k-$th matrix $B_{kk}$
on the block diagonal we have the first $k-$ rows from $B_{kk}$ removed and
also $k$ - last rows added, which should be taken from the below - diagonal
matrix $B_{k+1,k}$. Since according to (C62) the $n\times (n-k+1)$ matrix $%
B_{sk\text{ }}$has a nonzero $(s-k+1)$ column, the matrix $B_{k+1,k}$ will
have a nonzero $k+1-k+1=2$ column. Therefore the new transformed in this way
$n\times (n-k+1)$ matrix, which will be denoted as $\widetilde{B}_{kk}$ ,
will have the following structure:\
\begin{equation}
\widetilde{B}_{kk}\equiv \left(
\begin{array}{ccccccc}
g^{k+1,k} & g^{k+1,k+1} & .. & .. & ... & .. & g^{k+1,n} \\
g^{k+2,k} & g^{k+2,k+1} & .. & .. & . & . & g^{k+2,n} \\
.. & .. & .. & .. & . & .. & ... \\
g^{nk} & g^{n,(k+1)} & .. & .. & .. & .. & g^{nn} \\
0 & g^{k1} & .. & 0 & .. & 0 & 0 \\
.. &  & .. & .. & .. & .. & .. \\
0 & g^{kk} & .. & 0 & 0 & 0 & 0
\end{array}
\right)  \tag{C87. }
\end{equation}
In the same way, the below - diagonal transformed matrix $\widetilde{B}%
_{k+1,k}$, obtained from $B_{k+1,k}$ by removing its first $k$ and adding $k
$ rows from $B_{k+2,k}$ , will be of the following kind:

\begin{equation}
\widetilde{B}_{k+1,k}\equiv \left(
\begin{array}{cccccc}
0 & g^{k,(k+1)} & 0 & .. & 0 & 0 \\
.. & .. & .. & .. & .. & .. \\
0 & g^{k,n} & 0 & .. & 0 & 0 \\
0 & 0 & g^{k,1} & .. & 0 & 0 \\
.. & .. & .. & ... & .. & .. \\
0 & 0 & g^{k,k} & .. & 0 & 0
\end{array}
\right) \text{ \ \ \ \ .}  \tag{C88. }
\end{equation}
Since the block matrix (C59) has a \textbf{triangular structure, }for our
further purposes only the structure of the block - diagonal matrices $%
\widetilde{B}_{kk}$ will be relevent.

Next, our \ goal will be to divide the the block - matrix (C59) into
elemetary block matrices with an \textbf{equal number (}$n$\textbf{)} of
rows and columns. Let us remind once again that the block - matrice (C59)
contained elementary block - matrices with an \textbf{unequal number of
columns} - $n,(n-1),(n-2)......$correspondingly. For the purpose, we shall
take one (left) column from the block matrix with $(n-2)$ columns and
transfer it to the left to the block matrice with $(n-1)$ columns. As a
result, the block matrices on the second block matrix column (B. M. C.) will
already contain $n$ columns. Since in the block matrix column one column has
been transfered, one has to add $3$ columns from the $(n-3)-$rd block matrix
column to the $(n-2)-$nd \ block matrix column in order to obtain again a
block matrix column, consisting of elementary $n\times n$ matrices.
Continuing in the same manner with the $(n-3)-$rd \ B. M. C., one has to add
to its right end $6$ columns from the $(n-4)-$th B. M. C. Now let us write
down the numbers of the corresponding block columns and below with a $(-)$
sign the number of columns, transferred to the (neighbouring) B. M.\ C. ;
with a $(+)$ sign the number of columns, joined to the B. M.C. (to the right
side) from the neighbouring (right) B. M. C.
\begin{equation*}
(n-1)\text{ \ \ \ \ \ ; \ \ \ \ \ }(n-2)\text{ \ \ \ \ ; \ \ \ \ \ \ \ \ \ \
\ \ \ \ }(n-3)\text{ ; \ \ \ \ \ \ \ \ \ \ \ \ \ \ \ \ \ }(n-4)\text{ \ \ \
; \ \ \ \ \ \ \ \ \ \ \ \ \ \ \ \ }(n-5)\text{ \ \ \ \ \ \ \ \ \ \ \ \ \ \ \
\ }
\end{equation*}
\begin{equation}
+1\text{; \ \ \ \ \ \ \ \ \ \ \ \ \ }-1\text{ \ \ \ \ \ \ }+3\text{ \ \ \ \
\ \ \ ; \ \ }-3\text{ \ \ \ \ \ \ }+6\text{ \ \ \ \ ; \ }-6\text{ \ \ \ \ \
\ \ }+10\text{ \ \ \ ; \ \ \ \ }-10\text{ \ \ \ \ \ \ \ \ }+15\text{ \ \ \ \
\ \ \ \ \ \ .}  \tag{C89. }
\end{equation}
Now let us look at the numbers with a minus sign, which form the following
number sequence (with the corresponding number in the sequence denoted):

\begin{equation*}
1,3,6,10,15,21......
\end{equation*}
\begin{equation}
\text{\ \ \ \ \ \ \ \ \ \ \ \ \ \ \ \ \ \ \ \ \ \ \ \ \ \ \ \ }\ \ \text{\ \
\ \ \ \ }1,2,3,4,5,6,......\text{\ \ \ \ \ \ \ \ \ \ \ \ \ \ \ \ \ \ \ \ \ \
\ \ \ \ \ \ \ \ \ \ \ \ \ \ \ }  \tag{C90. }
\end{equation}

It is trivial to note that each number in the sequence (upper row) is in
fact a sum of the corresponding numbers in the lower row up to that number.
For example, the number $21$ in the sequence (upper row) can be represented
as a sum of the numbers in the sequence (lower row): $21=1+2+3+4+5+6$. The
same with the number $10$ $\Longrightarrow 10=1+2$\bigskip $+3+4$.
Therefore, to the sequence number $k$ in the low row will correspond the
number $\frac{k(k+1)}{2}$ in the upper row, which is the sum of the first $k$
numbers in the low row. Since the number $k$ in the lower (C90) corresponds
to the $(n-k)-$th block column, the corresponding number will be $\frac{%
k(k-1)}{2}$, and it will correspond to the number of columns, which have to
be removed from the diagonal block matrix $B_{k+1,(k+1)}.$ At the same time,
to the right end one should add $k+\frac{k(k-1)}{2}$ $=\frac{k(k+1)}{2}$
columns. This number is exactly equal to the number of left columns, removed
from the (right) neighbouring matrix $B_{(k+1),(k+2)}.$ This can serve also
as a consistency check that the performed calculation is consistent and
correct.

From the matrix (C87) for $\widetilde{B}_{k+1,(k+1)}$ we have to delete the
first $\overline{s}+1$ left columns, the first (upper) elements of which
begin with the elements $g^{(k+2),(k+1)};$ $%
g^{(k+2),(k+2)}.............,g^{(k+2),(\overline{s}+k)}$, where $\overline{s}%
=$ $\frac{k(k-1)}{2}$. Now let us denote by $\overline{B}_{(k+1),(k+1)\text{
}}$ the matrix $\widetilde{B}_{k+1,(k+1)}$ with $\frac{k(k+1)}{2}$ left
columns removed and $\frac{k(k+1)}{2}$ right columns added. The upper
elements of the (left) remaining columns will be $g^{(k+2),(\overline{s}%
+k+1)},g^{(k+2),(\overline{s}+k+2)},........,g^{(k+2),n},$ where
\begin{equation}
\overline{s}+k+1=\frac{k(k-1)}{2}+k+1=\frac{k(k+1)}{2}+1\text{ }  \tag{C91. }
\end{equation}
and the matrix $\overline{B}_{(k+1),(k+1)\text{ }}$ will contain $n-\frac{%
k(k+1)}{2}$ left nonzero columns. Therefore the $n\times n$ matrix $%
\overline{B}_{(k+1),(k+1)\text{ }}$ will have the following structure:
\begin{equation}
\overline{B}_{(k+1),(k+1)\text{ }}\equiv \left(
\begin{array}{cc}
L_{1} & L_{2} \\
L_{3} & L_{4}
\end{array}
\right) \text{ \ \ \ ,}  \tag{C92. }
\end{equation}
where the $(n-k+1)\times \left[ n-\frac{k(k+1)}{2}\right] $ matrix $L_{1}$
is
\begin{equation}
L_{1}\equiv \left(
\begin{array}{cccc}
g^{(k+2),(\frac{k(k+1)}{2}+1)} & ... & ... & g^{(k+2),n} \\
g^{(k+3),(\frac{k(k+1)}{2}+1)} & ... & .... & g^{(k+3),n} \\
.... & ... & ... & ... \\
g^{n,\frac{k(k+1)}{2}+1} & ... & ... & g^{n,n}
\end{array}
\right)  \tag{C93. }
\end{equation}
and $L_{2},L_{3}$ and $L_{4}$ are zero matrices of dimensions $(n-k+1)\times
\left[ \frac{k(k-1)}{2}\right] ,(k+1)\times \left[ n-\frac{k(k+1)}{2}\right]
$ and $(k+1)\times \left[ \frac{k(k+1)}{2}\right] $ correspondingly. Note
also that the block matrices $L_{1}$ and $L_{3}$ contain $\left[ n-\frac{%
k(k+1)}{2}\right] $ nonzero columns, and the remaining $\left[ \frac{k(k+1)}{%
2}\right] $ columns of the matrices $L_{2}$ and $L_{4}$ are exactly equal to
the number of zero columns, added to the right side of the matrix $\overline{%
B}_{(k+1),(k+1)\text{ }}$ from the neighbouring matrix $\overline{B}%
_{(k+1),(k+2)\text{ }}.$ Since this result depends on the initial structure
of the matrix $B_{(k+1),(k+1)\text{ }}$ and on the expression (C91) (which
are both independent on the number of removed right columns), this also
confirms the consistency of the calculation.

Note that the block structure \ of the matrix (C92) has been revealed on the
base of the assumption that the elements $g^{(k+1),1},...g^{(k+1),(k+1)}$ in
the last $(k+1)$ rows and the second column in the matrix (C87) will be
among the first removed to the left (and outside the matrix) columns.
However, for $\overline{s}=0$ (i.e. $k=1$) the elements in the second column
of the last two rows of the matrix $\overline{B}_{2,2}$ will contain the
elements $g^{21},g^{22}.$ Therefore, after removing the first column in the
matrix (C87) $\widetilde{B}_{(k+1),(k+1)\text{ }}($for $k=1$) and adding to
the right one zero column, the obtained structure of the $n\times n$ matrix $%
\overline{B}_{2,2}$ will be the following:
\begin{equation}
\overline{B}_{2,2}=\left(
\begin{array}{cccccc}
g^{33} & g^{34} & ... & g^{3n} & 0 & 0 \\
g^{43} & g^{44} & ... & g^{4n} & 0 & 0 \\
.. & ... & .. & ... & .. & .. \\
g^{n3} & g^{n4} & ... & g^{nn} & 0 & 0 \\
g^{21} & 0 & .. & 0 & 0 & 0 \\
g^{22} & 0 & ... & 0 & 0 & 0
\end{array}
\right) \text{ \ .}  \tag{C94. }
\end{equation}
Having established the block structure of the matrix of coefficients of the $%
n-$dimensional predetermined system $g_{ij}g^{jk}=0$ with a homogeneous zero
R. H. S., it is now easy to show that $\frac{n(n+1)}{2}$ equations can be
chosen so that the determinant of coefficients will be zero. Let us take for
example the first $\frac{n(n+1)}{2}$ equations from the system of $(n^{2}-n)$
equations with a zero R. H. S. with the corresponding $\frac{n(n+1)}{2}%
\times \frac{n(n+1)}{2}$ dimensional block matrix. In the particular case
the dimension of the block matrix is determined by the number of block
matrices on the horizontal ($\frac{n(n+1)}{2}$)t and on t he vertical ($%
\frac{n(n+1)}{2})$. Therefore, outside this matrix will remain a matrix of $%
\frac{n(n+1)}{2}$ matrix block columns and $(n-1)-\frac{n(n+1)}{2}=\frac{%
(n-3)}{2}$ block rows. The last assumption presumes that the spacetime
dimension number $n$ is an \textbf{odd }one, so that $(n+1)$ and $(n-3)$ are
dividable by two. Otherwise, if $n$ is an \textbf{even number, one may
consider }$\frac{n}{2}\times \frac{n}{2}$ dimensional elementary block
matrices $\overline{B}_{k,k}$ . Then the full block matrix of the system
will have $(n+1)$ block matrices on the block horizontal (i.e. $(n+1)$ block
columns) and $2n^{2}$ matrices on \ the block vertical (i. e. $2n^{2}$ block
rows). The block matrix of the homogeneous system of equations (with a zero
R. H. S. ) will be $2(n-1)\times (n+1)$ ''block'' dimkensional. The chosen
block matrix will be $(n+1)\times (n+1)$ ''block'' dimensional. Outside this
matrix there will remain a block matrix of $2n-2-n-1=n-3$ block rows and $%
(n+1)$ block columns.

Let us now compute the determinant of the triangular matrix (C59), from
which we take the first $\frac{(n+1)}{2}$ (or $\frac{n}{2}$) block rows.
This $\frac{(n+1)}{2}\times \frac{(n+1)}{2}$ (or $\frac{n}{2}\times \frac{n}{%
2})$ block determinant $\widetilde{S}$ will be equal to
\begin{equation}
\widetilde{S}=\prod\limits_{i=1}^{\frac{n+1}{2}}(det\overline{B}_{ii})\text{
\ \ \ .}  \tag{C95. }
\end{equation}
But for $i\neq 1,2$ the expression for $det\overline{B}_{ii}$ has to be
found from formulae (C91) for the block matrix $\overline{B}_{kk}$. Since
only one of the submatrices $L_{1}$ is different from zero, it is clear that
$det\overline{B}_{ii}=detL_{1}.0=0$, therefore the whole expression (C95)
equals zero.

Thus we have proved that by taking $\frac{(n+1)}{2}$ (or $\frac{n}{2})$
consequent block rows \ from the initial (quadratic and triangular) block
matrix, the obtained block matrix will have a zero determinant.

\section{\protect\bigskip APPENDIX \ D: ANALYTICAL \ CALCULATION \ OF \
INTEGRALS \ $I_{N-1}\equiv \int e^{2\overline{k}\protect\varepsilon %
y}y^{n-1}dy$}

\bigskip Performing a simple integration by parts
\begin{equation}
I_{n-1}=\frac{\varepsilon }{2k}\int y^{n-1}d(e^{2\overline{k}\varepsilon y})%
\text{ \ \ \ ,}  \tag{D1 }
\end{equation}
it is easy to find that
\begin{equation*}
I_{n-1}=\frac{\varepsilon }{2\overline{k}}y^{n-1}e^{2\overline{k}\varepsilon
y}-\left( \frac{\varepsilon }{2\overline{k}}\right) ^{2}(n-1)y^{n-2}e^{2%
\overline{k}\varepsilon y}+
\end{equation*}
\begin{equation}
+\left( \frac{\varepsilon }{2\overline{k}}\right) ^{2}(n-1)(n-2)I_{n-3}\text{
\ \ \ \ .}  \tag{D2 }
\end{equation}
Expressing further $I_{n-3}$,$..$in the same way, the formulae can be
generalized as
\begin{equation*}
I_{n-1}=\frac{\varepsilon }{2k}y^{n-1}e^{2\overline{k}\varepsilon
y}+\sum\limits_{k=1}^{s}(-1)^{k}\left( \frac{\varepsilon }{2\overline{k}}%
\right) ^{2}(n-k)y^{n-k-1}e^{2\overline{k}\varepsilon y}+
\end{equation*}
\begin{equation}
+(-1)^{s+1}(n-1)(n-2)...(n-s-1)I_{n-s-2}\text{ \ \ .}  \tag{D3 }
\end{equation}
Let us assume that the final term will be
\begin{equation}
I_{n-s-2}=I_{1}=\frac{\varepsilon }{2\overline{k}}ye^{2\overline{k}%
\varepsilon y}-\left( \frac{\varepsilon }{2\overline{k}}\right) ^{2}e^{2%
\overline{k}\varepsilon y}\text{ \ \ \ ,}  \tag{D4 }
\end{equation}
which means that
\begin{equation}
n-s-2=1\text{ }\Longrightarrow s=n-3  \tag{D5 }
\end{equation}
and therefore formulae (D3) acquires the form
\begin{equation*}
I_{n-1}=\sum\limits_{k=1}^{n-3}(-1)^{k}\left( \frac{\varepsilon }{2\overline{%
k}}\right) ^{2}(n-k)y^{n-k-1}e^{2\overline{k}\varepsilon y}+
\end{equation*}
\begin{equation}
+\frac{\varepsilon }{2\overline{k}}y^{n-1}e^{2\overline{k}\varepsilon
y}+(-1)^{n-2}(n-1)!I_{1}\text{ \ \ .}  \tag{D6 }
\end{equation}
Now it remains only to calculate the first term with the sum. Let us denote
this term by $\widetilde{I}_{n-1}$. If we differentiate this term by $y$, we
may obtain
\begin{equation}
\frac{\partial \widetilde{I}_{n-1}}{\partial y}=2k\varepsilon \widetilde{I}%
_{n-1}+\widetilde{I}_{n-2}\text{ \ \ \ ,}  \tag{D7 }
\end{equation}
where in general $\widetilde{I}_{n-s}$ will denote
\begin{equation}
\widetilde{I}_{n-s}\equiv \sum\limits_{k=1}^{n-3}(-1)^{k}\left( \frac{%
\varepsilon }{2\overline{k}}\right) ^{2}(n-k)(n-k-1)....(n-k-s+1)\text{ }%
y^{n-k-s}e^{2\overline{k}\varepsilon y}\text{ \ \ .}  \tag{D8 }
\end{equation}
Note that the same equality as (D7)\ holds for arbitrary $s$, i.e.
\begin{equation}
\frac{\partial \widetilde{I}_{n-s}}{\partial y}=2k\varepsilon \widetilde{I}%
_{n-s}+\widetilde{I}_{n-s-1}\text{ \ \ \ .}  \tag{D9 }
\end{equation}
From this recurrent differential equality it is evident that if one knows $%
\widetilde{I}_{1}$, then $\widetilde{I}_{2}$ can be found as a solution of
an diffrential equation; in the same way, if $\widetilde{I}_{2}$ is known, $%
\widetilde{I}_{3}$ can be found and etc. Finally, $\widetilde{I}_{n-1}$ can
be found. Of course, one can try a certain starting ''probe'' function in
the form of a polynomial for $\widetilde{I}_{1}$ with unknown coefficients
and then try to find the recurrent relations for these coefficients, so that
(D9) is fulfilled.

This purely technical calculations shall not be performed here, because the
final answer for the integral $I_{n-1}=\int e^{2\overline{k}\varepsilon
y}y^{n-1}$ is already known and can be found too in the monograph by
Timofeev [75]
\begin{equation}
I_{n-1}=\frac{e^{2\overline{k}\varepsilon \text{ }y}}{(2\overline{k}%
\varepsilon )^{n}\text{ }}\sum\limits_{p=0}^{n-1}(-1)^{p}\left(
\begin{array}{c}
n-1 \\
p
\end{array}
\right) p!(2\overline{k}\varepsilon )^{n-1-p}y^{n-1-p}\text{ \ \ \ \ .}
\tag{D10 }
\end{equation}
\

\end{document}